\documentclass[fleqn,a4paper,12pt]{report}
\usepackage{setspace}
\usepackage{rotating}
\usepackage{amsmath,amsthm,amssymb,epsf,epsfig,subfigure}
\usepackage[small,center,compact,outermarks]{titlesec}
\usepackage{graphicx}
\usepackage{placeins}
\usepackage{hyperref}
\hypersetup{backref,colorlinks=true,linkcolor=red,citecolor=blue,linktocpage=true,breaklinks}
\newcommand*{\justifyheading}{\raggedright}
\let\oldfootnotesize\footnotesize
\renewcommand*{\footnotesize}{\oldfootnotesize\small}

\titleformat{name=\chapter}[display]{\normalfont \bfseries \large\filcenter}
{\vspace{-1.4 cm}\titlerule[1pt]\vspace{2pt}\titlerule\vspace{0ex}\large\chaptername\
\thechapter}{1ex} {\Large\bfseries}[\vspace{0ex}\titlerule]

\titleformat{\section} {\normalfont\large\bfseries\justifyheading}{\thesection}{1em}{}
\titleformat{\subsection}{\normalfont\bfseries\justifyheading}{\thesubsection}{1em}{}

\titlespacing{\chapter}{0pt}{1em}{2em}
\titlespacing{\section}{0pt}{2em}{1 em}
\titlespacing{\subsection}{0pt}{2em}{1 em}

\newpagestyle{Intro}[\small\sffamily]{
  \headrule
 \sethead{\textup{Chapter \thechapter} }{\textup{Introduction}}{\usepage}
    \setfoot[][][]{}{}{}}

\newpagestyle{References}[\small\sffamily]{
  \headrule
 \sethead{\textup{References} }{\textsl{}}{\usepage}
    \setfoot[][][]{}{}{}}

\newpagestyle{Chapter}[\small\sffamily]{

\headrule
\sethead{\textup{Chapter \thechapter }}{\textup{\chaptertitle}}{\usepage}
 \setfoot[][][]{}{}{}}

\newpagestyle{Acknowledgement}[\small\sffamily]{

\headrule
\sethead{\textup{Acknowledgement}}{\textup{}}{\usepage}
 \setfoot[][][]{}{}{}}

\newpagestyle{Abstract}[\small\sffamily]{

\headrule
\sethead{\textup{Abstract}}{\textup{}}{\usepage}
 \setfoot[][][]{}{}{}}

\newpagestyle{none}[\small\sffamily]{\sethead{\textup{}}{\textup{}}{\usepage}
 \setfoot[][][]{}{}{}}

\newcommand{\sech}{\operatorname{sech}}
\newcommand{\ef}{\operatorname{Erf}}

\begin{document}
\begin{titlepage}
\begin{center}
{ \hspace{-.35 cm} \large A STUDY OF SOME SOLVABLE QUANTUM MECHANICAL MODELS\\ AND\\ THEIR SYMMETRIES}
\par
\vspace{1.25in}
{\bf ( Ph.D. dissertation was defended on May 20, 2013 for the award of the Degree of Doctor of Philosophy in Science under University of Calcutta )}

\par
\vspace{1.3 in}

{\bf By}
\vspace{.15 cm}
\par
{\bf BIKASHKALI ~ MIDYA}
\vspace{0.25 in}
\par
Physics and Applied Mathematics Unit
 \par
 Indian Statistical Institute
 \par
 Kolkata-700108, India.
 
\end{center}

\end{titlepage}


\newpage
\pagenumbering{roman}
.
\vspace{4 cm}
\begin{center}
{ \textmd{\bf{I Dedicate This Thesis}\\
 To\\
 My Parents\\
 For\\
 Their Loving Understanding When I Spent Time On Research Instead Of Being With Them}.... }
\end{center}

\newpage
\hrule
\begin{center}
{\bf \large {Acknowledgement}}
\end{center}
\hrule
\vspace{2 cm}

\noindent {First of all, I gratefully acknowledge the financial support in the form of research fellowship, received from the Indian Statistical Institute, Kolkata. It has given me the opportunity to work in the wonderful atmosphere of Physics and Applied Mathematics Unit of this Institute.}\\

\noindent {I am sincerely and heartily grateful to my advisor, Barnana Roy, for her continuous encouragement, support and guidance throughout my research work.} \\

\noindent {I am indebted to Professor Rajkumar Roychoudhury of Physics and Applied Mathematics Unit, Indian Statistical Institute for some fruitful discussions during my research period.}\\

\noindent { I am thankful to all other faculty members, research scholars, and official staffs of Physics
and Applied Mathematics Unit, Indian Statistical Institute, for their kind help and cooperation in various
aspects.}\\

\noindent { I would like to thank google.com for providing huge information resources. 
Also thanks to the Wolfram research which provides the scientific software like Mathematica.}\\

\vspace{3 cm}
\hfill Bikashkali Midya

\newpage
\hrule
\begin{center}
{\large \bf {Abstract}}
\end{center}
\hrule

\vspace{1 cm}

The importance of exactly solvable quantum mechanical systems are ubiquitous in modern physics. This is because, the qualitative understanding of a complicated realistic system can be acquired by analyzing exactly solvable simplified
model that portrays the purport of a physical reality. On the other hand, discovery of new
symmetry enlarges our ability and possibility to describe physical phenomena and to
construct their mathematical formulations. In the present thesis, studies on some solvable quantum mechanical systems and the underlying symmetry of their solutions have been addressed.\\

The problems dealt with in the context of quantum systems characterized by position dependent mass (PDM) are: {{(A)}} The modified factorization technique based on excited state wave function(s) has been extended to these systems. It has been shown that the singular superpotential defined in terms of a mass function and an excited state wave function of a given PDM Hamiltonian can be used to construct non-singular isospectral Hamiltonians.  {{(B)}} Some exactly solvable potentials in the
PDM background are generated whose bound
states are given in terms of recently discovered Laguerre- or Jacobi-type $X_1$
exceptional orthogonal polynomials. These potentials are shown, with the help of supersymmetric quantum mechanics, to
possess shape invariant symmetry. {{(C)}} First and second-order
intertwining approach to generate isospectral PDM Hamiltonians with pre-determined spectral characteristics, have been formulated in a systematic way. The proposed techniques allow one to generate isospectral
potentials with the following spectral modifications: (i) to add
new bound state(s), (ii) to remove bound state(s) and (iii) to
leave the spectrum unaffected.\\

Quantum systems with non-Hermitian Hamiltonians are considered for the following problems: {{(A)}} The $\mathcal{PT}$-symmetric 
Hamiltonian with the periodic potential $V(x) = 4 \cos^2 x + 4 i V_0 \sin 2x$ has been shown to map into a Hermitian Hamiltonian for 
$V_0<.5$, by a similarity transformation: $ e^{- \theta p} ~x ~e^{\theta p} = x + i \theta,$ $\theta \in \mathbb{R}$. The corresponding 
energy band structure has been studied using Floquet analysis. It is also shown that, in addition to the $\mathcal{PT}$-threshold at $V_0 = 0.5$, there 
exist a second critical point near $V_0^c\sim .888437$ after which no part
of the eigenvalues and the band structure remains
real. {{(B)}} Using the method of point canonical transformation, some exactly solvable
rationally extended quantum Hamiltonians which are non-Hermitian in nature and whose
bound state wave functions are associated with Laguerre or Jacobi-type $X_1$ exceptional
orthogonal polynomials, have been derived. These Hamiltonians are shown to be quasi Hermitian so that the corresponding energy spectra is entirely real. {{(C)}} The generalized Swanson Hamiltonian $H_{GS} = w (\tilde{a}\tilde{a}^\dag+ 1/2) + \alpha \tilde{a}^2 + \beta \tilde{a}^{\dag^2}$ with $\tilde{a} = A(x)d/dx + B(x)$, has been shown to transform into the harmonic oscillator like Hamiltonian so long as $[\tilde{a},\tilde{a}^\dag]=$ constant. This reveals an intriguing result in the sense that in this case the resulting Hamiltonian does not always possess the spectrum of the harmonic oscillator. Reason for this anomaly is discussed in the frame work of position dependent mass models by choosing $A(x)$ as the inverse square root of the mass function.\\

The aspects of quantum nonlinear oscillator (QNLO) studied in this thesis are the following: {{(A)}} The coherent state of QNLO are constructed using the Gazeau-Klauder formalism. The details of the revival structures arising from different
time scales underlying the quadratic energy spectrum of the above mentioned quantum system are investigated by the phase analysis of the autocorrelation function. {{(B)}} Various generalizations e.g, exactly solvable, quasi exactly solvable and
non Hermitian PT-symmetric variants of the quantum nonlinear oscillator have been examined. Supersymmetric approach has been used to show that all the exactly solvable potentials for the QNLO, are shape invariant.

\newpage
\pagestyle{none}
\hrule
\begin{center}
{\bf \large{Publications}}
\end{center}
\hrule

\vspace{.5 cm}

\noindent The thesis is based on the following published papers:

\begin{itemize}

\item [M1.] B. Midya\\
 \textsl{Nonsingular potentials from excited state factorization of a quantum system with position dependent mass}\\
  {\bf Journal of Physics A44 (2011) 435306}.

\item[M2.] B. Midya and B. Roy,\\
 \textsl{Exceptional orthogonal polynomials and exactly solvable potentials in position dependent mass Schr\"{o}dinger Hamiltonians},\\
  {\bf Physics Letters A373 (2009) 4117}.

 \item[M3.] B. Midya, B. Roy and R. Roychoudhury,\\
  \textsl{Position Dependent Mass Schr\"{o}dinger Equation and Isospectral Potentials : Intertwining Operator approach },\\
   {\bf Journal of Mathematical Physics 51 (2010) 022109}.

\item [M4.] B. Midya, B. Roy and R. Roychoudhury,\\
 \textsl{A note on the $\mathcal{PT}$-invariant periodic potential $V(x)=4 \cos^2 x + 4 i V_0 \sin 2x$ },\\
  {\bf Physics Letters A374 (2010) 2605}.

\item[M5.] B. Midya,\\
 \textsl{Quasi-Hermitian Hamiltonians associated with exceptional orthogonal polynomials},\\
  {\bf Physics Letters A (2012)} in press; arXiv: 1205.5860.

\item[M6.] B. Midya, P. Dube and R. Roychoudhury,\\
\textsl{Non-isospectrality of the generalized Swanson Hamiltonian and harmonic oscillator}\\
{\bf Journal of Physics A44 (2011) 062001}.

\item[M7.] B. Midya, B. Roy and A. Biswas,\\
 \textsl{Coherent state of a nonlinear oscillator and its revival dynamics},\\
 {\bf Physica Scripta 79 (2009) 065003}.

\item[M8.] B. Midya and B. Roy,\\
 \textsl{A generalized quantum nonlinear oscillator },\\
  {\bf Journal of Physics A42 (2009) 285301}.
\end{itemize}

\vspace{-2 cm}
\tableofcontents
\listoffigures
\listoftables

\setlength{\topmargin} {0.5 cm}
\chapter[Introduction]{Introduction}\label{ch:intro}
\pagestyle{Intro}
\pagenumbering{arabic}
\noindent
The search for exactly solvable and integrable models in quantum mechanics have great significance in modern physics.
This is because, the qualitative understanding of a complicated realistic system can be acquired by analyzing exactly solvable simplified
model that portrays the purport of a physical reality.\\

~ The study of such systems in non-relativistic quantum
mechanics has a long and cherished history. By exactly solvable,
one means that the spectral properties, that is, the eigenvalues and eigenfunctions of the Hamiltonian characterizing the quantum system under consideration, can be given in an explicit and closed form. The most important examples are the harmonic oscillator and the hydrogen atom. First unified venture in finding these systems was made by Schr\"{o}dinger \cite{Sc40} who initiated
the {\it factorization} method. The main idea of the factorization method is to replace a given Schr\"{o}dinger equation,
which is a second order differential equation, by an
equivalent pair of first-order differential equations. This enables one to find the eigenvalues and the normalized eigenfunctions in a far easier manner than solving the
original Schr\"{o}dinger equation directly. As has been emphasized in the literature, the factorization method is a special case of Darboux construction
\cite{Da82}. Darboux's method has been generalized by Crum \cite{Cr55} to the case of an arbitrary number of eigenfunctions.
Also, the Schr\"{o}dinger factorization technique was generalized by Infeld and Hull \cite{IH51} and Mielnik \cite{Mi84} to obtain a wide family of
exactly solvable Hamiltonians. Another method for constructing exactly solvable systems is the method of integral
transformation, which forms the basis of the inverse scattering approach \cite{Ni84,Su85d}. These transformations are derived from the solution of the Gelfand-Levitan equation \cite{GL51} and are
known as Abraham-Moses-Pursey transformations \cite{AM80,Pu85}.\\

 ~ The concept of solvability has also been extended to
{\it quasi-exactly solvable} (QES) \cite{Fl79,Ra79} and {\it conditionally exactly solvable}
(CES)\cite{Du93a,JR98} models. Quantum mechanical Hamiltonians are said to be QES if a finite portion of the energy spectrum and associated eigenfunctions can be found exactly and in closed form. On the other hand, principal
 feature of CES Hamiltonians is that they have exact solutions only when certain conditions are satisfied by the potential parameters.\\

~ Besides their role in describing physical problems of the microscopic world, solvable quantum mechanical models also represent an interesting direction of research due to the mathematical elegance associated with the symmetries of the systems. Discovery of new symmetry enlarges our ability and possibility to describe new physical phenomena and to construct their mathematical formulations. The most widely known symmetries of quantum mechanical Hamiltonians are based on group theory (in particular Lie algebra). For example, Pauli \cite{Pa26} considered $\mathcal{SO}(4)$ symmetry in quantum mechanics and obtained hydrogen atom spectrum by only algebraic
methods, Fock \cite{Fo35} considered the Schr\"{o}dinger equation of the Kepler
problem in momentum space and had shown that the spectrum has $\mathcal{SO}(4)$ symmetry in energy-momentum space.
The usual approach \cite{Tu88,Sh89,Us93} to the analysis of QES quantum systems
is an algebraic one in which the Hamiltonian is expressed as a nonlinear combination of
generators of a Lie algebra like $\mathcal{SL}(2,Q)$, where $Q=R,C$, etc.. The other familiar symmetries of quantum mechanical systems are 
based on {\it supersymmetry, $\mathcal{PT}$-symmetry,
quasi} and {\it pseudo-Hermiticity}.

\section{Supersymmetric quantum mechanics and its application}
\noindent
{\underline{\bf Supersymmetry}:} The concept of Supersymmetry (SUSY) owes its origin to the remarkable papers by Ramond \cite{Ra71},
Neveu and Schwarz \cite{NS71}, Gelfand and Likhtam \cite{GL71}. Subsequent to these works, various models embedding SUSY were proposed within a
field theoretical framework \cite{VA73,WZ74}. The basic composition rules of SUSY contain both commutators and anti-commutators which enable ones
to circumvent the powerful "no-go" theorem of Coleman and Mandula \cite{CM67}. It should be mentioned here that, if supersymmetry exists in
nature then it must be spontaneously broken, because there is no evidence for degenerate boson-fermion pairs in nature, and this
fact has generated interest in supersymmetry breaking mechanism. In an attempt to construct a theory of SUSY that is unbroken at the tree-level
but could be broken by small non-perturbative corrections, Witten \cite{Wi81,Wi82} proposed a class of grand unified models within a field theoretical
framework. Specifically, he considered models (in less than four dimensions) in which SUSY could be broken dynamically. This led to the remarkable discovery
of supersymmetric quantum mechanics \cite{CR83,CF83,KNT85,So85,HR84,Ra86,Su85e,AB+84}
\footnote{There are some excellent books and review articles on Supersymmetric quantum mechanics e.g.
the books \cite{CK00,Ba00,Ju96,GMR11} and reviews \cite{CK95,Ro02}.}.
Historically, however, it was Nicolai \cite{Ni76} who sowed the seeds of
SUSY in non-relativistic quantum mechanics.\\

\noindent A quantum system, described by a Hamiltonian $H$ and a set of $N$ self-adjoint charge operators $Q_i$, is called supersymmetric if the following
 anti-commutation and commutation relations are valid for all $i,j = 1,\ldots,N$: $\{Q_i,Q_j\} = H \delta_{ij}$, $[H,Q_i] =0$ \cite{Ju96,Ro02}.
 However, Witten's supersymmetry corresponds to $N=2$ (referred to in the literature as SUSYQM\footnote{Throughout the thesis, we use the abbreviation SUSYQM to refer $N=2$ supersymmetric quantum mechanics.}) in which the two complex supercharges
 $Q^\pm = (Q_1 \pm i Q_2)/\sqrt{2}$ satisfy the linear super-algebra \cite{CK95,CK00}:
 $$(Q^\pm)^2 = 0,~~~ [H,Q^\pm] = 0,~~~ \{Q^+,Q^-\} = H,$$ i.e. the charge operators are nilpotent and commute with the super Hamiltonian $H$.
 The one-dimensional realization of this super algebra
 is usually given in terms of $2\times2$ matrices:
 $$Q^- = \begin{pmatrix}
0 & 0 \\
A^- & 0
\end{pmatrix}
, ~~~ Q^+ = \begin{pmatrix}
0 & A^+ \\
0 & 0
\end{pmatrix}, ~~~ H = \begin{pmatrix}
H^- & 0\\
0 & H^+
\end{pmatrix}.$$ This implies that the super Hamiltonian $H$ consists of two Schr\"{o}dinger type operators $H^{\pm}$,
which are known as supersymmetric partner Hamiltonians. $H^\pm$ can be factorized as:
 $$H^\pm = A^\mp A^\pm = = -\frac{\hbar^2}{2 m}\frac{d^2}{d x^2} + V^\pm (x), ~~~~ V^\pm(x) = W^2(x) \pm \frac{\hbar}{\sqrt{2 m}}W'(x),$$ where the operators
 $A^-$ and $A^+ = (A^-)^\dagger$ are given by $A^\pm = \mp \frac{\hbar}{\sqrt{2 m}}\frac{d}{dx} + W(x)$. $W(x)$ is known as the superpotential.
 The Hamiltonians $H^\pm$ are isospectral and differ at most
 in the ground-state energy level. An important aspect of
SUSYQM is that a hierarchy of isospectral solvable Hamiltonians, with a relationship between the eigenvalues and eigenfunctions of
the different members
of the hierarchy, can be constructed by repeated factorization \cite{Su86,AB+85,Su87,Su85a}.
SUSY is said to be unbroken when
$Q^+ |0\rangle = Q^- |0\rangle = 0$, where $|0\rangle = (\psi_0^- ~~ \psi_0^+)^T$ is the ground state of $H$ \cite{CK00}.
Thus the ground state energy must be zero in case of unbroken SUSY and non-zero for broken SUSY. For unbroken SUSY the energy eigenvalues $E_n^\pm$ and the eigenfunctions $\psi_n^\pm$
of $H^\pm$ are related in the following way:
$$E_n^+ = E_{n+1}^-, ~~~ E_0^- = 0, ~~ n= 0,1,2...$$
$$ \psi_n ^+ = [E_{n+1}^-]^{-\frac{1}{2}} A^- \psi_{n+1}^-, ~~~ \psi_{n+1}^- = [E_n^+]^{-\frac{1}{2}} A^+ \psi_n^+$$
 Witten \cite{Wi82} proposed an index $\Delta_F$, known as Witten index, to determine whether SUSY is broken or unbroken. This index, defined by the difference between the number of zero-energy states of
the partner Hamiltonians, is non-zero (zero) for unbroken (broken) SUSY.
However,
certain supersymmetric periodic quantum systems \cite{DF98} may produce a zero-energy doublet of the ground states, resulting
in a completely isospectral pair of partner Hamiltonians. If this happens, one has $\Delta_F = 0$ even in the
case of unbroken supersymmetry. Such special property is named as self-isospectrality in the literature \cite{CJ+08,CJP08}.\\

~ In a nonlinear generalization\footnote{In the literature, there are several synonyms for the nonlinear SUSY:
the higher order SUSY \cite{Fe10,FG04}, the polynomial SUSY \cite{AICD95,AIS93,AIN95} or the $\mathcal{N}-$fold SUSY
\cite{AST01,AST01a}.} of
SUSYQM, two supercharges are higher-order $(n \ge 1)$ differential operators, which satisfy the nonlinear
super algebra
\cite{AICD95,AS03,AIS93,AIN95,CNP07,Pl04,KP01,AST01,AST01a,FG04,Fe10}:
$(Q^\pm)^2 = 0,~ [H,Q^\pm] = 0,~ \{Q^+,Q^-\} = \mathcal{P}_n(H)$, where $\mathcal{P}_n(H)$ is a $n$-th degree
polynomial of super-Hamiltonian $H$. The generalization of one-dimensional SUSYQM to higher space dimension
$d \ge 2$ is made in ref.\cite{An+85,An+84,AIN95a,IGV06}. The attempt to preserve the one-dimensional SUSYQM algebra
but with superpotentials depending on multidimensional coordinate leads to potentials which are amenable to standard separation of
variables \cite{KTV01}. Among multidimensional SUSY QM models, those most developed are the two-dimensional ones
\cite{AI88,Ca+04a,IN03,Io04}.
In the two-dimensional case
the supersymmetrization of a given Schr\"{o}dinger Hamiltonian involves not only a
second Hamiltonian but also a $2 \times 2$ matrix Schr\"{o}dinger operator. The spectra of
these three components of the super Hamiltonian are interrelated, and their wave functions
are connected by the components of supercharges. Other generalization of the one-dimensional SUSYQM are para \cite{Kh93,DV93,HS10}, ortho \cite{KMR93} and fractional
\cite{Du93,Qu03} supersymmetry.\\

\noindent {\underline{\bf Shape Invariance}:} A conceptual breakthrough in understanding the connection between solvable
 potentials and SUSY was made by Gendenshtein \cite{Ge83} who introduced a discrete re-parametrization invariance
called {\it shape invariance} (SI).
Two supersymmetric partner potentials $V^\pm$ are said to be shape invariant if they satisfy \cite{CK95}:
$$V^+(x,a_0) = V^-(x,a_1) + R(a_0),$$ where the parameters $a_0$,
 $a_1 = f(a_0)$ represent characteristics of the potentials, such as its strength, depth or width.
 Gendenshtein showed that whenever the shape invariance relationship is satisfied by the two supersymmetric
 potentials, the energy spectra can be determined in purely algebraic fashion.
 Specifically, energy eigenvalues are given by $$E_n^- = \sum_{k=1}^n R(a_k), E^-_0 =0.$$ Subsequently, it has been shown \cite{DKS86,DKS88} that the wave functions
 $\psi_n^-(x;a_0)$ of the potential $V^-$ can also be obtained in a straightforward way by consecutive application of ladder operator $A^+(x;a_k)$ as
 $$ \psi_n^-(x,a_0)=
 A^+(x;a_0) A^+(x;a_1) .... A^+(x;a_{n-1}) \psi_0^-(x;a_0).$$
 A systematic study of the relationship between solvability and shape invariance was carried out in \cite{CGK87}. This study has revealed that most of the known solvable potentials
(such as Coulomb, harmonic oscillator, Morse, Eckart, Poschl-Teller, Scarf, Rosen-Morse etc)
 are shape invariant. The explicit expressions of the wave functions of these known SI potentials
 are obtained by Dabrowska et al \cite{DKP88}. However, the exactly solvable Ginocchio class \cite{Gi84} of potentials and more generally the
 class of Natanzon potentials \cite{Na79}, for which the corresponding Schr\"{o}dinger equation can be reduced to confluent or general
 hypergeometric differential equation,
 are in general not shape invariant. This implies that shape invariance integrability condition is a sufficient, but not necessary, condition for exact solvability.
 Mainly four classes of shape invariant potentials (SIP) are reported in the literature: (i) the translational class \cite{CGK87,Le89},
 where the parameters $a_0$ and $a_1$
 of the two
supersymmetric partners are related to each other by translation $a_1 = a_0 + \alpha$; (ii) the scaling class \cite{KS93,Ba+93}, where
$a_1 = q a_0, (0<q<1)$; (iii) the cyclic class \cite{SRK97}, characterized by the parameters which repeat after a
cycle of $p$ iterations i.e. $a_0=a_p, a_1=f(a_0)=a_{p+1}$ and (iv) the exotic class \cite{Ba+93}, where $a_1 = q a_0^p$
and $a_1 = q a_0/(1+ p a_0)$, $p= 2,3...$. The potential algebra of these systems have also been identified
\cite{EQ87,WA90,Ba98,Ch+98,GRS99,BRA99}, which provides an alternate methods of finding exact solutions. The idea of SI has been extended to
the more general concept of SI in two steps and even multi-steps \cite{Ba+93,Su08}. The two-step SI approach was utilized for dealing with 
problems with spontaneously broken
SUSY \cite{Du+93,GMS01}. Very recently \cite{BG+10,BGM11,Ra11}, it has been shown that all the additive
SI potentials can be generated from the Euler equation of momentum conservation for inviscid fluid flow.\\

\noindent{\underline{\bf Application of SUSYQM and SI approach}:} The idea of of SUSYQM and shape invariance approach have profitably been applied to enlarge the families of exactly
\cite{Le92,GM+99,Le04,TF01,Le89}, quasi-exactly \cite{Tk99,QT03,Tk01,TV02,KS99} and
conditionally exactly \cite{JR98,LR99,NRV94} solvable potentials in non-relativistic quantum mechanics.
SUSYQM allows one to derive potentials which provide the same phase shifts as a given potential, even if the number of bound
states is modified \cite{ACD88,Ba87b,Ba87a,KS89b,SB97}.
Starting from a potential with a continuum of energy eigenstates, the
methods of supersymmetric quantum mechanics has been used to generate families of
potentials which have bound states embedded in the continuum \cite{PSP93,St95}. The relationships between reflection and
transmission amplitudes of two partner potentials supporting scattering states are obtained by SUSYQM \cite{KS88,ACDI95}.
In order for
scattering to take place, it is necessary that both the partner
potentials $V^\pm$ are finite as $x \rightarrow +\infty$ or as $x \rightarrow -\infty$ or both \cite{CK95}.
Using the formalism of supersymmetric quantum mechanics, a large number of new
analytically solvable one-dimensional periodic potentials have been obtained and their band structures are studied
\cite{KS04a,IGV08,FNN00,SK02}.
In particular the energy band structure of lame potential $V(x) = m a(a + 1) ~sn^2(x, m)$ and associated Lame potential $V(x) = p m ~sn^2(x,m) + q m ~cn^2(x, m)/dn^2(x,m)$, both of which
involve Jacobi elliptic function and modulation parameter $m$, are discussed. The supersymmetric partners of these potentials and
corresponding band edge energies and the wave functions are also derived \cite{KS04a}. Application of irreducible second order SUSYQM
to the Schr\"{o}dinger operators, with Lame and periodic piece-wise transparent potentials, reveal that the pairs of
factorization energies inside the same forbidden band can create new non-singular potentials with periodicity
defects and bound states embedded into the spectral gaps \cite{Fe+02}.\\

~ Approximate methods based on SUSYQM, like large-$N$ expansion \cite{IS85,RV88}, $\delta$-expansion
for the superpotential \cite{CR90}, a variational method \cite{GRT93,CDS94,FR00}, and WKB
approximation
(SWKB) for the case of unbroken SUSY \cite{Kh85,DKS91,HKS97} have also been developed.
The large-$N$ expansion method has been formulated \cite{IS85} in the framework of SUSYQM to study the spherically
symmetric potentials, e.g. Hulthen, screened Coulomb potentials etc., in $N$ spatial
dimensions. It is shown that the supersymmetric partner potential of a given potential can be effectively
treated as being in $N+2$ dimensions. This fact is responsible for improving the convergence of large-$N$ expansions.
The $\delta$ expansion method has been used in ref.\cite{CR90} to solve the Riccati equation for the superpotential
involved in factorization of the Hamiltonian for the anharmonic oscillator with $V(x) = \lambda x^{2 + 2 \delta}.$
Variational method based on the hierarchy of Hamiltonians \cite{GRT93} permits one to evaluate the excited states for
one-dimensional systems like anharmonic \cite{CDS94} and Morse \cite{FR00} oscillator. SWKB reveals several interesting features. Unlike the
standard WKB method, the leading order SWKB formula yields exact analytic
expressions for the energy eigenvalues.
It is not only exact for large $n$ (like WKB approximation) but by construction it is also exact for the ground
state of the partner potentials \cite{BM91}. Besides, it has been proved \cite{DKS86} that the lowest-order SWKB
quantization relation reproduces the exact bound-state
spectra for several analytically solvable potentials. The reason behind obtaining
the exact analytic results for these potentials is
that they satisfy the shape-invariance condition
and for shape-invariant potentials (with translation), lowest-order SWKB is
necessarily exact \cite{DKS86}. A systematic higher order SWKB expansion has also been
developed to obtain an explicit expression for the quantization condition which contains all terms up
to order $\hbar^6$, which vanishes for all
known shape-invariant potentials \cite{Ad+86}. Energy eigenvalue spectrum has also been
obtained for several non-SIPs \cite{DKV87,Va92} and it turns out that in many
of the cases the SWKB gives better results than the usual WKB approximation.
In ref.\cite{Fr+88,SP90}, some attempts have been made to obtain the bound state eigenfunctions within SWKB
formalism.  The lowest order WKB
quantization condition, in case of broken SUSY, has also been derived \cite{Du+99,IJ94}. It is worth mentioning here that the ideas of supersymmetric
quantum mechanics are applied to the tunneling problem for both the symmetric \cite{KKS88} and asymmetric \cite{GPS93} double-well
potentials. The tunneling time is evaluated by developing a systematic perturbation expansion whose leading
term is an improvement over the standard WKB result.\\

~ Several aspects of the Dirac equation have been studied within SUSYQM formalism \cite{HKN86,NT93,Co+88} as well.
In particular, it has been shown \cite{Co+88} that corresponding to an analytically
solvable one-dimensional Schr\"{o}dinger problem with potential $V(x)$ there always exists
a corresponding Dirac problem with static scalar potential $\phi(x)$ in $(1+1)$ dimensions
which is also analytically solvable, $V(x)$ and $\phi(x)$ are related by $V(x) = \phi^2(x) + \phi'(x)$.
The existence of SUSY for massless Dirac equation in two as well as in four Euclidean dimensions has also been shown.
The problem of the Dirac particle in a Coulomb field
has been solved algebraically by using the concepts of SUSY and shape
invariance \cite{Su85f}. The SUSY of the Dirac electron in the field of a magnetic
monopole has been studied \cite{HV84}. The level degeneracies of a Dirac electron
in a constant magnetic field \cite{KM84b} and the accidental degeneracy of systems with spin-orbit coupling \cite{Ui84} have been
interpreted using SUSYQM. A supersymmetric analysis is presented for the $d$-dimensional Dirac equation with central potentials
under spin-symmetric and pseudo spin-symmetric regimes \cite{HY10}.
In particular, such Dirac Hamiltonian with Coulomb and Kratzer potentials have been solved. Exactly solvable scalar and vector potentials for $(1+1)$
dimensional Klein-Gordon equation has been obtained using supersymmetric shape invariance approach \cite{JR07}.
Supersymmetry in quantum mechanics is also extended from square-integrable states to quasi-normal modes (which satisfy the outgoing-wave boundary condition),
in a Klein-Gordon formulation \cite{Le+01,JR07a}.\\

~ The supersymmetric generalization of the exactly solvable $N$-particle
Calogero model was considered and related spectrum
was found exactly in \cite{FM90,IN02,Gh12,GKS98}.
The intimate connection
between the soliton solutions of the KdV equation and SUSYQM was studied \cite{SG95,GR94}. Supersymmetry has been extended to the non-commutative plane \cite{HSK04,Gh05} as well.
It has been shown \cite{HSK04} that the Supersymmetry algebra for Pauli equation on non-commutative plane holds for all orders of the noncommutative
parameter $\theta$ provided the gyro-magnetic ratio is $2$. The energy spectrum
of this model is also obtained when the corresponding magnetic field is uniform. Supersymmetric quantum mechanics has also been formulated \cite{GS09} on a two-dimensional noncommutative plane. This approach has been applied to factorize the non-commutative harmonic oscillator. It turns out that the supersymmetry is partially broken and the number of supercharges decreases
in the presence of non-commutativity. Recently, supersymmetry is formulated for one-dimensional quantum mechanical
systems with reflections \cite{PVZ11,Ta12}. One of its characteristic features is that both a
supersymmetric Hamiltonian and a supercharge component involve reflection operator $\mathcal{R}$ (which is defined by $\mathcal{R} f(x) = f(-x)$ ). An
important aspect shown in ref. \cite{PVZ11} is that exact eigenfunctions of such a system are expressed
in terms of little $-1$ Jacobi polynomials which is one of the missing families of classical
orthogonal polynomials.\\

 \noindent{\underline{\bf Factorization based on Excited state}:} The factorization of Schr\"{o}dinger Hamiltonian in SUSYQM, discussed earlier,
 gives rise a Riccati differential equation for the superpotential \cite{CK00,Ba00}. The superpotential is usually defined in terms of ground state of a Hamiltonian. But the factorization is not unique in the sense that two
different superpotentials can give rise to the same potential. One can try to construct new isospectral potentials exploiting non-uniqueness of factorization and
obtain a one-parameter family of potentials with the parameter arising as an integration constant \cite{Mi84,Fe84}. In the standard unbroken SUSYQM, it is impossible
to use an excited state of the initial Hamiltonian and at the same time avoid creating singularities
 in the partner potential \cite{Ro97}. Singularities, which appear at the nodes of excited state wave function, are responsible for
 both the destruction of degeneracy of the spectrum of the partner Hamiltonians and creation of negative energy state(s)
 \cite{JR84,Ca95,DP99}. Physically, the destruction of degeneracy with a singular superpotential occurs because the singularity
imposes additional boundary conditions on the wave functions and breaks the real axis into two or more disjoint intervals.
As a result, some or all of the wave functions of the SUSY partner Hamiltonian may not belong to the same Hilbert space of square integrable functions.
In that case, the usual proof of degeneracy between the excited states of the partner Hamiltonians
does not hold, since the supersymmetric operators map square integrable functions to state outside the
 Hilbert space \cite{PS93}. Very recently, the non-uniqueness of the factorization has been exploited in ref. \cite{BU10a,BU10b,DR11} to construct nonsingular isospectral partner Hamiltonians
 by using excited state wave function of a given nonsingular Hamiltonian. The algorithm of this `modified factorization' is
 independent of the choice of wave functions used to define the superpotential and removes the ambiguity about the non degeneracy
 between the spectrum of isospectral partners.\\

\noindent{\underline{\bf Exceptional Orthogonal Polynomials and SUSYQM}:} Recently, SUSYQM has also attracted some attention in the construction of exactly solvable rationally extended potentials
\cite{Qu08,Qu09a,Gr11a,Gr11b,OS09a,OS10,Qu11,BQ10,OS10d,Qu11c} whose bound state wave functions
are given in terms of recently discovered exceptional orthogonal polynomials (EOPs). Exceptional $X_l$ orthogonal polynomials are the infinite sequence of eigenfunctions, ${\{\widehat{P}_n\}}^\infty_{n\ge l}$, $l=1,2,3...$, which satisfy a class of Sturm-Liouville problems:
 $$p(x) \widehat{P}''_n(x) + q(x) \widehat{P}'_n(x) + r(x) \widehat{P}_n(x) = \lambda_n \widehat{P}_n(x),$$
 where $p(x), q(x)$ and $r(x)$ are rational functions of $x$ \cite{UK09}. In contrast to the families of (Jacobi, Laguerre and Hermite) classical orthogonal
polynomials, which start with a constant, the $X_l$ EOP families begin with
polynomial of degree $l\ge1$, but still form complete orthogonal
sets with respect to some positive-definite measure. The first examples of EOP, the so-called Laguerre- and Jacobi-type $X_1$ families,
corresponding to $l = 1$, are proposed in \cite{UK09,UK10}. These $X_1$ EOPs are then shown to appear in the bound state wave functions of
rational extensions of the radial oscillator and Scarf I potential respectively \cite{Qu08}. Such rational extensions are also re-constructed in
the framework of SUSYQM, where they give
rise to a new class of translationally shape invariant potentials \cite{BQR09}. The most general $X_l$, $l=1,2,3...$, Laguerre and Jacobi type exceptional
polynomials and associated shape invariant
potentials are obtained by Sasaki and Odake \cite{OS09a} (the case of $l = 2$ was also discussed in Ref.\cite{Qu09a}).
Some important properties of the $X_l$ EOPs, such as Gram-Schmidt orthonormalization, Rodrigues formulas,
generating functions, the actions of the forward and
backward shift operators on these polynomials \cite{HOS11} and structure of their zeros \cite{HS12} are studied. The $X_l$ EOP and associated shape invariant potentials are also shown to be obtainable through several equivalent approaches
to the SUSYQM procedure, such as the Darboux-Crum transformation \cite{UK10,STZ10},
the Darboux-B\"{a}cklund transformation \cite{Gr11a} and the pre-potential approach \cite{Ho11d,Ho11a}. EOPs are generalized to multi-indexed
families by making use of multi-step Darboux transformations \cite{UKM12},
the Crum-Adler mechanism \cite{OS11a}, and higher-order SUSYQM \cite{Qu11}. Exceptional $X_l$ Laguerre polynomials
are shown to constitute the main part of the solutions to the Dirac equation coupled minimally and non-minimally with
some external fields and also to the Fokker-Planck equation \cite{Ho11b}. Conditionally exactly solvable partners of the
radial and linear oscillators, with broken as well as unbroken supersymmetry, are shown to be associated with exceptional Laguerre
and exceptional Hermite polynomials \cite{DR10}. Also the EOPs have been studied comprehensively within the
framework of the supersymmetric WKB quantization condition \cite{Sr12}, quasi-exactly solvable problems and
 $\mathcal{N}$-fold
supersymmetry \cite{Ta10}. Recently, EOPs are studied in connection with Wilson, Askey-Wilson,
Racah and $q$-Racah polynomials used in the framework
of discrete quantum mechanics \cite{OS09b,OS10,OS11c}.\\

\noindent{\underline{\bf Spectral design via SUSYQM}:}
There is a growing interest nowadays to design systems whose Hamiltonians have predetermined spectral characteristics.
This is useful mostly in producing artificial low dimensional
 structures with controllable physical properties e.g., quantum dots, quantum wells \cite{GG04}, traps in atomic physics \cite{Ph98},
spatially confined quantum systems \cite{WJP95} etc. In the quantum well profile optimization
isospectral potentials are generated through SUSYQM \cite{To01}.
These are necessary because, for instance, intersubband optical transitions in a quantum well, may be grossly enhanced by achieving the resonance
conditions i.e. appropriate spacings between the most relevant states and also by tailoring the wave functions so that the
matrix elements relevant for this particular effect are maximized \cite{In00}. In this
context, the idea of designing potentials with prescribed quantum spectra is worth investigating. Some progress in this
area has been made by restricting the construction of potentials isospectral to a given
initial one except for a few energy values through the usage of SUSYQM, Darboux transformation, factorization
method \cite{Mi84,Fe84,Su85e,BS95,BS97,Ha11a,Ha10a,Ba93,Am88,AC04,KS89}
and Abraham-Moses-Pursey procedure \cite{AM80,Pu85}. The underlying idea
of most of these procedures has been summarized in an algebraic scheme known as {\it intertwining} approach.
In the $n$-th order intertwining approach, two different Hamiltonians $H_0$ and $H_1$,
with nearly same spectrum, are intertwined by a $n$th order linear differential operator $\mathcal{L}$
 as \cite{LS75,CRF01}: $$\mathcal{L} H_0= H_1 \mathcal{L}.$$
$\mathcal{L}$ is known as the intertwiner. This intertwining relation allows one to construct a new Hamiltonian $H_1$ isospectral to the initial Hamiltonian $H_0$.
Moreover, if $\psi$ is an eigenfunction of $H_0$ with eigenvalue $E$ then $\phi = \mathcal{L} \psi$ is also an eigenfunction of $H_1$ with the same eigenvalue $E$. When $H_0$ and $H_1$ are Hermitian
operators, $\mathcal{L}^\dag$ acts as another intertwiner which intertwines in the reverse way: $H_0 \mathcal{L}^\dag = \mathcal{L}^\dag H_1$ and
this in turn implies the existence of a hidden symmetry $[H_0,\mathcal{L}^\dag \mathcal{L}] = 0 = [ \mathcal{L} \mathcal{L}^\dag,H_1]$ \cite{Ca+98}.
The ingredients to implement
the first order intertwining, which is equivalent to first order SUSYQM, are nodeless seed solutions of $H_0$ associated to
factorization energies less than or equal to the ground state energy \cite{FG04,Or98}. Thus first order intertwining allows one
to modify only the ground-state energy level of the initial Hamiltonian $H_0$ to obtain new non-singular partner Hamiltonian $H_1$.
The second order intertwining, which is equivalent to second order SUSYQM \cite{FMR03,FNN00,AF08},
has been proved to be very powerful technique to build a new family of isospectral
Hamiltonians with desired spectral modifications. In this approach one has the freedom of choosing any 
seed solution not necessarily nodeless,
as in the case of first order intertwining, to generate non-singular SUSY partners.
 Second order intertwining \cite{FNN00,AF08} can be implemented by either (i) iteration of first order Darbaux transformation twice or
(ii) looking for the intertwining operator directly as a second order differential operator.
The latter option offers several interesting possibilities of spectral manipulation e.g. \cite{FH05}:
 \begin{itemize}
 \item[--]{ creation of two new levels}
between two neighboring energies,
\item[--]{ deletion of two consecutive bound states}
\item[--]{ leave the spectrum unchanged.}
\end{itemize}
It is possible to generate families of real isospectral partner potentials by using two complex 
factorization energies as well \cite{FMR03}.
The higher order differential operators for the intertwiners leads in a natural way to the nonlinear \cite{AC04} or
higher order SUSY QM \cite{Fe10} (see also the references cited there in). The method of intertwining has also been extended to higher dimensional time independent \cite{KTP01}
and time dependent \cite{Ha10,HPS09} Schr\"{o}dinger equation. For the latter case, the conditions for the existence of a
Darboux transformation in $(n+1)$ dimensions are analyzed and compared to their $(1+1)$ dimensional counterparts.
A complete solution of these conditions is given for $(2+1)$ dimensions. The differential-matrix intertwining operator has been used to
construct one-dimensional
electric potentials or one-dimensional external scalar fields for which the Dirac equation is exactly solvable \cite{An91,SP00,NPS03}. Quite recently, the method of intertwining has been used
\cite{SH08} to generate families of isospectral position dependent mass Hamiltonians.\\

\noindent Other areas where the method of SUSYQM has been extended and used are classical
mechanics \cite{Ba00,JM94,SWY00,KN08}, condensed matter \cite{LN03,JP12},
statistical \cite{Ju96,MP06}, nuclear \cite{BS94,SSB07,LSSB00,BCA99,ABC00a} and atomic \cite{KN84} physics.

\section{$\mathcal{PT}$-symmetry and Pseudo-Hermiticity in Quantum mechanics}

\noindent {\underline{\bf $\mathcal{PT}$-symmetry}:} The interest in non-Hermitian Hamiltonians was stepped 
up by a conjecture of Bender and Boettcher \cite{BB98} that some $\mathcal{PT}$-symmetric
Hamiltonians could possess real bound-state eigenvalues. A non-Hermitian Hamiltonian $H$ is said to be $\mathcal{PT}$-symmetric if it is invariant under
the combined transformation of parity $(\mathcal{P}: \hat{x},\hat{p} \rightarrow -\hat{x},-\hat{p})$ and time-reversal $(\mathcal{T} :
\hat{x},\hat{p},i \rightarrow \hat{x},-\hat{p},-i)$ operators i.e.
$$H^{\mathcal{PT}} = \mathcal{PT} H \mathcal{PT} = H.$$
Because the operator $\mathcal{PT}$ is not linear, $[H,\mathcal{PT}] = 0$ does not always signify
that the Hamiltonian operator and $\mathcal{PT}$ operator has simultaneous eigenstates. If every eigenfunction of a $\mathcal{PT}$-symmetric 
Hamiltonian is also an eigenfunction
of the $\mathcal{PT}$ operator, it is said that the $\mathcal{PT}$-symmetry of $H$ is unbroken. Conversely, if some of the eigenfunctions of a
$\mathcal{PT}$-symmetric Hamiltonian are not simultaneously eigenfunctions of the $\mathcal{PT}$ operator, the$\mathcal{PT}$-symmetry of $H$ is said to be broken \cite{Be05,Be07}.
Thus to establish that the eigenfunctions of a particular $\mathcal{PT}$-symmetric Hamiltonian are real, it is necessary to prove that the $\mathcal{PT}$-symmetry of
the Hamiltonian is unbroken. This is difficult to show but a complete and rigorous proof has been given in \cite{DDT01}.\\

~ A $\mathcal{PT}$-symmetric
Hamiltonian need not be Hermitian, thus it is essential to have a fully consistent
quantum theory whose dynamics is described by a non-Hermitian Hamiltonian. In the case of a non-Hermitian $\mathcal{PT}$-symmetric Hamiltonian
it is crucial that the boundary conditions be imposed
properly to solve the corresponding Schr\"{o}dinger eigenvalue problem. Boundary conditions $\psi(x)_{|x|\rightarrow \infty} = 0$ in this context, are in general located
within wedges bounded by Stokes lines in the complex $x$-plane \cite{Be07}. The eigenfunction $\psi(x)$ vanishes most rapidly at the center of the wedges.
The natural choice of the inner product suitable for $\mathcal{PT}$-symmetric quantum mechanics is the $\mathcal{PT}$-inner product which is defined by
$$\langle\phi| \psi\rangle^{\mathcal{PT}} = \int_\Gamma ~[\phi(x)]^{\mathcal{PT}} \psi(x)~ dx =
\int_{\Gamma} ~[\phi(-x)]^* \psi(x) ~dx,$$
$\Gamma$ being the contour in the Stokes wedges. However, this guess for an inner product
is not acceptable for formulating a valid quantum theory because the norm of a
state is not always positive \cite{BBJ02}. One of the ways to circumvent this difficulty is to introduce $\mathcal{CPT}$ inner product
\cite{BBJ03,BBRR04}:
$$\langle\phi| \psi\rangle^{\mathcal{CPT}} = \int_{\Gamma} ~ [\phi(x)]^{\mathcal{CPT}} \phi(x)~ dx,$$ where $\mathcal{C}$ is the charge conjugation operator which
commute with both $H$ and $\mathcal{PT}$. The $\mathcal{C}$ operator is defined in co-ordinate space as a sum over
the $\mathcal{PT}$ normalized eigenfunctions $\phi_n$ of the Hamiltonian: $\mathcal{C}(x,y) = \sum_{n=0}^\infty \phi_n(x) \phi_n(y)$.
The $\mathcal{C}$ operator has eigenvalues $\pm1$ and it measures the sign of the $\mathcal{PT}$-norm
of an eigenstate. As a result this inner product is positive definite, as states with negative norm are multiplied by $-1$ when acted on by the $\mathcal{C}$
operator. It is not always easy to obtain a closed form expression of the $\mathcal{C}(x, y)$ operator and often one
has to rely on various approximation techniques \cite{BMW03,BJ04,BT06}. In some of the exactly solvable $\mathcal{PT}$-symmetric systems, like Scarf I,
the closed form expression of the operator $\mathcal{C}(x,y)$ has been constructed algebraically \cite{RR07a}. The completeness relation for $\mathcal{PT}$-symmetric system has been derived in terms of the $\mathcal{CPT}$ conjugate \cite{Wi03}: $$\sum_{n=0}^\infty \phi_n(x) [\mathcal{CPT}
\phi_n(y)] = \delta(x-y).$$
For a $\mathcal{PT}$-symmetric Hamiltonian the potential satisfies $V^*(-x) = V(x)$. This allows one to derive the
modified continuity equation for $\mathcal{PT}$-symmetric quantum mechanics which is given by \cite{BQZ01,Be+10} $\rho_t(x,t) + j_x(x,t) = 0$,
where $\rho(x,t)= \psi^*(-x,t) \psi(x,t)$ is the probability density and $j(x,t) = i \psi^*_x(-x,t) \psi(x,t)- i \psi^*(-x,t) \psi_x(x,t)$
is the current density. The $\mathcal{PT}$-symmetric extension of quantum mechanics is recently shown \cite{Be+10a} to exhibit a correspondence to
complex classical mechanics \cite{BBM99} via complex correspondence principle.
Other mathematical developments of the theory of $\mathcal{PT}$-symmetry can be found in
\cite{Sh++,PD98,We06,SG06}. All these studies on $\mathcal{PT}$-symmetric quantum mechanics make it clear that the requirement of Dirac Hermiticity, $H=H^\dag$,
for a Hamiltonian to possess real eigenvalues may be relaxed. However,$\mathcal{PT}$-symmetry of a non-Hermitian Hamiltonian
is neither a necessary nor a sufficient condition for the reality of its spectrum, only unbroken $\mathcal{PT}$-symmetry is a sufficient condition.\\

\noindent{\underline{\bf Pseudo-Hermiticity}:} In a parallel development \cite{Mo02a,Mo02b,Mo02c,Mo10}, it has been clarified that energy spectrum
of a non-Hermitian
Hamiltonian $H$ acting on the Hilbert space $\mathcal{H}$, is real if and only if the following three equivalent conditions (i), (ii) and (iii) holds:
 \begin{itemize}
 \item[(i)] there exist a positive definite, Hermitian operator $\eta: \mathcal{H}\rightarrow \mathcal{H}$ that fulfills $H^\dag = \eta H \eta^{-1}$ i.e. $H$ is {\it pseudo-Hermitian},
     \item[(ii)] $H$ is Hermitian with respect to some positive-definite inner product $\langle .|. \rangle_+$ on $\mathcal{H}$ (which is usually different from the standard inner product $\langle .|. \rangle$). A specific choice for $\langle .|. \rangle_+$ is $\langle \psi| \phi\rangle_\eta = \langle \psi|\eta| \phi\rangle$,
          \item[(iii)] $H$ may be mapped to a Hermitian Hamiltonian $h$ by a similarity transformation: $h= \rho H \rho^{-1}$, where  $\rho$ is linear, invertible and unitary operator.
 This in turn implies that $H$ is {\it quasi-Hermitian} \cite{SGH92,KS04,MB04}.
 \end{itemize}
  One can relax $H$ to be
weak pseudo-Hermitian \cite{So02} by not restricting
$\eta$ to be Hermitian. It has been shown \cite{BQ02} that weak pseudo-Hermiticity is not more general than pseudo-Hermiticity,
in fact these are essentially complementary concepts. A pseudo-Hermitian quantum system is defined by a (quasi-Hermitian)
 Hamiltonian operator and an associated metric
operator $\eta$. So the construction of $\eta$ is important from a physical point of view.
There
are various methods of calculating a metric operator e.g. spectral method \cite{Mo05a,Mo06}, perturbation technique \cite{Jo05,Mo05b} etc. For potentials with unbroken $\mathcal{PT}$-symmetry it has been shown that the operator $\mathcal{P}$ plays
the role of $\eta$ \cite{Mo02a}. In ref.\cite{Ah01b}, the Hermitian linear automorphism $\eta = e^{- \theta p}$ is introduced as a pseudo-Hermitian operator
which affects an
imaginary shift of the co-ordinate:
$$ e^{- \theta p} ~x ~e^{\theta p} = x + i \theta.$$
For real potential $V(x)$ and real $\beta$, the non-Hermitian
Hamiltonians $H = [p + i \beta \nu(x)]^2 + V(x)$ are shown \cite{Ah02}
to be pseudo-Hermitian under the gauge like transformation
$e^{- 2 \beta \int \nu(x) dx} ~[p + i \beta \nu(x)]~ e^{ 2 \beta \int \nu(x) dx} = p - i \beta \nu(x)$. In an interesting paper \cite{Mo02d}, it has been shown that every pair of diagonalizable (not necessarily
Hermitian) Hamiltonians with discrete spectra possessing real or complex-conjugate pairs of
eigenvalues are isospectral and have identical degeneracy structure except perhaps for
the zero eigenvalue if and only if they are pseudo-supersymmetric partners. These pseudo-supersymmetric partner Hamiltonians $H_\pm$ may be factorized
as $H_+= L L^\sharp$ and $H_- = L^\sharp L$. The operator $L^\sharp = \eta^{-1} L^\dag \eta$ is the pseudo-adjoint of the operator $L$. In particular cases, this
factorization applies to $\mathcal{PT}$-symmetric and Hermitian Hamiltonians as well. Spontaneous $\mathcal{PT}$-symmetry breaking is shown to be accompanied by
the explicit breaking of pseudo-supersymmetry \cite{SR07a}.\\

~ The spectrum of non-Hermitian Hamiltonians differ from the
 spectrum of Hermitian ones by two essential features. These are the possible presence of spectral singularities \cite{Na54,Sa05,Mo11,Ah09}
 and exceptional points \cite{He04,Be04,Sm09,MR08}. In the case of continuous spectra the lack of digonalizability of
 the Hamiltonian is responsible for the presence of spectral singularity. At the spectral singularity both the reflection and transmission
coefficient diverge. Physically, a spectral
singularity is the energy of a scattering state that behaves
exactly like a zero-width resonance \cite{Mo09a,Ah09}. This resonance phenomena may
be used as a method for amplifying guided electromagnetic waves \cite{RDM05}. The spectral singularity of a one-dimensional non-Hermitian potential
has been discussed in connection with the supersymmetric quantum mechanics \cite{Sa05} and with the completeness of the bi-orthogonal basis \cite{MD09}.
The existence of a spectral singularity in the case of a delta function potential with complex coupling \cite{Mo11}, complex barrier potentials \cite{MS11} and complex Scarf II potential \cite{Ah12},
has been shown.
 On the other hand, exceptional points (EP) are branch point singularities of
the spectrum of a Hamiltonian. When a physical system is characterized by
$H= H_0 + \lambda H_1$ with $\lambda$ being a strength parameter, the spectrum $E_n$ and eigenfunctions $\psi_n$ are in general analytic function
of $\lambda$. At certain points in the
complex $\lambda$-plane two energy levels coalesce. Such points are called exceptional points \cite{Ka66,He04,Be04,Sm09}. Physical systems in which such a phenomenon occur are open quantum systems with decaying
unbound states \cite{Ro09}, atomic spectra \cite{CMW07}, $\mathcal{PT}$-symmetric waveguides \cite{KGM08} and chaotic optical microcavity \cite{Le+09} etc.
Quasi exactly solvable complex $\mathcal{PT}$-symmetric $\pi$ periodic potential $V (x) = (i \xi \sin 2x + N)^2$
where $\xi$ is real and $N$ is a positive integer, is shown to have exceptional points for odd
values of $N \ge 3$ \cite{BQR08}.\\

\noindent{\underline{\bf Application of $\mathcal{PT}$-symmetry and Pseudo-Hermiticity}:}
The first example of $\mathcal{PT}$-symmetric
potential $V(x)=x^2(i x)^\epsilon$, $\epsilon\in \mathbf{R}$, possesses real
energy eigenvalues for $\epsilon \ge 0$, which however, turns into complex conjugate pairs
as the potential parameter $\epsilon$ exceeds a certain critical value at which $\mathcal{PT}$-symmetry is spontaneously broken \cite{BB98}.
This study was performed using numerical technique and
WKB approximation. Subsequently,
some of the exactly solvable $\mathcal{PT}$-symmetric potentials are obtained by simply setting
the coupling constant of the well known potential to imaginary values, while in some other cases the co-ordinate $x$ is shifted to $x-i \epsilon$.
Among the one-dimensional exactly solvable $\mathcal{PT}$-symmetric potentials, harmonic oscillator \cite{Zn99a}, anharmonic oscillator \cite{Mo05d},
 Morse potential \cite{Zn99}, Eckert, Poschl-Teller and Hulthen potentials \cite{Zn00b}, square well potential \cite{Zn01,BMQ02d},
 double well potential with point-interactions \cite{ZJ05}, Scarf I, II \cite{BR00,Ah01f,Le06,Le06a} and Rosen-Morse \cite{Le08,LM09} potentials are worth to be quoted.
 The $\mathcal{PT}$-symmetric versions of shape-invariant potentials
have been systematically constructed and conditions have been formulated for having real \cite{LZ00} or complex
 \cite{LZ01} energy eigenvalues in their spectra. These results follows from the method of variable transformations
 in which imaginary shift of the co-ordinate appears in a natural way. The Coulomb-harmonic oscillator correspondence in
 $\mathcal{PT}$-symmetric quantum mechanics has been shown in \cite{ZL00}.
 Exactly solvable models with $\mathcal{PT}$-symmetry
 with an asymmetric coupling of channels \cite{Zn06} as well as the coupled channel version of $\mathcal{PT}$-symmetric square
 well potential \cite{Zn06a} are also studied. Group
theoretical formalism of potential algebras
has been extended to non-Hermitian Hamiltonians by complexifying the algebras $so(2,1)$, thereby getting the algebra $sl(2,C)$ \cite{BQ00}.
The algebra $sl(2,C)$ enables one to construct both $\mathcal{PT}$-symmetric and non $\mathcal{PT}$-symmetric Hamiltonians with real eigenvalues.
$\mathcal{PT}$-symmetric potentials like, complex quartic
potential \cite{BB98a}, quartic potentials with centrifugal
and Coulombic terms \cite{Zn00d}, complex sextic potential \cite{BCQ00,BM05} and the Khare-Mandal potential $V(x) = -(z \cosh 2x -i M)^2$ \cite{KM00} are
shown to be QES.
The QES solvable $\mathcal{PT}$-symmetric partner potential of the Khare-Mandal potential has been obtained in \cite{Ba+00}.
It has been pointed out recently that even after suitably perturbing either a Hermitian or a complex $\mathcal{PT}$-invariant QES potential,
one can still obtain another Hermitian or complex, $\mathcal{PT}$-invariant, QES potential \cite{KM09}.\\

~ The energy band formations due to one-dimensional $\mathcal{PT}$-symmetric periodic potentials $V(x) = i (\sin x)^{2N+1}$ \cite{BDM99} and infinite chain of
delta functions with different couplings \cite{Ce03}, have been calculated numerically. It has appeared from the graph of the discriminant $\Delta(E)$, of the energy momentum $(E-k)$ dispersion relation, that the band structure of complex periodic potentials
appear and disappear under perturbation \cite{BDM99}. Later it was shown that \cite{Sh04} the the appearance and disappearance of such energy bands
imply the existence
of non real spectra due to complex deformation of real intervals. The energy-momentum dispersion relation of another
periodic potential, e.g. complex Kroning-Penney type potential, reveals unusual rounded (continuous)
double band (without band gap) structure near Brillouin zone boundaries \cite{Jo99,Ah01e}. Complex $\mathcal{PT}$-invariant periodic potentials have been
obtained from the associated Lame potentials using the anti-isospectral transformation $x \rightarrow ix + \beta$,
where $\beta$ is any non-zero real number \cite{KS06,KS04d}. These periodic potentials are analytically solvable and have finite
number of band gaps, when potential parameters are integers. Explicit expressions for the band edges and dispersion relation of some of
these potentials are also given.\\

~ A general formalism is worked out for the description of one-dimensional
scattering in non-hermitian quantum mechanics and constraints on transmission and reflection coefficients are derived in the cases of $\mathcal{P}$, $\mathcal{T}$ or $\mathcal{PT}$
invariance of the Hamiltonian \cite{CDV07}. For a non-Hermitian (complex) scattering potential the probability of reflection $(R)$
and transmission $(T)$ do not always add to $1$ and one instead has $R+T+A =1$, where $A$ is the probability of absorption \cite{Ah01i}.
Also it has been proved
that $R_{right} \ne R_{left}, T_{left} = T_{right}$. The reflectivity turns out to be symmetric i.e. $R_{right} = R_{left}$, if
the complex potential is
spatially symmetric. One-dimensional scattering by Complex Coulomb potential \cite{LSZ09},
 complex square well \cite{CDV07}, complex delta potential \cite{Mo06,Jo08} and complex Scarf II potential \cite{Ah12} are studied in details.
 Recently, it has been conjectured that all these $\mathcal{PT}$-symmetric scattering potentials yields the following invariance:
$R_{left}(-k) = R_{right}(k), T (-k) = T(k) $ \cite{Ah12}. \\

 ~ The idea of supersymmetric quantum mechanics with some modifications can be utilized to study non-Hermitian Hamiltonians having $\mathcal{PT}$-symmetry
 \cite{Zn01b,An+99b,Zn02c,DDT01,ZC+00}. The key feature of $\mathcal{PT}$-symmetric SUSYQM is the existence of a complex superpotential $W(x)$. $W(x)$ has to be broken into real and imaginary parts which are subsequently splitted into even and odd components. The PT-symmetric requirement $V^*(-x) = V(x)$ then imposes
conditions on these components of $W(x)$, which are expressed in terms of a system of first-order linear differential equations. This equation is
homogeneous when the factorization energy is real and inhomogeneous when it
is complex \cite{Le04a}. Realizations of $\mathcal{PT}$-symmetry
in terms of second-order SUSYQM and para-supersymmetric quantum
mechanics have been constructed \cite{An+99b} and a quasi-parity-dependent factorization energy has
been introduced in order to account for the dual structure of the energy levels \cite{LZ02}.
SUSYQM has also been used to explain the spontaneous $\mathcal{PT}$-symmetry breaking \cite{DDT01,BMQ02d}.
Spontaneous $\mathcal{PT}$-symmetry breaking has been studied in the case of exactly solvable $\mathcal{PT}$-symmetric square-well
\cite{ZL01d}, Scarf II \cite{Ah01f,LCV02}, purely imaginary delta function potential \cite{JZ05d}, Coulomb potential \cite{Le09},
as well as for the quasi exactly solvable Khare-Mandal potential \cite{MK00} and sextic potential \cite{BM05}.
 $\mathcal{PT}$-symmetric quantum brachistochrone problem has been studied in refs.\cite{Mo07c,AF08b,GS08}. It has been shown \cite{BBJB07} that the minimum time,
required to reach the final quantum state from an initial state, can be made arbitrarily small without violating time-energy uncertainty principle.
 The idea of $\mathcal{PT}$-symmetry has been applied to relativistic quantum mechanics,
where the time-reversal operator $\mathcal{T}$ reverses the sign of the time operator $x^0$ \cite{BM11}. Apart from the above mentioned $\mathcal{PT}$-symmetric
models, the other solvable models discussed in the literature are multi-dimensional $\mathcal{PT}$-symmetric models \cite{Na02,Le07,BDMS01},
$\mathcal{PT}$-symmetric deformation of Calogero model \cite{ZT01,FZ08}, non-Hermitian quantum graphs \cite{Zn09}, quantum lattice models \cite{Zn11a,Zn12}.\\

A major impetus for the study of non-Hermitian Hamiltonians has come from recent experimental observation of $\mathcal{PT}$-symmetry breaking
in  structured optical waveguides \cite{Gu09,Ru10}, where the complex index of refraction is manipulated by introducing loss and gain terms.
These experiments of optical waveguides make use of the quantum-optical analogy that the wave equation for the transverse electric field mode is
formally equivalent to the one-dimensional Schr\"{o}dinger equation.
$\mathcal{PT}$-symmetry demands that the complex refractive index obeys the condition $n(\vec{r}) = n^*(-\vec{r}),$ in other words, the real part of
the refractive
index should be an even function of position, whereas the imaginary part must be odd. This analogy has led to several studies on $\mathcal{PT}$-symmetric optical
lattices \cite{Ma08,Be08,Mu08a} in order to synthesize a new class of materials with properties and characteristics that have no analog in Hermitian gain/loss structures.
The stationary properties of such $\mathcal{PT}$-symmetric optical lattices are theoretically studied in
both ID and 2D spatial dimensions. In particular, beam dynamics \cite{Ma08,Ma10} and Bloch oscillations \cite{Lo09a} in these new type of optical structures
reveal some of the exotic structures associated with these periodic potentials, such as skewness of the associated Floquet-Bloch modes,
non-reciprocity of light propagation, band merging, double refraction, abrupt phase transition \cite{Ma08}, power oscillations \cite{Ru10,Ma08},
unidirectional invisibility near the spontaneous $\mathcal{PT}$-symmetry breaking point \cite{Li+11,Jo12,Lo11} etc. Quantum scenarios analogous to the waveguide
experiments in optics have been reported in ref.\cite{CW12} where a Bose-Einstein condensate is placed either in a double well potential \cite{KGM08} or into
two $\mathcal{PT}$-symmetric $\delta$-function traps \cite{CW12} and the effects of the non-linearity, in the Gross-Pitaevskii equation, on the
$\mathcal{PT}$-properties of the
condensate were studied. In the nonlinear domain, one and two dimensional soliton solutions were found to exist below and above the
phase transition point in nonlinear $\mathcal{PT}$-lattices \cite{Ma08,Mu08a}. The interplay of the Kerr nonlinearity with the $\mathcal{PT}$-threshold was analyzed \cite{Mu08a}
and analytical periodic solutions were also derived \cite{Mu08b}.
Also there exists a vast amount of literature on solitons and Bose-Einstein condensates in periodic optical and nonlinear
lattices with $\mathcal{PT}$-symmetry and their nonlinear optical analogs \cite{AAG07,BK10,DM11,GKN08} (and references cited there in).
The notion of non-Hermiticity in quantum mechanics has also been discussed in quantum field theory \cite{FF07,BCM+05,BJR05,Be05a,BM08}, cosmology \cite{Mo05,Mo04a}, classical mechanics \cite{BBM99,Na04}, graphene \cite{Fa11},
classical optics \cite{Ko10}, spin chain \cite{AF09}, many particle quantum systmes \cite{BM01,Ba02,Ba+04} and in nonlinear dynamical systems \cite{BHH07,CFB11} etc.

\section[Position-dependent mass Schr\"{o}dinger equation]{Position-dependent mass Schr\"{o}dinger equation (PDMSE)}

\noindent The concept of position dependent mass (PDM) comes from the effective mass approximation \cite{Tr88} which is an useful tool for studying the motion of carrier electrons in pure crystals \cite{GK93} and the virtual-crystal approximation in the
  treatment of homogeneous alloys as well as in semiconductors. The attention to the effective mass approach stems
  from the extraordinary development in crystallographic growth techniques like molecular beam epitaxy \cite{Jo85} which allow the production
of non uniform semiconductor specimen with abrupt heterojunctions \cite{Ba88}. The presence of position dependent mass in quantum mechanical problems gives rise to the following basic problems:
\begin{itemize}
\item Ordering ambiguity \cite{EHT90,Sh59} arises in defining a unique Hermitian Hamiltonian operator, due to non-commutativity between the
momentum operator $p$ and the mass function $m(x)$. There are many prescriptions for the suitable kinetic energy operator in this scenario,
e.g. Gora and Williams \cite{GW69}, Bastard \cite{Ba81} $(r=-1,s=t=0)$, Zhu and Kroemer \cite{ZK83} $(r=t=-\frac{1}{2},s=0)$, Li and Kuhn \cite{LK83}
$(r=0,s=t=-\frac{1}{2})$ etc,
all of which are reducible from a general form of kinetic energy operator given by Von Roos \cite{Ro83}:
$$H =-\frac{1}{4}\left(m(x)^r p m(x)^s p m(x)^t + m(x)^t p m(x)^s p m(x)^r\right) + V(x),$$
where the ambiguity parameters $r, s, t$ are related by $r+s+t=-1$. For an another choice $r=t=0, s=-1$, the Von-Roos Hamiltonian reduces to
$H= -\frac{1}{2} \left(p \frac{1}{m} p\right) + V$. This form of the Hamiltonian had been first proposed by Ben-Daniel and Duke \cite{BD66}
and later it has been reconsidered by Levy-Leblond \cite{Le95} in connection with the
concept of instantaneous Galilean invariance of the problem of particle with position-dependent mass.\\

\item The presence of discontinuity in the mass function gives rise to the problem of determining appropriate matching condition, on the effective-mass
wave function $\psi$ and its spatial derivative, for which the probability current is continuous across the  the heterojunction.
The matching condition in one-dimension requires that both $m^r \psi$ and $m^{r+t} \frac{d\psi}{dx}$ must be continuous across the junction \cite{MB84}. The matching conditions to be
obeyed by a time-dependent wave function, at a potential and
mass discontinuity when the particle interacts with an oscillating electric
field, has recently been investigated \cite{ML01}.

\item  Apart from the usual square integrability condition on the bound state wave function, an additional restriction
$|\psi(x)|^2/ m(x) \rightarrow 0$ at the end points of $\psi(x)$, has to be imposed for the physical systems with
position dependent mass which vanishes at
 one or both end points \cite{BBQT05,QT04}.
\end{itemize}

~ Much attention has been given to seek exact, quasi-exact or conditionally exact solutions of Schr\"{o}dinger equations
with PDM due to their usefulness in physical applications. The one dimensional PDMSE with smooth mass and potential steps was solved exactly by Dekar
et al.\cite{DC99,DCH98}. The wave functions of this system was obtained in terms of Heun's function. Dutra and Almeida \cite{DA00} discussed the relationship between exact solvability of PDMSE and the ordering ambiguity. In the context of discussing the exact solvability of Morse potential in PDMSE, it is shown \cite{BGQ06} how the map from the full real line $(-\infty,\infty)$ to the half line $(0,\infty)$ connects an exactly solvable Morse potential into an exactly solvable Coulomb one. In this context it is appropriate to mention that quantum nonlinear oscillator (QNLO) can be considered as a system with position dependent mass $m(x) = (1+\lambda x^2)^{-1}$. The Schr\"{o}dinger
equation for QNLO is exactly solved as a Sturm-Liouville problem and the
eigenfunctions are obtained in terms of $\lambda$-deformed
Hermite polynomials \cite{Ca+04,Ca+07}.
The formalism of supersymmetric quantum mechanics
was extended to include position
dependent mass \cite{PR99,DHA03,GO02,GG02,GN07,BGS10,SL03} and it has been shown that any
one-dimensional quantum system with a given PDM and a potential has a supersymmetric partner characterized 
by the same PDM but with a new potential.
The form of this supersymmetric partner potential depends on the form of the original potential 
and the form of the position dependent mass \cite{PR99}.
A procedure has been given for the generation of isospectral combinations of the potential and the 
effective mass variations by supersymmetric quantum mechanics,
accompanied by the co-ordinate transformation method \cite{MI99}. The framework of $\mathcal{N}$-fold 
supersymmetry in one body quantum
mechanical systems has been extended to position dependent mass scenario \cite{Ta06}. Complete classification 
and general form of effective potentials
for type-$A$ $\mathcal{N}$-fold SUSY in the PDM case are given. The effect of position dependent mass profiles on dynamical breaking
of $\mathcal{N}$-fold SUSY in several type-B and $X_2$ models has been investigated in \cite{MRT12}. The SI integrability condition in the PDM scenario leads
to certain relationship between the superpotential and the mass function \cite{PR99}. The method of deformed shape invariance \cite{BBQT05}, which is based on the
equivalence between the PDMSE and the deformed canonical commutation relation \cite{QT04}, has been used to generate pairs of potential
for which the PDMSE is exactly solvable. This approach not only gives the bound-state energy spectrum, but also the ground-state
and the first few excited-state wave functions. The determination of general wave functions, of these systems, requires to solve a differential-difference equation, which can be avoided by combining the deformed
shape-invariance technique with the point canonical transformation method (PCT) \cite{Qu09b}.\\

~ The method of PCT \cite{Le89,BS62}, which consists in transforming a PDMSE into a second-order differential
equation of some special functions, has been used to obtain wide class of both exactly and quasi-exactly solvable PDM systems \cite{Al02,BQ05a,Qu09b,GIN06}.
It is shown that \cite{BQ05a} the nature of the potentials, namely exact or quasi-exact solvability, depend on the relationship between the
mass function and the change of variable used
in point canonical transformation. The interest in quadratic algebra approach to PDMSE is highlighted by
constructing spectrum generating algebras for a class of $d$-dimensional radial harmonic oscillators 
with $d\ge2$ \cite{Qu07e}.
The $d$-dimensional generalization of the point canonical transformation \cite{CC04,MM06} and first order 
intertwining \cite{Qu07e} for a quantum
particle endowed with a position-dependent mass is also described. Quasi-exactly solvable PDM systems like, sextic oscillator, QES Coulomb
and QES Morse potentials, whose solutions are given in terms of orthogonal polynomials satisfying mass dependent recurrence relation,
are constructed using $sl(2,R)$ Lie algebra \cite{KKK02}. The transmission probabilities of the scattering problem with various position dependent mass are calculated using variable phase method
\cite{BV95} and co-ordinate transformation \cite{KSK05}.
Other methods which have been used to study one-dimensional systems with non-constant mass are Green's function approach \cite{CDH95},
path integral approach \cite{YY94}, Darboux transformation \cite{Ha08,CH11} and Lie algebraic approach \cite{RR02a,RR05}.
The last one is particularly interesting since it produces not only the solutions but also reveals the symmetry of the problem. In this context, it may be mentioned that
for a class of PDMSE the shape invariance condition is equivalent to a potential symmetry algebra \cite{JR09}.
The investigations concerning PDM problems have also been extended to relativistic \cite{Al+07} and non-Hermitian $\mathcal{PT}$-symmetric or more generally
pseudo-Hermitian \cite{JYJ05,BQR06} systems as well.

\section{Coherent States and revival dynamics}

\noindent {\underline{\bf Coherent state}:}
Another important line of research related to exactly solvable quantum mechanical models is the construction of coherent states (CS).
Coherent states were first proposed by Schr\"{o}dinger \cite{Sc26} in 1926 in connection with the most classical states of the quantum harmonic oscillator. However, CS became widely recognized after the work of Glauber \cite{Gl63} and
Sudarshan \cite{Su63} in 1963. Glauber \cite{Gl63}, constructed the eigenstates of the annihilation operator of the harmonic oscillator in order to study the correlation function of the electromagnetic field, a subject of great importance in quantum optics.
Sudarshan \cite{Su63} showed that in the coherent states basis $\{|\alpha \rangle \}$, it is always possible to express the density operator in the diagonal form:
$\rho = \int P(\alpha) | \alpha \rangle \langle \alpha | d^2\alpha$,
where $P(\alpha)$  is known as $P$-representation of
the phase space distribution of the density matrix. Roughly at the same time as Glauber and Sudarshan, Klauder \cite{Kl63} developed a set
of continuous states in which the basic ideas of coherent states for arbitrary Lie groups were contained. Coherent states are usually defined in one of the the following three equivalent ways \cite{Zh+90}:
\begin{itemize}
 \item The coherent states $|\alpha \rangle$ are eigenstates of the annihilation operator $a$ with eigenvalue $\alpha$:
 ~$a |\alpha \rangle = \alpha |\alpha \rangle,$ where
 $\alpha$ is a complex number,
 \item The coherent states $| \alpha \rangle$ can be obtained by applying a displacement operator $D(\alpha) = \exp(\alpha a^\dag - \alpha^* a)$
 on the vacuum state of the harmonic oscillator, i.e. $| \alpha \rangle = D(\alpha) |0\rangle$,
 \item These coherent states saturate the Heisenberg's uncertainity relation i.e. $(\Delta p) (\Delta x) = \frac{\hbar}{2}$, where $\Delta p ={[\langle p^2 \rangle - \langle p \rangle ^2]}^{1/2}.$
\end{itemize}
From a group theoretic point of view, the harmonic oscillator coherent states arise
in systems whose dynamical symmetry group is associated with the Heisenberg-Weyl algebra. Gilmore \cite{Gi72,Gi74} and Perelomov \cite{Pe72,Pe86} defined a set of generalized coherent states for any semi-simple
group $G$. Barut and Girardello \cite{BG71}, constructed a set of generalized coherent states for the group $SU(1,1)$
as eigenstates of the non-compact operator of $SU(1,1)$. Coherent states for general potentials were constructed in a series of papers by
Nieto and co-authors \cite{NS78,NS79,GNS80,GNB79}. Their results are valid only to the lowest order of $\hbar$, i.e.,
they are semi-classical in nature.
Indeed, it is not always easy
to obtain a closed form expressions for coherent states
for arbitrary potentials\footnote{A large body of work on the coherent states has by now appeared. This vast literature was exhaustively collected,
catalogued and classified by Klauder and Skagerstam in the book `Coherent states: Applications in Physics and Mathematical Physics' \cite{KS85}.}.\\

~ For quantum mechanical systems with known discrete spectrum a very elegant method of constructing
coherent states has been suggested by Klauder \cite{Kl96}. Later, this method has been extended to systems with
 discrete as well as continuous spectrum energy by Gazaeu and Klauder \cite{GK99}.
The Gazaeu-Klauder (G-K) coherent states of a Hamiltonian $H$ are a two parameter set of
coherent states, $\{ |j,\gamma \rangle:$ $j \ge 0$, $-\infty  \le \gamma \le \infty \}$,
which satisfy the following requirements:
\begin{itemize}
 \item Continuity of labelling-- $(j',\gamma ') \rightarrow (j,\gamma) \Rightarrow |j', \gamma '\rangle \rightarrow |j, \gamma\rangle$
 \item Resolution of unity-- $ \mathbf{I} = \int ~|j,\gamma \rangle  ~\langle j,\gamma | ~ d\mu(j,\gamma)$
 \item Temporal stability-- $ e^{-i H t} ~|j,\gamma \rangle  = |j, \gamma + \omega t\rangle$, ~$\omega$= constant.
 \item Action identity-- $ \langle j, \gamma|H|j, \gamma\rangle = \omega j$.
\end{itemize}
The G-K coherent states have been studied for a variety of quantum systems: Poschl-Teller \cite{An+01,ED02} and
Morse potentials \cite{RR02,Po03}, trigonometric Rosen-Morse \cite{CF08}, anharmonic oscillator \cite{Ro03}, pseudo-harmonic oscillators \cite{PSZ08}, quantum particles confined
by a double-well potential \cite{NAH03}, and non-Hermitian $\mathcal{PT}$-symmetric Scarf I potential \cite{RR06}.
The Morse and anharmonic oscillator coherent states possess both classical like feature (approximate minimum uncertainty property) as well as nonclassical feature (squeezing property) in different parameter regimes. In ref. \cite{Ho01}, G-K coherent states for one mode periodic potentials have been constructed and
compared with the canonical (Glauber) coherent state. The Mandel parameter of the G-K coherent 
states corresponding to a class of shape invariant potentials (like Eckart and Rosen-Morse II) are 
shown \cite{CF02} to be non-negative, so that the corresponding weighting distribution of these states describe 
either the Poissonian or the sub-Poissonian statistics. Generalized coherent state which minimize the 
Robertson-Schr\"{o}dinger uncertainty relation are constructed \cite{ED01a}. G-K coherent states are shown to
 be a special case of these generalized coherent states. An extension of the G-K coherent states for higher dimension 
was made in \cite{NG03}, where
the formalism was applied to a two-dimensional fermion gas in a constant magnetic field. The vector coherent states
 of the G-K type, have also been
constructed and some physical applications of these have been addressed \cite{AB05}. It is worth mentioning that, 
nowadays, coherent states are being widely used in quantum information processing and quantum computing with
 continuous variables \cite{JKL01,BL05,Sa12}.
\\

\noindent{\underline{\bf Revival dynamics}:} In addition to exhibiting non-classical features such as squeezing, Poissonian and/or sub-Poissonian statistics etc, coherent states of systems possessing nonlinear energy spectra also exhibit revival and fractional revival, leading to the formation of Schrodinger cat-like states (a quantum superposition
of macroscopically distinguishable states). The phenomenon of wave packet revivals \cite{Ro04}, arises when a well-localized wave packet is produced and initially exhibits
a short-term time evolution with almost classical periodicity ($T_{cl}$) and then spreads significantly
after a number of orbits, entering a so-called collapsed phase when the probability is
spread (not uniformly) around the classical trajectory. On a much longer time scale after the
initial excitation, however, called the revival time (with $T_{rev} >> T_{cl}$), the packet relocalizes,
in the form of a quantum revival, in which the spreading reverses itself and the classical
periodicity is once again apparent. Even more interestingly, an additional temporal structures with smaller periodicities (fractions of $T_{cl}$) are found
at times equal to rational fractions of the revival time ($p T_{rev}/q$). These structures have been
interpreted as the temporary formation of a number of mini-packets or clones of the
original packet, exhibiting local periodicities $T_{cl}/q$,
and have come to be known as fractional revivals \cite{AP89,YMS90,YS91,Av92}. \\
For a general localized wave packet formed as a one-dimensional superposition of energy eigenstates $| n \rangle$:
$$\psi(x,t) = \sum_{n=0}^\infty c_n e^{-i E_n t} | n \rangle$$ with  $\sum_{n\ge0} |c_n|^2 = 1$, the 
concept of revival arises from the weighting probabilities $|c_n|^2$. The weighting factors $c_n$ in the 
superposition are related to the initial wave function as $c_n = \langle  n | \psi(x,0) \rangle$. Assuming the 
wave packet is strongly weighted around the mean value $\langle n \rangle = \bar{n}$ and the spread 
$\Delta n = [\langle n^2 \rangle - \langle n \rangle ^2]^{1/2}$ is small compared to $\bar{n}$, the energy $E_n$ can be expanded in Taylor series \cite{AS97}:
$$ E_n \simeq E_{\bar{n}} + (n-\bar{n}) E_{\bar{n}}' + \frac{1}{2} (n-\bar{n})^2 E_{\bar{n}}'' + \frac{1}{6}(n-\bar{n})^3 E_{\bar{n}}''' + ...,$$
 where each prime on $E_{\bar{n}}$ denotes a derivative with respect to quantum number $n$. 
These derivatives define distinct time scales that depend on
$n$. The first time scale, $T_{cl} = \frac{2 \pi}{|E_{\bar{n}}'|}$ is the classical period. It controls the initial
behavior of the packet. The second time scale, $T_{rev} =  \frac{4 \pi}{|E_{\bar{n}}''|}$ is the revival time, 
which governs the appearance of the fractional and full revival. The behavior of the wave packet at a time 
larger than $T_{rev}$ is controlled by the third time scale $T_{sr} =  \frac{12 \pi}{|E_{\bar{n}}'''|}$ and is 
called super-revival time. The revival of wave packets may be studied\footnote{ One of the early papers to point 
out the possibility of revival behavior in the context of infinite well was by Segre and Sullival \cite{SS76}.
 They studied the motion of the bound state wave packets in such a system using a simple ``sum over energy eigenstates"
 method.}
by examining the absolute square of the autocorrelation
function \cite{AP89,Na90} $$|A(t)|^2= |\sum_{n=0}^\infty |a_n|^2 e^{-i E_n t}|^2.$$
The quantity $|A(t)|^2$ measures the overlap of the initial state wave packet $\psi(x,0)$ with the state $\psi(x,t)$ at later time $t$.
Numerically, $|A(t)|^2$ varies between $0$ and $1$. The maximum value $1$ is reached when $\psi(x,t)$ exactly matches the initial wave packet
$\psi(x,0)$, and minimum value $0$ corresponds to $\psi(x,t)$ whose shape is distinct from initial wave packet.
 On the other hand, fractional
revival and super-revival appear when periodic peaks in $|A(t)|^2$ occur at the rational fractions of the classical period and revival
time respectively.\\

~ Recently, the sum of Shannon information entropies in
position and momentum spaces has been shown \cite{RS07} to
provide a useful tool for describing fractional revivals,
complementary to the usual approach in terms of the above mentioned autocorrelation
function. The underlying idea is that the
position-space Shannon entropy measures the uncertainty in
the localization of the particle in space. The lower is
this entropy, the more concentrated is the probability density $|\psi(x)|^2$. This in turn implies that the uncertainity is smaller and the accuracy in predicting the localizatiobn of the particle is higher.
Equivalently, the momentum-space entropy measures the uncertainty
in predicting the momentum of the particle. Thus,
the sum of Shannon entropies in position and momentum spaces gives an account of the spreading
(high entropy values) and the regeneration (low entropy
values) of initially well localized wave packets during the
time evolution. The temporary formation of fractional
revivals of the wave packet correspond to the relative minima
of the sum of these Shannon information entropies. Similar studies  associated with Renyi entropy has also been reported \cite{RS08}. \\

~ It has been shown that the phenomena of full and fractional revival occur in the wave packet dynamics of various atomic,
  molecular, optical and condensed matter systems such as Rydberg atom \cite{YMS90,BKT95,YS91}, Jaynes-Cummings model
  \cite{Av92,RA86,BMK92}, molecular vibrational states \cite{EVU94}, atom bouncing vertically in a gravitational cavity \cite{CM95},
  graphene \cite{KK09}, Morse like anharmonic system \cite{VE94}, and system with position dependent mass \cite{Sc06}.
  The simplest quantum systems for which one would
see fractional and full revivals are those for which the energy
spectrum depends  as $n^2$, e.g., the potential well \cite{AS97,Ro00,RH02}, the Poschl-Teller potential
\cite{An+01,RBP05} and the rigid rotator in two dimensions (a quantum system closely related to the infinite square well)
\cite{BKP96}. Revival phenomena have also been observed for systems 
with energy spectra depending on two quantum numbers \cite{BA97,BKT97}. These phenomena has been 
experimentally observed in both atomic \cite{Wa+94} and molecular systems \cite{VVS96}.\\

\newpage

\section{{Outline of the Thesis}}
\noindent
In the light of previously mentioned developments, we have studied some solvable quantum systems and explored the underlying
symmetry of their solutions. The result of these studies are discussed in the remaining part of the thesis in a systematic way.\\

\noindent {\bf Chapter \ref{c2}} deals with the quantum systems which are characterized by position dependent mass
Schr\"{o}dinger Hamiltonians. The
modified factorization technique for a PDM Hamiltonian to construct nonsingular isospectral Hamiltonians have been proposed in section \ref{s2.1}.
It should be mentioned here that due to the presence of position dependent mass, the algorithm is different and more
complicated compared to the constant mass case.

In section \ref{s2.2}, the method of point canonical transformation has been used to
 generate some exactly solvable shape invariant PDM Hamiltonians associated with Laguerre- or Jacobi-type $X_1$ EOPs.
The results are analyzed in the light of SUSYQM approach.

 A complete scheme for the first and second order
intertwining approach have been then developed, in section \ref{s2.3}, to construct isospectral Hamiltonians with several possible spectral modifications e.g. deletion of bound state(s), creation of bound state(s) and leave the spectrum unchanged. The methods have also been explained with suitable examples.\\

\noindent In {\bf Chapter \ref{c3}}, the quantum systems with some non-Hermitian Hamiltonians are considered. It is shown that the $\mathcal{PT}$-symmetric Hamiltonian with the optical potential $V (x) = 4 \cos^2 x + 4i V_0 \sin 2x$ can be mapped into a Hermitian Hamiltonian for $V_0 < .5$, by a similarity transformation. The energy band structure of this Hamiltonian has been studied using the Floquet analysis. Existence of a second critical point ($V_0^c \sim .888437$) of the potential $V(x)$, apart from the known
critical point $V_0 = 0.5$, has also been identified. All these results are reported in section \ref{s3.1}.

 In the section \ref{s3.2} , we have used the method of point canonical transformation to construct some non-Hermitian Hamiltonians whose bound state wave functions are given in terms of exceptional $X_1$ Laguerre- or Jacobi- type polynomials. These Hamiltonians are shown to be both quasi- and pseudo-Hermitian.

 In the last part of this chapter, the generalized Swanson Hamiltonian $H_{GS} = w(\tilde{a}\tilde{a}^\dag + 1/2) + \alpha \tilde{a}^2 + \beta \tilde{a}^{\dag 2}$ with $\tilde{a} = A(x)d/dx + B(x)$, $\alpha,\beta, w \in \mathbb{R}$, has been considered.
 It is shown that the
equivalent Hermitian Hamiltonian of $H_{GS}$ can be transformed into the harmonic
oscillator type Hamiltonian so long as $[\tilde{a}, \tilde{a}^\dag]$ = constant. It has also been shown that, though the commutator of $\tilde{a}$ and $\tilde{a}^\dag$
is constant, the generalized
Swanson Hamiltonian is not necessarily isospectral to the harmonic oscillator. Reason
for this anomaly is discussed in the frame work of position dependent mass models by
choosing $A(x)$ as the inverse square root of the mass profile.\\

\noindent Studies of quantum nonlinear oscillator has been addressed in {\bf Chapter \ref{c4}}. In the first part of this chapter, the coherent state of the nonlinear oscillator is constructed
using the Gazeau-Klauder formalism. The nature of the weighting distribution and the Mandel parameter are
discussed. The details of the revival structure arising from different time scales underlying
the quadratic energy spectrum are also investigated by the phase analysis of the autocorrelation
function. These studies

 In section \ref{s4.2}, various generalizations e.g, exactly solvable Hermitian and non-Hermitian variants of the quantum nonlinear oscillator
are discussed. SUSYQM has also been used to show that these generalized potentials possess shape invariance symmetry. \\

\noindent Finally in {\bf Chapter \ref{c5}}, some future issues related to the works presented in this thesis have been mentioned.


\chapter[Quantum Systems with Position Dependent Mass]{Quantum Systems with Position Dependent Mass \footnote{\noindent This chapter is based on the following three papers:\\
(i) B. Midya, `` Nonsingular potentials from excited state factorization of a quantum system with position-dependent mass''
{ J. Phys.A} 44 (2011) 435306; (ii) B Midya and B. Roy,``Exceptional orthogonal polynomials and exactly solvable potentials in 
position dependent mass Schroedinger Hamiltonians", { Phys. Lett. A} 373 (2009) 4117-4122;
(iii) B. Midya, B. Roy and R. Roychoudhury, ``Position Dependent Mass Schroedinger Equation and Isospectral Potentials : Intertwining Operator approach'',
{J. Math. Phys.} 51 (2010) 022109.}}\label{c2}
\pagestyle{Chapter}
\noindent The one-dimensional time-independent Schr\"{o}dinger equation associated with a particle endowed with a position-dependent
effective mass reads \cite{BD66,Le95}:
\begin{equation}
H \psi(x) = \left[-\frac{d}{dx}
\frac{1}{M(x)} \frac{d}{dx} +V(x)\right] \psi(x) = E \psi(x)\label{e1.1}
\end{equation}
where $M(x)$ is the dimensionless
form of the particle's effective mass $m(x) = m_0 M(x)$ (we have considered $\hbar = 2m_0 = 1$), $V(x)$ denotes the potential, $\psi(x)$ is the particle's wave function corresponding to the energy $E$.

In this chapter, we have studied several aspects of the quantum systems characterized by the above mentioned position-dependent mass Schrodinger equation (PDMSE). First, in section \ref{s2.1} we have extended the notion of modified factorization technique based on excited state wave function to the PDMSE. Section \ref{s2.2} deals with the PDM Hamiltonians associated with exceptional orthogonal polynomials. Finally in section \ref{s2.3}, first and second order intertwining method have been studied systematically to generate isospectral PDM Hamiltonians with several spectral modifications.

\section{Modified factorization of quantum system with position dependent mass}\label{s2.1}
\noindent The modified factorization technique for constant mass Schr\"{o}dinger Hamiltonian \cite{BU10b,BU10a} is based on the non-uniqueness of the superpotential used to
factorize the Hamiltonian. In this technique, excited state wave functions of a given Hamiltonian can be used to construct non-singular isospectral partner
potentials. In this section, we extend the modified factorization technique with suitable examples to handle cases with position-dependent mass. It is to be noted here that due the position dependent mass function, the modified factorization technique of the PDMSE is different and more involved than constant mass case. In order to make this section self contained we have first briefly recalled the notion of
usual factorization technique based on ground state wave function of PDM Hamiltonian.

\subsection{Ground state factorization}\label{s2.1.1}
\noindent
 The minimal version of PDM SUSYQM is based on ground state wave function and the charge operators are represented by
 $Q^+ = A_0^+ \sigma_-$ and $Q^- = A_0^- \sigma_+$, where $\sigma_{\pm}= \sigma_1 \pm i \sigma_2$ are the combinations
 of pauli matrices $\sigma_{1,2}$ and the associated two first order linear operators $A_0^\pm$ are given by \cite{PR99}
\begin{equation}
A^\pm_0 = \pm\frac{1}{\sqrt{M}}\frac{d}{dx} - \left(\frac{1 \mp 1}{2}\right) \left(\frac{1}{\sqrt{M}}\right)' + W_0 .\label{e0.8}
\end{equation}
where `prime' denotes the differentiation with respect to $x$. The components of the supersymmetric PDM Hamiltonian ${\bf{H}} = diag(H^-,H^+)$ can be
factorized in terms of $A_0^+$ and $A_0^-$ to obtain
\begin{equation}
\displaystyle H_0^\mp = A_0^\mp A_0^\pm = -\frac{1}{M}\frac{d^2}{dx^2} + \frac{M'}{M^2} \frac{d}{dx}+ V_0^\mp (x)
\end{equation}
where without loss of generality we have considered zero ground state energy of $H_0^-$ and the two partner potentials $V_0^\mp$ can be
written in terms of superpotential $W_0$ and mass function, as
\begin{equation}
\displaystyle V_0^\mp (x) = W_0^2 \mp \left(\frac{W_0}{\sqrt{M}}\right)' + \left (\frac{1 \mp 1}{2}\right) \left[ \frac{W_0 M'}{M^{\frac{3}{2}}}-
\frac{1}{\sqrt{M}}\left(\frac{1}{\sqrt{M}}\right)''\right].\label{e0.1}
\end{equation}
Clearly, $V_0^+ = V_0^- + \frac{2 W_0'}{\sqrt{M}} - \frac{1}{\sqrt{M}}\left(\frac{1}{\sqrt{M}}\right)''$.
The ground state wave function $\psi_0^{-}$ of the Hamiltonian $H_0^-$ determines the superpotential
\begin{equation}
W_0(x) = - \frac{\psi_0^{{-'}}}{\sqrt{M} \psi_0^{-}}~,\label{e0.6}
\end{equation}
which implies that for a physically acceptable mass function\footnote{The mass function should be positive and without any
singularity through out its domain of definition.} the superpotential and hence two partner potentials $V_0^\mp$ are free from singularity
(provided the given potential $V_0^-$ is without any singularity). It is well established that when SUSY is unbroken then the spectrum of the two
PDM partner Hamiltonians $H_0^\mp$ are degenerate except for the lowest energy of $H_0^-$. Moreover, like constant mass case, the energy eigenvalues and normalized eigenfunctions
of the two PDM SUSY partner Hamiltonians $H_0^\pm$ are related by \cite{PR99}
\begin{equation}\begin{array}{lll}
\displaystyle E_n^{(0)+} = E_{n+1}^{(0)-},~~~~E_0^{(0)-} = 0\\
\displaystyle \psi_n^{+} = (E_{n+1}^{(0)-})^{-\frac{1}{2}}~A_0^+ \psi_{n+1}^{-}\\
\displaystyle \psi_{n+1}^{-} = (E_{n}^{(0)+})^{-\frac{1}{2}}~A_0^- \psi_{n}^{+},~ n= 0,1,2...
\end{array}
\end{equation}
For a well behaved mass function the usual factorization technique for PDMSE allows us to construct nonsingular
isospectral potentials of
a given nonsingular potential using equations (\ref{e0.6}) and (\ref{e0.1}). It should be noted here that the factorization technique mentioned
above is not unique and can be extended to use excited state wave functions. This generalization is mathematically straightforward but physically
very nontrivial since it yields singular partner potentials. In the following we shall show that the excited state of a given position dependent mass
Hamiltonian can be used to construct nonsingular isospectral potentials.

\subsection{Excited state factorization and its modification}
\noindent
In this case the superpotential is defined using the excited state wave functions, as
\begin{equation}
W_n(x) = - \frac{\psi_n^{-'}}{\sqrt{M}\psi_n^{-}}~~,n>0 \label{e0.2}
\end{equation}
where $\psi_n^{-}$ is the $n$-th excited state wave function of an exactly solvable nonsingular Hamiltonian $H_0^-$.
We subtracting the energy $E_n^{(0)-}$ of the excited state from the Hamiltonian $H_n^-$ so that the resulting Hamiltonian can be factorized. In this case
two partner potentials $V_n^\pm$ are given by
\begin{equation}\begin{array}{ll}
\displaystyle V_n^-(x) =  W_n^2 - \left(\frac{W_n}{\sqrt{M}}\right)'  = V^-_0(x) -E_n^{(0)-}\\
\displaystyle V_n^+ (x) = V_n^-(x) + \frac{2 W'_n}{\sqrt{M}}- \frac{1}{\sqrt{M}}\left(\frac{1}{\sqrt{M}}\right)''
\end{array}\label{e0.5}
\end{equation}
The singular superpotential $W_n$ contributes nothing new to $V_n^-$ except a constant energy shift to $V^-_0$.
In this case energy eigenvalues of the Hamiltonians $H^-_n$ and $H_0^-$ are related by $E_k^{(n)-} = E_k^{(0)-} - E_n^{(0)-}, k=0,1,2...$
($k$ denotes the energy level and $n$ refers to the $n$th eigenfunction of $H_n^-$ used in factorization). But the partner potential $V_n^+$
becomes singular at the node(s) of the wave function $\psi_n^{-}$. This singularity is responsible for both the destruction of degeneracy of the
spectrum and creation of negative energy state(s). The singularity breaks the real axis into more than one disjoint intervals and imposes additional
boundary conditions on the wave functions. In this case, the supersymmetry operators map square integrable functions to states outside the Hilbert space.
As a result, some or all of the wave functions of the SUSY partner Hamiltonians may not belong to the same Hilbert space of square integrable functions and
usual proof of degeneracy between the excited states of the partner Hamiltonians does not hold \cite{JR84,Ca95}. This warrants the following modification
of the excited state factorization.

We define the operators \footnote{$A_n^\pm$ are obtained from equation (\ref{e0.8}) after replacing $W_0$ by $W_n$.} $A_n^\pm$ as
\begin{equation}
\tilde{A}_n^\pm = \pm\frac{1}{\sqrt{M}}\frac{d}{dx} - \left(\frac{1 \mp 1}{2}\right) \left(\frac{1}{\sqrt{M}}\right)' + W_n + f_n
\end{equation}
where $W_n$ is given by equation (\ref{e0.2}) and the unknown function $f_n(x)$ will be evaluated shortly.
Using this operators $\tilde{A}_n^\pm$ we first obtain
\begin{equation}
\tilde{H}^+_n = \tilde{A}_n^+\tilde{A}_n^- = -\frac{1}{M}\frac{d^2}{dx^2} + \frac{M'}{M^2} \frac{d}{dx}+ \tilde{V}^+_n (x)
\end{equation}
where
\begin{equation}
\tilde{V}_n^+ = V_n^+ + \frac{f'_n}{\sqrt{M}} + \left(2 W_n + \frac{M' }{2 M^{\frac{3}{2}}}\right) f_n + f_n^2.
\end{equation}
At this point we assume
\begin{equation}
\frac{f'_n}{\sqrt{M}} + \left(2 W_n + \frac{M' }{2 M^{\frac{3}{2}}}\right) f_n + f_n^2 = \beta, ~~\beta \in \mathbb{R}\label{e0.3}
\end{equation}
 so that $\tilde{H}_n^+ = H_n^+ + \beta$ is solvable. This assumption also ensures the non uniqueness of the factorization viz.
  \begin{equation}
  A_n^+ A_n^- + \beta = \tilde{A}_n^+ \tilde{A}_n^-.
  \end{equation}
 For $\beta=0$, the Riccati equation (\ref{e0.3}) reduces to the Bernoulli equation whose solution is given by
\begin{equation}
f_n(x) = \frac{(\psi^{-}_n)^2}{\sqrt{M}\left(\lambda + \int (\psi^{-}_n)^2 dx\right)},~~~~\lambda \in \mathbb{R}.\label{e0.9}
\end{equation}
For nonzero $\beta$ the Riccati equation (\ref{e0.3}) is not always solvable but it can be transformed into the following
second order linear differential equation
\begin{equation}
-\frac{1}{M}\psi^{+''} +\frac{M'}{M^2} \psi^{+'} + \left[W_n^2 + \left(\frac{W_n}{\sqrt{M}}\right)'+ \frac{W_n M'}{M^{\frac{3}{2}}}-
\frac{1}{\sqrt{M}}\left(\frac{1}{\sqrt{M}}\right)''\right]\psi^{+} =-\beta \psi^+ \label{e0.4}
\end{equation}
with the help of following transformation
\begin{equation}
f_n(x) = \frac{1}{\sqrt{M}} (log \chi_n)',~~~~ \chi_n = \frac{e^{-\int \sqrt{M} W_n dx}}{\sqrt{M}} \psi^+.\label{e0.16}
\end{equation}
Using the relation (\ref{e0.5}) it is clear that the equation (\ref{e0.4}) is (nearly) isospectral to the following solvable equation
\begin{equation}
-\frac{1}{M}\psi^{-''} +\frac{M'}{M^2} \psi^{-'} + V_0^-(x) \psi^- = ( E_n^{(0)-}- \beta) \psi^-\label{e0.7}
\end{equation}
where the solutions $\psi^\pm (x)$ corresponding to two equations (\ref{e0.4}) and  (\ref{e0.7}) are related by
\begin{equation}
\psi^+ \sim A_n^+ \psi^-.\label{e0.17}
\end{equation}
By changing the order of operations between $\tilde{A}_n^+$ and $\tilde{A}_n^-$ and using relations (\ref{e0.5}), (\ref{e0.3})
we obtain the partner Hamiltonian $\tilde{H}_n^-$ of $\tilde{H}_n^+$ as
\begin{equation}
\tilde{H}_n^- = \tilde{A}_n^- \tilde{A}_n^+ = -\frac{1}{M} \frac{d^2}{dx^2} + \frac{M'}{M^2}\frac{d}{dx} + \tilde{V}^-_n(x)
\end{equation}
where
\begin{equation}
\tilde{V}_n^- = V_n^- - \frac{2 f_n'}{\sqrt{M}} + \beta = V_0^- - \frac{2 f_n'}{\sqrt{M}} - E_n^{(0)-} + \beta.\label{e0.14}
\end{equation}
At this point few comments are worth mentioning. For a given nonsingular potential $V_0^-$ and a mass function the new potential
$\tilde{V}_n^+ (= V_n^+ +\beta)$ is singular but the potential $\tilde{V}_n^-$ has no singularity if $f_n(x)$ is nonsingular in which case
the corresponding spectrum of the two isospectral Hamiltonians $\tilde{H}_n^-$ and $H_n^-$ are related by
\begin{equation}
\tilde{E}_k^{(n)-} = E_k^{(n)-} + \beta,~~~k=0,1,2....
\end{equation}
For $\beta = 0$, the singularity of $f_n$ given in (\ref{e0.9}) can be controlled by the arbitrary constant $\lambda$
and in all other cases it depends on the nature of solution of the equation (\ref{e0.7}).
 Moreover, the eigenfunctions of the two nonsingular Hamiltonians $H_n^-$ and $\tilde{H}_n^-$ are related by
\begin{equation}
\begin{array}{ll}
\displaystyle \tilde{\psi}^{-}_k~ \sim ~\tilde{A}_n^- A_n^+ \psi_k^{-}\\
\displaystyle \psi^{-}_k ~\sim ~ A_n^- \tilde{A}_n^+ \tilde{\psi}_k^{-}
\end{array}\label{e0.10}
\end{equation}
except for the zero energy state of $\tilde{H}_n^-$.
\begin{figure}[]
\centering
\includegraphics[width=2.65in, height=2.75in]{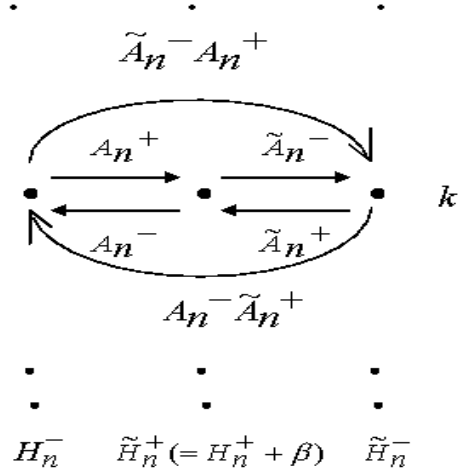}
\caption[Schematic diagram of energy levels of two partner Hamiltonians $H_n^-$, $\tilde{H}_n^-$ and the intermediate Hamiltonian $\tilde{H}_n^+$.]{\label{fig}Energy levels of two partner Hamiltonians $H_n^-$, $\tilde{H}_n^-$ and the intermediate Hamiltonian $\tilde{H}_n^+$.
Combined actions of the operator $\tilde{A}_n^- A_n^+$ and $A_n^- \tilde{A}_n^+ $ are shown.}\label{f0.1}
\end{figure}
\FloatBarrier
The proof of the result \ref{e0.10} is as follows: we start with
\begin{equation}\begin{array}{llll}
\tilde{H}_n^- (\tilde{A}_n^- A_n^+ \psi_k^-) = \tilde{A}_n^- (\tilde{A}_n^+ \tilde{A}_n^-) A_n^+ \psi_k^- \\
= \tilde{A}_n^- (A_n^+ {A}_n^- +\beta) A_n^+ \psi_k^- \\
= \tilde{A}_n^- A_n^+ ({A}_n^-A_n^+) \psi_k^- +\beta \tilde{A}_n^- A_n^+ \psi_k^- \\
= \tilde{A}_n^- A_n^+ (H_n^- \psi_k^-) +\beta \tilde{A}_n^- A_n^+ \psi_k^- = (E_k^{(n)-}+\beta)\tilde{A}_n^- A_n^+\psi_k^-.
\end{array}\label{e0.11}
\end{equation}
In the second step we have used the non-uniqueness of the factorization of $H_n^+$ and in the last step we have used the
PDMSE $H_n^-\psi_k^- = E_k^{(n)-}\psi_k^-$. Hence, if $\psi_k^-$ is a eigenfunction of $H_n^-$ corresponding to eigenvalue $E_k^{(n)-}$ then
$\tilde{A}_n^- A_n^+\psi^-_k$ is a eigenfunction of $\tilde{H}_n^-$ with eigenvalue $(E_k^{(n)-} + \beta).$ The second relation of equation (\ref{e0.10})
can be proved in a analogous way. The wave function $\tilde{\psi}_{k'}^-$ (say) corresponding to zero energy of the Hamiltonian $\tilde{H}_n^-$
can be obtained by directly solving $\tilde{A}_n^+ \tilde{\psi}^{-}_{k'} =0$.

 Hence, excited state wave function of a given position dependent mass Hamiltonian can be used to construct nonsingular partner
Hamiltonian $\tilde{H}_n^-$ using equation (\ref{e0.14}) provided the function $f_n$ is nonsingular. This has been made possible in
two step factorization. In the first step, an intermediate singular Hamiltonian $H_n^+$ has been created and in the second step non-uniqueness
of the factorization technique has been used to remove the singularities. The operator $-\tilde{A}_n^- A_n^+$ is similar to the second order reducible
intertwining operator ${\cal{L}} =\frac{1}{M(x)} \frac{d^2}{dx^2} + \eta(x) \frac{d}{dx} + \gamma(x)$ mentioned in ref.\cite{MRR10} which intertwines
two Hamiltonians $H_n^-$ and $\tilde{H}_n^-$
\begin{equation}
\tilde{H}_n^- \tilde{A}_n^- A_n^+ = \tilde{A}_n^- A_n^+ H_n^-,
\end{equation}
if one identifies $-\eta(x) = \frac{f_n}{\sqrt{M}} + \frac{M'}{M^2}$ and $-\gamma(x) = V_n^- + f_n W_n.$
In this regard the present algorithm is equivalent to the confluent second order SUSY transformation \cite{MN00} in
constant mass case but the relation between the two isospectral partner potentials is different in the present approach due to
the position dependence of the mass function. Moreover, the energy levels of the two nonsingular isospectral partner Hamiltonians
$H_n^-$ and $\tilde{H}_n^-$ are degenerate. In figure \ref{f0.1}, we have shown schematically the combined action of the operators
$\tilde{A}_n^- A_n^+$ and $A_n^- \tilde{A}_n^+ $ on the wave function of two isospectral partner Hamiltonians
$H_n^-$ and $\tilde{H}_n^-$ respectively. The operators $A_n^+$ or $\tilde{A}_n^+$ destroy a node but $\tilde{A}_n^-$ or $A_n^-$
create the same in the eigenfunctions so that the overall number of nodes remains the same.\\

\subsection{Examples of modified factorization}
\noindent It is worth mentioning here that the proposed technique mentioned in the preceding section is more general and can be applied
to any solvable PDM Hamiltonian to obtain some acceptable isospectral Hamiltonians which might be useful in various fields of condensed matter physics.
For illustration purpose we are considering here two exactly solvable position dependent mass Hamiltonians. In example 1.1, we have considered $\beta=0$ while
non zero $\beta$ has been considered in example 1.2.

\begin{figure}[]
\centering
\includegraphics[height=6 cm, width=7.25 cm]{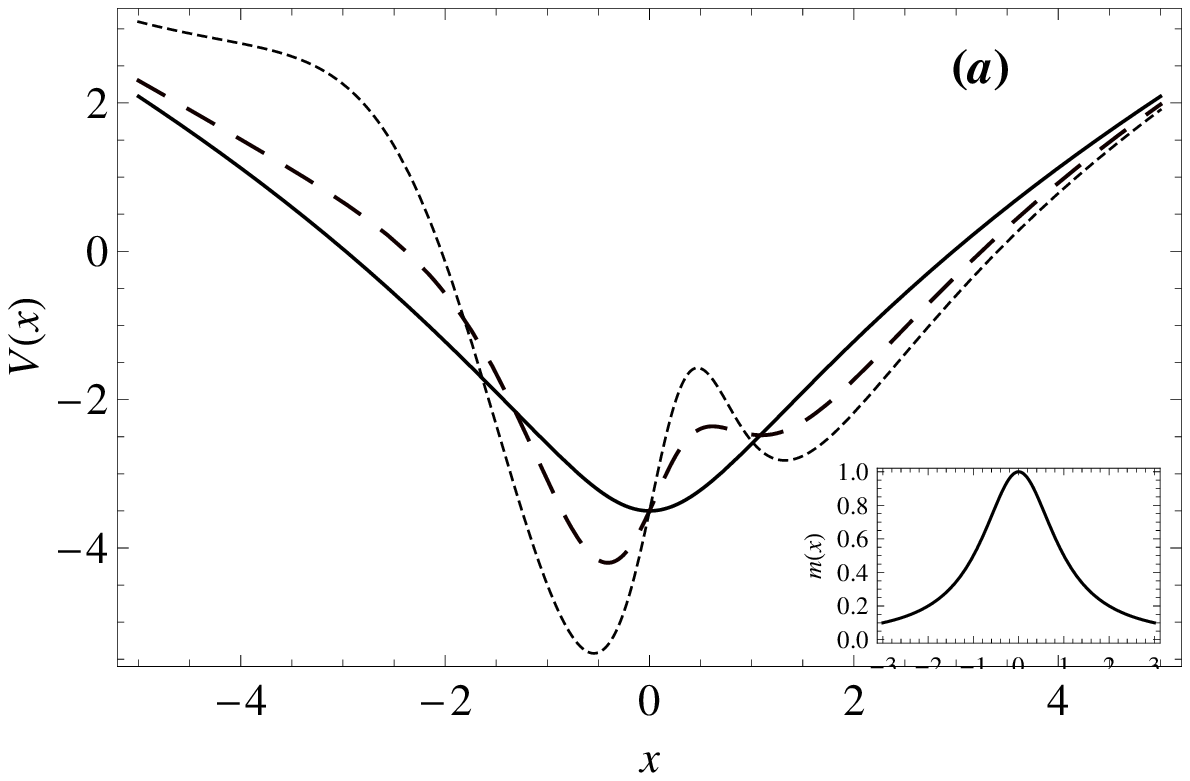}~~~~ \includegraphics[height=6 cm, width=7.25 cm]{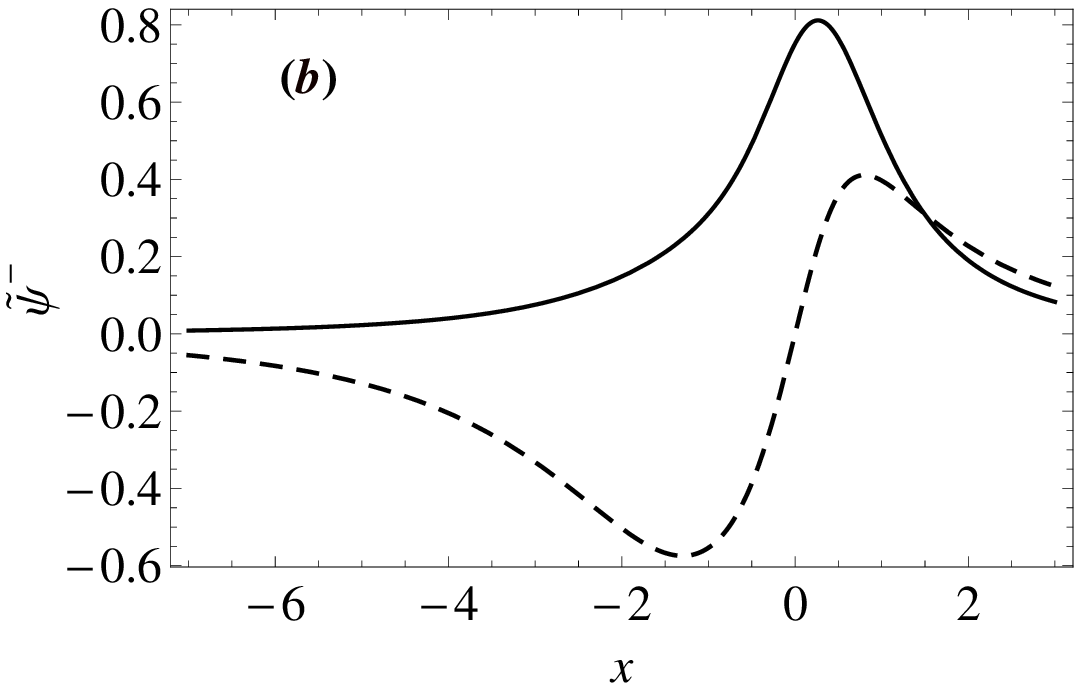}
\caption{(a) Plots of the given potentials $V_1^-$ (solid line) and $\tilde{V}_1^-$ for $\lambda =1$ (dashed line), $\lambda=.7$ (dotted line) and
the mass function $m(x)$. (b) The lowest (solid line) and first excited state (dashed line) wave functions of the Hamiltonian $\tilde{H}_1^-$ for $\lambda =1.$}
\label{f0.2}
\end{figure}

\begin{figure}[]
\centering
\includegraphics[height=6 cm, width=7.25 cm]{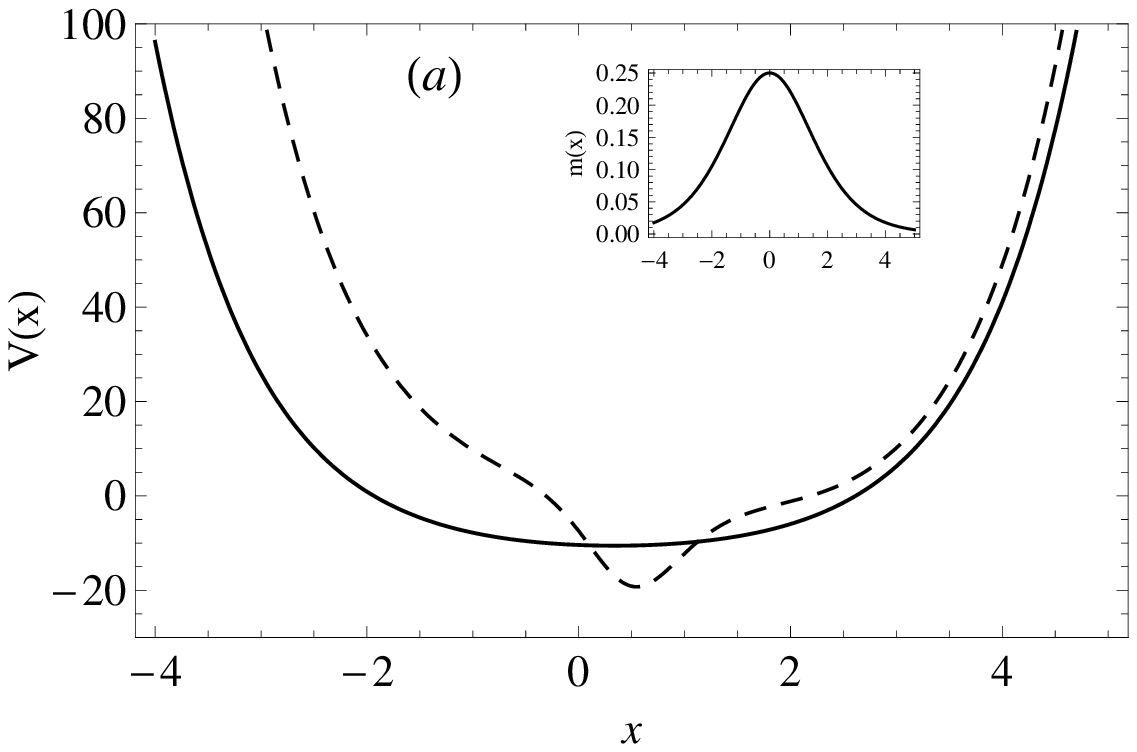}~~~\includegraphics[height=6 cm, width=7.25 cm]{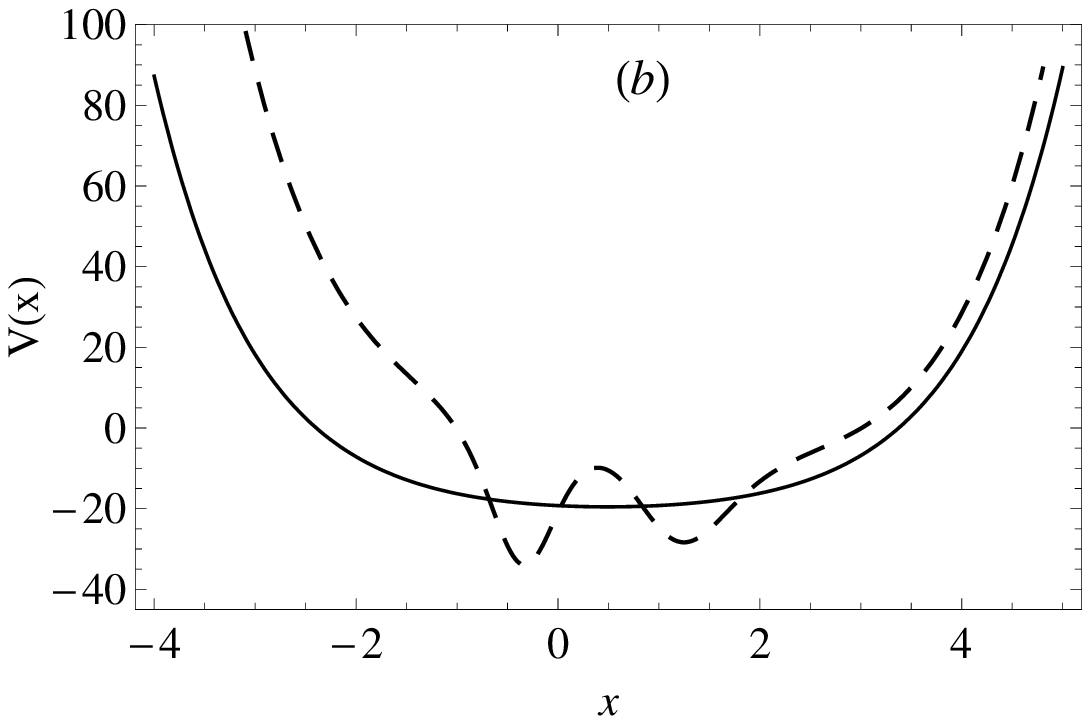}
\caption{Plots of (a) the potential $V_1^-=V_0^- - 13$ (solid line),
its nonsingular partner $\tilde{V}_1^-$ (dashed line) and the mass function $m(x)$,
(b) the potential $V_2^-=V_0^- -22$ (solid line) and its partner $\tilde{V}_2^-$ (dashed line) for $a=1, b=5, c=4, \beta=1$.}\label{f0.3}
\end{figure}

\noindent{\bf Example 1.1 ($\beta = 0$):~~} We consider the following mass function and potential possessing harmonic oscillator like spectra \cite{Pe08}:
\begin{flalign}
M(x) = \frac{1}{1 + \alpha^2 x^2}
\end{flalign}
\begin{align}
V_0^- (x) = \left(\frac{\sinh^{-1} (\alpha x)}{\alpha}\right)^2 - \frac{\alpha^2}{4} \left(\frac{2 + \alpha^2 x^2}
{1 + \alpha^2 x^2}\right),~~~ x\in(-\infty,\infty).
\end{align}
The bound state solution are given by
\begin{equation}
\psi_k^{-} = \sqrt{\frac{1}{2^k k!}} \left(\frac{1}{\pi}\right)^{\frac{1}{4}} \frac{e^{-\frac{1}{2}~
\left(\frac{\sinh^{-1} (\alpha x)}{\alpha}\right)^2}}{(1 + \alpha^2 x^2)^{1/4}}~ H_k\left(\frac{\sinh^{-1} (\alpha x)}{\alpha}\right)
\end{equation}
\begin{equation}
E_k^{(0)-} = 2 k +1, ~~~~k=0,1,2...
\end{equation}
where $H_k(x)$ denotes the Hermite polynomial. Now we consider here first excited state $\psi_1^{-} = \frac{2\sinh^{-1} x ~e^{-\frac{1}{2}
\left(\sinh ^{-1} x\right)^2}}{(1 + x^2)^{1/4}}$, $\beta=0,$ and $\alpha=1$ for which
the partner potential $V_1^-$ and $\tilde{V}_1^-$ are obtained using equation (\ref{e0.5}), (\ref{e0.9}) and (\ref{e0.14}) as
\begin{equation}
V_1^- = \left({\sinh^{-1} ( x)}\right)^2 - \frac{1}{4} \left(\frac{2 +  x^2}{1+  x^2}\right) -3
\end{equation}
{\small\begin{equation}
\tilde{V}_1^- = -\frac{2 + x^2}{4 + 4 x^2} + (\sinh^{-1} x )^2 -3+ 2 \sqrt{(1+x^2)} \frac{d}{dx} \left[\frac{4 \left(\sinh^{-1} x\right)^2}{\sqrt{\pi}e^{(\sinh^{-1} x)^2}
 (2 \lambda+ \ef(\sinh^{-1} x))-2 \sinh^{-1} x}\right]
\end{equation}}
respectively. The energy eigenvalues of the Hamiltonian $\tilde{H}_1^-$ are given by $\tilde{E}^{(1)-}_k = 2k -2, ~k=0,1,2...$
The normalized first excited state wave function corresponding to zero energy of $\tilde{H}_1^-$ is evaluated by solving
$\tilde{A}_1^+ \tilde{\psi}^{-}_1 = 0$ as
\begin{equation}
\tilde{\psi}_1^{-} = \frac{\pi^{\frac{1}{4}} \sqrt{8\lambda^2 -2}~~ e^{\frac{(\sinh^{-1} x)^2}{2}}
\sinh^{-1} x }{(1+x^2)^{\frac{1}{4}}\left[2\sqrt{\pi}\lambda e^{(\sinh^{-1} x)^2} - 2 \sinh^{-1} x +
\sqrt{\pi} e^{(\sinh^{-1} x)^2} \ef(\sinh^{-1} x)\right]}.
\end{equation}
Hence, the new potential $\tilde{V}_1^{(1)-}$ is nonsingular for $\lambda>1/2$.
The lowest energy wave function corresponding to negative energy $-2$ is obtained through $\tilde{A}_1^{-} A_1^+ \psi^{-}_0$ as
\begin{equation}
\tilde{\psi}^{-}_0 = \frac{\pi^{\frac{1}{4}}[2 \lambda + \ef(\sinh^{-1} x)] e^{\frac{(\sinh^{-1} x)^2}{2}}}{(1+x^2)^
{\frac{1}{4}}\left[2\sqrt{\pi} \lambda e^{(\sinh^{-1} x)^2} - 2 \sinh^{-1} x + \sqrt{\pi} \ef(\sinh^{-1} x) e^{(\sinh^{-1} x)^2}\right]}.
\end{equation}
In figure \ref{f0.2}(a), we have plotted the potential $V_1^-(x)$, its isospectral partner $\tilde{V}_1^-$
for two parameter values $\lambda=1, .7$ and the mass function. In figure \ref{f0.2}(b),
lowest and first excited state wave functions of the Hamiltonian $\tilde{H}_1^-$ for $\lambda=1$ have been plotted.
\FloatBarrier
\noindent{\bf Example 1.2 ($\beta \ne 0$):~~} Here we consider the following mass function and the potential
\begin{equation}
M(x) = \frac{1}{4} \sech^2 ~(\frac{x}{2})\label{e0.18}
\end{equation}
\begin{equation}
V_0(x) = \frac{(a+b-c)^2 - 1}{4} e^x + \frac{c(c-2)}{4} e^{-x},~~c>\frac{1}{2}, a+b-c+\frac{1}{2}>0,~x\in(-\infty,\infty),
\end{equation}
for which one linearly independent solution of the equation (\ref{e0.7}) is known to be (see appendix of ref.\cite{MRR10})
\begin{equation}
\psi^-(x) = C_1 ~e^{c x/2}(1+e^x)^{\frac{P-1}{2}}  ~~{_2}F_1\left(\frac{a+b+P}{2},\frac{-a-b+2c+P}{2},c,e^{-x}\right),\label{e0.19}
\end{equation}
where
\begin{equation}\begin{array}{ll}
P^2 = (a+b)^2 -2 c(a+b-c+1) + 4 (E_n^{(0)-}-\beta),\\
E_n^{(0)-} = n^2 + n(a+b) + \frac{c(a+b-c+1)}{2}.
\end{array}
\end{equation}
The bound state wave functions of the Hamiltonian $H_0^-$ are given \cite{MRR10}, in terms of Jacobi polynomial ${P}_n^{(\sigma,\delta)}(x)$, as
\begin{equation}
\psi_n^- \sim \frac{e^{cx/2}}{(1+e^x)^{(a+b+1)/2}}~~ {P}_n^{(c-1,a+b-c)}\left(\frac{1-e^x}{1+e^x}\right),~~n=0,1,2...\label{e0.15}
\end{equation}
Now we consider first excited state i.e. $n=1$ for which the function $f_n$ can be obtained using equations (\ref{e0.16}) and (\ref{e0.17}) as
{\small \begin{equation}
 f_n= \frac{1}{\sqrt{M}}\frac{d}{dx} [\log(A_1^+ \psi^-(x))]+\frac{\cosh \left(\frac{x}{2}\right)}{4} \left[\frac{2 c}{(a+b-c+1)e^x -c} + \frac{a+b+3}{1+e^x}+c-a-b-1\right] + \sinh \left(\frac{x}{2}\right)
\end{equation}}
where $A_1^+ = \frac{1}{\sqrt{M(x)}}\left(\frac{d}{dx}-\frac{\psi_1^{-'}}{\psi_1^-}\right)$,
$M(x), \psi^-$ and $\psi_1^-$ are given by equations (\ref{e0.18}), (\ref{e0.19}) and (\ref{e0.15}) respectively. In order to obtain
new nonsingular potential $\tilde{V}_1^-$ we have to choose $a,b,c, \beta$ such that $f_n$ has no singularity. The analytical expression
of $f_n$ is too involved so it is very difficult to find the range of parameters values for which $f_n$ is nonsingular. In figure \ref{f0.3}(a),
we have plotted the potential $V_1^-(x) = V_0^- - E_1^{(0)-}$ and its nonsingular partner $\tilde{V}_1^- = V_1^- -2f_n/\sqrt{M}-E_1^{(0)-}+\beta$
for particular values of $a=1,b=5, c=4$ and $\beta=1$. For the same set of parameter values, isospectral partner potential $\tilde{V}_2^-$
has been drawn in figure \ref{f0.3}(b) by considering second excited state factorization i.e. $n=2$. The energy eigenvalues of the new Hamiltonians $\tilde{H}_1^-$ and $\tilde{H}_2^-$ are given by $\tilde{E}_k^{(1)-} = k^2+ 6k -6$ and $\tilde{E}_k^{(2)-}
 = k^2 + 6 k-15$ respectively. In both cases the wave functions can be obtained using equations (\ref{e0.10}) and (\ref{e0.15}).
\FloatBarrier

\section[Position-dependent mass Hamiltonians associated with exceptional $X_1$ orthogonal polynomials]
{New exactly solvable PDM Hamiltonians associated with exceptional $X_1$ orthogonal polynomials}\label{s2.2}
\noindent Here, some new exactly solvable potentials, whose bound state solutions are written in terms of recently discovered \cite{UK09,UK10}
exceptional $X_1$ Laguerre or Jacobi type orthogonal polynomials,
are obtained using the method of point canonical transformation (PCT). In the constant-mass case, i.e., for $M(x) = 1$,
this procedure has been thoroughly
investigated \cite{BS62,Le89}. A similar study in the PDM context is more involved as has been discussed in the below.

\subsection[The method of PCT in position-dependent mass background]{The method of PCT in PDM background}\label{s2.2.1}
 \noindent In PCT approach \cite{Le89,BS62}, we search for the solution of equation (\ref{e1.1}) in the form
 \begin{equation}
 \psi(x)= f(x)~F\left(g(x)\right)\label{e1.3}
 \end{equation}
 where $F(g)$ is some special function which satisfies the second order differential equation
 \begin{equation}
 \frac{d^2F}{dg^2}+Q(g) \frac{dF}{dg}+R(g) F=0. \label{e1.4}
 \end{equation}
and $f(x)$ and $g(x)$ are unknown functions of $x$ to be determined.
It should be mentioned here that in order to be physically acceptable, $\psi(x)$ has to satisfy the following two conditions:
\begin{itemize}
\item[(i)]{ Like quantum systems with constant mass it will be square integrable over domain of definition {\bf{D}} of $M(x)$
 and
 $\psi(x)$ i.e.,$$\int_D |\psi_n(x)|^2 dx<\infty$$}
 \item[(ii)]{ The Hermiticity of the Hamiltonian in the Hilbert
  space spanned by the eigenfunctions of the potential $V(x)$ is ensured by the following extra
 condition {\cite{BBQT05}}
 $$\frac{|\psi_n(x)|^2}{\sqrt{M(x)}}\rightarrow 0$$ at the end points of the interval where $V(x)$ and $\psi_n(x)$ are
  defined. This condition imposes an additional restriction whenever the mass function $M(x)$ vanishes at any one or both
  the end points of $\mathcal {D}$.}
\end{itemize}
Now, substituting $\psi(x)$ from equation (\ref{e1.3}) in equation (\ref{e1.1})
and comparing the coefficients of different order derivatives appear in the resulting equation, one obtains
\begin{equation}
Q(g(x)) = \frac{g''}{g'^2}+\frac{2f'}{fg'}-\frac{M'}{Mg'},  ~~~~ R(g(x)) = \frac{f''}{fg'^2}+(E-V)\frac{M}{g'^2}-\frac{M'f'}{Mfg'^2} \label{e1.6}
\end{equation}
After some algebraic manipulations we obtain explicit expressions for $f(x)$ and $E-V(x)$ as
\begin{equation}
f(x)\propto \sqrt{\frac{M}{g'}}
~~~~\exp\left(\frac{1}{2}\int^{g(x)}Q(t)dt\right)\label{e1.7}
\end{equation}
and
\begin{equation}
E-V =\frac{g'''}{2Mg'}-\frac{3}{4M}\left(\frac{g''}{g'}\right)^2+\frac{g'^2}{M}
\left(R-\frac{1}{2}\frac{dQ}{dg}-\frac{Q^2}{4}\right)-\frac{M''}{2M^2}+\frac{3M'^2}{4M^3}
\label{e1.8}
\end{equation}
respectively. For equation (\ref{e1.8}) to be satisfied, we need to find some functions $M(x)$, $g(x)$
 ensuring the presence of a constant term on the right hand side of equation (\ref{e1.8}) to
  compensate $E$ on its left hand side and giving rise to an potential $V(x)$ with well behaved wave functions involving special function $F(g)$. In PCT
  approach there are many options for choosing $M(x)$,
for example $M(x) = \lambda g'^2(x)$ \cite{BG04,Al02}, $M =
\lambda g'(x)$ , $M = \displaystyle \frac{\lambda}{g'(x)}$ \cite{BQ05a}, $\lambda$
 being a constant. Some of these choices give rise to exactly solvable potentials while some other correspond to quasi-solvable potentials.
 Here, we choose $M(x)=\lambda g'(x)$ so that the expression $E-V(x)$ reduces to
\begin{equation}
E-V=\frac{g'}{\lambda}\left(R-\frac{1}{2}\frac{dQ}{dg}-\frac{Q^2}{4}\right)\label{e1.9}
\end{equation}
and we are now in a position to choose the form of $F(g)$, $Q(g)$ and $R(g)$ which satisfy equation (\ref{e1.4}). It is to be noted here that
different choices for the function $F(g)$ will give rise to different solvable Hamiltonians whose solutions are given by (\ref{e1.3}) and (\ref{e1.7}).

\subsection{{Hamiltonian associated with exceptional $X_1$ Laguerre polynomial}}
\begin{figure}[]
\centering
\includegraphics[height=6 cm, width=7.25 cm]{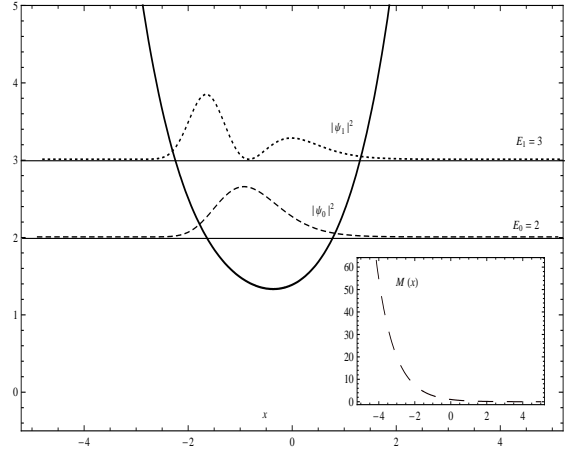}
\caption[Plot of the potential associated with Laguerre type $X_1$ EOP] { Plot of the potential $V$ (solid line)
given in equation (\ref{e1.12}), square of its first two bound state
wave functions $|\psi_o(x)|^2$ (dashed line) and $|\psi_1(x)|^2$
(dotted line), for the mass function $M(x)$ (long dashed line)
given in equation (\ref{e1.11}). Here we have considered $b=1,\alpha=2.$}\label{f1.1}
\end{figure}

\noindent We consider the special function $F_n(g) \propto \hat{L}_n^{(\alpha)}$, where
$\hat{L}_n^{(\alpha)},~n=1,2,3,...,~ \alpha>0$ is the Laguerre type $X_1$ exceptional orthogonal polynomials \cite{UK09} for
which $Q(g)=-\frac{(g-\alpha)(g+\alpha+1)}{g(g+\alpha)}$ and $R(g)=\frac{n-2}{g}+\frac{2}{g+\alpha}$.
For these values of $Q(g)$ and $R(g)$ the equation (\ref{e1.9}) reeds
\begin{equation}
E-V  =\frac{g'}{\lambda g}\left(\frac{2\alpha
n+\alpha^2-\alpha+2}{2\alpha}\right)-\frac{g'}{g^2}
\left(\frac{\alpha^2-1}{4\lambda}\right)-\frac{2g'}{\lambda(g+\alpha)^2}-\frac{g'}{\lambda
\alpha(g+\alpha)}-\frac{g'}{4\lambda}. \label{e1.10}
\end{equation}
Taking $\frac{g'}{\lambda g}= \mbox{Constant}$ (say $C$), a
constant term can be created on the right-hand side of the above
equation ($C$ must be positive in order to get increasing energy eigenvalues for successive $n$ values).
Consequently we have the expression of mass function, energy eigenvalues and potential as
\begin{equation}
M(x)=e^{-bx}~,~~~~~~~ -\infty<x<\infty\label{e1.11}
\end{equation}
\begin{equation}
E_m=b^2 \left(m+\frac{\alpha+1}{2}\right)+\frac{b^2}{\alpha} + \bar{V}_0 \label{e1.43}
\end{equation}
\begin{equation}
V(x) = \left[\frac{b^2}{4}\left((\alpha^2-1)e^{bx}+e^{-bx}\right)\right]+\frac{b^2}{4}\left(\frac{4}{\alpha(1+\alpha
e^{bx})}+\frac{8e^{bx}}{(1+\alpha
e^{bx})^2}\right)+\bar{V}_0 \label{e1.12}
\end{equation}
respectively, where $C\lambda=-b,b>0$ and $\bar {V_0}$ are constants.
In the expression (\ref{e1.43}) we have reset the quantum number $n=m+1$. The normalized eigenfunctions of the Hamiltonian are obtained using equations (\ref{e1.3}) and (\ref{e1.7}) as
\begin{equation}
\psi_m(x)=\left(\frac{b~
m!}{(m+\alpha+1) \Gamma(m+\alpha)}\right)^{1/2}
~~\frac{exp\left[-\frac{1}{2}\{(\alpha+1)bx+e^{-bx}\}\right]}{\alpha+e^{-bx}}~~\hat{L}_{m+1}^{(\alpha)}
(e^{-bx}),~~~~~m=0,1,2...\label{e1.14}
\end{equation}
It is to be noted that the square integrability condition (i) stated
earlier does not impose any additional restriction on $\alpha$ but for condition (ii) to be satisfied, $\alpha$ should be greater than $-1$.
In figure \ref{f1.1}, we
have plotted the mass function $M (x)$ given in (\ref{e1.11}), potential $V(x)$
given in (\ref{e1.12}) and square of first two bound state wave functions.
\FloatBarrier

\subsection{{Hamiltonian associated with exceptional $X_1$ Jacobi polynomial}}
\begin{figure}[]
\centering
\includegraphics[height=6 cm, width=8.25 cm]{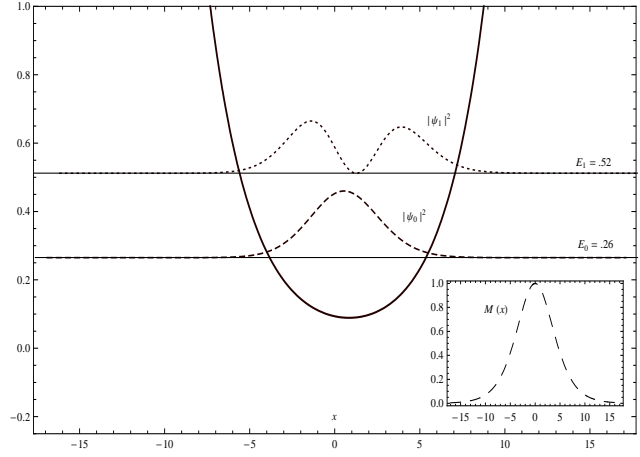}\caption[Plot of the potential associated with Jacobi type $X_1$ EOP]{ Plot of the
potential $V$ (solid line) given in equation (\ref{e1.15}),
square of its first two bound state wave functions $|\psi_o(x)|^2$
(dashed line) and $|\psi_1(x)|^2$ (dotted line), for the mass
function $M(x)$ (long dashed line) given in equation (\ref{e1.13}).
We have considered here $a=.2, \alpha=2, \beta=2.5.$}\label{f1.2}
\end{figure}

 \noindent Next we choose $F_n(g)\propto
\widehat{P}_n^{(\alpha,\beta)}(x)$, where
$\widehat{P}_n^{(\alpha,\beta)}$, $n=1,2,3,...\cdots$,
$\alpha,\beta>-1,~\alpha\neq\beta$ is the Jacobi type $X_1$
exceptional orthogonal polynomial, for which we have \cite{UK09}
\begin{equation}
\begin{array}{ll}
\displaystyle Q(g)=-\frac{(\alpha+\beta+2)g-(\beta-\alpha)}{1-g^2}-\frac{2(\beta-\alpha)}{(\beta-\alpha)g-(\beta+\alpha)},\\
\displaystyle R(g)=-\frac{(\beta-\alpha)g-(n-1)(n+\alpha+\beta)}{1-g^2}-\frac{(\beta-\alpha)^2}{(\beta-\alpha)g-(\beta+\alpha)}.
\end{array}
\end{equation}
For these $Q(g)$ and $R(g)$ equation (\ref{e1.8}) becomes
\begin{equation}
E-V = \frac{g'}{\lambda}\left[\frac{A_1g+A_2}{1-g^2}
+\frac{A_3g+A_4}{(1-g^2)^2}+\frac{A_5}{(\beta-\alpha)g-(\beta+\alpha)
}+\frac{A_6}{[(\beta-\alpha)g-(\beta+\alpha)]^2}\right]\label{e1.42}
\end{equation}
where
\begin{equation*}
\begin{array}{ll}
\displaystyle A_1=\frac{\beta^2-\alpha^2}{2\alpha \beta}~,~~~~A_2=n^2+(\beta+\alpha-1)n+
\frac{1}{4}\{(\beta+\alpha)^2-2(\beta+\alpha)-4\}+\frac{\beta^2+\alpha^2}{2\alpha\beta}\\
A_3=\frac{\beta^2-\alpha^2}{2}~,~~~~A_4=-\frac{\beta^2+\alpha^2-2}{2}~,
\displaystyle ~~~~~A_5=\frac{(\beta+\alpha)(\beta-\alpha)^2}{2\alpha \beta}
~,~~~A_6=-2(\beta-\alpha)^2.
\end{array}
\end{equation*}
A constant term on the
right-hand side of (\ref{e1.42}) can be generated by making $\frac{g'}{\lambda(1-g^2)}=$ constant (say $C_1>0$). Consequently we have
\begin{equation}
g(x)=\tanh(ax)~~,~~~~~~~~~~M(x)=\sech^2(ax)~,~~~~~~~-\infty<x<\infty.\label{e1.13}
\end{equation}
For this choice of $g(x)$  and mass function we obtain the potential
\begin{equation}
V(x) = \left[\frac{a^2}{4}\left((\alpha^2-1)e^{2ax}+
(\beta^2-1)e^{-2ax}\right)\right]+\frac{a^2}{4}\left(\frac{4(\alpha-\beta)(\alpha-3\beta)}{\alpha(\beta+\alpha
e^{2ax})}-\frac{8\beta(\alpha-\beta)^2}{\alpha(\beta+\alpha
e^{2ax})^2}\right)+\hat{V_0},\label{e1.15}
\end{equation}
corresponding bound state energy eigenvalues and eigenfunctions are given by
\begin{equation}
E_m=a^2\left(m+\frac{\alpha+\beta}{2}\right)\left(m+\frac{\alpha+\beta+2}{2}\right)
+a^2\left(\frac{\beta}{\alpha}-\frac{\alpha^2+\beta^2-2}{4}\right)+\hat{V}_0
\label{e1.16}
\end{equation}
and
{\small \begin{equation}
\begin{array}{ll}
\displaystyle \psi_m(x)=\left[\frac{a(\alpha-\beta)^2~ m!
(2m+\alpha+\beta+1) \Gamma(m+\alpha+\beta+1)}{2^{\alpha+\beta-1}
(m+\alpha+1)(m+\beta+1)\Gamma(m+\alpha)\Gamma(m+\beta)}\right]^{1/2}~\\\\

\displaystyle ~~~~~~~~~~ \times \frac{(1-\tanh(ax))^{\frac{\alpha+1}{2}}
(1+\tanh(ax))^{\frac{\beta+1}{2}}}{\alpha+\beta+(\alpha-\beta) \tanh(ax)}~~
\widehat{P}_{m+1}^{(\alpha,\beta)}(\tanh(ax)) ~~~~~m=0,1,2,...\label{e1.17}
\end{array}
\end{equation}}
respectively, where $\hat{V}_0$ and $C_1\lambda=a>0$ are constants. In expressions (\ref{e1.16}) and (\ref{e1.17}) we have reset the quantum number $n=m+1$.
An additional restriction $\alpha, \beta > -1/2$ is to be imposed to satisfy
condition (ii) stated before whereas the square integrability condition does not require any extra restriction on the parameters. In
figure \ref{f1.2} we have plotted the mass function given in (\ref{e1.13}) and potential $V(x)$ given in (\ref{e1.15}) and square of its first two bound state
wave functions.

\FloatBarrier
It should be mention here that the new potentials (\ref{e1.12}) and (\ref{e1.15}) are isospectral to the potentials
 previously obtained in \cite{BQ05a} whose bound state wave functions are given in terms of classical Laguerre and Jacobi
orthogonal polynomials respectively. More interestingly in both cases the expression within the square bracket coincides
 with the potentials associated with classical Laguerre and Jacobi orthogonal polynomials. So the new potentials obtained here can be considered as rationally
extended version of the standard potentials, of PDM Hamiltonians, which are associated with classical Laguerre or Jacobi polynomials.

\subsection[Supersymmetry and Hamiltonians associated with exceptional orthogonal polynomials]{Supersymmetry and Hamiltonians associated with EOP}
\noindent Here we combine the results of Point canonical transformation with supersymmetric quantum mechanics (SUSY QM) method described in section \ref{s2.1.1}.
The superpotentials $W_0(x)$, corresponding to the potentials obtained in (\ref{e1.12}) and (\ref{e1.15}), are determined by the equation (\ref{e0.6})
and the ground state wave functions of both the potentials, as
\begin{equation}
 W_0(x)=\frac{b}{2}\left[(\alpha+1)e^{\frac{bx}{2}}-e^{-\frac{bx}{2}}\right]
 -\frac{b~e^{\frac{3}{2}bx}}{\alpha(\alpha+1)e^{2bx}+(2\alpha+1)e^{bx}+1}\label{e1.46}
 \end{equation}
 and
 \begin{equation}
\begin{array}{ll}
\displaystyle
 W_0(x)=\frac{a}{2}\left[(\alpha-\beta)\cosh(ax)+(\alpha+\beta+2) \sinh(ax)\right]\\
 \displaystyle ~~~~~~~~~~~~+\frac{2a(\alpha-\beta)}{\left[(\alpha+\beta)\cosh(ax)
 +(\alpha-\beta)\sinh(ax)\right]\left[(\alpha+\beta+2)\cosh(ax)+(\alpha-\beta)\sinh(ax)\right]}
 \end{array}\label{e1.47}
 \end{equation}
 respectively. Using equations (\ref{e0.1}) and (\ref{e1.46}) one obtains
 \begin{equation}
 V_0^-(x) = \frac{b^2}{4}\left[(\alpha^2-1)e^{bx}+e^{-bx}+\frac{4}{\alpha(1+\alpha e^{bx})} +
 \frac{8e^{bx}}{(1+\alpha e^{bx})^2}\right]-b^2\left(\frac{\alpha+1}{2}+\frac{1}{\alpha}\right) \label{e1.52}
\end{equation}
and
\begin{equation}
\displaystyle V_0^+=
\frac{b^2}{4}\left[\alpha(\alpha+2)e^{bx}+e^{-bx}+\frac{4}{(\alpha+1)(1+(\alpha+1)
e^{bx})}+\frac{8e^{bx}}{(1+(\alpha+1)
e^{bx})^2}\right]-\frac{b^2}{2}\left(\frac{\alpha^2+\alpha+2}{\alpha+1}\right) .
  \label{e1.48}
 \end{equation}
Also using equations (\ref{e0.1}) and (\ref{e1.47}) we have
\begin{equation}\begin{array}{ll}
\displaystyle
 V_0^- = \frac{a^2}{4}\left[(\alpha^2-1)e^{2ax}+
(\beta^2-1)e^{-2ax}+\frac{4(\alpha-\beta)(\alpha-3\beta)}{\alpha(\beta+\alpha
e^{2ax})}-\frac{8\beta(\alpha-\beta)^2}{\alpha(\beta+\alpha
e^{2ax})^2}\right]\\ \\ \displaystyle ~~~~~~~~~~~~~~~~ -\frac{a^2}{4}\left(2\alpha
\beta+2\alpha+2\beta+2+\frac{4\beta}{\alpha}\right)\label{e1.53}
\end{array}
\end{equation}
\begin{equation}\begin{array}{ll}
 \displaystyle
V_0^+ = \frac{a^2}{4}\left[(\alpha(\alpha+2))e^{2ax}+
(\beta(\beta+2))e^{-2ax}+\frac{4(\alpha-\beta)(\alpha-3\beta-2)}{(\alpha+1)\left((\alpha+1)
e^{2ax}+\beta+1\right)}\right]\\ \\
\displaystyle ~~~~~~~~~~~~~~~-\frac{a^2}{4}\left(\frac{8(\beta+1)(\alpha-\beta)^2}{(\alpha+1)\left((\alpha+1)
e^{2ax}+\beta+1\right)^2}+2\alpha
\beta+\frac{4(\beta+1)}{\alpha+1}\right).
\end{array}\label{e1.49}
\end{equation}
The potentials $V_0^-$ obtained in (\ref{e1.52}) and (\ref{e1.53})
are same with the potentials given in (\ref{e1.12}) and (\ref{e1.15}) for
$\bar{V_0}=-b^2(\frac{\alpha+1}{2}+\frac{1}{\alpha})$ and
$\widehat{V}_0=-\frac{a^2}{4}\left(2\alpha
\beta+2\alpha+2\beta+2+\frac{4\beta}{\alpha}\right)$ respectively. From the equations (\ref{e1.52}) and (\ref{e1.48}), we observe that the
the potential $V_0^-$ and it's supersymmetric partner potentials
$V_0^+$ satisfy the following relation
\begin{equation}
V_0^+(x,\alpha)=V_0^- (x,\alpha+1)+b^2.\label{e1.54}
\end{equation}
Also from equation (\ref{e1.53}) and (\ref{e1.49}) we have the relation
\begin{equation}V_0^+ (x,\alpha,\beta) = V_0^-(x,\alpha+1,\beta+1) + a^2(\alpha+\beta+2).\label{e1.55}
\end{equation}
Hence, we conclude that the new potentials obtained here have translational shape invariant symmetry.

\section[Intertwining operator approach to spectral design of a position-dependent mass Hamiltonian]
{Intertwining operator approach to spectral design of a PDM Hamiltonian}\label{s2.3}

\noindent
In this section, systematic procedures of first and second order intertwining
operator approach to generate the isospectral Hamiltonians in PDM scenario have been studied with the following spectral modifications: (i) to add new bound state(s) below the ground state
(ii) to remove lowest bound state(s) and (iii) to leave the spectrum unaffected. Second order intertwiner for PDM Hamiltonian has been constructed earlier \cite{SH08}
by iterating two first order intertwiner. But this approach failed to perform some of the above mentioned spectral modifications.
This shortcoming can be removed by
considering directly a second order linear differential operator as the intertwiner. In the following we have first discussed the first order intertwining
which uses a first order differential operator as intertwiner.

\subsection{First-order intertwining}\label{s2.3.1}
\noindent We consider the following two one-dimensional PDM Schr\"{o}dinger
Hamiltonians with same
spectrum but with different potential
\begin{equation}
{H}_0\psi(x)=E^{(0)} \psi(x)~,~~{H}_0 \equiv -\left[\frac{d}{dx}\left(\frac{1}{M(x)}\right)\frac{d}{dx}\right]+V_0(x)\label{e2.15}
\end{equation}
and
\begin{equation}
    {{H}}_1 \phi(x)=E^{(1)} \phi(x)~,~~{{H}}_{1}
     \equiv -\left[\frac{d}{dx}\left(\frac{1}{M(x)}\right)\frac{d}{dx}\right]+ V_1(x).\label{e2.16}
\end{equation}
The solution of one of the above equation (say (\ref{e2.15})) is known. We connect above two Hamiltonians (\ref{e2.15}) and (\ref{e2.16}) by means
of the intertwining relation
\begin{equation}
{\cal{L}} {{H}}_0 = {{H}}_1 {\cal{L}}, ~~~~~~~~~\phi(x) = {\cal{L}} \psi(x).\label{e2.60}
\end{equation}
Without loss of generality let us consider the first order
intertwining operator as
\begin{equation}
\mathcal{L}=\frac{1}{\sqrt{M(x)}}\frac{d}{dx}+A(x)\label{e2.17}
\end{equation}
 Now using the intertwining relation (\ref{e2.60}) and equating the coefficients of
like order of derivatives we obtain
\begin{equation}
V_1 = V_0 +\frac{2A'}{\sqrt{M}}-\frac{3M'^2}{4M^3}+\frac{M''}{2M^2}\label{e2.29}
\end{equation}
and
\begin{equation}
A(V_1-V_0)=-\frac{A'M'}{M^2}+\frac{V_0'}{\sqrt{M}}+\frac{A''}{M}\label{e2.30}
\end{equation}
where `prime' denotes differentiation with respect to $x$. Using
(\ref{e2.29}), equation (\ref{e2.30}) reduces to
\begin{equation}
\frac{A''}{\sqrt{M}}-\frac{A'M'}{M^{\frac{3}{2}}}-\frac{AM''}{2M^{\frac{3}{2}}}+\frac{3AM'^{2}}{4M^{\frac{5}{2}}}
-2A A'+V_0'=0\label{e2.36}
\end{equation}
Integrating equation (\ref{e2.36}) we get
\begin{equation}
\frac{A'}{\sqrt{M}}-\frac{A M'}{2M^{\frac{3}{2}}}-A^2+ V_0 =\mu\label{e2.31}
\end{equation}
where $\mu$ is a constant of integration. After substituting
\begin{equation}
A(x)=-\frac{K}{\sqrt{M}},\label{e2.57}
\end{equation}
 $K=K(x)$ being an auxiliary function, in equation (\ref{e2.31})
we obtain the following Riccati  differential equation
\begin{equation}
-\frac{K'}{M}+\frac{KM'}{M^2}-\frac{K^2}{M} + V_0=\mu\label{e2.58}
\end{equation}
The equation (\ref{e2.58}) can be linearized by the assumption
$K(x)=\frac{\mathcal{U'}(x)}{\mathcal{U}(x)}$. Substituting this
value of $K(x)$ in equations (\ref{e2.57}) and (\ref{e2.58}) we get
\begin{equation}
A(x)=-\frac{\mathcal{U'}}{\sqrt{M}\mathcal{U}}\label{e2.59}
\end{equation}
and
\begin{equation}
-\frac{1}{M}\mathcal{U}''-\left(\frac{1}{M}\right)'\mathcal{U}'+V_0 ~\mathcal{U}=\mu ~
\mathcal{U}\label{e2.18}
\end{equation}
respectively. The equation (\ref{e2.18}) is similar to equation
(\ref{e2.15}) with $E^{(0)} =\mu$. $\mu$ is the factorization
energy and $\mathcal{U}(x)$ is the seed solution. It should be
noted here that $\mathcal{U}(x)$ need not be normalizable solution
of (\ref{e2.18}). However, for an acceptable $V_1(x)$ (without
singularity), $\mathcal{U}(x)$ must not have any node on the real
line. For this, we shall restrict ourself on $\mu \leq E^{(0)}_0$ throughout, $E^{(0)}_0$ being the ground state energy eigenvalues
of the Hamiltonian $H_0$. For a given potential $V_0(x)$ the first order isospectral partner potential $V_1(x)$ and the solution of
the corresponding Hamiltonian ${H}_1$ can be obtained using equations (\ref{e2.60}). The solution of ${H}_1$ at the factorization energy
$\mu$ can not be obtained with the help of (\ref{e2.60}) because $\mathcal{L} \mathcal{U} = 0$. It can be obtained by directly solving
$\mathcal{L}^\dag \phi_\mu = 0$ and is given by \cite{SH08}
\begin{equation}
\phi_\mu \sim \frac{\sqrt{M}(x)}{\mathcal{U}(x)}.\label{e2.72}
\end{equation}
At this point following conditions of spectral modification are worth mentioning.
\begin{itemize}
\item[$\circ$]{
{\underline{Condition for deletion of ground state} :} For a suitably chosen mass function $M(x)$, if
$\mathcal{U}(x)$ corresponds to the ground state wave function of
${{H}}_0$ then the potential $V_1(x)$ has no new
singularity, except the singularity due to $V_0(x)$ and the solution $\phi_{\mu}(x)$
defined in (\ref{e2.72}) is not normalized so that $\mu$ does not
belong to the bound state spectrum of ${{H}}_1$.}

\item[$\circ$]{{\underline{Condition for isospectral transformation} :} If $\mathcal{U}(x)$ is nodeless
and unbounded at one end point of the domain of definition of $V_0(x)$ then $\phi_{\mu}(x)$ is not normalized.
So we can generate ${{H}}_1$ strictly isospectral to ${{H}}_0$.}

\item[$\circ$]{{\underline{Condition for creation of bound state below the ground state} :} If $\mathcal{U}(x)$ is nodeless and unbounded at the
both end points then $\phi_{\mu}(x)$ defined in (\ref{e2.72})
is normalizable, so that $\mu$ can be included
 in the bound state spectrum of ${{H}}_0$ to generate ${H}_1$. In this case maximal set
of bound state wave functions of ${{H}}_1$ are given by
$\{\phi_{\mu}, {\mathcal{L}}_1\psi\}$.}
\end{itemize}

\subsection{Examples of First order Intertwining}
\noindent It may be emphasized that the results mentioned in section \ref{s2.3.1} are
most general and valid for any potential $V_0(x)$. However to
illustrate the above procedure with the help of an example we
shall need non-normalizable solutions of equation (\ref{e2.18}) corresponding
to a particular mass function $M(x)$. The general solution of this equation can be obtained using PCT approach mentioned in
section \ref{s2.2.1} by assuming $F(g) \sim ~ _2F_1(a,b,c,g)$, where $ ~_2F_1(a,b,c,g)$ is the Hypergeometric function. In such case, the potential,
mass function and the general non-normalized solutions of the PDMSE (\ref{e2.18}) are given by (detail calculation has been given in the appendix
of ref.\cite{MRR10}, we
have considered $p=\lambda=1$ for simplicity)
\begin{equation}
V_0(x)= \frac{[(a+b-c)^2-1]}{4} e^{x} + \frac{c(c-2)}{4} e^{-
x}\label{e2.10}
\end{equation}
 \begin{equation}
 M(x)=\frac{1}{4} \sech^2\left(\frac{ 1}{2}x\right)\label{e2.92}
 \end{equation}
 \begin{equation}\begin{array}{ll}
 \displaystyle \mathcal{U}(x)=\alpha\frac{e^{\frac{c}{2}x}}{(1+e^{
x})^{\frac{a+b+1}{2}}}~ _2F_1\left(a,b,c,\frac{e^{ x}}{1+e^{
x}}\right)\\
~~~~~~~~~~ \displaystyle +\beta\frac{e^{\left(1-\frac{c}{2}\right) x}}{(1+e^{
x})^{\frac{a+b-2c+3}{2}}}~
 _2F_1\left(a-c+1,b-c+1,2-c,\frac{e^{ x}}{1+e^{
 x}}\right)
 \end{array}\label{e2.13}
 \end{equation}
\begin{equation}
 \mu=-ab+\frac{(a+b+1)c}{2}-\frac{c^2}{2}\label{e2.9}
\end{equation}
respectively. For the potential $V_0(x)$, the bound state solutions and eigenstates of the
equation (\ref{e2.15}) are obtained as \cite{MRR10}
\begin{equation}
 \psi_n(x)=\left(\frac{ (2n+\sigma+\delta+1) n!~\Gamma(n+\sigma+\delta+1)}
 {\Gamma(n+\sigma+1)\Gamma(n+\delta+1)}\right)^{1/2} \frac{e^{\frac{(\sigma+1)}{2}}x}{\left(1+e^{
 x}\right)^{\frac{\sigma+\delta+2}{2}}} {P}_n^{(\sigma,\delta)}
 \left(\frac{1-e^{x}}{1+e^{x}}\right)\label{e2.14}
 \end{equation}
 and
 \begin{equation}
 E^{(0)}_n=n^2+n(\sigma+\delta+1)+\frac{(\sigma+1)(\delta+1)}{2}~,~~~n=0,1,2,...\label{e2.26}
 \end{equation}
 respectively, where $b=1-a+\sigma+\delta,~c=1+\sigma$ with $c>\frac{1}{2}~\mbox{and}~a+b-c+\frac{1}{2}>0$ and ${P}_n^(\sigma,\delta)$ is the Jacobi polynomial.
 The asymptotic behavior of the solution $\mathcal
{U}(x)$ given in (\ref{e2.13}), at both end points $\pm \infty$ are
given by \cite{AS65}
\begin{equation}
 \mathcal{U}(x)\sim (A_1\alpha+B_1\beta) e^{-\frac{a+b-c+1}{2}x}+(A_2\alpha+B_2
 \beta)e^{-\frac{c-a-b+1}{2}x}~~~\mbox{as}
 ~x\rightarrow\infty\label{e2.22}
 \end{equation}
 where $$A_1=\frac{\Gamma(c)\Gamma(c-a-b)}{\Gamma(c-b)\Gamma(c-a)}~,
 ~~B_1=\frac{\Gamma(2-c)\Gamma(c-a-b)}{\Gamma(1-a)\Gamma(1-b)}$$
 $$A_2=\frac{\Gamma(c)\Gamma(a+b-c)}{\Gamma(a)\Gamma(b)}~,
 ~~B_2=\frac{\Gamma(2-c)\Gamma(a+b-c)}{\Gamma(a-c+1)\Gamma(b-c+1)}$$
 and
 \begin{equation}
\mathcal{U}(x)\sim \alpha e^{\frac{c}{2}x}+ \beta~
e^{(1-\frac{c}{2})x}~~~\mbox{as}
 ~x\rightarrow -\infty\label{e2.23}
 \end{equation}
 From these asymptotic behaviors it is clear that $\mathcal {U}(x)$ will unbounded
 at $x \rightarrow \infty$ if $|a+b-c|>1$ and it is unbounded at $x \rightarrow -\infty$ if $c<0 ~\rm or~ c>2$.
Therefore $\mathcal{U}(x)$ will nodeless at the finite part of the
$x$ axis if  $A_1\alpha+B_1\beta$, $A_2\alpha+B_2
 \beta$,$\alpha$ and $\beta$ are all positive and $|a+b-c|>1$, $c<0 ~\rm or~
 c>2$.

 Now we are going to generate isospectral potentials of the
 potential (\ref{e2.10}) with various possible spectral
 modifications.\\

\noindent {\underline{Deletion of the initial ground state} :}
In this case the  factorization energy $\mu$ is equal to the
ground state energy $E_0$ giving $ a~ \rm {and/or}~ b = 0$ and
$\mathcal {U}(x)$ becomes the ground state wave function
$\psi_0(x)$ which is obtained from (\ref{e2.14}) as
 \begin{equation}
\mathcal{U}(x)=\psi_0(x)\propto
\frac{e^{\frac{c}{2}x}}{(1+e^x)^{\frac{a+b+1}{2}}}\label{e2.93}
\end{equation}
The isospectral partner of $V_0(x)$ given in equation (\ref{e1.10}),
is obtained using equation (\ref{e2.29}), (\ref{e2.10}), (\ref{e2.92})
and (\ref{e2.93}) and is given by
\begin{equation}
V_1(x)=\frac{c^2-1}{4}e^{-x}+\frac{(a+b-c)(2+a+b-c)}{4}e^x+\frac{a+b}{2}
\label{e2.28}
\end{equation}
The above potential (\ref{e2.28}) can also be obtained from the
initial potential (\ref{e2.10}) by making the changes $a\rightarrow
a+1, b\rightarrow b+1, c\rightarrow c+1$. This in turn imply that the potentials $V_0(a)$ and $V_1(x)$ are shape invariant.
\begin{figure}[]
\epsfxsize=3 in \epsfysize=2 in \centerline{\epsfbox{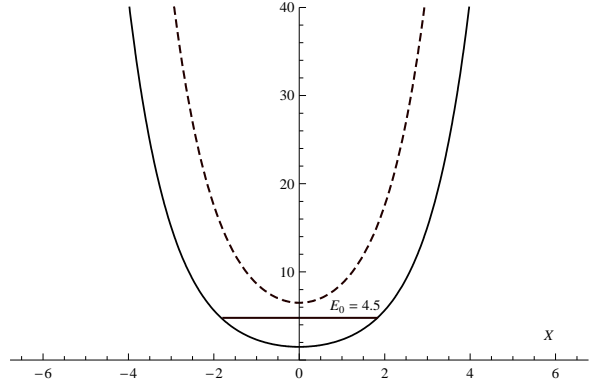}}
\label*{}\caption{Plot of the potential $V_0(x)$ (solid line) given
in (\ref{e2.10}) and its first order isospectral partners
$V_1(x)$ (dashed line) given in (\ref{e2.28}) by deleting the
ground state $E_0^{(0)}=4.5$, we have considered here $a=5, b=0, c=3,
\alpha=1,\beta=0.$ }\label{f2.1}
\end{figure}
\FloatBarrier
Since $\psi_0(x)$ is a bounded
solution, $\phi_{\mu}(x)=\frac{\sqrt{M}}{\psi_0}$ is
unbounded at $x\rightarrow\pm \infty$ so that $\phi_{\mu}$ can not be the ground state wave function of $V_1$. Thus we delete the ground state of
$V_0$ to obtain $V_1$. Therefore the eigenvalues of $H_1$ are given by
 \begin{equation}
 E^{(0)}_n=E^{(1)}_{n+1}=(n+1)^2+(n+1)(a+b)
 +\frac{c(a+b-c+1)}{2}~,~~~n=0,1,2...
 \end{equation}
 Corresponding bound
 state wave functions of $V_1(x)$ are obtained using equations (\ref{e2.60}) and (\ref{e2.14}) as
\begin{equation}
\phi_n(x)\propto
\frac{e^{(1+\frac{c}{2})x}}{(1+e^x)^{(\frac{a+b+5}{2})} \sech(\frac{x}{2})}~
{P}_{n}^{\left(c,a+b-c+1\right)}\left(-\tanh\frac{x}{2}\right)~,~~
n=0,1,2...
\end{equation}
We have plotted the potentials $V_0(x)$ given in (\ref{e2.10}) and
$V_1(x)$ given in (\ref{e2.28}) for
$a=5,b=0,c=3,\alpha=1,\beta=0$ in figure \ref{f2.1}.\\

\noindent {\underline{Strictly isospectral potentials} :} The strictly (strict in
the sense that the spectrum of the initial potential and its
isospectral potential are exactly the same) isospectral potentials
can be generated with the help of those seed solutions which
vanish at one of the ends of the $x$-domain.  Now for $\beta=0~
\mbox{and}~ \alpha >0$, it is seen from (\ref{e2.22}) that
$\mathcal{U}$ is unbounded at $x\rightarrow\infty$ if $|a+b-c|>1$.
But the solution (\ref{e2.14}) become unbounded for $a+b-c<-1$. So
we must take $a+b-c>1.$ On the other hand
 from (\ref{e2.23}) it is observed that  $\mathcal{U}(x)\rightarrow 0$ at $x\rightarrow-\infty$ if
$c<2$ or $c>0$ but $\psi_n(x)$ are not normalizable for the values
of $c<2$ so we must take $c>0$. So ~$\mathcal{U}(x)$ vanishes at
$x\rightarrow-\infty$ and unbounded at $x\rightarrow\infty$~if $a+b-c>1 ~\mbox{and}~ c>0.$ In this case the spectrum of
the isospectral potential as well as original potential are
identical i.e. $E^{(0)}_n=E^{(1)}_n~,n=0,1,2...$ Considering the seed
solution as  $$\mathcal{U}(x)=\frac{e^{\frac{c}{2}x}}{(1+e^{
x})^{\frac{a+b+1}{2}}}~ _2F_1\left(a,b,c,\frac{e^{ x}}{1+e^{
x}}\right),~~~~~a+b-c>1, c>0$$ we have calculated the explicit
form of the partner potential using equations (\ref{e1.10}), (\ref{e2.29}), (\ref{e2.10}), and (\ref{e2.92})
  {\small \begin{equation}\begin{array}{llll}
  \displaystyle V_1(x)=\frac{1}{8}[2c(c-2) e^{-x}+2((a+b-c)^2-1)e^{x}\\\\
  ~~~~-
  \frac{\sech^2(\frac{x}{2})}{c^2(1+c)~ _2F_1\left(a,b,c,\frac{e^x}{1+e^x}\right)}
  \{-4a^2b^2(1+c) \left(_2F_1\left(1+a,1+b,1+c,\frac{e^x}{1+e^x}\right)\right)^2 + 4abc~ _2F_1\left(a,b,c,\frac{e^x}{1+e^x}\right)\\\\
  ~~\left((a+1)(b+1)
   \left(_2F_1\left(2+a,2+b,2+c,\frac{e^x}{1+e^x}\right)\right)^2
  -(1+c)\sinh x \left(_2F_1\left(a+1,b+1,c+1,\frac{e^x}{1+e^x}\right)\right)^2\right)\\\\
  ~~~~~~~ -8c^2(1+c)\cosh^3 x
  \left(_2F_1\left(a,b,c,\frac{e^x}{1+e^x}\right)\right)^2
  \left((a+b)\cosh(\frac{x}{2})+(1+a+b-2c) \sinh(\frac{x}{2})\right)\}
  \end{array}
  \end{equation}}
 \begin{figure}[]
\epsfxsize=2.75 in \epsfysize=2 in \centerline{\epsfbox{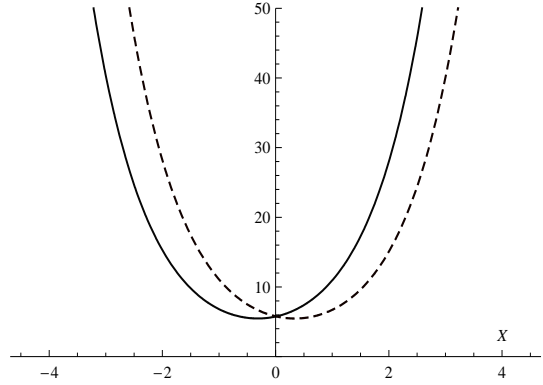}}
\label*{}\caption{Plot of the potential $V_0(x)$ (solid line) and
its first order isospectral partner (dashed line) given in
(\ref{e2.74}). }\label{f2.2}
\end{figure}
\FloatBarrier
 In particular for $a=3, b=5, c=4, \alpha=1, \beta=0$ we have
 \begin{equation}\begin{array}{ll}
 \displaystyle V_0(x)=\frac{1}{4}(23 \cosh x+7 \sinh x)\\
 V_1(x)=\frac{15e^{-x}}{4}+2e^x+\frac{4+e^x-3e^{2x}}{(4+3e^x)^2}.
 \end{array}\label{e2.74}
 \end{equation}
 These potentials are plotted in figure \ref{f2.2}.
  In this case eigenfunctions and eigenvalues of the above partner potential $V_1(x)$ are given by
 \begin{equation}\begin{array}{ll}
\displaystyle \phi_n(x)\propto
\frac{e^{3x}\left[(2+e^x)(n+7) {P}_{n-1}^{(4,4)}\left(-\tanh
\left(\frac{x}{2}\right)\right)+(1+e^x)(5+3e^x) {P}_n^{(3,3)}(-\tanh
(\frac{x}{2}))\right]}{(1+e^x)^6(2+e^x) \sech
\left(\frac{x}{2}\right)}\\
\displaystyle E^{(1)}_n=n^2+8n+10~,~~~~~n=0,1,2...
\end{array}
 \end{equation}
respectively.\\

\begin{figure}[]
\epsfxsize=4.25 in \epsfysize=2.5 in \centerline{\epsfbox{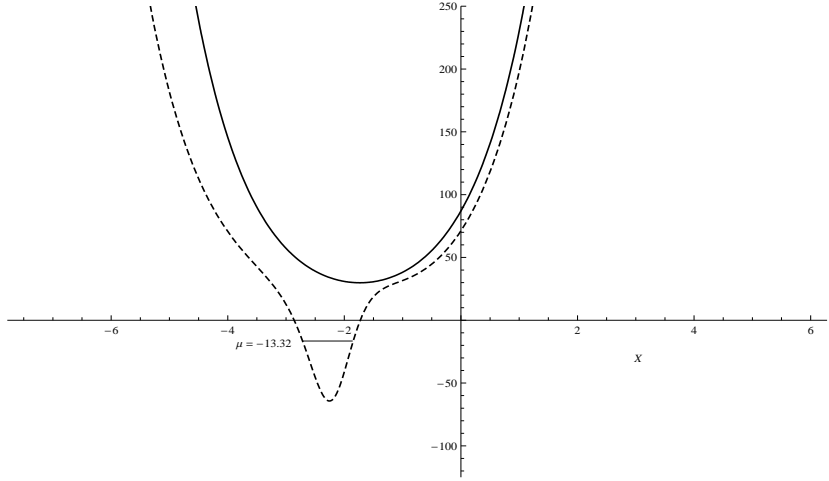}}
\label*{}\caption{Plot of the potential $V_0(x)$ (solid line) given
in (\ref{e2.10}) and its first order isospectral partner $V_1$
(dashed line) by inserting the state $\mu=-13.32$. We have
considered here $a=2.8,b=20, c=4.4, \alpha=\beta=1$.}\label{f2.3}
\end{figure}

\noindent{\underline{Creation of a new ground state} :} In this case we
consider $\mu<E_0$. The new state can be created below the ground
state of the initial potential with the help of those seed
solutions which satisfies the following two conditions: (i) it
should be nodeless throughout the $x$-domain and (ii) it should be
unbounded at both the end points of the domain of definition of
the given potential $V_0(x)$. From the asymptotic behaviors of the
seed solution $\mathcal{U}$, given in equations (\ref{e2.22}) and
(\ref{e2.23}) we have, for $|a+b-c|>1$ together with either
$c<0~\mbox{or}~c>2$, the above two conditions are satisfied. But
to get $\psi_n(x)$ as physically acceptable, we shall take
$c>2~\mbox{and}~a+b-c>1.$ In this case the spectrum of the partner
potential is $\{\mu, E_n^{(0)},n=0,1,2...\}$, $E_n^{(0)}$ being the energy
eigenvalues of the original potential $V_0(x)$ given in (\ref{e2.10}).
Corresponding bound state wave functions are
$\{\phi_{\mu}(x), \phi_n(x),n=0,1,2,...\}$. For $a+b-c>1, c>2$ and the seed
solution $\mathcal{U}$ given in (\ref{e2.13}), the general
expression of the isospectral potential becomes too involved so
instead of giving the explicit expression of the partner potential
we have plotted in figure \ref{f2.3} the original potential $V_0(x)$ given in
(\ref{e2.10}) and its partner potential $V_1(x)$ (which is
obtained using (\ref{e2.29})) considering the particular values $a=
2.8, b=20, c=4.4~ \mbox{and} ~\alpha=\beta=1$. In this case the
energy eigenvalues of $V_1(x)$ are given by
\begin{equation}
E^{(1)}_n=\{-13.32,E_n^{(0)},n=0,1,2...\}=\{-13.32,n^2+22.8n+42.68,n=0,1,2...\}
\end{equation}
\FloatBarrier

\subsection{Second-order intertwining}
\noindent Here we have constructed the second intertwiner directly by taking the following second-order differential operator
\begin{equation}
\mathcal{L}_2 \equiv \frac{1}{M}\frac{d^2}{dx^2}+\eta(x)\frac{d}{dx}+\gamma(x)\label{e2.37}
\end{equation}
where $\eta(x)$ and $\gamma(x)$ are unknown functions of $x$. Substitution of this intertwiner in the intertwining relation
(\ref{e2.60}) and comparing the coefficients of like order derivatives leads to a set of equations involving $V_0(x), V_2(x)\footnote{In second order intertwining we assume
$H_1$ to be $H_2$ with $V_1(x)$ as $V_2(x)$ and $E_n^{(1)}$ as $E_n^{(2)}$.}, \eta(x), \gamma(x)$ and
their derivatives
\begin{equation} V_2 =
V_0 +2\eta'+\frac{M'}{M}\eta-\frac{3M'^2}{M^3}+\frac{2M''}{M^2}\label{e2.38}
\end{equation}
\begin{equation}
(V_2-V_0)\eta =
\frac{2V_0'}{M}+\frac{2\gamma~'}{M}+\frac{\eta''}{M}-\frac{\eta'M'}{M^2}
+\frac{M''\eta}{M^2}-\frac{2M'^2\eta}{M^3}+\frac{6M'^3}{M^5}-\frac{6M'M''}{M^4}+\frac{M'''}{M^3}\label{e2.39}
\end{equation}
\begin{equation}
(V_2-V_0)\gamma =
\frac{V_0''}{M}+V_0'\eta+\frac{\gamma~''}{M}-\frac{M'\gamma~'}{M^2}\label{e2.40}
\end{equation}
Now using (\ref{e2.38}) the equations (\ref{e2.39}) and (\ref{e2.40})
reads
{\small \begin{equation}\displaystyle
 2\eta\eta'+\frac{M'\eta^2}{M}-\frac{3\eta
M'^2}{M^3}+\frac{2M''\eta}{M^2}-\frac{2\gamma~'}{M}+\frac{M'\eta'}{M^2}\\
\displaystyle +\frac{2M'^2\eta}{M^3}-\frac{\eta''}{M}
-\frac{M''\eta}{M^2}-\frac{2V_0'}{M}-\frac{6M'^3}{M^5}+\frac{6M'M''}{M^4}-\frac{M'''}{M^3}=0
 \label{e2.41}
\end{equation}}
and
\begin{equation}
\gamma\left(2\eta'+\frac{M'\eta}{M}-\frac{3M'^2}{M^3}+\frac{2M''}{M^2}\right)+\frac{M'\gamma~'}{M^2}-\frac{\gamma~''}{M}-\eta
V_0'-\frac{V_0''}{M}=0\label{e2.43}
\end{equation} respectively. Equation (\ref{e2.41}) can be integrated to obtain
\begin{equation}
\gamma=\frac{M\eta^2}{2}+\frac{M'\eta}{2M}-\frac{\eta'}{2}-V_0+\frac{M'^2}{M^3}-\frac{M''}
{2M^2}+C_1\label{e2.42}
\end{equation}
where $C_1$ is an arbitrary constant. Using (\ref{e2.42}) in
(\ref{e2.43}) we obtain
{\small \begin{equation}
\begin{array}{lll}
\displaystyle
\frac{\eta'''}{2M}+M\eta'\eta^2-\frac{\eta\eta'M'}{2M}-2\eta'^2-2\eta'V_0 +\frac{5\eta'M'^2}{M^3}-\frac{3\eta'M''}{M^2}
+\frac{\eta^3M'}{2}-\frac{\eta^2M'^2}{2M^2}-\frac{\eta V_0 M'}{M}-\frac{2\eta M'^3}{M^4}\\\\

\displaystyle +\frac{\eta^2M''}{2M}+\frac{5\eta
M'M''}{2M^3}-\frac{\eta''M'}{M^2}-\eta \eta''-\eta V'-\frac{\eta
M'''}{2M^2}+2C_1\eta'+\frac{C_1\eta
M'}{M}+\frac{3VM'^2}{M^3}-\frac{18M'^4}{M^6}\\\\

\displaystyle
+\frac{49M'^2M''}{2M^5}-\frac{3C_1M'^2}{M^3}-\frac{2V_0 M''}{M^2}-\frac{4M''^2}{M^4}+\frac{2C_1M''}{M^2}-\frac{V_0'M'}{M^2}
-\frac{9M'M'''}{2M^4}+\frac{M''''}{2M^3}=0\label{e2.44}
\end{array}
\end{equation}}
Multiplying  by $\left(\eta M+\frac{M'}{M}\right)$, above equation
(\ref{e2.44}) can be integrated to obtain
{\small \begin{equation}
\begin{array}{lll}
\displaystyle \frac{\eta
\eta''}{2}-\frac{\eta'^2}{4}-\eta'\eta^2M+\frac{\eta^4M^2}{4}-M\eta^2
V_0+C_1 M\eta^2+\frac{C_1M'^2}{M^3}+\frac{2C_1M'\eta}{M}-\frac{2M'V_0\eta}{M}-\frac{M'^2V_0}{M^3}\\\\

\displaystyle -\frac{M''\eta^2}{2M} + \frac{M'\eta^3}{2}
+\frac{5M'^3\eta}{m^4}-\frac{2M'\eta\eta'}{M}+\frac{5M'^2\eta^2}{4M^2}+\frac{M'\eta''}{2M^2}-\frac{M''\eta'}{2M^2}
+\frac{M'''\eta}{2M^2}-\frac{4M'M''\eta}{M^3}+\frac{3M'^4}{M^6}\\\\

\displaystyle -\frac{5M'^2M''}{2M^5} -\frac{M''^2}{4M^4}+\frac{M'M'''}
{2M^4}+C_2=0,\label{e2.45}
\end{array}
\end{equation}}
where $C_2$ is the constant of integration. For a given potential
$V_0(x)$, the new potential $V_2(x)$ and $\gamma(x)$ can be
obtained from (\ref{e2.38}) and (\ref{e2.42}) if the solution
$\eta(x)$ of (\ref{e2.45}) is known. To obtain $\eta(x)$ we take the
Ans\"{a}tz
\begin{equation}
\eta'=M\eta^2+2\left(\eta+\frac{M'}{M^2}\right)\tau+\frac{M'}{M}\eta+\frac{2M'^2}{M^3}-\frac{M''}{M^2}+\xi\label{e2.46}
\end{equation}
where $\xi$ is a constant to be determined and $\tau$ is a
function of $x$. Using above ans\"{a}tz in equation (\ref{e2.45}) we
obtain the following equation
{\small \begin{equation}\begin{array}{ll}\displaystyle
M\left(\frac{\tau'}{M}+\frac{\tau^2}{M}-\frac{M'\tau}{M^2}-V_0+C_1-\frac{\xi}{2}\right)\eta^2
+\frac{2M'}{M}\left(\frac{\tau'}{M}+\frac{\tau^2}{M}-\frac{M'\tau}{M^2}-V_0+C_1-\frac{\xi}{2}\right)\eta\\\\
\displaystyle
+\frac{M'^2}{M^3}\left(\frac{\tau'}{M}+\frac{\tau^2}{M}-\frac{M'\tau}{M^2}-V_0 +C_1-\frac{\xi}{2}\right)
+\left(C_2-\frac{\xi^2}{4}\right)=0\label{e2.47}
\end{array}
\end{equation}}
Since equation (\ref{e2.47}) is valid for arbitrary $\eta$, the
coefficients of each power of $\eta$ must vanish, which give
$\xi^2=4C_2$ and
\begin{equation}
\frac{\tau'}{M}+\frac{\tau^2}{M}-\frac{M'\tau}{M^2}-V_0+C_1-\frac{\xi}{2}=0\label{e2.48}
\end{equation}
Now defining $\mu=C_1-\frac{\xi}{2}$ , the above equation can be
written as
\begin{equation}
\frac{\tau'}{M}+\frac{\tau^2}{M}-\frac{M'\tau}{M^2}=V_0-\mu~,~~~~\mu=C_1-\frac{\xi}{2}=C_1\mp\sqrt{C_2}\label{e2.49}
\end{equation}
The equation (\ref{e2.49}) is a Riccati equation which can be
linearized by defining $\tau=\frac{\mathcal{U'}}{\mathcal{U}}$.
Making this change in equation (\ref{e2.49}) we obtain
\begin{equation}
-\frac{1}{M}\mathcal{U}''-\left(\frac{1}{M}\right)'\mathcal{U}'+V_0~\mathcal{U}=\mu~
\mathcal{U}\label{e2.54}
\end{equation}
Depending on whether $C_2$ is zero or not, $\xi$ vanishes or takes
two different values $\pm \sqrt{C_2}$. If $C_2 = 0$, we need to
solve one equation of the form (\ref{e2.49}) and then the equation
(\ref{e2.46}) for $\eta (x)$. If $C_2 \neq 0$, there will be two
different equations of type (\ref{e2.49}) for two factorization
energies $\mu_{1,2}=C_1\mp\sqrt{C_2}$. Once we solve them, it is
possible to construct algebraically a common solution $\eta (x)$
of the corresponding pair of equations (\ref{e2.46}). There is an
obvious difference between the (A) real case with $C_2 > 0$ and (B)
complex case $C_2 < 0$; thus there follows a natural scheme of
classification for the solutions $\eta (x)$ based on the sign of $C_2$.
Here, we shall not discuss the case $C_2=0.$\\

\noindent{\underline{(A) Real Case $(C_2>0)$} :} Here we have $\mu_{1,2}\in\mathbb{R}, \mu_1\neq\mu_2$. Let the
corresponding solutions of the  Riccati  equation (\ref{e2.49}) be
denoted by $\tau_{1,2}(x).$ Now the associated pair of equations
(\ref{e2.46}) become
\begin{equation}
\eta'=M\eta^2+2\left(\eta+\frac{M'}{M^2}\right)\tau_1+\frac{M'}{M}\eta+\frac{2M'^2}{M^3}-\frac{M''}{M^2}+\mu_2-\mu_1\label{e2.76}
\end{equation}
and
\begin{equation}
\eta'=M\eta^2+2\left(\eta+\frac{M'}{M^2}\right)\tau_2+\frac{M'}{M}\eta+\frac{2M'^2}{M^3}-\frac{M''}{M^2}+\mu_1-\mu_2\label{e2.77}
\end{equation}
respectively. Subtracting (\ref{e2.76}) from (\ref{e2.77}) and using
(\ref{e2.54}) we obtain $\eta(x)$ as
\begin{equation}
\eta(x)=\frac{\mu_1-\mu_2}{\tau_1-\tau_2}-\frac{M'}{M^2}=-\frac{W'(\mathcal{U}_1,\mathcal{U}_2)}{M
W(\mathcal{U}_1,\mathcal{U}_2)}\label{e2.55}
\end{equation}
where $\mathcal{U}_1 , \mathcal{U}_2$ are the seed solutions of
the equation (\ref{e2.54}) corresponding to the factorization energy
$\mu_1$ and $\mu_2$ respectively and
$W(\mathcal{U}_1,\mathcal{U}_2)=\mathcal{U}_1\mathcal{U}'_2-\mathcal{U}'_1\mathcal{U}_2$,
is the Wronskian of
$\mathcal{U}_1$ and $\mathcal{U}_2$. It is clear from (\ref{e2.38}) and (\ref{e2.55}) that  mass
function $M(x)$ is nonsingular and does not vanish at the finite
part of the $x$-domain, so that the new potential $V_2(x)$ has
no extra singularities (i.e. the number of singularities in $V_0$
and $V_2$ remains the same) if
$W(\mathcal{U}_1,\mathcal{U}_2)$ is nodeless there. The spectrum
of $H_2$ depends on whether or not its two
eigenfunctions $\phi_{\mu_{1,2}}$ which belongs as well to
the kernel of $\mathcal{L}^{\dagger}$ can be normalized \cite{FG04}, namely
\begin{equation}\mathcal{L}^{\dagger}{\phi}_{\mu_{j}}=0~~~~\mbox{and}~~~~
H_2\phi_{\mu_{j}}=\mu_{j}\phi_{\mu_{j}},~~~j=1,2\label{e2.56}
\end{equation}
where $\mathcal{L}^{\dagger}$ is the adjoint of $\mathcal{L}$ and
is given by
$$\mathcal{L}^{\dagger}=\frac{1}{M}\frac{d^2}{dx^2}-\left(\eta+\frac{2M'}{M^2}\right)\frac{d}{dx}+\left(\frac{2M'^2}{M^3}-\frac{M''}{M^2}
-\eta'+\gamma\right)$$ For $j=1$ the explicit expression of the
two equation mentioned in (\ref{e2.56}) are
\begin{equation}
\frac{1}{M}\frac{d^2\phi_{\mu_1}}{dx^2}-\left(\eta+\frac{2M'}{M^2}\right)\frac{d\phi_{\mu_1}}{dx}+\left(\frac{2M'^2}{M^3}-\frac{M''}{M^2}
-\eta'+\gamma\right)\phi_{\mu_1}=0\label{e2.78}
\end{equation}
and
\begin{equation}
-\frac{1}{M}\phi_{\mu_1}''-\left(\frac{1}{M}\right)'\phi_{\mu_1}'+(V_2-\mu_1)\phi_{\mu_1}=0\label{e2.79}
\end{equation}
respectively. Adding (\ref{e2.78}) from (\ref{e2.79}) we obtain
\begin{equation}
-\left(\frac{M'}{M^2}+\eta\right)
\frac{d\phi_{\mu_1}}{dx}+\left(V_2-\mu_1+\frac{2M'^2}{M^3}-\eta'+\gamma-\frac{M''}{M^2}\right)\phi_{\mu_1}=0\label{e2.80}
\end{equation}
Substituting the values of $V_2$ and $\gamma$ from (\ref{e2.38})
and (\ref{e2.42}) with $2C_1=\mu_1+\mu_2$, in the above equation
(\ref{e2.80}), we get
\begin{equation}
\frac{d}{dx}\left(log{\phi_{\mu_1}}\right)=\frac{\eta'+3\eta\frac{M'}{M}+\frac{M''}{M^2}+M\eta^2+2(C_1-\mu_1)}{2(\eta+\frac{M'}{M^2})}\label{e2.81}
\end{equation}
Now using our ans\"{a}tz (\ref{e2.46}) in (\ref{e2.81}) and then
integrating we obtain
\begin{equation}
\phi_{\mu_1}\propto\frac{M\left(\eta+\frac{M'}{M^2}\right)}{\mathcal{U}_1} \propto\frac{M\mathcal{U}_2}{W(\mathcal{U}_1,\mathcal{U}_2)}\label{e2.82}
\end{equation}
Above procedure can be applied to obtain  $\phi_{\mu_{2}}$
as
\begin{equation}
\phi_{\mu_2}\propto\frac{\eta
M+\frac{M'}{M}}{\mathcal{U}_2}\propto\frac{M\mathcal{U}_1}{W(\mathcal{U}_1,\mathcal{U}_2)}\label{e2.83}
\end{equation}
If both $\phi_{\mu_{1,2}}$ are normalizable then we get the
maximal set of eigenfunctions of $H_2$
as $\{\phi_{\mu_1},\phi_{\mu_2},\phi_{n}\propto\mathcal{L}\psi_n\}$.

Among the several spectral modifications which can be achieved
through the real second order intertwining of PDM Hamiltonians, some cases are worth to be mentioned.
\begin{itemize}
\item[$\circ$] {{\underline{ Condition of deletion of first two energy levels} :} For $\mu_1=E^{(0)}_0$
and $\mu_2=E^{(0)}_1$ the two solutions of equation (\ref{e2.54}) are the
normalizable solutions of equation (\ref{e2.15}) i.e,
$\mathcal{U}_1(x)=\psi_0(x)$ and $\mathcal{U}_2(x)=\psi_1(x)$
respectively. It turns out that the Wronskian is nodeless in the finite part of the real line and vanishes at the boundary of the domain of $V_0(x)$ but two
solutions $\phi_{\mu_1}$ and $\phi_{\mu_2}$ are
non-normalizable which is clear from the relation (\ref{e2.82}). Thus $Sp({{H}}_2) = Sp({{H}}_0)-\{E^{(0)}_0,E^{(0)}_1\}=\{E^{(0)}_2,E_3^{(0)},...\}$.}

\item[$\circ$] {{\underline{Condition of isospectral transformations} :}
If we take $\mu_2<\mu_1< E^{(0)}_0$ and choose two seed solutions $\mathcal{U}_1(x)$ and
$\mathcal{U}_2(x)$ such way that either $\mathcal{U}_{1,2}(x_l)=0$ or
$\mathcal{U}_{1,2}(x_r)=0$, $x_l$ and $x_r$ being the end points
of the domain of definition of $V_0(x)$, then the Wronskian
$W(\mathcal{U}_1,\mathcal{U}_2)$ vanishes at $x_l$ or $x_r$.
Hence $\phi_{\mu_1}$ and $\phi_{\mu_2}$ become
non-normalizable so that
$Sp({{H}}_0)=Sp({{H}}_2)$.}

\item[$\circ$]{{\underline{Condition of creation of two new levels below the ground state} :} For
$\mu_2 < \mu_1< E^{(0)}_0$ and choosing $\mathcal{U}_1(x)$ and
$\mathcal{U}_2(x)$ in such way that $\mathcal{U}_2(x)$ has exactly one
node and $\mathcal{U}_1(x)$ is nodeless on the domain of $V_0(x)$ then the Wronskian
$W(\mathcal{U}_1,\mathcal{U}_2)$ becomes nodeless and unbounded at the boundary so that two
wave functions $\phi_{\mu_1}$ and $\phi_{\mu_2}$ become
normalizable. Therefore the spectrum of ${{H}}_2$
becomes $Sp({{H}}_2)= Sp({{H}}_0) \bigcup \{\mu_1,\mu_2\}=\{\mu_1,\mu_2,E^{(0)}_n, n=0,1,2...\}.$}
\end{itemize}

\noindent{\underline{(B) Complex case $(C_2<0)$} :} For $C_2<0$ the two factorization energies $\mu_1$ and $\mu_2$
become complex. In order to construct real $\bar{V}$ we shall
choose $\mu_1$ and $\mu_2$ as complex conjugate to each other i.e,
$\mu_1=\mu\in \mathbb{C}$ and $\mu_2=\bar{\mu}$. For the same
reason we shall take $\tau_1(x)=\tau(x)$ and
$\tau_2(x)=\bar{\tau}(x).$ Hence the real solution $\eta(x)$ of
(\ref{e2.46}) generated from the complex $\tau(x)$ of (\ref{e2.49})
becomes
\begin{equation}
\eta(x)=\frac{\mu-\bar{\mu}}{\tau-\bar{\tau}}-\frac{M'}{M^2}=\frac{Im(\mu)}{Im(\tau)}-\frac{M'}{M^2}
=-\frac{W'(\mathcal{U},\bar{\mathcal{U}})}
{MW(\mathcal{U},\bar{\mathcal{U}})}\label{e2.91}
\end{equation}
Defining $w(x)=\frac{W(\mathcal {U},\bar{\mathcal
{U}})}{M(\mu-\bar{\mu})}$, $\eta(x)$ becomes
\begin{equation}
\eta(x)=-\frac{w'}{Mw}-\frac{M'}{M^2}
\end{equation}
For the factorization energies $\mu$ and $\bar{\mu}$ the equation
(\ref{e2.49}) becomes
$$-\frac{1}{M}\mathcal{U}''-\left(\frac{1}{M}\right)'\mathcal{U}'+V_0 \mathcal{U}=\mu~
\mathcal{U} ~~\mbox{and}~~
-\frac{1}{M}\mathcal{\bar{U}}''-\left(\frac{1}{M}\right)'\mathcal{\bar{U}}'+V_0\mathcal{\bar{U}}=\bar{\mu}~
\mathcal{\bar{U}}$$ Multiplying first equation by
$\mathcal{\bar{U}}$ and second equation by $\mathcal{U}$ and then
subtracting we obtain
\begin{equation}
\frac{W'(\mathcal{U},\mathcal{\bar{U}})}{M(\mu-\bar{\mu})}-\frac{M'W(\mathcal{U},\mathcal{\bar{U}})}{M^2(\mu-\bar{\mu})}=|\mathcal{U}|^2\label{e2.84}
\end{equation}
Using above relation (\ref{e2.84}) we have
\begin{equation}
w'(x)=\frac{W(\mathcal{U},\mathcal{\bar{U}})}{M(\mu-\bar{\mu})}-\frac{M'W(\mathcal{U},\mathcal{\bar{U}})}
{M^2(\mu-\bar{\mu})}=|\mathcal{U}|^2
\end{equation}
which implies that $w(x)$ is a non-decreasing function. So it is
sufficient to choose
\begin{equation}
\lim_{x\rightarrow x_l} \mathcal{U}=0 ~\mbox{or}
~~\lim_{x\rightarrow x_r} \mathcal{U}=0
\end{equation}
for the Wronskian $W$ to be nodeless. It is to be noted here that
in this case we can only construct potentials which are strictly
isospectral with the initial potential.

\subsection{Examples of second order intertwining}
\noindent To illustrate the second order intertwining with the help of an example we
shall need non-normalizable solutions of (\ref{e2.54}) (which is
similar to equation (\ref{e2.18}) but with two factorization
energies) corresponding to a particular mass function $M(x)$. To
illustrate the second order intertwining with an example we have
considered the potential (\ref{e2.10}) as an initial potential.
 Corresponding seed solution and the factorization energy $\mu =
 \mu_1$ are given by \cite{MRR10}
\begin{equation}
\mu_1=-ab+\frac{(a+b+1)c}{2}-\frac{c^2}{2}\label{e2.75}
\end{equation}
\begin{equation}\begin{array}{ll}
 \displaystyle \mathcal{U}_1(x)=\alpha\frac{e^{\frac{c}{2}x}}{(1+e^{
x})^{\frac{a+b+1}{2}}}~ _2F_1\left(a,b,c,\frac{e^{ x}}{1+e^{
x}}\right)\\
\displaystyle ~~~~~~~~~~~~~~~~~ +\beta\frac{e^{\left(1-\frac{c}{2}\right)
x}}{(1+e^{x})^{\frac{a+b-2c+3}{2}}}~
 _2F_1\left(a-c+1,b-c+1,2-c,\frac{e^{ x}}{1+e^{ x}}\right)
 \end{array}
 \label{e2.50}
 \end{equation}
We notice that the potential (\ref{e2.10}) and corresponding
Hamiltonian are invariant under the transformation $a\rightarrow
a+\nu$ and $b\rightarrow b-\nu$, $\nu\in\mathbb{R}-\{0\}$. But the
solution (\ref{e2.13}) of the corresponding Schr\"{o}dinger equation
changes to
\begin{equation}
 \begin{array}{ll}
 \displaystyle
 \mathcal{U}_2(x)=\alpha\frac{e^{\frac{c}{2}x}}{(1+e^{
x})^{\frac{a+b+1}{2}}}~ _2F_1\left(a+\nu,b-\nu,c,\frac{e^{
x}}{1+e^{
x}}\right)\\
\displaystyle ~~~~~~~~~~~
+\beta\frac{e^{\left(1-\frac{c}{2}\right) x}}{(1+e^{
x})^{\frac{a+b-2c+3}{2}}}~
 _2F_1\left(a+\nu-c+1,b-\nu-c+1,2-c,\frac{e^{ x}}{1+e^{
 x}}\right)\label{e2.51}
 \end{array}
 \end{equation}
 and the corresponding factorization energy is given by
 \begin{equation}
\mu_2=-ab+\frac{(a+b+1)c}{2}-\frac{c^2}{2}+\nu(a-b)+\nu^2\label{e2.52}
 \end{equation}
 Thus the general solutions of the equation (\ref{e2.54}) for the two
 factorization energies $\mu_1$ and $\mu_2$, are given by (\ref{e2.50}) and
 (\ref{e2.51})respectively.
 The asymptotic behaviors  of the seed solution $\mathcal{U}_2$ remains same as $\mathcal{U}_1$, which are given in
  (\ref{e2.22}) and (\ref{e2.23}). In the following we have constructed the second order isospectral partner potentials of the potential (\ref{e2.10}) with various possible spectral modifications.\\

\begin{figure}[]
\epsfxsize=2.75 in \epsfysize=2 in \centerline{\epsfbox{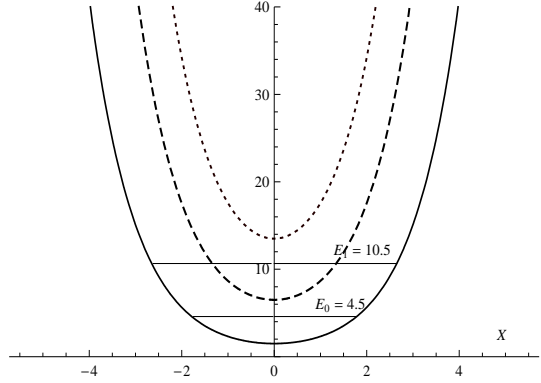}}
\caption{ Plot of the original potential (solid line) for
$a= 5, b=0, c=3$ and its first-order SUSY partner (dashed line) by
deleting the ground state $E_0=4.5$ and second-order SUSY partner
(dotted line) by deleting two successive states $E_0=4.5,
E_1=10.5.$}\label{f2.4}
\end{figure}

\noindent {\underline{Deletion of first two energy levels} :}
 Let us take $\mu_1=E_0^{(0)}$ and $\mu_2=E_1^{(0)}$,
 $\mathcal{U}_1=\psi_0(x)$
 and $\mathcal{U}_2=\psi_1(x)$ which are given in (\ref{e2.14}). The Wronskian
 $W(\mathcal{U}_1,\mathcal{U}_2)$ is given by
 \begin{equation}
 W(\mathcal{U}_1,\mathcal{U}_2)\propto \frac{e^{(c+1)x}}{(1+e^x)^{a+b+3}}\label{e2.53}
 \end{equation}
which is nodeless and bounded in $(-\infty,\infty)$ as
$c>-\frac{1}{2}$ and $a+b-c+\frac{1}{2}>0$. The second-order SUSY
partner of $V_0(x)$ is obtained using equation (\ref{e2.38}) and is
given by
\begin{equation}
V_2(x)=\frac{1}{4}\left[\left(c^2-2c(a+b+2)+(a+b+1)(a+b+3)\right)e^x+c(c+2)e^{-x}+4(a+b+1)\right]\label{e2.85}
\end{equation}
Here, the eigenfunctions
$\phi_{\mu_1}\propto\frac{M\mathcal{U}_2}{W(\mathcal{U}_1,\mathcal{U}_2)}$
and
$\phi_{\mu_2}\propto\frac{M\mathcal{U}_1}{W(\mathcal{U}_1,\mathcal{U}_2)}$
of $H_2$ associated to $\mu_1=E^{(0)}_0$ and $\mu_2=E^{(0)}_1$
are not normalizable since $\lim_{x\rightarrow-\infty,\infty}\bar{\psi}_{\mu_{1,2}}(x)=\infty$. Thus $Sp(H_2)=Sp(H_0)-\{E^{(0)}_0,E^{(0)}_1\}=\{E^{(0)}_2,E^{(0)}_3...\}.$\\
In particular taking $a=5,b=0,c=3$ we have plotted the potential
$V_0(x)$ and its second-order SUSY partner $V_2(x)$ given in
(\ref{e2.10}) and (\ref{e2.85}) respectively, in figure \ref{f2.4}.
\FloatBarrier

\begin{figure}[]
\epsfxsize=2.75 in \epsfysize=1.85 in \centerline{\epsfbox{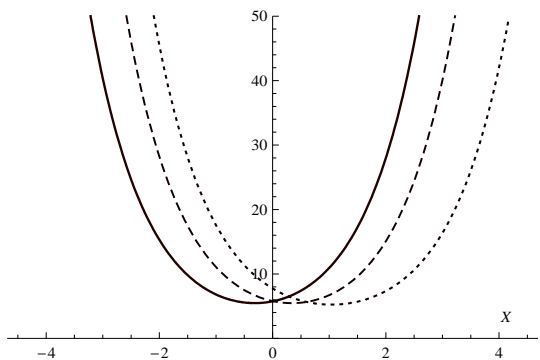}}
\label*{}\caption{ Plot of the original potential (solid line) and
its first-order (dashed line) and second-order (dotted line) SUSY
partner by making the isospectral transformation for $a=3,b=5
,c=4,\alpha=1, \beta=0,\nu=1$.}\label{f2.5}
\end{figure}

\noindent {\underline{Strictly isospectral potentials} :}
 The strictly isospectral partner potentials can be constructed by creating two new
 energy levels in the limit when each seed vanishes at one of the
 ends of the $x$-domain. Now from the asymptotic behaviors of
 the seed solutions, we note that both the seed solutions vanish at
 $x\rightarrow -\infty$ for $\beta=0, \alpha>0$ if $a+b-c>1$ and
 $c>0$. Considering $\beta=0, \alpha=1$ in (\ref{e2.50}) and (\ref{e2.51}) we take two seed solution as
 \begin{equation}
 \mathcal{U}_1(x)=\frac{e^{\frac{c}{2}x}}{(1+e^{
x})^{\frac{a+b+1}{2}}}~ _2F_1\left(a,b,c,\frac{e^{ x}}{1+e^{
x}}\right)
 \end{equation}
 and
 \begin{equation}
 \mathcal{U}_2(x)=\frac{e^{\frac{c}{2}x}}{(1+e^{
x})^{\frac{a+b+1}{2}}}~ _2F_1\left(a+\nu,b-\nu,c,\frac{e^{
x}}{1+e^{ x}}\right)
 \end{equation}
 Since $\mathcal{U}_{1,2}(x)\rightarrow 0$ at $x\rightarrow -\infty$, from the expressions (\ref{e2.82}) and
 (\ref{e2.83}) we can conclude that $\lim_{x\rightarrow -\infty}\phi_{\mu_{1,2}}(x)=\infty$ which
 implies that $\mu_{1,2}$ does not belongs to
 $Sp(H_2)$ i.e. $V_2(x)$ is strictly isospectral
 to $V_0(x).$ Here the general expression of the partner potential is too involved so instead of giving
 the explicit expression we have considered particular values $a=3 , b = 5, c=4, \nu
 =1, \alpha=1, \beta=0$. Corresponding expression of the partner potential and its energy spectrum
 are $$V_2(x)= 1+\frac{3}{4} (9 \cosh x- 7 \sinh x), ~~~~~E^{(2)}_n=E^{(0)}_n=n^2+8 n+10,~~n=0,1,2...$$
 respectively. In figure \ref{f2.5}, we have plotted the initial potential,
 its first and second-order strictly isospectral partner
 potentials for the parameter values $a=3,b=5 ,c=4,\alpha=1, \beta=0,\nu=1$.
\FloatBarrier

\begin{figure}[]
\epsfxsize=4.75 in \epsfysize=2.25 in
\centerline{\epsfbox{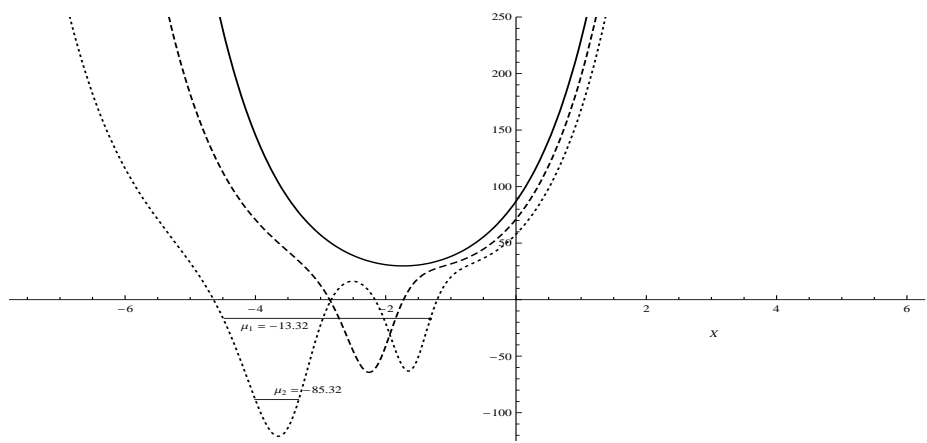}} \label*{}\caption{Plot of the
original potential (solid line) for $a= 2.8, b=20, c=4.4$ and its
first-order SUSY partner (dashed line) by creating a new level
$\mu_1=-13.32$ and second-order SUSY partner (dotted line) by
creating two new levels $\mu_1=-13.32, \mu_2=-85.32$}\label{f2.6}
\end{figure}
\FloatBarrier

\noindent{\underline{Creation of two new levels below the ground state} :} Two
energy levels can be created taking $\mu_2<\mu_1<E_0$ and using
those seed solutions $\mathcal{U}_1$ and $\mathcal{U}_2$ for which
the Wronskian become nodeless. In this case the expressions of the
Wronskian contains several Hypergeometric function, so it is very
difficult to mention the range of $a,b,c$ and $\nu$ for which it
is nodeless. In particular for $a= 2.8, b=20, c=4.4 , \alpha=1,
\beta=1$ we have the Wronskian is found to be nodeless. For the
same values of $a,b,c$ we have plotted the potential and its
second order partner in figure \ref{f2.6}. The second-order isospectral
partner is obtained using equation (\ref{e2.38}).

\begin{figure}[]
\epsfxsize=2.95 in \epsfysize=1.75 in \centerline{\epsfbox{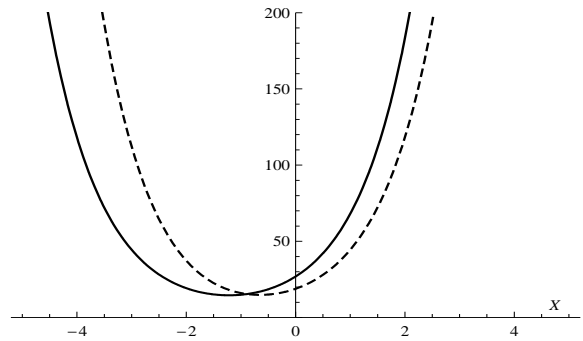}}
\caption{ Plot of the original potential (solid line) for
$a= 6.1-5 ~i,b=8+5 ~i,c=4.1, \nu=1.9, \mu_1=-51.25+9.5~i,
\mu_2=-51.25-9.5 ~i.$ and its second-order isospectral partner
(dashed line).}\label{f2.7}
\end{figure}

\noindent {\underline{Example of strictly isospectral transformation for complex factorization energies} :} The complex factorization
energy $\mu_1$ and $\mu_2$ given by equation (\ref{e2.75}) and
(\ref{e2.52}), can be made conjugate to each other in several ways.
One of the way is by making following restrictions on $a,b, c,\nu$
{\bf :} $c\in \mathbb{R}, ~Im(a)=-Im(b),~ \nu=Re(b)-Re(a)$. But in
order to keep the initial potential real we have to made two more
restrictions e.g. $Re(a)+Re(b)-c>1$ and $c>2.$ In particular
taking $a= 6.1-5 ~i,b=8+5 ~i,c=4.1 , \nu=1.9$ we have two
factorization energy $ \mu_1(=\mu) =-51.25+9.5~i$ and $
\mu_2(=\bar{\mu}) = -51.25-9.5~i.$  For these values of $a,b,c,
\nu$ and $\alpha=1, \beta=0$ the seed solution $\mathcal{U}$
becomes
\begin{equation}
\mathcal{U}(x)= \frac{e^{2.05 x}}{(1+e^x)^{7.55}}~
_2F_1\left(6.1-5~i,8+5 ~i,4.1,\frac{e^x}{1+e^x}\right)
\end{equation}
Clearly $\mathcal{U}(-\infty)=0$ and $|\mathcal{U}|\rightarrow
\infty$ as $x\rightarrow \infty$ so this seed $\mathcal{U}$ and
its conjugate $\mathcal{\bar{U}}$ can be used to obtain the
second-order SUSY partner potential ${V}_2(x)$ with the help of
equations (\ref{e2.38}) and (\ref{e2.91}). In figure \ref{f2.7} we have plotted
the initial potential $V_0(x)$  given in (\ref{e2.10}) and its
isospectral partner $V_2(x)$ for the parameter values
mentioned earlier.
\FloatBarrier

\section{Summary}
\noindent To summarize, we point out the following main results :
\begin{itemize}
 \item[$\blacktriangleright$]{We have generalized the modified factorization technique of a Schr\"{o}dinger Hamiltonian suggested in ref.\cite{BU10a,BU10b} to a quantum system characterized
by position-dependent mass Hamiltonian. This generalization is done in such a way that a given mass function and the
excited state wave function of a PDM Hamiltonian can be used to generate new solvable physical Hamiltonians. This method
is applied to a number of exactly solvable PDM Hamiltonians which leads to nontrivial isospectral Hamiltonians.}

  \item[$\blacktriangleright$]{The method of point canonical transformation has been used to generate some exactly solvable shape invariant potentials for position
dependent (effective) mass Schr\"{o}dinger equation whose bound
state solutions are given in terms of Laguerre or Jacobi type
$X_1$ exceptional orthogonal polynomials. The obtained potentials are in fact generalizations (by some rational
functions) of the potentials whose bound state solutions involve classical Laguerre or Jacobi
orthogonal polynomials. We have also shown that these two type of potentials, one is associated with exceptional orthogonal polynomials
and other is to classical orthogonal polynomials,
are isospectral. The method discussed here can be used for
other choices of the function $g(x)$ in order to generate other
type of exactly and quasi exactly solvable  potentials for the one
dimensional Schr\"{o}dinger equation with position dependent mass.}

   \item[$\blacktriangleright$]{We have discussed the possibilities for designing
quantum spectra of position dependent mass Hamiltonians offered by
the first and second order intertwining technique. To generate spectral
modifications by first order intertwining, we have used solutions
to the position dependent mass Schr\"odinger equation
corresponding to factorization energy (not belonging to the
physical spectrum of the initial problem) less than or equal the
ground state energy in order to avoid singularity in the
isospectral partner potential. Thus it is
possible to generate isospectral partner potentials (i) with the
ground state of the given potential deleted (ii) with a new
state created below the ground state of the given potential (iii)
with the spectrum of the given potential unaffected.
In the case of second order intertwining, instead of using the
iterative method used in \cite{SH08}, the second order intertwiner
is constructed directly by taking it as second order linear
differential operator with unknown coefficients. The main advantage of this construction is that
one can generate isospectral partner potentials
directly from the initial potential without generating first-order
partner potentials. The apparently intricate system of equations
arising from the second-order intertwining relationship is solved for the
coefficients by taking an ansatz. In this case the spectral
modifications are done by taking appropriately chosen
factorization energies which may be real or complex. For real
unequal factorization energies, it is possible to generate
potentials (i) with deletion of first two energy levels (ii) with
two new levels embedded below the ground state of the original
potential (iii) with identical spectrum as of the original
potential. For complex factorization energies, it is shown how to
obtain strictly isospectral potentials.}
\end{itemize}


\chapter[Quantum Systems with Non-Hermitian Hamiltonian]{Quantum Systems with Non-Hermitian Hamiltonian\footnote{This chapter is based on the following three papers:\\
(i) B. Midya, B Roy and R. Roychoudhury, ``A note on $\mathcal{PT}$ invariant periodic potential $V(x) = 4 \cos^2 x + 4 i V_0 \sin 2x$'',
Phys. Lett. A 374 (2010) 2605. (ii) B. Midya, ``Quasi-Hermitian Hamiltonians associated with exceptional orthogonal polynomials'', {Phys. Lett. A} (2012) in press; arXiv:1205.5860.
(iii) B Midya, P Dube and R Roychoudhury,`` Non-isospectrality of the generalized Swanson Hamiltonian and harmonic oscillator'',
J. Phys. A 44 (2011) 062001.}} \label{c3}
\pagestyle{Chapter}
\noindent
In this chapter,
we shall study some non-Hermitian Hamiltonians: (A) Hamiltonian corresponding to a complex periodic potential (\ref{e3.1}); (B) Hamiltonians which
are associated with
exceptonal $X_1$ orthogonal polynomials and (C) the non-Hermitian oscillator known as Swanson Hamiltonian (\ref{e3.01}).

\section{Non Hermitian Hamiltonian with $V(x) = 4 \cos^2 x + 4 i V_0 \sin 2x $}\label{s3.1}
\noindent Quite recently, the prospect of realizing
complex $\mathcal{PT}$-symmetric periodic potentials within
 the framework of optics has been suggested \cite{Gu09,Ru10,Mu08a,Ma08,Lo09a,Lo09b,Ma10}. The $\mathcal{PT}$-symmetric optical beam evolution is governed by the nonlinear Schr\"{o}dinger (NLS)-like
 equation \cite{Mu08b,Ma08}
 \begin{equation}
  i \frac{\partial \phi}{\partial z} + \frac{\partial^2 \phi}{\partial x^2} + V(x) \phi + g |\phi|^2 \phi = 0\label{e3.0}
 \end{equation}
where $\phi(x,z)$ is proportional to the electric field envelope,
$z$ is a scaled propagation distance and $g=\pm 1$ correspond to self-focusing or defocusing nonlinearity. The optical potential
$V(x) = V_R(x) + i V_I(x)$ plays the role of a complex refractive index, where $V_R(-x)= V_R(x)$ is the index guiding and
$V_I(-x) = - V_I(x)$ is the gain/loss distribution. The nonlinear stationary solution of equation (\ref{e3.0}) are sought in the form $\phi(x,z) = e^{i\beta z} \psi(x)$ where
$\psi(x)$ satisfies
\begin{equation}
 \frac{d^2\psi}{dx^2} + V(x) \psi(x) + g |\psi|^2 \psi = \beta \psi. \label{e3.00}
\end{equation}
As mentioned in chapter 1, many interesting results have already been obtained by considering the above equations with some $\mathcal{PT}$-symmetric optical potentials. Here we consider the $\mathcal{PT}$ invariant periodic potential
\begin{equation}
V(x) = 4 \cos^2 x + 4 i V_0 \sin 2x,~~~ V_0\in \mathbb{R} \label{e3.1}
\end{equation}
where $V(x + \pi) = V(x)$. The
reason for considering this particular potential (\ref{e3.1}) lies in
the fact that many physically interesting results have
been obtained while considering this potential in optical lattice.
 Specifically, it has been identified that the $\mathcal{PT}$ threshold
occurs at $V_0 = 0.5$ \cite{Mu08a,Ma08}. Below this threshold all the eigenvalues for every band structures are real
and all the forbidden gaps are open whereas at the threshold, the spectrum is whole real line and there are no band gap. On the other hand, when $V_0
> 0.5$ the first two bands (starting from the lowest band) start to merge together and in doing so they form oval-like structure with a
related complex spectrum. The soliton
solution of the nonlinear Schr\"{o}dinger equation corresponding to this potential has also been studied \cite{Mu08b}. But in all
these studies some numerical techniques have been used. Therefore it is interesting to reconsider this potential so that  the notion of pseudo-Hermiticity \cite{Mo10} can be used to obtain some interesting (except the existing one) results analytically.
Here, we shall argue that one can anticipate by analytical argument that when $V_0<.5$ the ${\cal{PT}}$-symmetry of the potential is unbroken. We shall first derive the
equivalent Hermitian Hamiltonian of the corresponding non-Hermitian Hamiltonian with the potential (\ref{e3.1}) for $V_0< 0.5$.
It will be
shown that there exist another critical point $V_0^c \sim 0.888437$ after which no part of the band structure remains real. The corresponding
 band structure will be obtained using the Floquet analysis.

\subsection[Equivalent Hermitian analogue]{Equivalent Hermitian analogue of $V(x) = 4 \cos^2 x + 4 i V_0 \sin 2x $}

\noindent We consider the following linear eigenvalue problem that correspond to the linearized version of equation (\ref{e3.00})
\begin{equation}
H\psi(x) = \frac{d^2\psi(x)}{dx^2}+V(x)\psi(x)=\beta\psi(x)\label{e3.4}
\end{equation}
where the potential $V(x)$ is given in equation (\ref{e3.1}). In order to find the equivalent Hermitian analogue of the above Hamiltonian, we re-write
potential $V(x)$ given in (\ref{e3.1}) as
\begin{equation}
V(x)=\left\{\begin{array}{lll}
\displaystyle 2+2\sqrt{1-4V_0^2}~
\cos\left(2x-i~ \tanh^{-1}2V_0\right),~~~~~~~~V_0< .5\\
\displaystyle 2+2e^{2 i x},~~~~~~~~~~~~~~~~~~~~~~~~~~~~~~~~~~~~~~~~~~~~~~~~~V_0=.5\\
\displaystyle 2+2i \sqrt{4V_0^2-1}~ \sin\left(2 x-i~
 \tanh^{-1}\frac{1}{2V_0}\right),~~~~~~V_0> .5
\end{array}\right.\label{e3.2}
\end{equation}
Due to the fact that for a $\theta \in \mathbb{R}$ the Hermitian linear automorphism $\eta=e^{-\theta p}$ affects an
imaginary shift of the co-ordinate \cite{Ah01b}: $e^{-\theta p} x e^{\theta p} = x + i \theta$, we have for $V_0< 0.5$
the Hamiltonian defined in (\ref{e3.4}) reduced to a equivalent Hermitian Hamiltonian $h$
relations
\begin{equation}
h = \rho H \rho^{-1}
\end{equation}
where
\begin{equation}
\rho= e^{-\frac{\theta}{2} p} ,~~~~~\theta= \tanh^{-1} 2V_0, ~p=-i \frac{d}{dx}
\end{equation}
is a unitary operator. The Hermitian Hamiltonian $h$ is given by
\begin{equation}
h\equiv\frac{d^2}{dx^2}+2+2\sqrt{1-4V_0^2}~ \cos~ 2x.
\end{equation}
Defining  $\eta= \rho^2 = e^{-\theta p}$, a positive definite operator, we have $H^\dagger = \eta H \eta^{-1}$.
Which implies that the Hamiltonian considered here is pseudo-Hermitian with respect to a positive definite metric operator $\eta$, so that the
spectrum of $H$ is entirely real for $V_0 < 0.5$. For $V_0=.5$, it was shown that \cite{Ga80} the spectrum of this potential $V(x) = 2+2 e^{2 i
x}$ is purely continuous and fills the semi axis $[0,\infty)$. While for $V_0> .5$, we have $\eta V(x) \eta^{-1} = V(x+i \theta) \neq
V^*(x)$, where $\eta= e^{-\theta p}, \theta= \tanh^{-1}\frac{1}{
2V_0}$ which indicates that in this case the Hamiltonian $H$ is
not $\eta$ pseudo-Hermitian so the reality of the eigenvalues is
not ensured.

 Hence, we propose that $V_0=V_0^{th}= 0.5$ is the transition
point after which the eigenvalues of equation (\ref{e3.4}) become
complex and spontaneous breaking of $\mathcal{PT}$ symmetry occurs. In the next section, we show that this proposition is indeed true.

\subsection[Band Structure and existence of second critical point]{Band structure of the potential (\ref{e3.1}) and existence of second critical point}\label{s3.1.2}

\noindent After making the change of variable $z=x-\frac{i}{2} \tanh^{-1}\frac{1}{2V_0}$ in case of $V_0> .5$,
the equation (\ref{e3.4}) transforms into
\begin{equation}
\frac{d^2\psi(z;\beta,V_0)}{dz^2}+\left(2-\beta+2i\sqrt{4V_0^2-1}~
\sin 2z\right)\psi(z;\beta,V_0)=0.\label{e3.5}
\end{equation}
Under simple coordinate translation $z \rightarrow
\frac{\pi}{4}-z$ the  equation (\ref{e3.5}) reduces to Mathieu
equation $ \frac{d^2\psi}{dz^2} + (a-2 i q \cos2z)\psi = 0$
\cite{AS65}, with characteristic value $a = 2-\beta$ and $q =
\sqrt{4V_0^2 - 1}$. Tabular
values from an old paper of Mulholland and Goldstein \cite{MG29}
and Bouwkamp \cite{Bo48} reflect that the characteristic values $q$
would be both real and complex conjugate: the transition occurring
in the neighborhood of $q= q_0= 1.46876852$. The same critical(exceptional) point was obtained in
ref.\cite{BQR08}, while considering the large $N$ limit of the quasi
exactly solvable $\mathcal{PT}$-symmetric potential $V(x) = -(i \xi
\sin 2x + N)^2$. Hence it follows that there
is a second critical point in the neighborhood of $V_0=V_0^{c}
\sim .888437$ after which all the eigenvalues of the equation
(\ref{e3.5}) becomes complex and occur in complex
conjugate pairs. Consequently no part of the band structure remain
purely real.\\

\begin{figure}[]
\begin{center}
\includegraphics[width=2.75in]{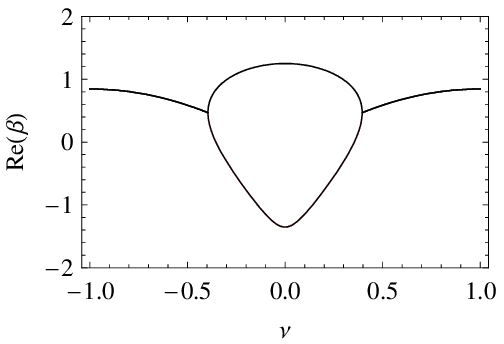} ~~~~~\includegraphics[width=2.75in]{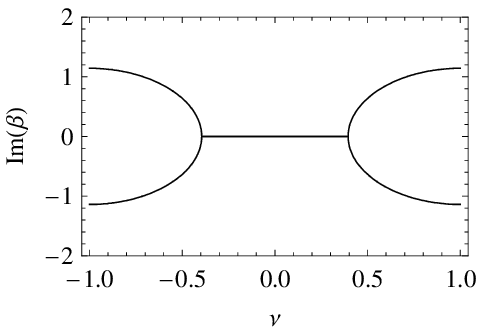}
 \end{center}
 \caption{Band structure for the $\mathcal{PT}$ invariant potential $V(x)=4 \cos^2 x + 4iV_0 \sin 2x$ for $V_0$=.75}\label{f3.2}
 \end{figure}

\begin{figure}[]
 \begin{center}
 \includegraphics[width=2.75in, height=1.75in]{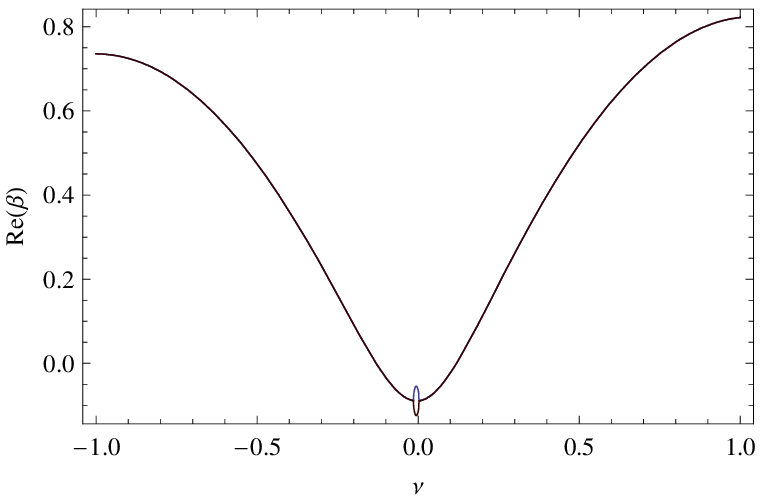}~~~~~\includegraphics[width=2.75in,height=1.75in]{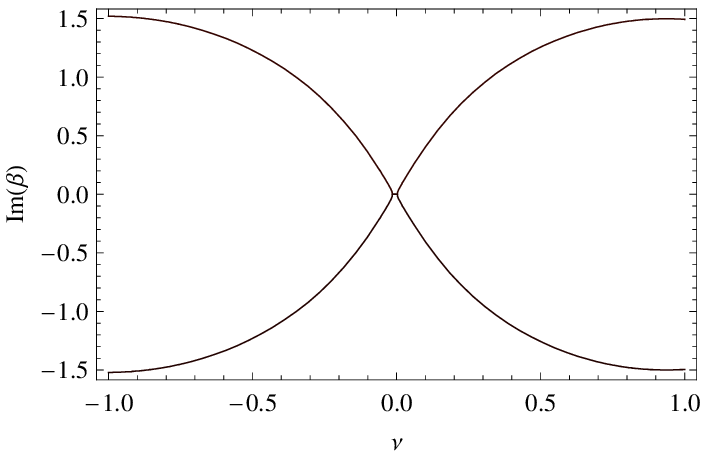}
 \end{center}
 \caption{Band structure for the $\mathcal{PT}$-invariant potential $V(x)=4 \cos^2 x + 4iV_0 \sin 2x$ for
$V_0$=.8885}\label{f3.3}
 \end{figure}

Now to determine the eigenvalues of the equation (\ref{e3.5}) and
correspondingly the band structure of the potential given in
(\ref{e3.1}), we proceed as follows: as the coefficient
of the eq.(\ref{e3.5}) are $\pi$ periodic, so according to
Floquet-Bloch theorem \cite{Fl83} the eigenfunctions are of the
form $\psi(z;\beta, V_0)=\phi_\nu(z;\beta, V_0) e^{i\nu z}$, where
$\phi_\nu(z+\pi)=\phi_\nu(z)$ and $\nu $ stands for the real Bloch
momentum. Since $\phi_\nu(z)$ is periodic so the entire solution
of equation (\ref{e3.5}) can be expanded in the Fourier series
\begin{equation}
\psi(z,\beta) =  \sum_{k\in \mathbb{Z}} c_{2k} (\nu;\beta,V_0)
e^{i(\nu+2k) z}\label{e3.6}
\end{equation}
which, inserted into eq.(\ref{e3.5}), gives the recurrence relation
\begin{equation}
 c_{2k}+\zeta_{2k}[c_{2(k-1)}-c_{2(k+1)}]
=0,~~\zeta_{2k}=\frac{-\sqrt{4V_0^2-1}}{[(\nu+2k)^2+\beta-2]},~~\forall
k\in \mathbb{Z}\label{e3.7}
\end{equation}
for the coefficients $c_{2k}(\beta,\nu)$. For any
finite(truncated) upper limit $n\in \mathbb{Z}$, eq. (\ref{e3.7})
can be written as a matrix equation
\begin{equation}
\mathcal{M}_n(\nu;\beta,V_0) \mathcal{C}_n^t = 0\label{e3.8}
\end{equation}
where $\mathcal{C}_n=(c_{-2n},...,c_{-2},c_0,c_2,...,c_{2n})$ and
$\mathcal{M}_n(\nu;\beta,V_0)$ is a $(2n+1)\times (2n+1)$ matrix given
by
\begin{equation}
\mathcal{M}_n(\nu;\beta,V_0) =\left(%
\begin{array}{ccccccccc}
  1 & -\zeta_{2n} &  &  &  &  &  & &  \\
  \zeta_{2n-2} & . & . & . & . & . & . & . &  \\
   & . & \zeta_2 & 1 & -\zeta_2 & 0 &0 & . &  \\
   & . & 0 & \zeta_0 & 1 & -\zeta_0 & 0 & . &  \\
   & . & 0 & 0 & \zeta{-2} & 1 & -\zeta_{-2} & . &  \\
   & . & . & . & . & . & . & . & -\zeta_{-2n+2} \\
   &  &  &  &  &  &  & \zeta_{-2n} & 1 \\
\end{array}%
\right).\label{e3.9}
\end{equation}
\begin{figure}[]
\centering
\subfigure[]{\includegraphics[width =2.75in] {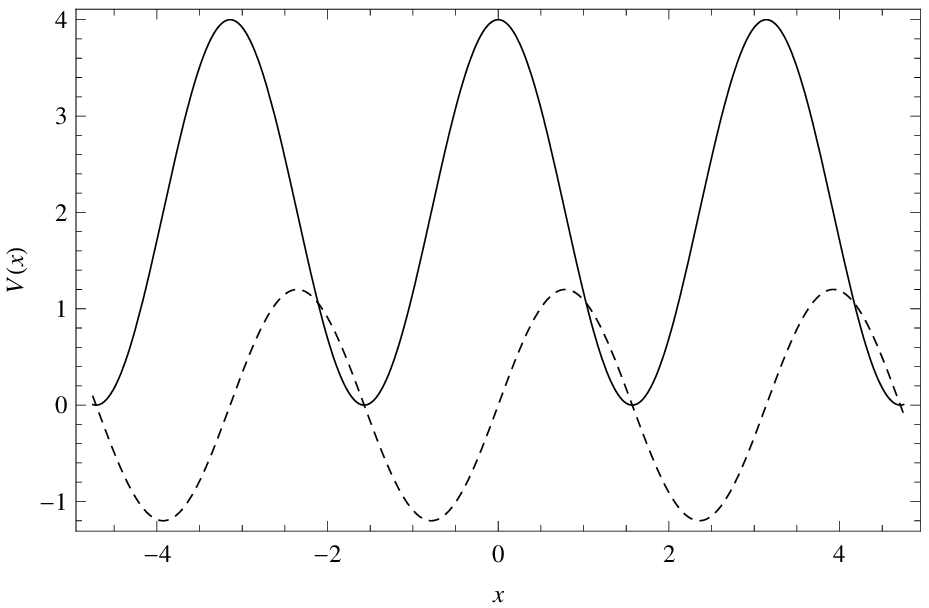}
}
 \subfigure[]{\includegraphics[width=2.75in] {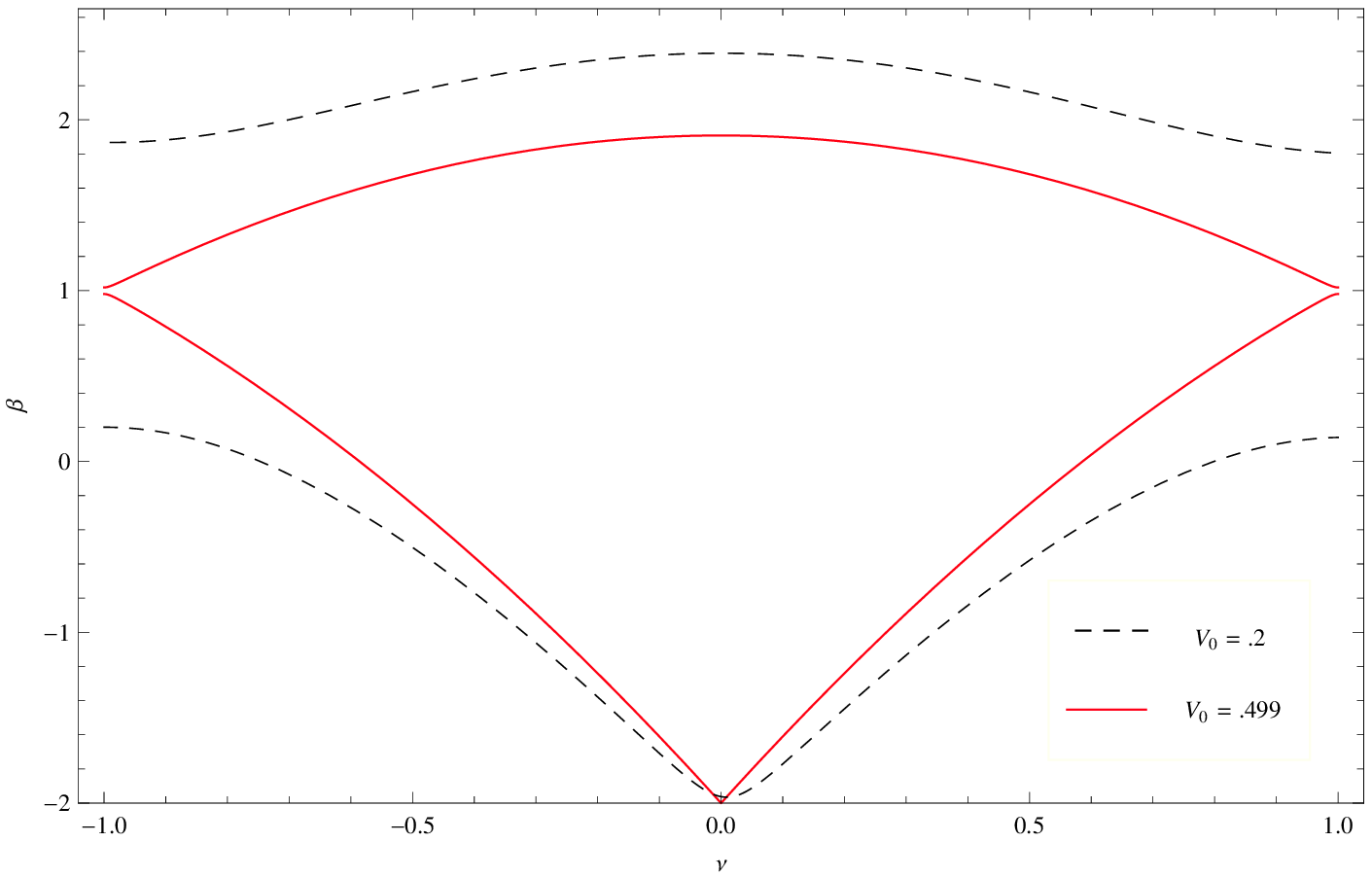}
}
\caption{(a) Real (solid line) and imaginary (dashed) part of the potential $V(x) = 4 \cos^2 x + 4 i V_0 \sin 2x$ for $V_0=0.3$.
(b) The band structure corresponding to $V_0 = 0.2$ (dashed line) and $V_0 =.499$ (solid red line).}
\label{f3.1}
\end{figure}
\FloatBarrier
For the finite dimensional case, finding the non-trivial solutions
of the homogeneous system of eqn.(\ref{e3.8}) is equivalent to solve
\begin{equation}
det(\mathcal{M}_n) = 0.\label{e3.10}
\end{equation}
The solution of the above eqn.(\ref{e3.10}) gives the relation
between the energy eigenvalues $\beta$ and bloch momentum $\nu$.
Figures \ref{f3.1}b, \ref{f3.2} and \ref{f3.3} have been plotted taking $n=3$. Figure \ref{f3.1}a shows the real and imaginary part of the
complex periodic potential given in
(\ref{e3.1}). In figure \ref{f3.1}b, the band structure of the potential $V(x)$ for $V_0=.2$ and .499 have been shown. In figures \ref{f3.2} and \ref{f3.3}, we have plotted the band structures for $V_0 =0.75$ and near the value of second critical point $V_0^c \sim .888437$ respectively. It is found that below the value $V_0 =0.5$ all bands are real and all the forbidden band gaps are open,
whereas after this point the bands become complex and start to merge together and form oval like structure.
After the second critical point this oval like structure vanishes consequently no part of the band structure remains real.
 Hence, we conclude that $V_0=V_0^{th} = 0.5$ is the
$\mathcal{PT}$ threshold and $V_0=V_0^{c} \sim .888437$ is the second critical point.
\FloatBarrier

\FloatBarrier

 \section[Quasi-Hermitian Hamiltonians associated with $X_1$ exceptional orthogonal polynomials]
{Quasi-Hermitian Hamiltonians associated with Laguerre or Jacobi type $X_1$ exceptional orthogonal polynomials }\label{s3.2}
\noindent Here we use the method of point canonical transformation (PCT) to derive some exactly solvable non-Hermitian Hamiltonians whose bound state wave functions are associated with Laguerre or Jacobi type $X_1$
 exceptional orthogonal polynomials. For this we first briefly recall the method of point canonical transformation.

 In PCT approach \cite{BS62,Le89}, the general solution of the Schr\"{o}dinger equation (with $\hslash = 2 m = 1$)
\begin{equation}
 H \psi(x) = -\frac{d^2\psi(x)}{dx^2} + V(x) \psi(x) = E \psi(x)\label{e8.1}
\end{equation}
can be assumed as
\begin{equation}
 \psi(x) \sim f(x) F(g(x)) \label{e8.18}
\end{equation}
where $F(g)$ satisfies the second order linear differential equation of a special function
\begin{equation}
 \frac{d^2 F}{dg^2} + Q(g) \frac{dF}{dg} + R(g) F(g) =0.\label{e8.2}
\end{equation}
Substituting the assumed solution $\psi(x)$ in equation (\ref{e8.1}) and comparing the resulting equation with the equation (\ref{e8.2})
one obtains the following two equations for $Q(g(x))$ and $R(g(x))$
\begin{subequations}
  \begin{equation}
    Q(g) = \frac{g''}{g'^2} + \frac{2 f'}{f g'}
  \end{equation}
  \begin{equation}
    R(g) = \frac{E-V(x)}{g'^2} + \frac{f''}{f g'^2}
  \end{equation}
\end{subequations}
respectively. After some algebraic manipulations above two equations reduces to
\begin{subequations}
\begin{equation}
 f(x) \approx g'(x)^{-1/2} ~ e^{^{\frac{1}{2} \int Q(g) dg}},\label{e8.7}
\end{equation}
\begin{equation}
 E-V(x) = \frac{g'''}{2g'} - \frac{3}{4} \left(\frac{g''}{g'}\right)^2 + g'^2 \left(R-\frac{1}{2} \frac{dQ}{dg}
 -\frac{1}{4} Q^2\right).\label{e8.3}
\end{equation}
\end{subequations}
Now we are in a position to choose the special function $F(g)$ (consequently $Q(g)$ and $R(g)$).
The equation (\ref{e8.3}) becomes meaningful for a proper choice of $g(x)$ ensuring the presence of a constant term in the right-hand
 side which connects the energy in the left-hand side. The remaining part of equation (\ref{e8.3}) gives the potential. Corresponding bound
state wave functions involving the special function $F(g)$ are obtained with the help of equations (\ref{e8.18}) and (\ref{e8.7}), as
\begin{equation}
\psi(x) \sim g'(x)^{-1/2} ~ e^{^{\frac{1}{2} \int Q(g) dg}}~ F(g(x)).\label{e8.24}
\end{equation}
Here, we choose the special function to be the exceptional $X_1$ Laguerre polynomial viz, $F(g) \propto \hat{L}_n^{(a)}(x)$.
For real $a>0$ and $n=1,2,3...$, these polynomials $\hat{L}_n^{(a)}(x)$ satisfy the differential equation \cite{UK09,UK10}
\begin{equation}
x \frac{d^2y}{dx^2} - \frac{(x-a)(x+a+1)}{x+a} \frac{dy}{dx} + \left(\frac{x-a}{x+a} +n-1 \right) y = 0.
\end{equation}
The polynomial $\hat{L}_n^{(a)}(x)$ has one zero in $(-\infty,-a)$,
remaining $n-1$ zeros lie in $(0,\infty)$ .
Moreover, these polynomials are orthonormal \cite{UK09} with respect to the rational weight $\widehat{W} = \frac{e^{-x} x^a}{(x+a)^2}$
\begin{equation}
\int_0^\infty \frac{e^{-x} x^a}{(x+a)^2} \hat{L}_n^{(a)}(x) \hat{L}_m^{(a)}(x) dx = \frac{(a+n)\Gamma(a+n-1)}{(n-1)!}\delta_{nm}.\label{e8.28}
\end{equation}
The expressions for $Q(g)$ and $R(g)$, corresponding to the choice $F(g) = \hat{L}_n^{(a)}(x)$, are given by
\begin{equation}
Q(g) = - \frac{(g-a)(g+a+1)}{g(g+a)}, ~~~~ R(g) = \frac{g-a}{g(g+a)} + \frac{n-1}{g}.\label{e8.27}
\end{equation}
Using them in equation (\ref{e8.3}), we have the expression for $E-V(x)$ as
\begin{equation}
E-V(x) = \frac{g'''}{2g'} -\frac{3}{4} \left(\frac{g''}{g'}\right)^2 + \frac{(2 n a + a^2 - a +2)g'^2}{2 a g} - \frac{g'^2}{a(g+a)}-\frac{(a^2-1) g'^2}{4 g^2} - \frac{2 g'^2}{(g+a)^2} - \frac{g'^2}{4}\label{e8.4}
\end{equation}
At this point we choose $g'^2/g = \mbox{constant} = k^2, k \in \mathbb{R}-\{0\}$, which is satisfied by
\begin{equation}
g(x) = \frac{1}{4} (k x +d)^2,\label{e8.29}
\end{equation}
where $d$ is an arbitrary constant of integration. Here two cases may arise, namely, $d=0$ and $d\ne0$. Without loss of generality we can choose, for the moment, $d=0$.
For this choice, substituting $g(x)$ in equation (\ref{e8.4}) and separating out the potential and the energy, we have
\begin{equation}\begin{array}{ll}
V(x) = \frac{k^4 }{16} x ^2 + \left(a^2-\frac{1}{4}\right) \frac{ 1}{ x^2} + \frac{4 k^2}{k^2 x^2 + 4 a} - \frac{32 a k^2}{(k^2 x^2 + 4 a)^2},
\\\\
E_n = \frac{k^2( 2n +a-1)}{2}.
\end{array}\label{e8.5}
\end{equation}
The potential $V(x)$ is singularity free in the interval $0<x<\infty$. The same potential has earlier been reported in ref.\cite{Qu08}.
It has been shown that the potential $V(x)$ is the extension of the standard radial oscillator by addition of last two rational terms.
Such terms do not change the behavior of the potential for large
values of $x$, while small values of $x$ produce some drastic effect on the minima of the potential. The normalized wave functions corresponding to the potential
can be determined, in terms of Laguerre $X_1$ EOPs, using equations (\ref{e8.24}), (\ref{e8.28}) and (\ref{e8.27}), as
\begin{equation}
\psi_n(x) = \left( \frac{(n-1)!~ k^{2 a +2}}{2^{2a-3} (a +n) \Gamma(a+n-1)}\right)^{\frac{1}{2}} ~\frac{ x^{a+\frac{1}{2}}}{k^2 x^2 + 4 a} ~~e^{-\frac{k^2 x^2}{8}} ~~ \hat{{L}}_n^{(a)}
\left(\frac{k^2 x^2}{4}\right), ~~ n=1,2,3...\label{e8.34}
\end{equation}
It is worth mentioning here that the choice $d=0$ in equation (\ref{e8.29}) always gives rise to Hermitian potential.
Nonzero real values of $d$ do not make any significant difference
in the potential and its solutions.
The non-Hermiticity can be invoked into the potential only if $d$ is purely imaginary. We set $d= i \epsilon$, $\epsilon \in \mathbb{R}-\{0\}$, and $g(x) = \frac{1}{4}(k x + i \epsilon)^2$
for which the potential reduces to
\begin{equation}
\widetilde{V}(x) = \frac{k^2(k x + i \epsilon)^2}{16} + \frac{k^2 (a^2-\frac{1}{4})}{(k x + i \epsilon)^2} + \frac{4 k^2}{(k x + i \epsilon)^2 + 4 a}
- \frac{32 a k^2}{ [(k x + i \epsilon  )^2 + 4 a]^2}\label{e8.33}
\end{equation}
The non-Hermitian potential (\ref{e8.33}) is free from singularity through out the whole real $x$ axis. Since the energy $E_n$ has no dependence on $d$,
the non-Hermitian Potential $\widetilde{V}(x)$ also shares the same real energy spectrum of $V(x)$. This requires further explanation. In the following, we show that the potential $\widetilde{V}(x)$ is actually quasi-Hermitian. For this we define the operator
\begin{equation}
\rho = e^{\frac{\epsilon}{k} p} , ~~p = -i \frac{d}{dx}\label{e8.9}
\end{equation}
which has the following properties
\begin{equation}
\rho x \rho^{-1} = x - \frac{i \epsilon}{k},~~ \rho p \rho^{-1} = p, ~~ \rho f(x) \rho^{-1} = f\left(x - \frac{i \epsilon}{k}\right). \label{e8.6}
\end{equation}
In other words, the operator $\rho$ has an effect of shifting the coordinate $x$ to $x - \frac{i \epsilon}{k}$.
For the proof of the results (\ref{e8.6}), follow the reference \cite{Ah01b}.
For this operator we have the following similarity transformation
\begin{equation}
\rho \widetilde{V}(x) \rho^{-1} = \widetilde{V} \left(x - \frac{i \epsilon}{k}\right) = V(x).
\end{equation}
This ensures that the non-Hermitian Hamiltonian corresponding to the potential $\widetilde{V}(x)$ is quasi-Hermitian.
 The equivalent Hermitian potential $V(x)$, which corresponds to $d=0$, is given in equation (\ref{e8.5}).
It is very easy to show that the positive definite operator $\eta = \rho^2$ satisfies
$\eta \widetilde{V}(x) \eta^{-1} = \widetilde{V}^\dag(x)$ ensuring the potential to be pseudo-Hermitian. The potential $\widetilde{V}(x)$ also satisfies
$\widetilde{V}^*(-x) = \widetilde{V}(x)$ and hence is $\mathcal{PT}$-symmetric. The wave functions of the potential $\widetilde{V}(x)$ can be determined by $\widetilde{\psi}_n (x) = \rho^{-1} \psi_n (x) =
\psi_n \left(x + \frac{i \epsilon}{k}\right)$.
\begin{figure}[]
\centering

\subfigure[]{
   \includegraphics[width=2.90in,height=1.85in] {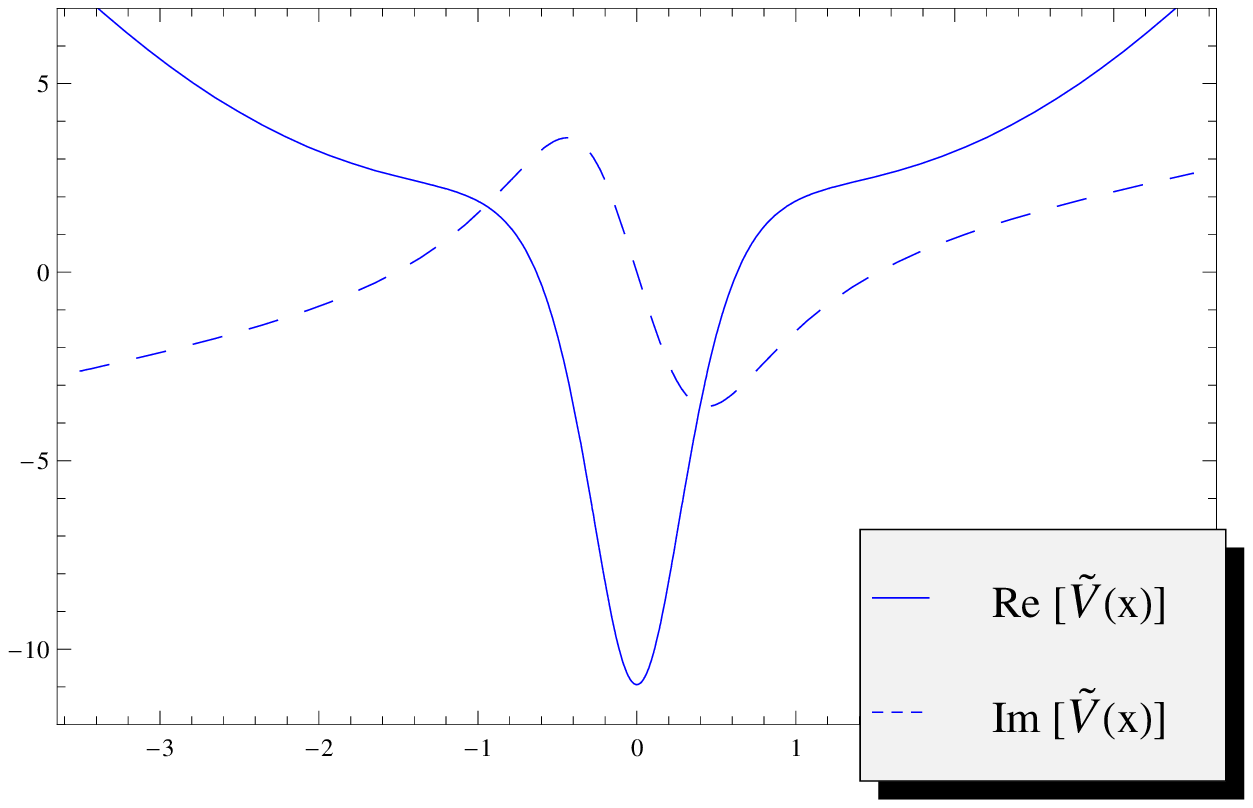}
   \label{f1a}
}
 \subfigure[]{
   \includegraphics[width =2.90in,height=1.85in] {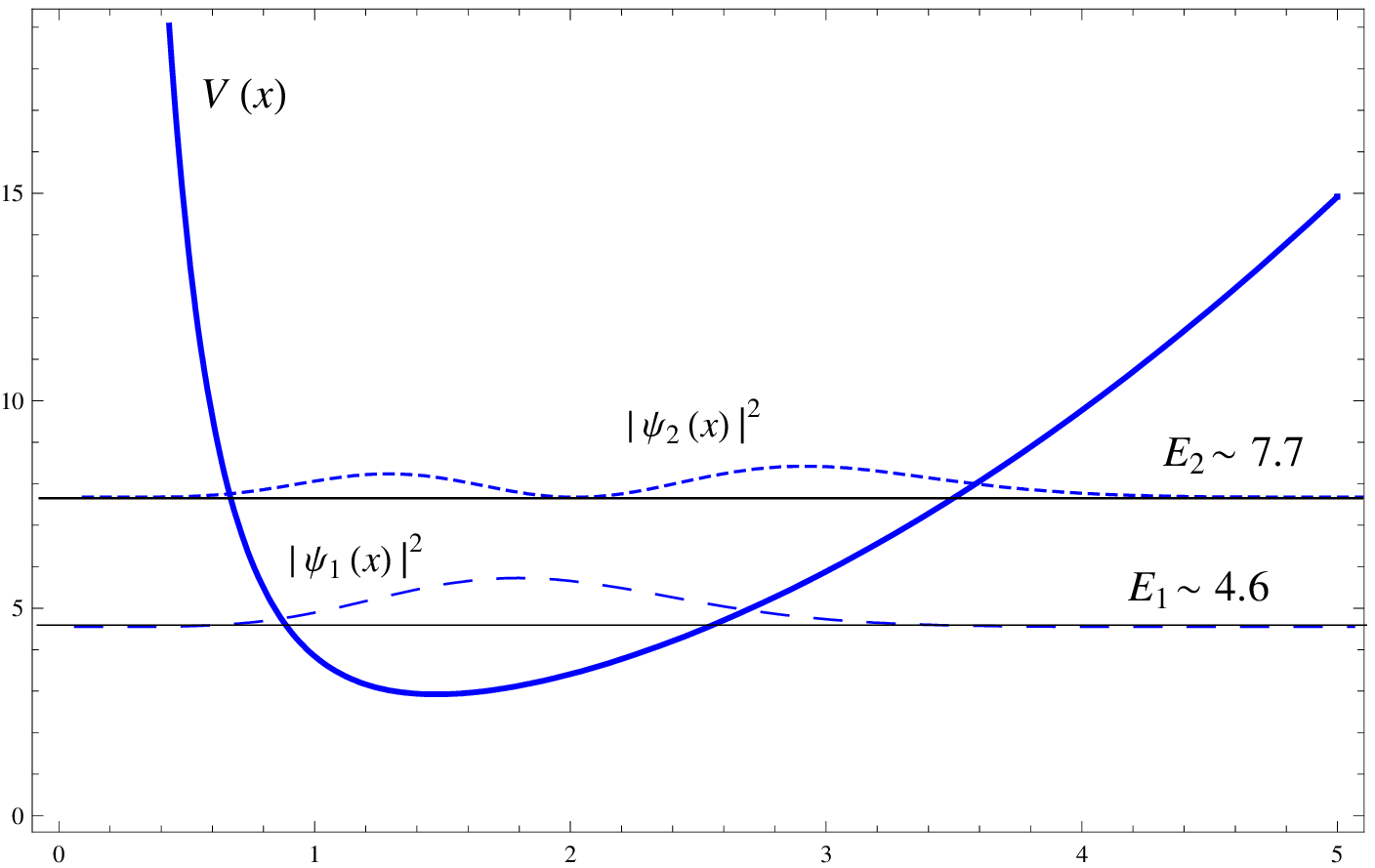}
   \label{f1b}
}
\caption{(a) Plot of the real (solid line) and imaginary (dashed line) parts of the quasi-Hermitian Potential $\widetilde{V}(x)$ associated with
 $X_1$ Laguerre EOPs. ~~(b). Plot of the corresponding equivalent Hermitian potential $V(x)$ (thick line) and square of the absolute value of its lowest
two wave functions. Here we have considered $\epsilon = 1.2,
a = 2, k = 1.75$.}\label{f3.4}
\end{figure}
\FloatBarrier
In figure \ref{f3.4}a, we have shown the real and imaginary parts of the potential $\widetilde{V}(x)$ given in (\ref{e8.33}), while figure \ref{f3.4}b
shows its equivalent Hermitian
analogue $V(x)$ given in (\ref{e8.5}). Using first two members of exceptional $X_1$ Laguerre polynomials
$\hat{L}_1^{(a)}(x) = -x-a-1, \hat{L}_2^{(a)}(x) = x^2 -a(a+2)$,
we have also plotted in figure \ref{f3.4}b
the absolute value of
first two wave functions given in (\ref{e8.34}).\\

Next we choose $F(g)$ to be Jacobi-type $X_1$ EOP, $\hat{P}_n^{(a,b)}$, which is defined for real $a,b >-1$, $a\ne b$ and $n=1,2,3..$.
In this case the expression for $Q(g)$ and $R(g)$ are given by \cite{UK09}
 \begin{equation}
 \begin{array}{ll}
\displaystyle Q(g) =-\frac{(a+b+2)g + a -b}{1-g^2} - \frac{2(b-a)}{(b-a)g - b -a}, \\
\displaystyle R(g) = -\frac{(b-a)g -(n+a+b)(n-1)}{1-g^2} - \frac{(a-b)^2}{(b-a)g -b-a}
\end{array}\label{8.00}
\end{equation}
Using these expressions in (\ref{e8.3}) and choosing $g'^2/(1-g^2) = \mbox{constant} = k^2 (k \ne 0)$, we have
\begin{equation}
g(x) = \sin (k x + d).\label{e8.38}
\end{equation}
Like the exceptional Laguerre polynomials, the choice $d=0$ gives rise to the potential, energies and corresponding bound state wave functions, as
\begin{equation}\begin{array}{ll}
V(x) =  \frac{k^2 (2 a^2 +2 b^2 -1)}{4} \sec^2 k x - \frac{k^2(b^2-a^2)}{2} \sec k x \tan k x + \frac{2 k^2(a+b)}{a+b - (b-a) \sin k x} - \frac{8 k^2 a b}{\left[a+b - (b-a) \sin k x\right]^2},\\ \\
E_n = \frac{k^2}{4} (2 n + a + b -1)^2,
\end{array}\label{e8.8}
\end{equation}
\begin{equation}
\psi_n(x) \approx ~ \frac{(1- \sin k x)^{\frac{a}{2}+\frac{1}{4}} (1+ \sin kx)^{\frac{b}{2}+
\frac{1}{4}}}{a+b- (b-a) \sin k x}~ \hat{P}_n^{(a,b)}(\sin k x), ~~~~ ~~~n= 1, 2, 3...\label{e8.39}
\end{equation}
respectively. The above periodic potential $V(x)$, which is free from singularity in the interval $-\frac{\pi}{2 k} <x < \frac{\pi}{2 k}$,
can be interpreted \cite{Qu08} as the rational extension of the standard trigonometric scarf potential which is
associated with classical Jacobi polynomials. The wave functions in equation (\ref{e8.39}) are regular iff $a,b>-1/2$.

Here, the non-Hermitian potential corresponding to the choice $d= i \epsilon$ is obtained as
\begin{equation}\begin{array}{ll}
\displaystyle \widetilde{V}(x) = \frac{k^2 (2 a^2 +2 b^2 -1)}{4} \sec^2 (k x + i \epsilon) - \frac{k^2(b^2-a^2)}{2} \sec (k x + i \epsilon) \tan (k x + i \epsilon)\\ \\
\displaystyle ~~~~~~~~~~~~~~+ \frac{2 k^2(a+b)}{a+b - (b-a) \sin (k x + i \epsilon)} +
 \frac{2 k^2[ (a-b)^2 - 4 a b]}{\left[a+b - (b-a) \sin (k x + i \epsilon) \right]^2}
\end{array}
\end{equation}
This potential $\widetilde{V}(x)$, which is defined on whole real line, also shares the same real eigenvalues of the potential
given in (\ref{e8.8}). Like the rationally extended radial oscillator the above non-Hermitian potential is also quasi-Hermitian under the the
operator $\rho$ defined in (\ref{e8.9}). The corresponding equivalent analogue is the one given in equation (\ref{e8.8}) which
corresponds to the choice $d=0$. The potential $\widetilde{V}(x)$ also fulfills the requirement of $\mathcal{PT}$-symmetry i.e.
$\widetilde{V}^*(-x) = \widetilde{V}(x)$, only if $a=b$. However, if we consider the other solution $g(x) = \cos(k x+d)$ of $g'^2/(1-g^2) = k^2$,
the corresponding potential $\widetilde{V}(x)$ becomes $\mathcal{PT}$-symmetric for all real values of $a$ and $b$.  The wave functions of $\widetilde{V}(x)$ can be determined by operating $\rho^{-1}$ on $\psi_n$ given in (\ref{e8.39}). Here, we have not considered the complex values of
$a,b$ because this will give rise to the exceptional Jacobi polynomials with complex indices and complex arguments. The orthogonality properties for such
 complex polynomials may depend on the interplay between integration contour and parameter values.
 \begin{figure}[]
\centering
\subfigure[]{
   \includegraphics[width =2.9in, height=1.95in] {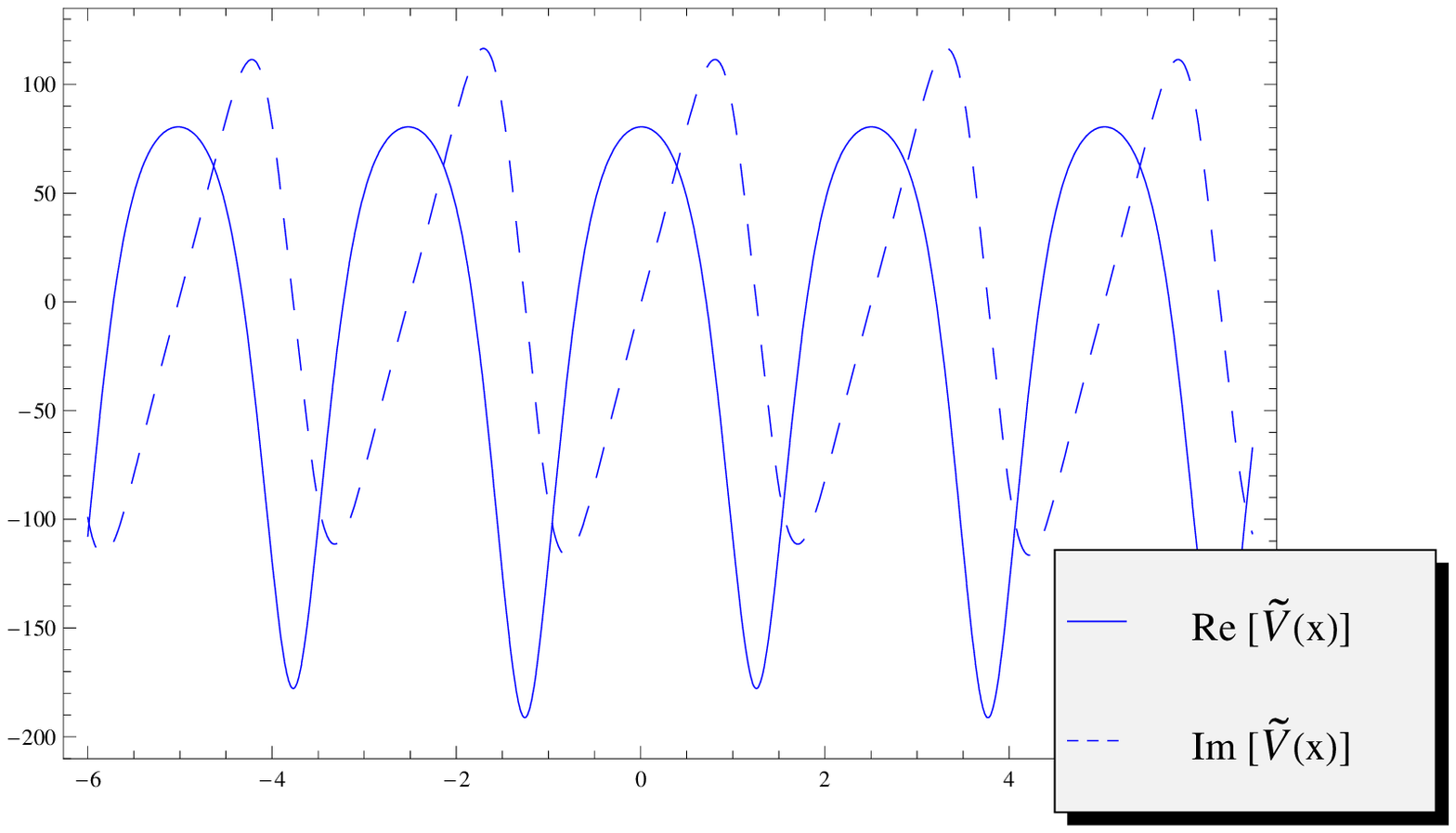}
   \label{f2a}
}
 \subfigure[]{
   \includegraphics[width =2.9in, height=1.95in] {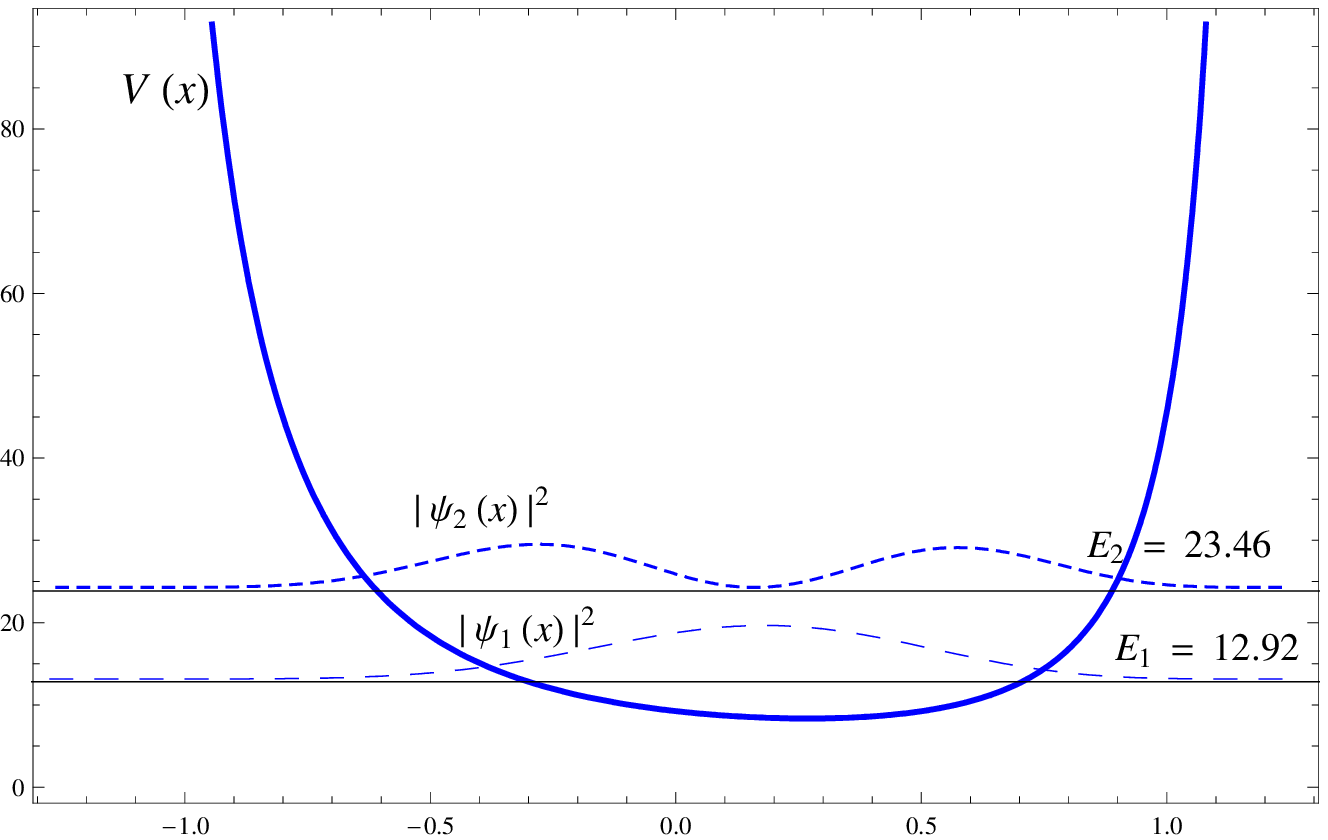}
   \label{f2b}
}
\caption[(a) Plot of the real (solid line) and imaginary (dashed line) parts of the quasi-Hermitian Potential $\widetilde{V}(x)$
 associated with $X_1$ Jacobi EOPs. ~~(b).
Plot of the corresponding equivalent Hermitian potential $V(x)$ (thick line)...]{(a) Plot of the real (solid line) and imaginary (dashed line) parts of the quasi-Hermitian Potential $\widetilde{V}(x)$
 associated with $X_1$ Jacobi EOPs. ~~(b).
Plot of the corresponding equivalent Hermitian potential $V(x)$ (thick line) and square of the absolute value of its lowest two wave functions.
Here we have considered $a = 1.75,
b = 3,
k = 1.25,
\epsilon = 1$.}\label{f3.5}
\end{figure}
\FloatBarrier
In figure \ref{f3.5}a, we have plotted the real and imaginary parts of the potential $\widetilde{V}(x)$. The corresponding equivalent
Hermitian analogue $V(x)$ and square of its first two wave functions are plotted in figure \ref{f3.5}b. We have used the expression
of first two members of Jacobi type $X_1$ EOPs, $\widehat{P}^{(a,b)}_1 = -\frac{x}{2} - \frac{2+a+b}{2(a-b)}$ and
$\widehat{P}^{(a,b)}_2 = -\frac{a+b+2}{4} x^2 - \frac{a^2+b^2+2(a+b)}{2(a-b)}x - \frac{a+b+2}{4}$ to plot the square of the wave functions.

\newpage
\section{Generalized Swanson Hamiltonian}\label{s3.3}

\subsection[Equivalent Hermitian analogue]{Generalized Swanson Hamiltonian and its equivalent Hermitian analogue}
\noindent The harmonic oscillator Hamiltonian augmented by a non-Hermitian ${\cal{PT}}$-symmetric part is one of the important example of pseudo-Hermitian Hamiltonian.
This non-Hermitian oscillator was first discussed by Swanson \cite{Sw04} who considered the Hamiltonian
 \begin{equation}
 H_{S} = w \left(a^{\dagger} a + \frac{1}{2}\right) + \alpha a^2 + \beta a^{\dagger^2},~~~~w, \alpha, \beta \in \mathbf{R}\label{e3.01}
 \end{equation}
 where $a, a^\dag$ are bosonic harmonic oscillator annihilation and creation operators satisfying usual commutation relationship $[a,a^\dag] = 1$.
If $\alpha \ne \beta$, $H_S$ is non-Hermitian ${\cal{PT}}$-symmetric (${\cal{P}}$-pseudo-Hermitian).
Swanson, with the help of a Bogoliubov type transformation, showed that the transformed Hamiltonian has the eigenvalue of a harmonic
 oscillator system with frequency $\sqrt{w^2-4\alpha \beta}$ so long as $w>\alpha+\beta$ and the eigenfunctions can be derived from
the eigenfunctions of the harmonic oscillator. Transition probabilities governed by the $H_{S}$ are also shown to be manifestly unitary \cite{Sw04}.
Consequently, many properties and generalizations of this Hamiltonian have been reported
 \cite{Jo05,BQR05,MGH07,Qu07,BT08,SR07}. In particular, it has been shown \cite{Jo05} that the non-Hermitian oscillator (\ref{e3.01}) can be mapped
to harmonic oscillator with the help of following similarity transformation,
\begin{equation}
h =\rho_{(\alpha,\beta)}~ H_S~ \rho_{(\alpha,\beta)}^{-1} = -\frac{1}{2} (w-\alpha-\beta)\frac{d^2}{dx^2} + \frac{1}{2} \frac{w^2-4 \alpha \beta}{w-\alpha -\beta} x^2
\end{equation}
 with $\rho_{(\alpha,\beta)} = \exp[-\frac{1}{2}\frac{\alpha-\beta}{w-\alpha-\beta}x^2]$. However, this choice for metric operator $\rho$ is not unique.
 A family of metric
 operators for the Hamiltonian $H_S$ have been constructed \cite{MGH07} and they are re-examined in the light of $su(1,1)$ \cite{Qu07}. In ref.\cite{BQR05}, the authors have explained the hidden
symmetry of the Hamiltonian $H_{S}$ and found the pseudo-Hermitian operators $\eta$ in the form $\eta= \rho^{-1}_{(\alpha,\beta)} \rho_{(\beta,\alpha)}$.
The positivity of $\eta$ is provided by the property satisfied by $\rho$ viz., $\rho_{(\alpha,\beta)} = \rho^{-1}_{(\beta,\alpha)}$.
It should be put emphasize here that if one writes $\eta= \rho^{-1}_{(\alpha,\beta)} \rho_{(\beta,\alpha)}$ then
the boundedness of the operator $\eta$ can not be guaranteed. If $\eta$ is not bounded metric operator an alternative
construction of $\rho$ can be made which is not based on the metric operator \cite{KS04}. The Hamiltonian given in (\ref{e3.01}) is generalized many
ways, for example, a general first order differential form of the annihilation and creation operators $\tilde{a}$ and $\tilde{a}^\dag$ are proposed \cite{BQR05}
\begin{equation}
\tilde{a}= A(x) \frac{d}{dx} + B(x)~~~~\tilde{a}^\dag = -A(x) \frac{d}{dx} + B(x) -A'(x)~~~~~ A(x), B(x) \in \mathbb{R}, \label{e4.3}
\end{equation}
where `prime' denotes derivative with respect to $x$ and $[\tilde{a},\tilde{a}^\dag] = 2A B'-A A''$ ($\ne 1$ in general).
After replacing $a, a^\dag$ by $\tilde{a}$ and $\tilde{a}^\dag$
 in equation (\ref{e3.01}) we have the the generalized Swanson Hamiltonian \cite{BQR05}
\begin{equation}
H_{GS} = w \left(\tilde{a}^{\dagger} \tilde{a} + \frac{1}{2}\right) + \alpha \tilde{a}^2 + \beta \tilde{a}^{\dagger^2},~~~~~~ w, \alpha, \beta \in \mathbb{R}.
\label{e4.00}
\end{equation}
 The Hamiltonian $H_{GS}$ is $\mathcal{PT}$-symmetric if $A(x)$ is an odd
function and $B(x)$ is an even function of $x$. For the general choice of $\tilde{a},\tilde{a}^\dag$ and $w-\alpha-\beta = 1$,
 the Hermitian equivalent form of the Hamiltonian $H_{GS}$ is obtained by a similarity transformation \cite{BQR05}
\begin{equation}
\tilde{h} = \rho_{(\alpha,\beta)} H_{GS}~ \rho_{(\alpha,\beta)}^{-1} = -\frac{d}{dx} A(x)^2 \frac{d}{dx}
+V_{eff} (x)\label{e4.6}
\end{equation}
where
\begin{equation}
\tilde{\rho}_{(\alpha,\beta)} = A(x)^{\frac{\alpha-\beta}{2}}
\exp\left(-(\alpha-\beta)\int^x \frac{B(x)}{A(x)}
dx'\right),\label{e4.7}
\end{equation}
\begin{equation}
\begin{array}{ll}
\displaystyle V_{eff}(x) = \frac{(\alpha+\beta) A A''}{2} + \left[\frac{\alpha+\beta}{2} + \frac{(\alpha-\beta)^2}{4}\right] A'^2 - 4 \tilde{w}^2 A'B \\
 \quad \quad \quad \quad \quad \displaystyle + 4 \tilde{w}^2 B^2 - (\alpha+\beta+1) A B' + \frac{\alpha+\beta+1}{2}\label{e4.15}
\end{array}
\end{equation}
\begin{equation}
\tilde{w} = \frac{\sqrt{1+ 2 (\alpha+\beta) + (\alpha-\beta)^2}}{2}.
\end{equation}
 There is an one to one correspondence between the energy eigenvalues of $\tilde{h}$ given in (\ref{e4.6}) and $H_{GS}$.
 If $\psi_n(x)$ are the wave functions of the equivalent Hermitian Hamiltonian $\tilde{h}$ then the wave functions of the
 Hamiltonian $H_{GS}$ are given by $\rho_{(\alpha,\beta)}^{-1} \psi_n(x)$.

Here our aim is to show that the equivalent Hermitian Hamiltonian (\ref{e4.6}) of the Hamiltonian (\ref{e4.00})
can be further transformed into a harmonic oscillator like Hamiltonian. It will also be shown that
though the commutator $[\tilde{a},\tilde{a}^\dag] = 1$, the generalized Swanson Hamiltonian is not
always isospectral to the harmonic oscillator. This anomaly will be explained in the frame work of position dependent mass scenario.\\

\subsection[Reduction of equivalent Hermitian Hamiltonian into harmonic oscillator]{Reduction of equivalent Hermitian Hamiltonian of $H_{GS}$ into harmonic oscillator}
\noindent We consider the commutator of $\tilde{a}$ and $\tilde{a}^{\dagger}$ as
\begin{equation}
[\tilde{a},\tilde{a}^\dagger] = 2 A(x) B'(x) - A(x) A''(x) = 1.
\end{equation}
On setting
\begin{equation}
z(x) = \int^x \frac{dx'}{A(x')}~~~~~~~~~~~~ B(x) =\frac{z''}{2z'^2} + \frac{1}{2} z, \label{e4.4}
\end{equation}
(for the sake of simplicity we assume integration constant to be zero), the Hamiltonian $\tilde{h}$ given in eq.(\ref{e4.6}) becomes
\begin{equation}
\tilde{h} = -\frac{d}{dx}A^2\frac{d}{dx} + V_{eff}(x),~~~~~V_{eff}(x) = \frac{1}{2} \frac{z'''}{z'^3} - \frac{5}{4} \frac{z''^2}{z'^4} +
\tilde{w}^2 z ^2.\label{e4.5}
\end{equation}
For the change of variable (\ref{e4.4}) we have
\begin{equation*}
A'= \frac{\dot{A}}{A},~~~~~~~ A'' = \frac{\ddot{A}}{A^2} -
\frac{\dot{A}^2}{A
^3}~~~~~ \mbox{etc.}
\end{equation*}
where `dot' represents derivative with respect to $z$.
Consequently the Schr\"{o}dinger eigenvalue equation for the
Hamiltonian $\tilde{h}$ reduces to
\begin{equation}
\left[-\frac{d^2}{dz^2} -\frac{\dot{A}}{A} \frac{d}{dz} +
\left(\frac{\dot{A}^2}{4A^2}-\frac{\ddot{A}}{2 A} + \tilde{w}^2 z
^2\right)\right]\psi(z) = E \psi(z)\label{e4.8}
\end{equation}
In order to eliminate the first derivative term from equation
(\ref{e4.8}) we assume
\begin{equation}
\psi(z) = A(z)^{-\frac{1}{2}} \phi(z)\label{e4.9}.
\end{equation}
For this assumption equation (\ref{e4.8}) reduces to
\begin{equation}
\left[-\frac{d^2}{dz^2} + \tilde{w}^2 z^2\right] \phi(z) = E \phi(z)\label{e4.10}
\end{equation}
This equation (\ref{e4.10}) is the Schr\"{o}dinger equation with harmonic oscillator like potential $V(z) = \tilde{w}^2 z^2$.
The important point is to note here that the co-ordinate transformation given by the first of the equation (\ref{e4.4}) has certain peculiarities.
The domain of $z(x)$ depends on the functional form of $A(x)$ and may not be the same as that of $x$.
This may lead to non-isospectrality of the original Hamiltonian $\tilde{h}$ and the transformed Hamiltonian.\\
Now the equation (\ref{e4.10}) can be transformed into confluent Hypergeometric equation
\begin{equation}
y \frac{d^2\chi}{dy^2} +\left(\frac{1}{2}-y\right) \frac{d\chi}{dy} + \left(\frac{E}{4\tilde{w}}-\frac{1}{4}\right) \chi =0\label{e14.2}
\end{equation}
by the following transformations
\begin{equation}
y= \tilde{w} z^2~~~~~~~~~~~~~ \phi(y) = e^{-\frac{y}{2}} \chi(y).\label{e4.11}
\end{equation}
Hence the general solution of the equation (\ref{e4.10}) are given by
\begin{equation}
\phi_{e} (z) \sim
e^{-\frac{\tilde{w}}{2} z^2} {_1}F_1
\left(\frac{1}{4}-\frac{E}{4 \tilde{w}},\frac{1}{2},\tilde{w}z^2\right)\label{e4.13}
\end{equation}
\begin{equation}
\phi_{o} (z) \sim
z e^{-\frac{\tilde{w}}{2} z^2} {_1}F_1
\left(\frac{3}{4}-\frac{E}{4 \tilde{w}},\frac{3}{2},\tilde{w}z^2\right),\label{e4.14}
\end{equation}
where $\phi_e$ and $\phi_o$ denote the even and odd solutions. The eigenfunctions of the Hamiltonian $\tilde{h}$ are given by
\begin{equation}
\psi(x)\sim A(z)^{-{\frac{1}{2}}} \phi(z),
\end{equation}
where $z(x)$ is given by equation (\ref{e4.4}).

It should be mentioned here that though the equivalent
Hermitian analogue of $H_{GS}$ is transformed into harmonic oscillator like Hamiltonian, they may not share the same eigenvalues. This apparent anomaly occurs because the domain of $x$
and the domain of the variable $z$ used in co-ordinate
transformation to obtain the harmonic oscillator are not necessarily same. In the next section, we determine the condition of
isospectrality between $H_{GS}$ and harmonic oscillator.

\subsection[Non-isospectrality between generalized Swansan Hamiltonian and harmonic oscillator]
{Non-isospectrality between generalized Swansan Hamiltonian and harmonic oscillator}\label{s3.6}
\noindent The condition of isospectrality between the the equivalent Hermitian Hamiltonian $\tilde{h}$ and harmonic oscillator can be obtained
by analyzing the behaviors of the eigenfunctions (\ref{e4.13}) and (\ref{e4.14}) which determine the eigenvalues of the equation (\ref{e4.10}).
If the domain of argument $\tilde{w} z^2$ is unbounded then $\phi_e(z)$  and $\phi_o(z)$ will not in general, square integrable because of
the asymptotic behavior of the confluent Hypergeometric function viz. \cite{AS65},
\begin{equation}
_1F_1(a,b,y) = \frac{\Gamma(b)}{\Gamma(a)} e^y y^{a-b} [1+O(y^{-1})],~~~~Re(y)>0.
\end{equation}
So in order to make the eigenfunctions square integrable one must take $a=-m$ $(m=0,1,2,3...)$ in which case $_1F_1(a,b,y)$ reduces to a polynomial. So from (\ref{e4.13}) $1/4 - E/(4 \tilde{w}) = -m$ or equivalently
\begin{equation}
E_{2m} = 2\tilde{w}\left(2 m + \frac{1}{2}\right)\label{e4.16}
\end{equation}
and from (\ref{e4.14}) $3/4-E/(4 \tilde{w}) = -m$ or equivalently
\begin{equation}
E_{2m+1}= 2\tilde{w}\left(2m+\frac{3}{2}\right)\label{e4.17}
\end{equation}
Combining (\ref{e4.16}) and (\ref{e4.17}) we have
\begin{equation}
E_n = 2 \tilde{w} \left(n + \frac{1}{2}\right),~~ n=0,1,2...
\end{equation}
Hence in this case the Hamiltonian $\tilde{h}$ (as well as $H_{GS}$) has the spectrum of a harmonic oscillator.

However, if the domain of $\tilde{w}z^2$ is finite then the required boundary condition to be satisfied by the eigenfunctions (\ref{e4.13}) and (\ref{e4.14}) is that they must vanish at the end points of the domain of $z$ and the eigenvalues are given by the zeroes of the confluent hypergeometric functions when the arguments attain their end points. The first approximation of the $m$'th ($m=1,2...$) positive zero $X_0$ of ${_1}F_1(a,b,y)$ is given by \cite{AS65}
\begin{equation}
X_0 = \frac{\pi^2\left(m+\frac{b}{2}-\frac{3}{4}\right)^2}{2 b -4 a} \left[1+ O\left(\frac{1}{(\frac{b}{2}-a)^2}\right)\right],~~~m=1,2...
\end{equation}
Proceeding in a similar way, as discussed just above, we have from eqns.(\ref{e4.13}) and (\ref{e4.14})
\begin{equation}
E_{2m}\approx \frac{\tilde{w}\pi^2}{4 z_{\pm}^2} (2m-1)^2,~~~E_{2m-1} \approx \frac{\tilde{w}\pi^2}{ z_{\pm}^2} m^2
\end{equation}
respectively. Combining these two one can write
\begin{equation}
E_n \approx \frac{\tilde{w} \pi^2}{4 z_{\pm}^2} (n-1)^2,~~n=1,2,3...\label{e4.18}
\end{equation}
where $z_{\pm}$ are the end points of the domain of definition of $V(z)$.
From the above fact it follows that, in this case the Hamiltonian $\tilde{h}$ (as well as $H_{GS}$) is not isospectral to the harmonic oscillator.

\subsection{Examples}
\noindent Here we give some concrete examples in favor of our obtained results.\\
 The generalization of the Swanson model enables us to connect those physical systems which are describable by a position dependent mass
 by choosing $ A(x) = M(x)^{-1/2}$ which is a strictly positive function. For this choice of $A(x)$ the equation (\ref{e4.6})
 reduces to the time independent PDMSE
\begin{equation}
\left(-\frac{d}{dx} \frac{1}{M(x)}\frac{d}{dx} + V_{eff} (x)
\right)\psi(x) = E \psi(x)
\end{equation}
with  $V_{eff}(x)$ is given by equation (\ref{e4.5}).  At this point we consider different mass profiles $M(x)$ to illustrate that the Hamiltonian given in (\ref{e4.6})
is not always isospectral to harmonic oscillator.\\

\noindent
{\bf Example 1: Isospectral case}\\
Let us consider the following mass function
\begin{equation}
M(x) = \frac{1}{1+x^2}~,~~x\in (-\infty,\infty)
\end{equation}
which has been considered in the study of quantum nonlinear
oscillator \cite{Ca+07}. For this choice of mass function, $z(x)$ is
given by
\begin{equation}
z(x) = \sinh^{-1} (x).
\end{equation}
It is clear that $z(x)\rightarrow \pm \infty$ as $x\rightarrow \pm\infty$.
So according to the condition obtained in section \ref{s3.6}, we have the Hamiltonian $\tilde{h}$ with the effective potential $V_{eff}(x)$
\begin{equation}
V_{eff} (x) = -\frac{2+x^2}{4(1+x^2)} + \tilde{w}^2 (\sinh^{-1}~ x)^2
\end{equation}
is isospectral to harmonic oscillator.
In table \ref{t3.1}, we have given a list of physically interesting mass functions, $V_{eff}(x)$, and eigen energies, for which the corresponding position dependent mass Hamiltonians are isospectral to the harmonic oscillator.\\
\begin{table}[h]
\begin{center}
\begin{tabular}{|c|c|c|}
  \hline
  $M(x)$ & $z(x)$ &$V_{eff}(x)$ \\
  \hline
   $\frac{1}{(1+x^2)}$&$\sinh^{-1}~x$  & $-\frac{2+x^2}{4(1+x^2)} + \tilde{w}^2 (\sinh^{-1}~ x)^2$ \\
   \hline
   $\cosh^2 x$&$\sinh~x$  & $\frac{1}{8} (7-3\cosh 2x) sech^4~x + \tilde{w}^2 \sinh^2~x$  \\
   \hline
   $\left(\frac{\gamma+x^2}{1+x^2}\right)^2$&$x+(\gamma-1)\tan^{-1}~x$  & $ \frac{(\gamma-1)(3x^4-2(a-2)x^2-a)}{(x^2+\gamma)^4}+\tilde{w}^2 [x+(\gamma-1)\tan^{-1}~x]^2$  \\
   \hline
   $e^{2 x} \sech^2~x$&$\log (1+e^{2 x})$  & $-\frac{3}{4} e^{-4 x}-\frac{1}{2}e^{-2 x}+\tilde{w}^2 [\log (1+e^{2 x})]^2  $   \\
   \hline
   $e^{-x}$& $-2e^{-\frac{x}{2}}$&$-\frac{3}{16} e^x+4 \tilde{w}^2 e^{-x}$\\
  \hline
  \end{tabular}
\end{center}
  \caption[Table of mass functions and effective potentials
for which the corresponding generalized Swanson Hamiltonians are isospectral to harmonic oscillator ]{Some physically interesting mass functions and corresponding effective potentials
for which the corresponding Hamiltonians $\tilde{h}$ are isospectral to harmonic oscillator with
$E_n=2\tilde{w}\left(n+\frac{1}{2}\right),n=0,1,2...$. In all these cases $z(x)$ are unbounded as $x\rightarrow\pm \infty.$ }\label{t3.1}
  \end{table}

\noindent {\bf Example 2: Non-isospectral case}\\
Now let us choose
\begin{equation}
M(x) = \sech^2(x),~~~x\in (-\infty,\infty)
\end{equation}
which depicts the solitonic profile \cite{Ba07}. For this choice, the function
$z(x)$ reeds
\begin{equation}
z(x) = \tan^{-1}(\sinh~ x).
\end{equation}
Here as $x\rightarrow\pm \infty, z(x)\rightarrow \pm \pi/2$, so the eigenvalues  of the Hamiltonian $\tilde{h}$ with
\begin{equation}
V_{eff}(x) = \frac{1}{4} - \frac{3}{4} \cosh^2~x +
\tilde{w}^2\left(\tan^{-1} (\sinh~x)\right)^2,
\end{equation}
will be given by zeroes of the functions given in eqns.(\ref{e4.13}) and (\ref{e4.14}) at $z=\pm \pi/2$. First approximate value of the energy eigenvalues are given by equation (\ref{e4.18})
\begin{equation}
E_n \approx n^2, ~n=1,2,3...
\end{equation}
In table \ref{t3.2}, we have given a list of physically interesting mass functions, $V_{eff}(x)$, and eigen energies,
 for which the corresponding Hamiltonians $\tilde{h}$ are not isospectral to the harmonic oscillator.

  \begin{table}[h]\begin{center}
  \begin{tabular}{|c|c|c|c|}
  \hline
  $M(x)$ & $z(x)$ &$V_{eff}(x)$ & $E_n \approx$\\
  \hline
  $\sech^2 x$& $\tan^{-1}(\sinh~x)$&$\frac{1}{4}-\frac{3}{4}\cosh^2 x +\tilde{w}^2 (\tan^{-1}(\sinh~x))^2$&$n^2$\\
  \hline
  $e^{-2x^2}$&$\frac{\sqrt{\pi}}{2} \ef~ x$&$-(1+3x^2)e^{2 x^2}+\frac{\pi\tilde{w}^2}{4} (\ef~ x)^2$ &$\pi n^2$\\
  \hline
  $\frac{1}{(1+x^2)^2}$& $\tan^{-1} ~x$ & $-(1+2 x^2)+\tilde{w}^2 (\tan^{-1}~x)^2$&$n^2$\\
  \hline
  \end{tabular}
\end{center}
  \caption[Table of mass functions and effective potentials
for which the corresponding generalized Swanson Hamiltonians are not isospectral to  harmonic oscillator ]{ Some physically interesting mass functions and corresponding
effective potentials for which the corresponding Hamiltonians $\tilde{h}$ are not isospectral to
harmonic oscillator. In all these cases $z(x)$ are finite as $x\rightarrow\pm \infty.$ Here $n$ takes the values $1,2,3...$ }\label{t3.2}
  \end{table}

\section{Summary}
\noindent To summarize, we point out the following main results :
\begin{itemize}
 \item[$\blacktriangleright$] {We have shown that for a positive definite operator $\eta$ the potential (\ref{e3.1}) is
$\eta$ pseudo-Hermitian for $V_0<.5$ which ensures that all the
band edges are real. It has also been that the eigenvalue problem (\ref{e3.4}) reduces to solving the Mathieu equation for all real $V_0$ .
 The band structure of the same potential has been studied using Floquet analysis. It is also shown that, in addition to the $\mathcal{PT}$-threshold
at $V_0 = 0.5$, there exist a second critical point near $V_0^c \sim .888437$ after which no part
of the eigenvalues and the band structure remains real.}

\item[$\blacktriangleright$]{Some exactly solvable non-Hermitian Hamiltonians
 whose bound state wave functions are associated with Laguerre and Jacobi-type $X_1$ exceptional orthogonal polynomials have been generated using the method of point canonical transformation.
 The Hamiltonians are shown to be quasi-Hermitian in nature so that the energy spectrum is real.
 It is to be noted here that the other choices of $g(x)$ in the expression $E-V(x)$ associated with Laguerre and Jacobi EOPs give
 rise to the several other exactly solvable Hermitian as well as quasi-Hermitian extended potentials.
 But in all these cases we have to redefine the parameters carefully so that $n$ dependent term appears only in the constant energy.}

\item[$\blacktriangleright$]{ We have studied a class of non-Hermitian Hamiltonians of the form
$H_{GS} = w(\tilde{a}\tilde{a}^\dag + 1/2) +\alpha \tilde{a}^2 + \beta \tilde{a}^{\dag^2}$,
where $w,\alpha,\beta$ are real constants and $\tilde{a},\tilde{a}^\dag$ are generalized annihilation
 and creation operators. When $\tilde{a}$ and $\tilde{a}^\dag$ satisfy the commutation relation $[\tilde{a},\tilde{a}^\dag] =$constant, 
 the Hamiltonian $H_{GS}$ has been shown to be transformed into
a harmonic oscillator like Hamiltonian. This reveals an intriguing result in the sense that in this case the resulting Hamiltonian does not always possess the spectrum of the
harmonic oscillator. Reason for this anomaly is discussed in the frame work of position dependent mass
models.}
\end{itemize}


\chapter[Quantum Nonlinear Oscillator]{Quantum Nonlinear Oscillator \footnote{This chapter is based on following two papers:\\
 (i) B. Midya, B. Roy and A. Biswas, ``Coherent state of nonlinear oscillator and its revival dynamics", Physica Scripta 79 (2009) 065003;
 (ii) B. Midya and B. Roy, ``A generalized quantum nonlinear oscillator", J. Phys. A42 (2009) 285301.}}  \label{c4}
\pagestyle{Chapter}

\noindent
The nonlinear differential equation
\begin{equation}
(1+\lambda x^2)\ddot {x} - (\lambda x)\dot {x}^2 + \alpha^2 x = 0,~~~\lambda > 0\label{e5.1}
\end{equation}
was first studied by  Mathews and Lakshmanan \cite{ML74} as an example of a non-linear oscillator and it was shown that the solution of (\ref{e5.1}) is
\begin{equation}
x = A ~\sin(\omega t + \phi)\label{e5.2}
\end{equation}
with the following additional restriction linking frequency and amplitude $\omega^2 = \frac{\alpha^2}{1+\lambda A^2}$.
Furthermore equation (\ref{e5.1}) can be obtained from the Lagrangian
\cite{ML74,LR03}
\begin{equation}
L = \frac{1}{2}\frac{1}{(1+\lambda x^2)}(\dot {x}^2 - \alpha^2
x^2)\label{e5.4}
\end{equation}
so that both the kinetic and the potential energy terms depend on the same
parameter $\lambda$. This $\lambda$-dependent system can be
considered as:
\begin{itemize}
\item[(i)]{ a deformation of the standard harmonic
oscillator in the sense that for $\lambda \rightarrow 0$ all the characteristics of the linear oscillator are recovered}
 \item[(ii)] {a particular case of a system with a position
dependent effective mass of the form $m(x) = \frac{1}{1+\lambda x^2}$.}
\end{itemize}
Recently, this particular nonlinear system has been
generalized to the higher dimensions and various properties of
this system have been studied \cite{Ca+04,Ca+05}. The classical Hamiltonian
corresponding to the $\lambda$-dependent oscillator is given by
\cite{ML74,LR03}
\begin{equation}
H_{cl} = \left(\frac{1}{2m}\right)P_x^2 +
\left(\frac{1}{2}\right)g\left(\frac{x^2}{1+\lambda
x^2}\right),~~~~~ P_x = \sqrt{1+\lambda
x^2}p_x,~~g=m\alpha^2,\label{e5.3}
\end{equation}
where $p_x$ is the canonically conjugate momentum defined by $p_x =
\frac{\partial L}{\partial \dot{x}}$, ${L}$ being the Lagrangian and $m$ is the mass. It has been shown \cite{Ca+07} that in the space ${\cal{L}}^2(\Re,d\mu)$ where $d\mu = \left(\frac{1}{\sqrt{1+\lambda x^2}}\right)dx$, the
differential operator $\sqrt{1+\lambda x^2}\frac{d}{dx}$ is skew
self adjoint. Therefore,
 contrary to the naive expectation of ordering ambiguities, the transition from the classical system to
 the quantum one is given by defining the momentum operator $P_x = -i \sqrt{1+\lambda x^2}\frac{d}{dx}$ so that
$$
(1+\lambda x^2)p_x^2 \rightarrow -\left(\sqrt{1+\lambda x^2}\frac{d}{dx}\right)
\left(\sqrt{1+\lambda x^2}\frac{d}{dx}\right).$$
Therefore the quantum version of the Hamiltonian (\ref{e5.3}) with
$\hbar=1$ becomes \cite{Ca+07}
\begin{equation}
H = -\frac{1}{2m}(1+\lambda x^2)\frac{d^2}{dx^2} - \left(\frac{1}{2m}\right)\lambda x\frac{d}{dx}
 + \frac{1}{2}g\left(\frac{1}{1+\lambda x^2}\right)\label{e5.00}
\end{equation}
where $g = \alpha (m\alpha + \lambda)$. It is to be
 noted here that in reference \cite{Ca+07} the value of the parameter $g$ has been slightly modified from that given in equation (\ref{e5.3}). After introducing adimensional variables $(y,\Lambda)$ as was done in \cite{Ca+07}
\begin{equation}
  x = \sqrt{\frac{1}{m\alpha}}y~~; ~~ \lambda = {m\alpha}\Lambda,\label{e5.21}
\end{equation}
  the Schr\"{o}dinger equation $H\psi = E\psi$ reduces to
\begin{equation}
\left[(1+\Lambda y^2)\frac{d^2}{dy^2} + \Lambda y \frac{d}{dy} -(1+\Lambda)\frac{y^2}{(1+\Lambda y^2)} + 2e\right]\psi = 0\label{e5.03}
\end{equation}
where $E = e( \alpha)$. The eigenvalues and eigenfunctions for $\Lambda < 0$ are given by
\cite{Ca+07}
\begin{equation}
\begin{array}{lcl}
\psi_n(y,\Lambda) &=& (1 - |\Lambda|y^2)^{\frac{1}{(2|\Lambda|)}} ~ {\cal {H}}_n(y,\Lambda)\\
\epsilon_n &=& (n+\frac{1}{2}) - \frac{1}{2}n^2 \Lambda~~~,~~~n =
0,1,2,...
\end{array} \label{e5.01}
\end{equation}
where ${\cal {H}}_n(y,\Lambda)$ is $\Lambda$-deformed Hermite
polynomial. For $\Lambda > 0$,
\begin{equation}
\begin{array}{lcl}
\psi_n(y,\Lambda) &=& (1+\Lambda y^2)^{-\frac{1}{2\Lambda}} ~ {\cal {H}}_n(y,\Lambda)\\
\epsilon_n &=& n +\frac{1}{2}) - \frac{1}{2}n^2\Lambda~~~,~~~n = 0,1,2
\cdots,N_{\Lambda}
\end{array}\label{e5.02}
\end{equation}
where $N_{\Lambda}$ denotes the greatest integer lower than
$n_{\Lambda}(=\frac{1}{\Lambda})$.

\newpage

\noindent Here we have first studied, in the next section, the coherent state of the above mentioned quantum nonlinear oscillator (QNLO)
and its revival dynamics.
In section \ref{s4.2}, we have examined various
Hermitian and non-Hermitian generalizations of QNLO.
In doing so a relationship between $\Lambda$-deformed Hermite polynomial and Jacobi polynomial has been established in section \ref{s4.3}.

\section {Gazeau-Klauder coherent state for nonlinear oscillator}\label{s4.1}
\noindent We have first constructed the Gazeau-Klauder coherent states \cite{Kl96,GK99} of the QNLO and then studied their revival dynamics to demonstrate the effect of mass parameter $\lambda$ on the temporal evaluation of these states. The motivation comes from the following two facts (i) the temporal evolution of the coherent states of systems, like  P\"oschl-Teller,
 Morse and Rosen-Morse, possessing nonlinear energy spectra can lead to revival and
 fractional revival, leading to Schr\"odinger cat and cat-like states, (ii) the study of temporal evolution of a free wave-packet with position dependent mass inside an infinite well \cite{Sc06} has revealed
that revival and partial revivals are not only different from the constant mass case but also they are very much dependent on the mass function $m(x)$. It should be noted here that, the mass parameter $\lambda$ of QNLO may be positive or negative. However for $\lambda>0$ the discrete energy spectrum of QNLO are finite and consequently for completeness property the continuum has to be taken into account. On the other hand for $\Lambda<0$ there are only
discrete energy states and henceforth we shall consider this choice. \\
From equation (\ref{e5.01}) the eigenvalues of the Hamiltonian $H^1 = H - \frac{\alpha}{2}$ are given by
\begin{equation}
E_n^1 = E_n - \frac{\alpha}{2} = \hbar\alpha\left[n+\frac{1}{2}n^2|\Lambda|\right]= \alpha\frac{n(n+\mu)}{\mu} = \alpha e_n^1\label{e5.9}
\end{equation}
where $\mu = \frac{2}{|\Lambda|}$.
The Gazeau-Klauder coherent state \cite{GK99} for this system is given by
\begin{equation}
|J,\gamma\rangle = \frac{1}{N(J)}\sum_n \frac{(J)^{\frac{n}{2}}exp(-i\gamma e_n^1)}{\sqrt{\rho_n}}|n\rangle \label{e5.10}
\end{equation}
where $\gamma = \alpha t$. The normalization constant $N(J)$ is given by
\begin{equation}
N(J) = \left[\sum_n \frac{J^n}{\rho_n}\right]^{\frac{1}{2}}~~~,0<J<R = \limsup_{n \rightarrow +\infty}\rho_n^{\frac{1}{n}}\label{e5.11}
\end{equation}
where $R$ denotes the radius of convergence and $\rho_n$ denotes the moments of a probability distribution $\rho(x)$
\begin{equation}
\rho_n = \int_0^R x^n \rho(x)dx = \prod_{i=1}^n e_i^1~~;~~\rho_0 = 1\label{e5.12}
\end{equation}
For the coherent state (\ref{e5.10})
\begin{equation}
\rho_n = \prod_{i=1}^n \frac{i(i+\mu)}{\mu} = \frac{\Gamma (n+1)\Gamma (n+1+\mu)}{\mu^n \Gamma(1+\mu)}~~,\rho_0 =1\label{5.13}
\end{equation}
so that $R$ is infinite and
\begin{equation}
\rho(J) = \frac{2\mu(J\mu)^{\frac{\mu}{2}}}{\Gamma (1+\mu)}K_{\mu}(2\sqrt{J\mu}),\label{5.14}
\end{equation}
$K_{\nu}(cx)$ being modified Bessel function \cite{AS65}. Also
\begin{equation}
N(J)^2 = \frac{\Gamma (1+\mu)}{(J\mu)^{\frac{\mu}{2}}}I_{\mu}(2\sqrt{J\mu})\label{e5.15}
\end{equation}
where $I_{\mu}(cx)$ is the modified Bessel function \cite{AS65}. So the coherent state (\ref{e5.10}) finally becomes
\begin{equation}
|J,\gamma\rangle = \frac{\sqrt{\Gamma (1+\mu)}}{N(J)}\sum_{n=0}^{\infty} \frac{(J\mu)^{\frac{n}{2}}
e^{-i\alpha (n+\frac{n^2}{\mu})t}}{\sqrt{n!\Gamma (n+1+\mu)}}|n\rangle\label{e5.16}
\end{equation}
Below we shall see that the coherent state (\ref{e5.16}) satisfies following four conditions:\\
{\bf 1. Continuity of labeling:} From the definition (\ref{e5.10}) it is obvious that\\
$$(J^{\prime} \gamma ^{\prime}) \rightarrow (J,\gamma) \Rightarrow |J^{\prime},\gamma^{\prime}\rangle \rightarrow |J,\gamma\rangle.$$\\
{\bf 2. Resolution of unity:} \\
{\small $$
\int|J,\gamma\rangle \langle J,\gamma|d\mu(J,\gamma) = \int|J,\gamma\rangle\langle J,\gamma|d\mu(J,\gamma) = \frac{1}{2\pi}
\int_{-\pi}^{\pi}d\gamma \int_0^{\infty}k(J)|J,\gamma \rangle \langle J,\gamma| ~dJ \label{e5.17}
$$}
where $k(J)$ is defined by
\begin{equation}
\begin{array}{lcl}
k(J) &=& N(J)^2\rho(J) \geq 0, ~~~~0 \leq J <R \\
     &=& \rho(J)\equiv 0,~~~ J>R.
\end{array}\label{e5.18}
\end{equation}
For the coherent state (\ref{e5.16})
\begin{equation}
k(J) = 2\mu I_{\mu}(2\sqrt{J\mu}) K_{\mu}(2\sqrt{J\mu})\label{e5.19}
\end{equation}
so that the resolution of unity is satisfied
$$
\int|J,\gamma\rangle \langle J,\gamma|d\mu(J,\gamma) = 1. \label{e5.20}
$$
{\bf 3. Temporal stability:} ~~~ $e^{-{i H^{1}t}} |J,\gamma \rangle = |J,\gamma + \alpha t \rangle.$\\
{\bf 4. Action identity:} ~~~ $\langle J, \gamma |H^{1}|J,\gamma \rangle = \alpha J .$\\
The overlapping of two coherent states is given by
\begin{equation}
 \langle J^{\prime},\gamma'|J,\gamma \rangle = \frac{\Gamma(\mu+1)}{N(J)N(J')}\sum_{n=0}^{\infty}
\frac{(JJ'\mu^2)^{\frac{n}{2}}e^{i(\gamma'-\gamma)e_n^1}}{n!\Gamma(n+1+\mu)}. \label{e5.23}
\end{equation}
If $\gamma = \gamma^{\prime}$, the overlapping is reduced to
\begin{equation}
 \langle J^{\prime},\gamma|J,\gamma \rangle = \frac{1}{\sqrt{I_{\mu}(2\sqrt{J\mu})I_{\mu}(2\sqrt{J'\mu})}} ~ I_{\mu}\left(2(JJ'\mu^2)^{\frac{1}{4}}\right). \label{e5.24}
\end{equation}

\subsection{Revival dynamics}\label{s4.1.2}
\begin{figure}[]
\epsfxsize=3.2 in \epsfysize=2.5 in
\centerline{\epsfbox{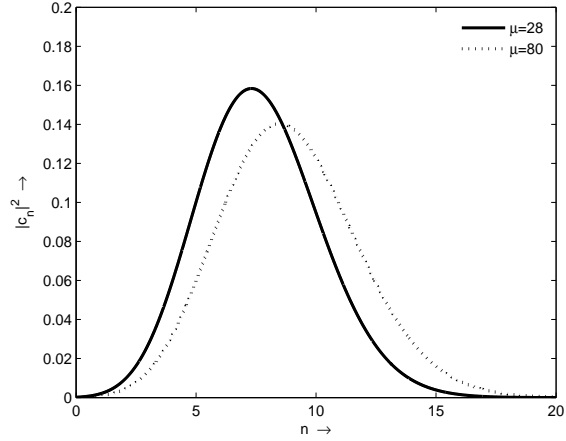}} \label*{}\caption{Plot of the
weighted distribution $|c_n|^2$ given in (\ref{e5.26}) for $J=10$ and
$\mu=28,80.$}\label{f5.1}
\end{figure}

\noindent In this section we shall study the revival dynamics of the coherent states (\ref{e5.16}) which are constructed in the previous section. To demonstrate the role of the mass parameter $\lambda$ on the revival dynamics, all the figures below are drawn for a fixed value of $J =10$
and two different values of $\mu$ which is inversely proportional to the mass parameter $\Lambda$ ($\mu = \frac{2}{|\Lambda|}$) .
For a general wave packet of the form
\begin{equation}
|\psi(t)\rangle = \sum_{n \geq 0}c_n e^{-iE_n t}|n \rangle \label{e5.25}
\end{equation}
with $\sum_{n \geq 0}|c_n|^2 = 1$, the concept of revival arises from the weighting probabilities $|c_n|^2$.
For the coherent state (\ref{e5.16}), the weighting distribution is given by
\begin{equation}
|c_n|^2 = \frac{(J\mu)^{n+\frac{\mu}{2}}}{n!\Gamma(n+1+\mu)I_{\mu}(2\sqrt{J\mu})}\label{e5.26}
\end{equation}
Since the weighting distribution $|c_n|^2$ is crucial for understanding the temporal behavior of
the coherent state (\ref{e5.16}), we show the curves of $|c_n|^2$ for $J=10$ and different $\mu$ in figure \ref{f5.1}.
\FloatBarrier

\begin{figure}[]
\epsfxsize=3.2 in \epsfysize=2.5 in
\centerline{\epsfbox{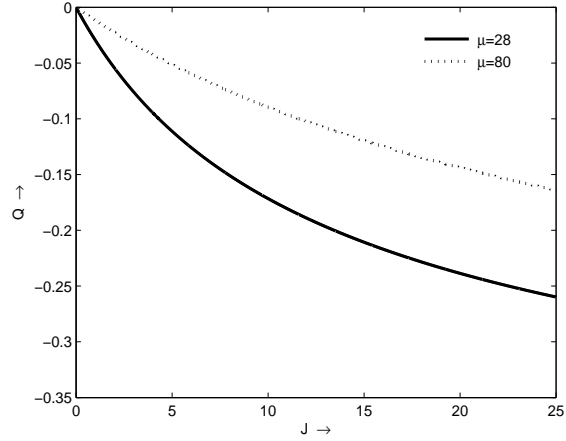}} \label*{}\caption[Plot of the Mandel parameter $Q$ ...]{ Plot of the
Mandel parameter $Q$ given in (\ref{e5.29}) for $J=10$ and $\mu=28,80.$}\label{f5.2}
\end{figure}

\noindent
The Mandel parameter $Q$ is defined by
\begin{equation}
Q = \frac{(\Delta n)^2}{\langle n \rangle} - 1 \label{e5.27}
\end{equation}
where
\begin{equation}
\begin{array}{lcl}
\langle n \rangle &=& \displaystyle \sum_{n=0}^{\infty} \frac{n J^n}{N(J)^2\rho_n}\\
\Delta n &=& [ \langle n^2 \rangle- \langle n \rangle^2]^{\frac{1}{2}}.
\end{array}\label{e5.28}
\end{equation}
The Mandel parameter determines the nature of the weighting distribution function $|c_n|^2$.
The case of $Q = 0$ coincides with the Poissonian weighting distribution $\displaystyle \frac{\langle n \rangle^n e^{- \langle n \rangle }}{n!}$ while the cases of
$Q>0$ and $Q<0$ correspond to the super-Poissonian or sub-Poissonian statistics respectively. For the coherent state (\ref{e5.16}),
\begin{equation}
Q = \sqrt{J\mu}\left[\frac{I_{\mu+2}(2\sqrt{J\mu})}{I_{\mu+1}(2\sqrt{J\mu})} - \frac{I_{\mu+1}(2\sqrt{J\mu})}{I_{\mu}(2\sqrt{J\mu})}\right] \label{e5.29}
\end{equation}
where
\begin{equation}
\langle n \rangle = \sqrt{J\mu} \frac{I_{\mu+1}(2\sqrt{J\mu})}{I_{\mu}(2\sqrt{J\mu})}. \label{e5.30}
\end{equation}
In figure \ref{f5.2}, we plot the Mandel parameter $Q$ for $J=10$ and $\mu = 28,80$.
It is evident from the figure that the Mandel parameter is sub-Poissonian and it has been observed that it remains so for all values $\mu$.

\FloatBarrier

Now assuming that the spread $\Delta n = [\langle n^2 \rangle -\langle n \rangle^2]^{\frac{1}{2}}$ is small compared to $\langle n \rangle \approx \bar {n}$,
we expand the energy $E_n^1$ in a Taylor series in $n$ around the centrally excited value $\bar {n}$:
\begin{equation}
E_n^1 \approx E_{\bar {n}}^1 + E_{\bar {n}}^{1'}(n-\bar {n}) + \frac{1}{2}E_{\bar {n}}^{1''}(n-\bar {n})^2 +
\frac{1}{6}E_{\bar {n}}^{1'''}(n-\bar {n})^3 + \cdots \label{e5.31}
\end{equation}
where each prime on $E_{\bar{n}}^1$ denotes a derivative with respect to $n$.
These derivatives define distinct time scales, namely the classical period $T_{cl} = \displaystyle \frac{2\pi}{|E_{\bar{n}}^{1'}|}$,
the revival time $t_{rev} = \displaystyle \frac{2\pi}{\frac{1}{2}|E_{\bar{n}}^{1''}|}$and so on. For $E_n^1$ in Eqn.(\ref{e5.9}),
$T_{cl} = \displaystyle \frac{2\pi}{\alpha(2\bar {n}+\mu)}$ and $t_{rev} = \displaystyle \frac{2\pi\mu}{\alpha}$. There is no
superrevival time here because the energy is a quadratic function in $n$. It is convenient to describe the wave packet
dynamics by an autocorrelation function
\begin{equation}
A(t) = \langle \psi(x,0)|\psi(x,t)\rangle = \sum_{n \geq 0}|c_n|^2 e^{{-iE_n^1t}} \label{e5.32}
\end{equation}
For the coherent state (\ref{e5.16}) the autocorrelation function (\ref{e5.32}) is given by
\begin{equation}
A(t) = \langle J,0|J,\alpha t \rangle = \frac{\Gamma (1+ \mu)}{N(J)^2}\sum_{n \geq 0}\frac{(J\mu)^n}{n!\Gamma(n+1+\mu)}e^{-i\alpha(n+\frac{n^2}{\mu})t}. \label{e5.33}
\end{equation}
\begin{figure}[]
\epsfxsize=6 in \epsfysize=4.5 in \centerline{\epsfbox{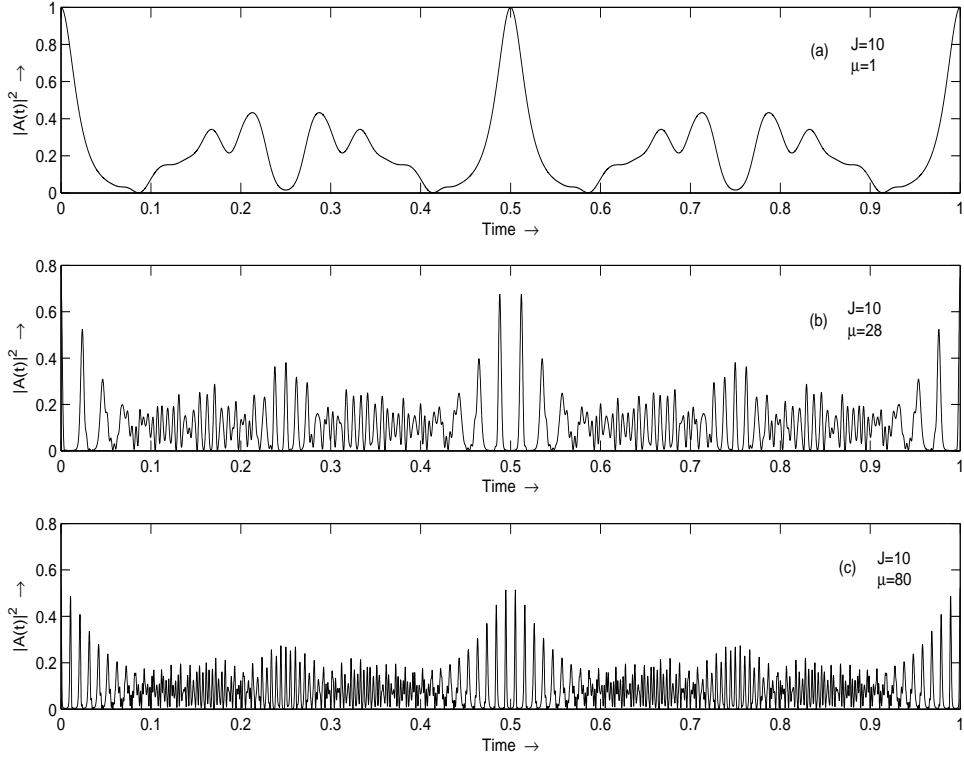}}
\label*{}\caption[Plot of the autocorrelation function $|A(t)|^2$... ]{Plot of the square modulus of the
autocorrelation function $|A(t)|^2$ given in (\ref{e5.33}) for $(a) J=10,
\mu=1, ~(b) J=10,\mu=28, (c) J=10, \mu=80.$} \label{f5.3}
\end{figure}
\FloatBarrier
From (\ref{e5.33}) it follows that $|A(t-t_{rev})|^2 = |A(t)|^2$ so that
the autocorrelation function is symmetric about
$\frac{t_{rev}}{2}$. In other words whatever happens in
$[0,\frac{t_{rev}}{2}]$ is repeated subsequently. We note that $0
\leq |A(t)|^2 \leq 1$ and it denotes the overlap between the
coherent state at $t=0$ and the one at time $t$. So, the larger
the value of $|A(t)|^2$, the greater the coherent state at time
$t$ will resemble the initial one. In figures \ref{f5.3}(a), (b) and (c) we plot the squared modulus of the autocorrelation function
(\ref{e5.33}) in units of $t_{rev}$ for $J=10$ and $\mu = 1,28,80$ respectively. Figure \ref{f5.3}(a) clearly shows the full revival. The sharp peaks in Figure \ref{f5.3}(b) and \ref{f5.3}(c) arise due to fractional revivals.
We can see from figures \ref{f5.3}(a) to \ref{f5.3}(c) that different values of $\mu$, which is inversely proportional to the mass parameter $\Lambda$, lead to qualitatively different types of motion of the coherent wave packet.\\

Now we shall study the mechanism of the fractional revival by phase analysis \cite{Ve+93,Ve+94}.
For this,we use the time scale determined by the period of the complete revival $t_{rev} = \frac{2\pi\mu}{\alpha} = 1$.
So the phase of the $n$-th stationary state is
\begin{equation}
\phi_n(t) = 2\pi(\mu n + n^2)t \label{e5.34}
\end{equation}
where $\mu = \frac{2}{|\Lambda|}$ is assumed to be an integer. At arbitrary moments of time, the phases (\ref{e5.34}) of
individual components of the packet are uniformly mixed so that it is not possible to make any definite conclusion about the value of the autocorrelation function. However, at specific moments, the distribution of phases can gain some order; for example, the phases can split into several groups of nearly constant values. The fractional revival of $q$-th order is defined as the time interval during which the phases are distributed among $q$ groups of nearly constant values. To consider fractional revival of order $q$, in the vicinity of time $t = \frac{1}{q}$, it is convenient to write $n$ as
$n = kq + \Delta$, where $k = 0,1,2,\cdots$ and $\Delta = 0,1,2,\cdots q-1$. Then the autocorrelation function can be written as
\begin{equation}
A(t) = \sum_{\Delta = 0}^{q-1}P_{\Delta}(t) \label{e5.35}
\end{equation}
where
\begin{equation}
P_{\Delta}(t) = \sum_k c_{kq+\Delta}e^{-i\phi_{kq+\Delta}(t)}, \label{e5.36}
\end{equation}
\begin{equation}
c_{kq+\Delta}=\frac{(J\mu)^{kq+\Delta+\frac{\mu}{2}}}{(kq+\Delta)!\Gamma(kq+\Delta+1+\mu)I_{\mu}(2\sqrt{J\mu})} \label{e5.37}
\end{equation}
and $\phi_{kq+\Delta}(t)$ is given by (\ref{e5.34}) replacing $n$ by $kq+\Delta$.
For quantum numbers that are multiples of the revival order i.e.$n = kq$, we have
\begin{equation}
\begin{array}{lcl}
\phi_{kq}(\frac{1}{q}) &=& 2\pi(k\mu + k^2q) = 0~ (mod 2\pi)\\
P_0(\frac{1}{q}) &=& \sum_k c_{kq}
\end{array} \label{e5.38}
\end{equation}
When $\Delta \neq 0$, the phases of the corresponding states are
\begin{equation}
\begin{array}{lcl}
\phi_{kq+\Delta}(\frac{1}{q}) &=& 2\pi(\mu\Delta+\Delta^2)/q\\
P_{\Delta}(\frac{1}{q}) &=& exp[-2\pi i(\mu\Delta q^{-1}+\Delta^2q^{-1})\sum_k c_{kq+\Delta}
\end{array} \label{e5.39}
\end{equation}
where in obtaining (\ref{e5.37}) and (\ref{e5.38}) we have made use of the result
\begin{equation}
2\pi(k(\mu+2\Delta)+k^2 q) = 0 (mod 2\pi) \label{e5.40}
\end{equation}
Thus from (\ref{e5.38}) and (\ref{e5.39}) it follows that, around time
$t=\frac{1}{q}$ the Gazeau-Klauder coherent state (10) splits into
$q$ packet fractions such that the $\Delta = 0$ packet fraction
has zero phase while relative to this the other packet fractions
have a constant phase $(\mu \Delta q^{-1} + q^{-1}{\Delta}^2)$.

\begin{figure}[]
\epsfxsize=7 in \epsfysize=3.5 in
\centerline{\epsfbox{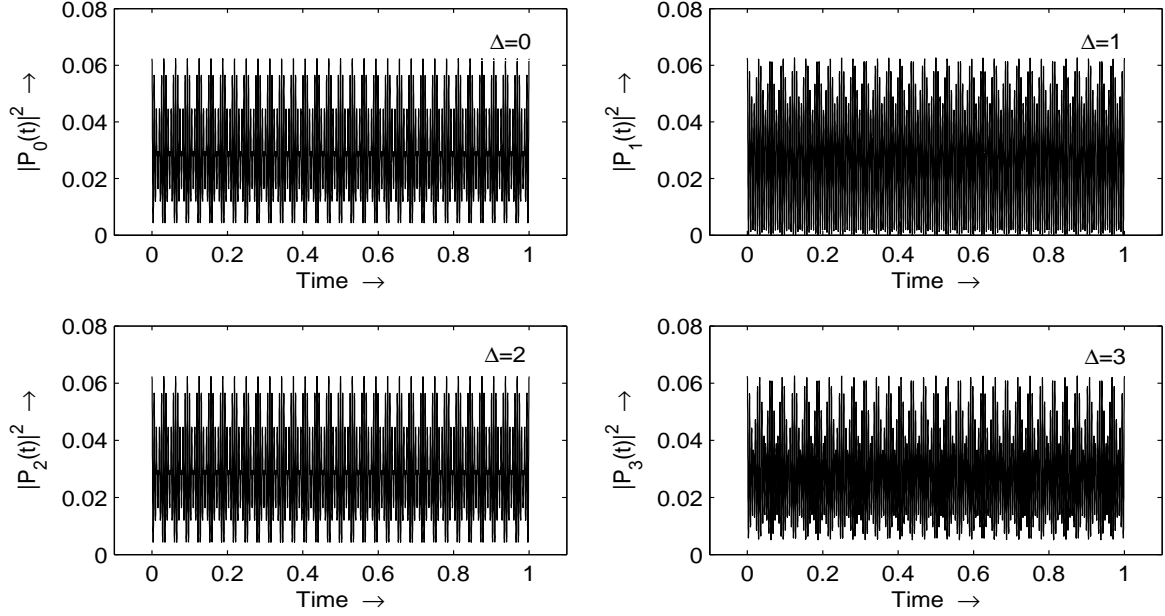}} \label*{}\caption{Plots of the
Survival functions $|P_{\Delta}(t)|^2$ given in (\ref{e5.37}) for
$\Delta=0,1,2,3$ and $J=10, \mu=28 .$}\label{f5.4}
\end{figure}
\FloatBarrier

\begin{figure}[]
\epsfxsize=7 in \epsfysize=3.5 in
\centerline{\epsfbox{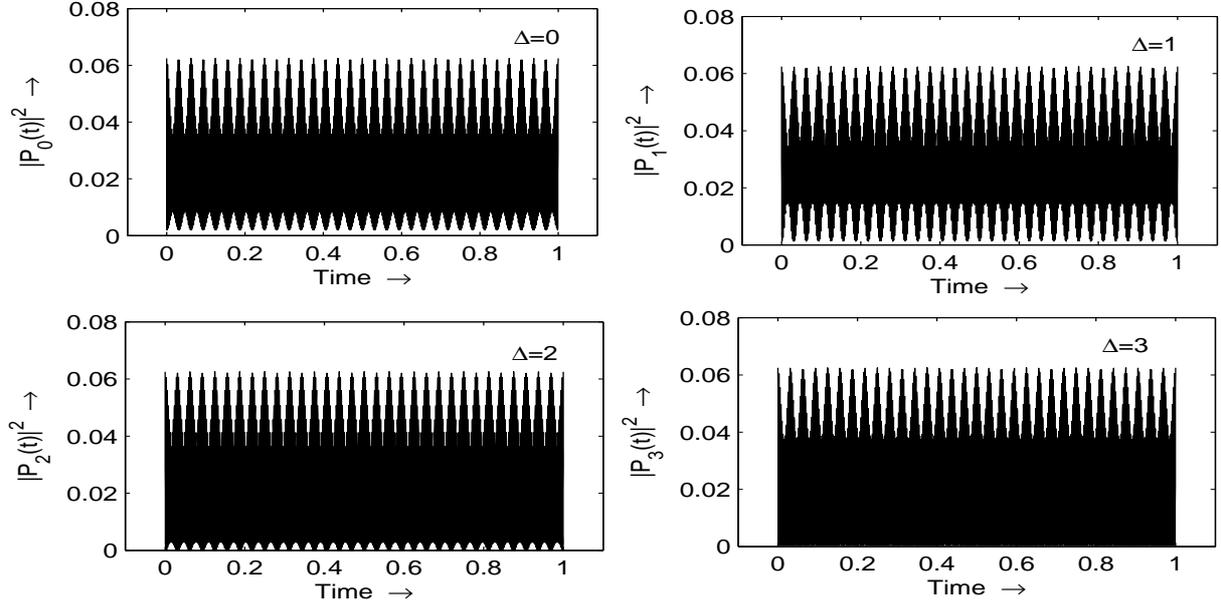}}\label*{} \caption{Plots of the
Survival functions $|P_{\Delta}(t)|^2$ given in (\ref{e5.37}) for
$\Delta=0,1,2,3$ and $J=10, \mu=80.$}\label{f5.5}
\end{figure}

It must be mentioned that the way the packet is divided into
fractions is determined only by the order of the revival analyzed.
Figure \ref{f5.4} and Figure \ref{f5.5} give the time dependence of the
survival functions $|P_{\Delta}(t)|^2$ in units $t_{rev}$ for
$\Delta = 0,1,2,3$ during the whole period of the complete revival
for fixed $J=10$ and $\mu = 28,80$ respectively.

\FloatBarrier

The intensity of the fractional revival can be defined as the value of the wave packet survival function at the moment of revival.
This value is composed of two terms: the sum of survival functions for the packet fractions ( the diagonal term) and the interference term
describing the interaction of the packet fractions
\begin{equation}
S\left(\frac{1}{q}\right) = \sum_{\Delta = 0}^{q-1} |P_{\Delta}\left(\frac{1}{q}\right)|^2 + \sum_{\Delta = 0}^{q-1}
\sum_{\Gamma \neq \Delta}P_{\Delta}\left(\frac{1}{q}\right)P_{\Gamma}^*\left(\frac{1}{q}\right) \label{e5.41}
\end{equation}

\begin{figure}[]
\epsfxsize=6 in \epsfysize=2.5 in
\centerline{\epsfbox{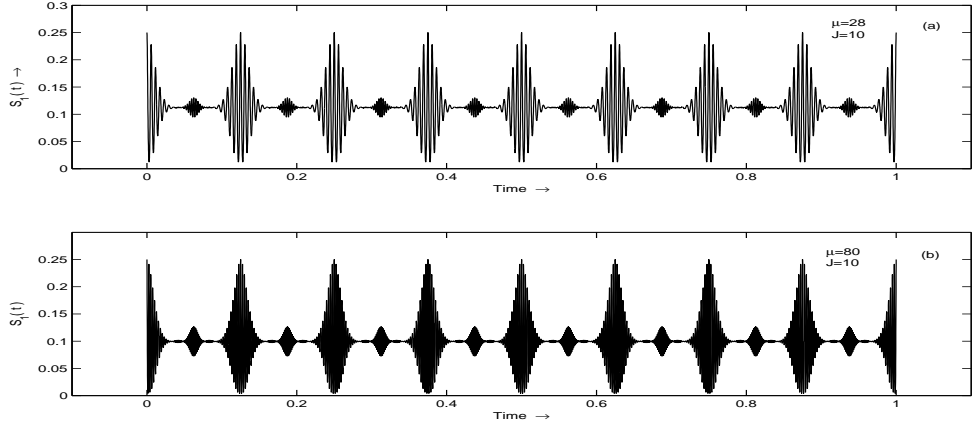}} \caption[Plots of the diagonal terms of the survival function ...]{Plots of the
diagonal term $S_1(t)=\Sigma_{\Delta=0}^3 |P_{\Delta}(t)|^2$ of the
survival function for (a)$J=10, \mu = 28$ and (b) $J=10, \mu=80.$}\label{f5.6}
\end{figure}
\FloatBarrier

\begin{figure}[]
\epsfxsize=6 in \epsfysize=2.5 in
\centerline{\epsfbox{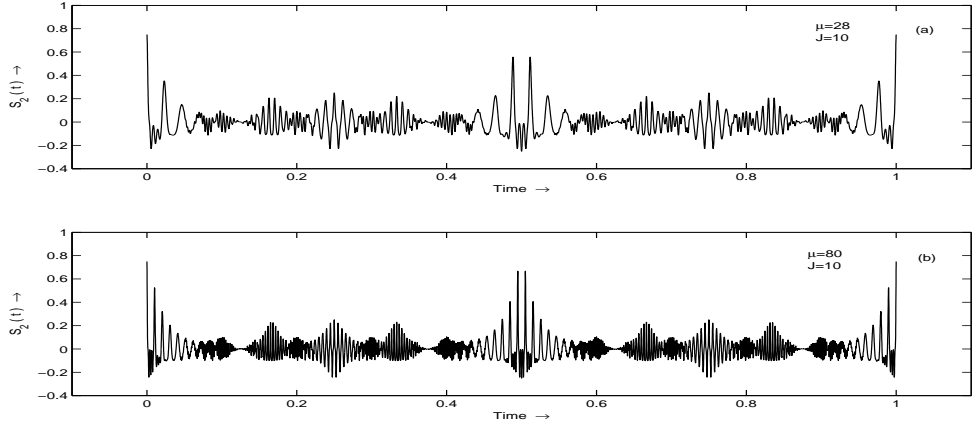}} \caption[ Plots of the interference terms of the survival function ...]{Plots of the
interference term $S_2(t)=\Sigma_{\Delta=0}^3 \Sigma_{\Gamma\neq
\Delta}^{3} P_{\Delta}(t)P_{\Gamma}^\ast(t)$ of the survival
function.(a) $J=10 , \mu=28$ ; (b) $J=10 , \mu=80.$}\label{f5.7}
\end{figure}

In Figure \ref{f5.6}(a) and Figure \ref{f5.6}(b) we present the time dependence of the diagonal
terms in units of $t_{rev}$ for fixed $J=10$ and $\mu = 28,80$ respectively. It is seen from the figures that
the diagonal term related to the individual packet fractions takes positive values only near the moments of fractional revival.
Figure \ref{f5.7}(a) and Figure \ref{f5.7}(b) show the time dependence of the
interference terms in units of $t_{rev}$ for fixed $J=10$ and $\mu
= 28,80$ respectively. It is seen from the figures that the
interference term plays a constructive, destructive or indifferent
role near the moments of fractional revival.

\FloatBarrier

\newpage

\section{Generalization of quantum nonlinear oscillator}\label{s4.2}

 \noindent Here we shall obtain a number of exactly solvable Hermitian, non-Hermitian $\mathcal{PT}$-symmetric and $\mathcal{PT}$-symmetric quasi exactly solvable potentials corresponding to the QNLO. In all these cases
 the same mass function $m(x) = \frac{1}{1+\lambda x^2}$ has been considered.

 \subsection{Hermitian generalization}\label{s4.2.1}
 \noindent We have the Schr\"{o}dinger equation corresponding to the Hamiltonian given in Eqn.(\ref{e5.00})
 with $m=1$ and $\lambda > 0$ as
\begin{equation}
\left[-(1+\lambda
    x^2)\frac{d^2\psi}{dx^2}-\lambda x
    \frac{d\psi}{dx} - \frac{g}{\lambda}\left(\frac{1}{1+\lambda x^2}\right)\right]\psi = E
    \psi\label{e6.5}
\end{equation}
\begin{equation}
E = 2e - \frac{g}{\lambda}\label{e6.43}
\end{equation}
where $e$ is the energy of the Hamiltonian (\ref{e5.00}). Now
expanding $(1+\lambda x^2)^{-1}$ for
$|x|<\frac{1}{\sqrt{\lambda}}$ we can write the potential of
equation (\ref{e6.5}) as
 \begin{equation}
 V(x)= -\frac{g}{\lambda} + g x^2-\lambda~ O(x^3)\label{e6.44}
\end{equation}
It is clear from (\ref{e6.44}) that the term
$(-\frac{g}{\lambda})$ in equation (\ref{e6.43}) cancels from both
sides of the equation (\ref{e6.5}), so that the new eigenvalues
(\ref{e6.43}) are actually the old eigenvalues $e$ of the
Hamiltonian (\ref{e5.00}).

Now generalizing the
potential of the  equation (\ref{e6.5}) as below, the corresponding
Schr\"{o}dinger equation now reads
    \begin{equation}
    -(1+\lambda
    x^2)\frac{d^2\psi}{dx^2}-\lambda x
    \frac{d\psi}{dx}+\left[\frac{B^2-A^2-A\sqrt{\lambda}}{1+\lambda x^2}+B(2A+\sqrt{\lambda}) \left(\frac{\sqrt{\lambda}x}
    {1+\lambda x^2}\right)+A^2\right]\psi=E\psi\label{e6.0}
    \end{equation}
It is seen from (\ref{e6.0}) that if we put $B=0$ then the
potential reduces to that of the nonlinear oscillator with
$\frac{g}{\lambda} = A^2 + A\sqrt{\lambda}$. This particular generalization is made
so that it corresponds to the hyperbolic Scarf II potential
\cite{CK95} in the constant mass case. In order to solve
(\ref{e6.0}), we now perform a transformation involving
change of variable given
by \cite{KO90}
\begin{equation} z = \int \frac{dx}{\sqrt{F(x)}} =
\frac{1}{\sqrt{\lambda}}~ \sinh^{-1}(\sqrt{\lambda}x)\label{t}
\end{equation}
where
\begin{equation}
\begin{array}{lcl}
F(x)&=&1+\lambda x^2~~,~~\lambda > 0\\
\end{array}\label{e6.6}
\end{equation}
Under the transformation (\ref{t}), Eqn.(\ref{e6.0}) reduces to a Schr\"{o}dinger equation
\begin{equation}
-\frac{d^2\psi}{dz^2} + V(z)\psi(z) = E\psi(z)\label{rsch}
\end{equation}
where the potential $V(z)$ is given by
\begin{equation}
V(z) =
(B^2-A^2-A\sqrt{\lambda}) \sech^{2}\left(z\sqrt{\lambda}\right) +
B(2A+\sqrt{\lambda})~\tanh\left(z\sqrt{\lambda}\right) \sech\left(z\sqrt{\lambda}\right)+A^2~~~\label{pot1}
\end{equation}
The potential (\ref{pot1}) is a standard solvable potential and
the solutions are given by \cite{CK95}
    \begin{equation}
     \psi_{n}(z)=N_n {i^n} \left(1+ \sinh
     ^{2}(z\sqrt{\lambda})\right)^{-\frac{s}{2}} e^{-r
     \tan^{-1}\left(\sinh(z\sqrt{\lambda})\right)}~P_{n}^{(ir-s-\frac{1}{2},-ir-s-\frac{1}{2})}
     \left(i~\sinh(z\sqrt{\lambda}\right)\label{e7}
     \end{equation}
 where $N_n$ is the normalization constant ,
  $s=\frac{A}{\sqrt{\lambda}},~ ~r=\frac{B}{\sqrt{\lambda}}$ and $P_n^{(\alpha,\beta)}(x)$ is
   the Jacobi Polynomial \cite{AS65}.
The normalization constants $N_n, n=0,1,2,...$ are given by \cite{Le01},
    \begin{equation}N_n=\left[\frac{ \sqrt{\lambda}~ n!~(s-n)\Gamma(s-ir-n+\frac{1}{2})\Gamma{(s+ir-n+\frac{1}{2})}}
    {\pi~2^{-2s}\Gamma(2s-n+1)}\right]^{1/2}\label{e6.40}
    \end{equation}
The eigenvalues $E_n$ are given by
    \begin{equation}
    E_n=n\sqrt{\lambda}(2A-n\sqrt{\lambda}),~~~~~
    n=0,1,2,...<s\label{en2}
    \end{equation}
 Subsequently by performing the inverse of the transformation (\ref{t}) we find the
 solution of eqn.(\ref{e6.0}) as
    \begin{equation}
    \begin{array}{lll}
    \displaystyle \psi_{n}(x)=\left[\frac{\sqrt{\lambda} ~n!~(s-n)\Gamma(s-ir-n+\frac{1}{2})\Gamma{(s+ir-n+\frac{1}{2})}}
    {\pi~2^{-2s}\Gamma(2s-n+1)}\right]^{1/2}\\
    \displaystyle ~~~~~~~~~~~~~~~~~~
    i^n (1+\lambda x^2)^{-\frac{s}{2}} ~e^{-r \tan^{-1}(x\sqrt{\lambda})}~
    P_n^{(ir-s-\frac{1}{2},-ir-s-\frac{1}{2})}(ix\sqrt{\lambda}),\\
    \displaystyle ~~~~~~~~~~~~~~~~~~~~~~~~~~~~~~~~~~~~~~~~~~~~~ n=0,1,2,\cdots <s~(=\frac{A}{\sqrt{\lambda}})
    \end{array}\label{e6.8}
    \end{equation}
 In the following, we show that the above mentioned generalization of QNLO
 produces all the results of a linear harmonic oscillator in the limit $\lambda\rightarrow 0$. For $B=0$,
$A=\frac{\alpha}{\sqrt{\lambda}}$ the potential of equation
(\ref{e6.0}) and it's energy eigenvalues (\ref{en2}) reduces to
\begin{equation}
V(x)=\left(-\frac{\alpha^2}{\lambda}-\alpha\right) (1+\lambda
x^2)^{-1}+\frac{\alpha^2}{\lambda} \label{I1}
\end{equation}
\begin{equation}
E_n=2n\alpha-n\lambda.\label{I2}
\end{equation}
respectively. For $|x| < \frac{1}{\sqrt{\lambda}}$, the potential (\ref{I1}) can be written as
\begin{equation}
 \begin{array}{ll}
\displaystyle V(x)=\left(-\frac{\alpha^2}{\lambda}-\alpha\right)
(1-\lambda x^2
 + \lambda^2 x^4-\lambda^3 x^6+...) +\frac{\alpha^2}{\lambda}\\
 \displaystyle
 ~~~~~~~=\alpha^2 x^2-\lambda(\alpha^2 x^4-\lambda \alpha^2
 x^6+...)+\lambda (\alpha x^2-\lambda \alpha x^4+...)-\alpha
 \end{array}\label{I3}
 \end{equation}
For $\lambda\rightarrow 0$ the potential reduces to
\begin{equation}
 V(x)=\alpha^2 x^2-\alpha \label{I4}
\end{equation}
It is clear from (\ref{I4}) and (\ref{I2}) that for
$\lambda\rightarrow 0$ the potential of the equation (\ref{e6.0}) and the energy
eigenvalues (\ref{en2}) reduces to those of a simple harmonic
oscillator. Now, for $A=\frac{\alpha}{\sqrt{\lambda}}, B=0$ and using the relation
(\ref{e9}) the expression for the wavefunction (\ref{e6.8}) is reduced to
\begin{equation}
 \psi_n(x)=N'_n (1+\lambda x^2)^{-\frac{2\alpha}{\lambda}}~
\mathcal{H}_n\left(\sqrt{\alpha}
x,\frac{\lambda}{\alpha}\right)\label{I5}
\end{equation}
where
\begin{equation}
N'_n=\frac{1}{2^n
n!}\left(\frac{\alpha}{\lambda}\right)^{\frac{n}{2}}N_n ~~=\left[\frac{\alpha^n (\frac{\alpha}{\lambda}-n)
\Gamma(\frac{\alpha}{\lambda}-n+\frac{1}{2})~\Gamma(\frac{\alpha}{\lambda}-n+\frac{1}{2})}{\pi
n!~
2^{2n-\frac{2\alpha}{\lambda}}\lambda^{n-\frac{1}{2}}\Gamma(\frac{2\alpha}{\lambda}-n+1)}\right]^{1/2}
\label{I6}
\end{equation}
When $\lambda\rightarrow 0$ the $\lambda$-deformed Hermite
polynomial becomes the conventional Hermite polynomial $H_n$
{\cite{Ca+07}}. Consequently at $\lambda\rightarrow 0$ limit the
unnormalized wave function given in equation (\ref{I5}) reduces
to
\begin{equation}
\psi_n(x) \propto e^{-\frac{\alpha x^2}{2}} H_n(\sqrt{\alpha}x)\label{I7}
\end{equation}
Using the asymptotic formula  $\Gamma(a z + b) \sim
\sqrt{2\pi}e^{-az}(az)^{az+b-\frac{1}{2}}$ (see 6.1.39 of the
ref. \cite{AS65}) in (\ref{I6}) we have
\begin{equation}
N'_n=\left(\frac{\sqrt{\alpha}-\frac{n\lambda}{\sqrt{\alpha}}}{\sqrt{\pi}2^n n!}\right)^{1/2}. \label{I8}
\end{equation}
Therefore from equations (\ref{I7}) and (\ref{I8}) it follows
that for $\lambda\rightarrow 0$ the wave function given in
equation (\ref{e6.8}) reduce to that of simple harmonic oscillator.\\

At this point it is natural to ask the following question :
 Are there other solvable potentials for the QNLO corresponding to the mass function $m(x)=\left(\frac{1}{1+\lambda x^2}\right)$?
  The answer to this question is in the affirmative. The procedure to obtain these potentials is similar to the one already described and so instead of treating each case separately we have presented the potentials and the corresponding solutions
 in Table \ref{t6.1}. The first two and the last two potentials in Table \ref{t6.1}
 are actually the generalizations of the nonlinear oscillator potential. Although the other two potentials in the Table are not
 generalizations of the nonlinear oscillator potential, nevertheless they are exactly solvable potentials with the same mass
 function. Incidentally all the potentials in table \ref{t6.1} are shape invariant as shown in section \ref{s4.3} and so the energy eigenvalues and eigenfunctions can also be obtained algebraically.

\subsection {Non-Hermitian $\mathcal{PT}$-symmetric generalization}\label{s4.2.2}

\noindent Here we shall find some complex potentials for which the QNLO is exactly solvable. For a constant mass Schr\"odinger Hamiltonian the condition for $\mathcal{PT}$-symmetry reduces to $V(x)=V^*(-x)$. However, in the case of position dependent mass an additional condition is required. To see this, we note that in the
present case the Hamiltonian is of the form
\begin{equation}
H = -\frac{1}{2m(x)}\frac{d^2}{dx^2}-\frac{m^\prime(x)}{2m^2(x)}\frac{d}{dx}+V(x),\label{hamil3}
\end{equation}
with $m(x) = \frac{1}{1+\lambda x^2}$. The conditions for the Hamiltonian (\ref{hamil3}) to be $\mathcal{PT}$-symmetric are given by \begin{equation}
m(x) = m(-x)~~,~~V(x) = V^*(-x). \label{mv}
\end{equation}
It may be pointed out that here we are working with a mass profile
$m(x)=(1+\lambda x^2)^{-1}$ which is an even function and
consequently satisfies the first condition of (\ref{mv}). To
generate non-Hermitian interaction in the present case we introduce complex coupling constant. As an example let us first consider the
potential appearing in (\ref{e6.0}). It can be seen from
(\ref{en2})
 that the energy for this potential does not depend on one of the potential parameters, namely $B$.
  Thus we make the generalized nonlinear oscillator potential of the equation (\ref{e6.0}) by replacing $B$ in terms of $i B$
\begin{equation}
V(x) = \left[\frac{B^2-A^2-A\sqrt{\lambda}}{1+\lambda x^2}+iB(2A+\sqrt{\lambda}) \left(\frac{\sqrt{\lambda}x}
    {1+\lambda x^2}\right)+A^2\right]\label{ptv}
\end{equation}
From (\ref{ptv}) it can be easily verified that $V(x)=V^*(-x)$ so
that the Hamiltonian (\ref{hamil3}) with this potential is
$\mathcal{PT}$-symmetric. In this case the spectrum is real and given
by (\ref{en2}). Proceeding in a similar way we have obtained the complex $\mathcal{PT}$-symmetric version of the potentials given in table \ref{t6.1}. These complex potentials for which the QNLO is exactly solvable are given in tabular form in the table \ref{t6.2}.

{\small \begin{sidewaystable}
    \centering
    {\tiny{
    \begin{tabular}{|@{}l@{}|@{}l@{}|@{}l@{}|@{}l@{}|@{}l@{}|@{}l@{}|} 	
    \hline

     $~~V(x)$ & $~~W(x)$& $~~E_n$ & $~~\psi_n(x)$ & $~~a_{i},i=0,1,..$ & $~~R(a_i)$\\

    \hline
    \hline

    $\frac{B^2-A^2-A\sqrt{\lambda}}{{1+\lambda x^2}}+B(2A+\sqrt{\lambda})\frac{\sqrt{\lambda}x}{1+\lambda
    x^2}+A^2$~~~
    &$A\frac{\sqrt{\lambda}x}{\sqrt{1+\lambda x^2}}+B\frac{1}{\sqrt{1+\lambda
    x^2}}$~~~& $n\sqrt{\lambda}(2A-n\sqrt{\lambda})$ ~~~& $i^n (1+\lambda x^2)^{-\frac{s}{2}}e^{-r~tan^{-1}
    (x\sqrt{\lambda})}$& $(A-i\sqrt{\lambda},B)$&$\sqrt{\lambda}\left[2A-(2i+1)\sqrt{\lambda}\right]$\\
    $$&$$&$$ &$P_n^{(ir-s-\frac{1}{2},-ir-s-\frac{1}{2})}(ix\sqrt{\lambda})$&$$&$$\\

    \hline

    $A^2+\frac{B^2}{A^2}-\frac{A(A+\sqrt{\lambda})}{1+\lambda x^2}+2B\frac{\sqrt{\lambda}x}{\sqrt{1+\lambda x^2}}~,~ B<A^2$
    &$A\frac{\sqrt{\lambda }x}{\sqrt{1+\lambda x^2}}+\frac{B}{A}$~~~& $A^2+\frac{B^2}{A^2}-
    (A-n\sqrt{\lambda})^2-\frac{B^2}{(A-n\sqrt{\lambda})^2}$ ~~~
    &$\left(1-\frac{x\sqrt{\lambda}}{\sqrt{1+\lambda
    x^2}}\right)^{\frac{s_1}{2}}\left(1+\frac{x\sqrt{\lambda}}{\sqrt{1+\lambda
    x^2}}\right)^{\frac{s_2}{2}}
    $&$(A-i\sqrt{\lambda},B)$&$A^2-\left[A-(i+1)\sqrt{\lambda}\right]^2$\\$$&$$ &$$ & $P_n^{(s_1,s_2)}\left(\frac{x\sqrt{\lambda}}{\sqrt{1+\lambda
    x^2}}\right)$&$$&$+\frac{B^2}{A^2}-\frac{B^2}{\left[A-(i+1)\sqrt{\lambda}\right]^2}$\\

    \hline

    $A^2+\frac{B^2}{A^2}-2B\frac{\sqrt{1+\lambda x^2}}{\sqrt{\lambda}x}+\frac{A(A-\sqrt{\lambda})}{\lambda
    x^2}~,~B>A^2$
    &$\frac{B}{A}-A\frac{\sqrt{1+\lambda x^2}}{x\sqrt{\lambda}}$
    & $A^2+\frac{B^2}{A^2}-
    (A+n\sqrt{\lambda})^2-\frac{B^2}{(A+n\sqrt{\lambda})^2}$& $\left(\frac{\sqrt{1+\lambda x^2}}{x\sqrt{\lambda}}
    -1\right)^
    {\frac{s_3}{2}}\left(\frac{\sqrt{1+\lambda x^2}}{x\sqrt{\lambda}}+1\right)^{\frac{s_4}{2}}$
    &$(A+i\sqrt{\lambda},B)$&$A^2-\left[A+(i+1)\sqrt{\lambda}\right]^2$\\
    $0\leq x\sqrt{\lambda}\leq \infty$& $$ &$$ &$P_n^{(s_3,s_4)}\left(\frac{1+\lambda x^2}{x\sqrt{\lambda}}\right)$&$$
    &$+\frac{B^2}{A^2}-\frac{B^2}{\left[A+(i+1)\sqrt{\lambda}\right]^2}$\\

    \hline

     $\frac{A^2+B^2+A\sqrt{\lambda}}{\lambda x^2}-B(2A+\lambda)\frac{\sqrt{1+\lambda x^2}}{\lambda
    x^2}+A^2,$&$A\frac{\sqrt{1+\lambda
    x^2}}{x\sqrt{\lambda}}-B\frac{1}{x\sqrt{\lambda}}$&$n\sqrt{\lambda}(2A-n\sqrt{\lambda})$&$(\sqrt{1+\lambda
    x^2}-1)^{(\frac{r-s}{2})}
    (\sqrt{1+\lambda x^2}+1)^{-(\frac{r+s}{2})}$&$(A-i\sqrt{\lambda},B)$&$\sqrt{\lambda}\left[2A-(2i+1)\sqrt{\lambda}\right]$\\
    $~A<B, ~~0\leq x\sqrt{\lambda}\leq \infty$&$$&$$&$P_n^{(r-s-\frac{1}{2},-r-s-\frac{1}{2})}\left(\sqrt{1+\lambda
    x^2}\right)$&$$&$$\\

    \hline

    $\frac{A^2+B^2-A\sqrt{|\lambda|}}{1+\lambda x^2}-B(2A-\sqrt{|\lambda|})
    \frac{x\sqrt{|\lambda|}}{1+\lambda x^2}-A^2$&$A\frac{x\sqrt{|\lambda|}}{\sqrt{1+\lambda x^2}}
    -B\frac{1}{\sqrt{1+\lambda
    x^2}}$&$n\sqrt{|\lambda|}(2A+n\sqrt{|\lambda|})$&$(1-x\sqrt{|\lambda|})^{(\frac{s'-r'}{2})}
    (1+x\sqrt{|\lambda|})^{(\frac{r'+s'}{2})}$&$(A+i\sqrt{|\lambda|},B)$&$\sqrt{|\lambda|}
    \left[2A+(2i+1)\sqrt{|\lambda|}\right]$\\
    $\frac{1}{-\sqrt{|\lambda|}}\leq
    x\leq\frac{1}{\sqrt{|\lambda|}}$&$$&$$&$P_n^{(s'-r'-\frac{1}{2},s'+r'-\frac{1}{2})}\left(x\sqrt{|\lambda|}\right)$&$$&$$\\

    \hline

    $\frac{A(A-\sqrt{|\lambda|})}{1+\lambda x^2}-2B\frac{x\sqrt{|\lambda|}}{1+\lambda
    x^2}-A^2+\frac{B^2}{A^2}$&$A\frac{x\sqrt{|\lambda|}}{\sqrt{1+\lambda
    x^2}}-\frac{B}{A}$&$\frac{B^2}{A^2}-A^2+(A+n\sqrt{|\lambda|})^2-\frac{B^2}{(A+n\sqrt{|\lambda|})^2}$
    &$\left(\frac{\lambda x^2}{1+\lambda
    x^2}-1\right)^{-(\frac{s'+n}{2})}e^{-a\sqrt{|\lambda|}x}$&$(A+i\sqrt{|\lambda|},B)$&$\frac{B^2}{A^2}-\frac{B^2}{\left[A+(i+1)\sqrt{|\lambda|}\right]^2}$\\
    $\frac{1}{-\sqrt{|\lambda|}}\leq
    x\leq\frac{1}{\sqrt{|\lambda|}}$&$$&$$
    &$P_n^{(-s'-n-ia,-s'-n+ia)}\left(-i\frac{x\sqrt{|\lambda|}}{\sqrt{1+\lambda x^2}}\right)$&$$&$-A^2+\left[A+(i+1)\sqrt{|\lambda|}\right]^2$\\

    \hline

      \end{tabular}
      }}
       \caption[Table excatly solvable Hermitian potentials corresponding to QNLO]{\small{Shape invariant potentials $V(x)$ for which the QNLO is exactly solvable. Corresponding, superpotential $W(x)$ , energy eigenvalue
      $E_n$ and wave functions $\psi_n(x)$ are also given. Here $s=\frac{A}{\sqrt{\lambda}}, r=\frac{B}{\sqrt{\lambda}},
       r_1=\frac{B}{\lambda}, a=\frac{r_1}{s-n}, s_1=s-n+a,s_2=s-n-a, s_3=a-n-s,s_4=-(s+n+a), s'=\frac{A}
       {\sqrt{|\lambda|}} ~\&~r'=\frac{B}{\sqrt{|\lambda|}}.$The first four entries correspond to $\lambda>0$ and the last two
        correspond to $\lambda<0.$}}\label{t6.1}

    \end{sidewaystable}}

{\begin{sidewaystable}
    \centering
    {\tiny{
    \begin{tabular}{|l c| c |c |c|}
    \hline

    & $V(x)$ & $W(x)$& $E_n$ & $\psi_n(x)$ \\

    \hline
    \hline

    &$\frac{-B^2-A^2-A\sqrt{\lambda}}{1+\lambda x^2}+iB(2A+\sqrt{\lambda})\frac{\sqrt{\lambda}x}{1+\lambda
    x^2}+A^2$~~~
    &$A\frac{\sqrt{\lambda}x}{\sqrt{1+\lambda x^2}}+iB\frac{1}{\sqrt{1+\lambda
    x^2}}$~~~& $n\sqrt{\lambda}(2A-n\sqrt{\lambda})$ ~~~& $i^n (1+\lambda x^2)^{-\frac{s}{2}}e^{-r~tan^{-1}
    (x\sqrt{\lambda})}$\\&$$&$$&$$ &$P_n^{(ir-s-\frac{1}{2},-ir-s-\frac{1}{2})}\left(ix\sqrt{\lambda}\right)$\\

    \hline

    &$A^2-\frac{B^2}{A^2}-\frac{A(A+\sqrt{\lambda})}{1+\lambda x^2}+i2B\frac{\sqrt{\lambda}x}{\sqrt{1+\lambda x^2}}~,~ B<A^2$
    &$A\frac{\sqrt{\lambda }x}{\sqrt{1+\lambda x^2}}+i\frac{B}{A}$~~~& $A^2-\frac{B^2}{A^2}-
    (A-n\sqrt{\lambda})^2+\frac{B^2}{(A-n\sqrt{\lambda})^2}$ ~~~
    &$\left(1-\frac{x\sqrt{\lambda}}{\sqrt{1+\lambda
    x^2}}\right)^{\frac{s_1}{2}}\left(1+\frac{x\sqrt{\lambda}}{\sqrt{1+\lambda
    x^2}}\right)^{\frac{s_2}{2}}
    $\\&$$&$$ &$$ & $P_n^{(s_1,s_2)}\left(\frac{x\sqrt{\lambda}}{\sqrt{1+\lambda
    x^2}}\right)$\\

    \hline

    &$A^2-\frac{B^2}{A^2}-2iB\frac{\sqrt{1+\lambda x^2}}{\sqrt{\lambda}x}+\frac{A(A-\sqrt{\lambda})}{\lambda
    x^2}~,~B>A^2$
    &$i\frac{B}{A}-A\frac{\sqrt{1+\lambda x^2}}{x\sqrt{\lambda}}$
    & $A^2-\frac{B^2}{A^2}-
    (A+n\sqrt{\lambda})^2+\frac{B^2}{(A+n\sqrt{\lambda})^2}$& $\left(\frac{\sqrt{1+\lambda x^2}}{x\sqrt{\lambda}}
    -1\right)^
    {\frac{s_3}{2}}\left(\frac{\sqrt{1+\lambda x^2}}{x\sqrt{\lambda}}+1\right)^{\frac{s_4}{2}}$\\
    &$0\leq x\sqrt{\lambda}\leq \infty$& $$ &$$ &$P_n^{(s_3,s_4)}\left(\frac{1+\lambda x^2}{x\sqrt{\lambda}}\right)$\\

    \hline

    & $\frac{A^2-B^2+A\sqrt{\lambda}}{\lambda x^2}-iB(2A+\lambda)\frac{\sqrt{1+\lambda x^2}}{\lambda
    x^2}A^2~,~A<B$&$A\frac{\sqrt{1+\lambda
    x^2}}{x\sqrt{\lambda}}-iB\frac{1}{x\sqrt{\lambda}}$&$n\sqrt{\lambda}(2A-n\sqrt{\lambda})$&$(\sqrt{1+\lambda
    x^2}-1)^{(\frac{r-s}{2})}
    (\sqrt{1+\lambda x^2}+1)^{-(\frac{r+s}{2})}$\\
    &$0\leq x\sqrt{\lambda}\leq \infty$&$$&$$&$P_n^{(r-s-\frac{1}{2},-r-s-\frac{1}{2})}\left(\sqrt{1+\lambda
    x^2}\right)$\\

    \hline

    &$\frac{A^2-B^2-A\sqrt{|\lambda|}}{1+\lambda x^2}-iB(2A-\sqrt{|\lambda|})
    \frac{x\sqrt{|\lambda|}}{1+\lambda x^2}-A^2$&$A\frac{x\sqrt{|\lambda|}}{\sqrt{1+\lambda x^2}}
    -iB\frac{1}{\sqrt{1+\lambda
    x^2}}$&$n\sqrt{\lambda}(2A+n\sqrt{|\lambda|})$&$(1-x\sqrt{|\lambda|})^{(\frac{s'-r'}{2})}
    (1+x\sqrt{|\lambda|})^{(\frac{r'+s'}{2})}$\\
    &$\frac{1}{-\sqrt{|\lambda|}}\leq
    x\leq\frac{1}{\sqrt{|\lambda|}}$&$$&$$&$P_n^{(s'-r'-\frac{1}{2},s'+r'-\frac{1}{2})}\left(x\sqrt{|\lambda|}\right)$\\

    \hline

    &$\frac{A(A-\sqrt{|\lambda|})}{1+\lambda x^2}-2iB\frac{x\sqrt{|\lambda|}}{1+\lambda
    x^2}-A^2-\frac{B^2}{A^2}$&$A\frac{x\sqrt{|\lambda|}}{\sqrt{1+\lambda
    x^2}}-i\frac{B}{A}$&$-\frac{B^2}{A^2}-A^2+(A+n\sqrt{|\lambda|})^2+\frac{B^2}{(A+n\sqrt{|\lambda|})^2}$
    &$\left(\frac{\lambda x^2}{1+\lambda
    x^2}-1\right)^{-(\frac{s'+n}{2})}e^{-a\sqrt{|\lambda|}x}$\\
    &$\frac{1}{-\sqrt{|\lambda|}}\leq
    x\leq\frac{1}{\sqrt{|\lambda|}}$&$$&$$
    &$P_n^{(-s'-n-ia,-s'-n+ia)}\left(-i\frac{x\sqrt{|\lambda|}}{\sqrt{1+\lambda x^2}}\right)$\\

    \hline

      \end{tabular}
      }}
      \caption[Table of exactly solvable non-Hermitian $\mathcal{PT}$-symmetric potentials corresponding to QNLO]{\small{Non-Hermitian $\mathcal{PT}$-symmetric potentials corresponding to QNLO. Here, $s=\frac{A}{\sqrt{\lambda}}, r=i\frac{B}{\sqrt{\lambda}},
       r_1=i\frac{B}{\lambda}, a=\frac{r_1}{s-n}, s_1=s-n+a, s_2=s-n-a, s_3=a-n-s, s_4=-(s+n+a),
       ~~~~s'=\frac{A}{\sqrt{|\lambda|}} ~\&~
       r'=i\frac{B}{\sqrt{|\lambda|}}.$ The first four entries correspond to $\lambda>0$ and the last two
        correspond to $\lambda<0.$}}\label{t6.2}
    \end{sidewaystable}}

\newpage
\subsection{Quasi exactly solvable $\mathcal{PT}$-symmetric generalization}\label{cqes}
\noindent The quasi-exactly solvable complex sextic potential in the constant mass Schr\"{o}dinger
equation has been discussed in ref.\cite{Ba+00a}. By using the
transformations (\ref{t}) for $\lambda > 0$ we obtain the corresponding quasi exactly solvable potentials of the QNLO.\\
For $\lambda > 0$, we consider the potential
\begin{equation}
V(x)=\sum_{k=1}^{6}\frac{c_k}{\lambda^{\frac{k}{2}}} ~ {\left(\sinh(x\sqrt{\lambda})\right)^{-k}}\label{qesp}
\end{equation}
where $V(x)$ is $\mathcal{PT}$-symmetric if $c_1, c_3, c_5$ are purely imaginary and $c_2, c_4, c_6$ are real. Following ref.{\cite{Ba+00a}}, the ansatz for the wave function is
taken as
\begin{equation}
\psi(x)=f(x)~ exp\left(-\sum_{j=1}^{4}
\frac{b_j}{\lambda^{\frac{j}{2}}}
(\sinh(x\sqrt{\lambda}))^{-j}\right) \label{qesw}
\end{equation}
where $f(x)$ is some polynomial function of $x$. We focus on the following choices of $f(x)$:
$$ \hspace{-7.5 cm}(a)~~~ f(x) = 1$$
$$ \hspace{-4 cm}(b)~~~ f(x) = \frac{\left[\sinh(x\sqrt{\lambda})\right]^{-1}}{\sqrt{\lambda}} + a_0$$
$$(c)~~~f(x)=\frac{\left[\sinh(x\sqrt{\lambda})\right]^{-2}}{\lambda}
+a_1\frac{\left[\sinh(x\sqrt{\lambda})\right]^{-1}}{\sqrt{\lambda}}+a_0$$
For complex potentials, $a_0$ is purely imaginary in (b), but in (c) $a_1$ is purely imaginary, but $a_0$ is real. Without going into the details of calculation (which are done in ref.\cite{Ba+00a} for constant mass case), which are quite straightforward, let us summarize our results.\\

\noindent {\bf Case 1:} $f(x) = 1$\\
In this case the relation between the parameters $c_i$ and $b_i$
are found to be $
c_1 = -3b_3+2b_1b_2,~~c_2 = -6b_4+3b_1b_3+2b_2^2,~~c_3 = 4b_1b_4+6b_2b_3, c_4 = 8b_2b_4+ \frac{9}{2}b_3^2,~~ c_5 = 12b_3b_4,~~c_6 =
8b_4^2$ and energy is given by
\begin{equation}
E=b_2-\frac{1}{2}b_1^2\label{e6.37}
\end{equation}
Without loss of generality, we can choose $c_6 = \frac{1}{2}$
which fixes the leading coefficient of $V(x)$. It gives $b_4 = \pm
\frac{1}{4}$. Taking the positive sign to ensure the
normalizability of the wave function we obtain
{\small \begin{equation}
\psi(x)=\exp~\left(-\frac{b_{1}\left[\sinh(x\sqrt{\lambda})\right]^{-1}}{\sqrt{\lambda}}
-\frac{b_{2}\left[\sinh(x\sqrt{\lambda})\right]^{-2}}{\lambda}-\frac{b_{3}\left[\sinh(x\sqrt{\lambda})\right]^{-3}}
{\lambda\sqrt{\lambda}}
-\frac{\left[\sinh(x\sqrt{\lambda})\right]^{-4}}{4\lambda^2}\right)\label{e6.36}
\end{equation}}
Now if $b_1$ and $b_3$ are purely imaginary then $c_1, c_3, c_5$ are also purely imaginary.
In that case $V(x)$ in Eqn.(\ref{qesp}) and $\psi(x)$ in Eqn.(\ref{qesw}) are $\mathcal{PT}$-symmetric and $E$ is real.\\

\noindent {\bf{Case 2:}} $f(x)=\displaystyle \frac{\left[\sinh(x\sqrt{\lambda})\right]^{-1}}{\sqrt{\lambda}} + a_0$, where $a_0$ is
purely imaginary.\\
In this case wave function is of the form
{\small \begin{equation}
\begin{array}{ll} \displaystyle
\psi(x)=\left(\frac{\left[\sinh(x\sqrt{\lambda})\right]^{-1}}{\sqrt{\lambda}}
+a_0\right)\\
\displaystyle
~~~~~~~~~\exp\left(-\frac{b_{1}\left[\sinh(x\sqrt{\lambda})\right]^{-1}}{\sqrt{\lambda}}
-\frac{b_{2}\left[\sinh(x\sqrt{\lambda})\right]^{-2}}{\lambda}-\frac{b_{3}\left[\sinh(x\sqrt{\lambda})\right]^{-3}}
{\lambda\sqrt{\lambda}}
-\frac{\left[\sinh(x\sqrt{\lambda})\right]^{-1}}{4\lambda^2}\right)
\end{array}\label{e6.35}
\end{equation}}
In this case the relation between the parameters are given by
\begin{equation*}
c_1=-6b_3+2b_1 b_2+a_0~~,~~c_2=-\frac{5}{2}+3b_1b_3+2b_2^2~~,~~c_3=b_1+6b_2b_3\\
c_4=2b_2+\frac{9}{2}b_3^2~~,~~c_5=3b_3~~,~~c_6=\frac{1}{2}\label{e6.34}
\end{equation*}
and $a_0$ satisfies the condition
\begin{equation}
 a_0^3-3b_3a_0^2+2b_2a_0-b_1=0\label{qesc}
\end{equation}
The corresponding energy is given by
\begin{equation}
E=-\frac{1}{2}b_1^2+3b_2-3a_0 b_3+a_0^2\label{e6.33}
\end{equation}

\noindent We now consider following two special cases.\\
\noindent (a) $b_1=b_3=0$ and $a_0^2<0$.

In this case $c_1$ is purely imaginary and $c_3 = c_5 = 0$. Moreover, $c_1=a_0=\pm i\sqrt{2b_2}$.
So we get two different complex potentials corresponding to above
two values of $c_1$ with same real energy eigenvalues. The
potential, energy values and the eigenfunctions are given by
{\small \begin{equation}
\begin{array}{lcl}
V(x) &=& \frac{1}{2}\frac{[\sinh(x\sqrt{\lambda})]^{-6}}{\lambda^3}+\frac{2b_2}
{\lambda^2}[\sinh(x\sqrt{\lambda})]^{-4}+\frac{(2b_2^2-\frac{5}{2})}{\lambda}[\sinh(x\sqrt{\lambda})]^{-2}\pm
\frac{i \sqrt{2b_2}}{\sqrt{\lambda}}\left[\sinh(x\sqrt{\lambda})\right]^{-1}\\
E &=& b_2>0\\
\psi(x) &=&
\left(\frac{[\sinh(x\sqrt{\lambda})]^{-1}}{\sqrt{\lambda}}\pm i
\sqrt{2b_2}\right)
\exp\left(-\frac{b_2}{\lambda}\left[\sinh(x\sqrt{\lambda})\right]^{-2}
-\frac{1}{4\lambda ^2}
\left[\sinh(x\sqrt{\lambda})\right]^{-4}\right)
\end{array}\label{e32}
\end{equation}}
It can be easily seen from the above equations that the potential
is $\mathcal{PT}$-symmetric,
 while the wave function is odd under $\mathcal{PT}$-symmetry.\\

\noindent (b) $b_1=0, b_3\neq 0$

Then from (\ref{qesc}) we get
\begin{equation}
a_0=\frac{1}{2} (3b_3 \pm \sqrt{9b_3^2-8b_2})\label{e6.31}
\end{equation}
So in order to make $a_0$ imaginary we must
have $9b_3^2-8b_2<0.$ or $b_3^2 = -|b_3|^2 \leq \frac{8}{9}b_2$.\\
In this case also there exist two different complex potentials
corresponding to two values of $b_3$ with the same real energy
eigenvalues $E=3b_2-3a_0b_3+a_0^2.$\\

\noindent {\bf{Case 3:}}
$f(x)=\frac{\left[\sinh(x\sqrt{\lambda})\right]^{-2}}{\lambda} +a_1
\frac{\left[\sinh(x\sqrt{\lambda})\right]^{-1}}{\sqrt{\lambda}}+a_0$,
where $a_1$ is imaginary and $a_0$ is real.\\
In this case the relation between the parameters is given by
 \begin{equation}
a_1 = 2b_3~~,~~ a_0 = \frac{1}{2} \left(2b_2 - b_3^2 \pm
\sqrt{(2b_2-3b_3^2)^2 + 2} \right)\label{qesc1}
 \end{equation}
The wave function, energy and the potential are of the form
\begin{subequations}
{\small
\begin{equation}
    \begin{array}{ll}
    \psi_\pm(x)=\left[\frac{\left[\sinh(x\sqrt{\lambda})\right]^{-2}}{\lambda}+2b_3 \frac{\left[\sinh(x\sqrt{\lambda})\right]^{-1}}{\sqrt{\lambda}}+\frac{1}{2}
    (2b_2-b_3^2\pm \sqrt{(2b_2-3b_3^2)^2+2}) \right]\\
    \displaystyle
      ~~~~~~\exp \left(-2b_3(b_2-b_3^2)\frac{\left[\sinh(x\sqrt{\lambda})\right]^{-1}}{\sqrt{\lambda}}
    -b_2 \frac{\left[\sinh(x\sqrt{\lambda})\right]^{-2}}{\lambda}
    -b_3\frac{\left[\sinh(x\sqrt{\lambda})\right]^{-3}}{\lambda\sqrt{\lambda}}
    -\frac{1}{4}\frac{\left[\sinh(x\sqrt{\lambda})\right]^{-4}}{\lambda^2}\right)
    \end{array},\label{e30}
    \end{equation}}
    \begin{equation}
    E_\pm=-2b_3^2(b_2-b_3^2)^2+3b_2-b_3^2\pm
    \sqrt{(2b_2-3b_3^2)^2+2},\label{e29}
    \end{equation}
   {\small  \begin{equation}
    \begin{array}{lll}
    \displaystyle
    V(x) =\frac{1}{2\lambda^3}[\sinh(x\sqrt{\lambda})]^{-6}+\frac{3b_3}{\lambda^2 \sqrt{\lambda}}[\sinh(x\sqrt{\lambda})]^{-5}
    +\frac{(2b_2+\frac{9}{2})}{\lambda ^2}[\sinh(x\sqrt{\lambda})]^{-4}\\
   \displaystyle ~~~~~~~ +\frac{2b_3}{\lambda \sqrt{\lambda}}(4b_2-b_3^2)[\sinh(x\sqrt{\lambda})]^{-3}
+\frac{[2(b_2^2+3b_2b_3^2-3b_3^4)-\frac{7}{2}]}{\lambda}[\sinh(x\sqrt{\lambda})]^{-2}\\
~~~~~~~~\displaystyle +\frac{b_3(4b_2^2-4b_2b_3^2-7)}{\sqrt{\lambda}}[\sinh(x\sqrt{\lambda})]^{-1},
\end{array}\label{qesv2}
\end{equation}}
\end{subequations}
respectively. The result (\ref{qesc1}) to (\ref{qesv2}) are valid both for real
and purely imaginary $b_i$. When $b_i$ are purely imaginary the
potential and wave function are $\mathcal{PT}$-symmetric while for
real $b_i$ $\mathcal{PT}$-symmetry is broken. In particular when
$b_3$ is purely imaginary we have a complex $\mathcal{PT}$-symmetric
two parameter family of potentials corresponding to two values of
$a_0$ with two distinct
real eigenvalues.

\section[Supersymmetric Shape invariance approach to generalized quantum nonlinear oscillator]
{Supersymmetric Shape invariance approach to generalized QNLO}\label{s4.3}
\subsection{Unbroken supersymmetry} \label{s4.3.1}
\noindent We consider two first order operators of the form
\begin{equation}
A = P_x - i W(x), ~~~~ A^\dagger = P_x + i W(x),~~~~ P_x =
\frac{1}{\sqrt{m(x)}}\left(-i\frac{d}{dx}\right) \label{op}
\end{equation}
to factorize the non-linear oscillator Hamiltonians $H_\pm$ as
\begin{equation}
H_{\pm} = -\frac{1}{m(x)}\frac{d^2}{dx^2} +
\left(\frac{m^\prime}{2m^2}\right)\frac{d}{dx} + W^2 \pm
\frac{W^\prime}{\sqrt{m}}\label{e6.12}
\end{equation}
For unbroken supersymmetry (SUSY), the
ground state of $H_-$ has zero energy, $E_0^{(-)} = 0$, which implies that
the ground state wave function $\psi_0^{(-)}(x)$ given by
($A\psi_0^{(-)} = 0$)
\begin{equation}
\psi_0^{(-)}(x) = N_0 ~\exp~\left[-\int^{x}~\sqrt{m(y)}~
W(y)dy\right] \label{gs}
\end{equation}
is normalizable. Hence, the superpotential $W(x)$ could be generated from the ground state solution of $H_-$ i.e. $W(x) = - \frac{\psi_0^{(-)}}{\sqrt{m}\psi_0^{(-)}}$. For the generalized nonlinear oscillator potential appears in equation(\ref{e6.0}) the superpotential is given by
\begin{equation}
W = A \frac{\sqrt{\lambda}x}{\sqrt{1+\lambda x^2}} +
B\frac{1}{\sqrt{1+\lambda x^2}}\label{e6.17}
\end{equation}
Therefore the Hamiltonians $H_\pm$ given in (\ref{e6.12}) reduced to
\begin{equation}
\begin{array}{lcl}
H_- &=& A^{\dagger}A\\
          &=& -(1+\lambda x^2)\frac{d^2}{dx^2} - \lambda x \frac{d}{dx} + \frac{B^2-A^2-A\sqrt{\lambda}}{1+\lambda x^2}+
           B(2A + \sqrt{\lambda})(\frac{\sqrt{\lambda}x}{1+\lambda x^2}) + A^2\\
H_+ &=& AA^{\dagger}\\
          &=& -(1+\lambda x^2)\frac{d^2}{dx^2} - \lambda x \frac{d}{dx} + \frac{B^2-A^2+A\sqrt{\lambda}}{1+\lambda x^2}+ B(2A - \sqrt{\lambda})(\frac{\sqrt{\lambda}x}{1+\lambda x^2}) + A^2
\end{array}\label{e18}
\end{equation}
The Hamiltonian $H_-$ is the Hamiltonian of the generalized QNLO with the potential of the Schr\"{o}dinger equation (\ref{e6.0}). Also these two Hamiltonians $H_\pm$ are related by
\begin{equation}
H_+(x;A,B) = H_-(x;A-\sqrt{\lambda},B) +
\sqrt{\lambda}(2A-\sqrt{\lambda})\label{e6.19}
\end{equation}
so that they satisfy shape invariance condition
\begin{equation}
H_+(x,a_0) = H_-(x,a_1) + R(a_0)\label{e6.20}
\end{equation}
where $\{a_0\} = (A,B), \{a_1\} = (A-\sqrt{\lambda},B)$ and
$R(a_0) = \sqrt{\lambda}(2A-\sqrt{\lambda}).$
The zero energy ground state $\psi_0(x,a_0)$ of the Hamiltonian $H_-$ is
found by solving $A\psi_0(x,a_0)=0$ i.e.
\begin{equation}
H_-(x,a_0)\psi_0(x,a_0)=0\label{e6.47}
 \end{equation}
 Now using (\ref{e6.20}) we
can see that $\psi_0(x,a_1)$ is an eigenstate of $H_+$ with
the energy $E_1=R(a_0)$, because
\begin{equation}\begin{array}{ll}\displaystyle
H_+(x,a_0)\psi_0(x,a_1)=H_-(x,a_1)\psi_0(x,a_1)+R(a_0)\psi_0(x,a_1)\\
\displaystyle
~~~~~~~~~~~~~~~~~~~~~~~~~~~ =R(a_0)\psi_0(x,a_1)~,~~~\mbox{[using
({\ref{e6.47}})]}\label{e6.46}
\end{array}
\end{equation}
Next, using the intertwining relation
$H_-(x,a_0)A^{\dagger}(x,a_0)=A^{\dagger}(x,a_0)H_+(x,a_0)$
and equation ({\ref{e6.20}}), we see that
\begin{equation*}
H_-(x,a_0)A^{\dagger}(x,a_0)\psi_0(x,a_1)=A^{\dagger}(x,a_0)H_+(x,a_0)\psi_0(x,a_1)=A^{\dagger}
\left[H_-(x,a_1)+R(a_0)\right]\psi_0(x,a_1)\label{e6.48}
\end{equation*}
and hence using (\ref{e6.47}) we arrive at
\begin{equation}
H_-(x,a_0)A^{\dagger}(x,a_0)\psi_0(x,a_1)=
R(a_1)A^{\dagger}(x,a_0)\psi_0(x,a_1)\label{e6.49}
\end{equation}
This indicates that $A^{\dagger}(x,a_0)\psi_0(x,a_1)$ is an
eigenstate of $H_-$ with an energy $E_1=R(a_0)$. Now iterating this
process we will find the sequence of energies for $H_-$ as
\begin{equation}
E^{(-)}_n = \sum_{i=0}^{n-1} R(a_i) =
n\sqrt{\lambda}(2A-n\sqrt{\lambda})~, ~~~E^{(-)}_0=0\label{e6.21}
\end{equation}
and corresponding eigenfunctions being
\begin{equation}
\psi_n(x,a_0)=A^{\dagger}(x,a_0)A^{\dagger}(x,a_1)...A^{\dagger}(x,a_{n-1})\psi_0(x,a_n)\label{e6.50}
\end{equation}
where $a_i=f(a_{i-1})=\underbrace{f(f(....(f(a_0)))}_{i \mbox{
times}}=\left(A-i\sqrt{\lambda},B\right)$ and
$R(a_i)=\sqrt{\lambda}\left[2\left(A-i\sqrt{\lambda}\right)-\sqrt{\lambda}\right]$.
In a similar way it can be shown that other potentials of table \ref{t6.1} are also shape invariant.
For all these potentials the energy, wave functions and other parameters related to shape
invariance property are given in the same table.

\subsection{Broken supersymmetry}

\noindent When supersymmetry is broken neither of the wave functions
$\psi_0^{(\pm)}(x)\approx exp[\pm\int^x\sqrt{m(y)}W(y)dy]$ are
normalizable and in this case all the energy values are degenerate
i.e, $H_+$ and $H_-$ have identical energy eigenvalues
\cite{CK00}
\begin{equation}
E_n^{(-)} = E_n^{(+)}\label{e6.22}
\end{equation}
with ground state energies greater than zero. So far as we know,
little attention has been paid till now to study problems
involving broken SUSY in the case of PDMSE. Broken supersymmetric
shape invariant systems in the case of constant mass Schr\"odinger
equation has been discussed in ref.\cite{Ju96,Ba00}. Below we
illustrate the two step procedure discussed in \cite{Du+93,Ga+01} for
obtaining the energy spectra of generalized QNLO when the SUSY is broken. \
For this, we consider the superpotential as
    \begin{equation}
    W(x,A,B)=A\sqrt{|\lambda|}\frac{x}{\sqrt{1+\lambda
    x^2}}-\frac{B}{\sqrt{|\lambda|}}\frac{\sqrt{1+\lambda x^2}}{x}, ~~~~~ 0<x
    <\frac{1}{\sqrt{|\lambda|}},~~\lambda<0. \label{sp}
    \end{equation}
 Then the supersymmetric partner potentials of the Hamiltonians $H_\pm$ are obtained using (\ref{e12}) as
    \begin{equation}
    \begin{array}{ll}
    \displaystyle
    V_-(x,A,B)=\frac{A(A-\sqrt{|\lambda|})}{1+\lambda
    x^2}-\frac{B(B-\sqrt{|\lambda|})}{\lambda x^2}-(A+B)^2   \\
    \displaystyle
    V_+(x,A,B)=\frac{A(A+\sqrt{|\lambda|})}{1+\lambda
    x^2}-\frac{B(B+\sqrt{|\lambda|})}{\lambda x^2}-(A+B)^2
    \end{array}\label{bsp}
    \end{equation}
The ground state wave function is obtained from (\ref{gs}) as
    \begin{equation}
    \psi_0^{(-)}\sim x^{\frac{B}{\sqrt{|\lambda|}}}~ (1+\lambda
    x^2)^{\frac{A}{2\sqrt{|\lambda|}}}\label{e23}
    \end{equation}
For $A>0,~B>0$ the ground state wave function $\psi_0^{(-)}$ is
normalizable which means the SUSY is unbroken. But for $A>0,~
B<0$  and $A<0,~ B>0$ , neither of $\psi_0^{(\pm)}$ are normalizable. Hence SUSY  is broken in both cases. \\
For $A>0,~ B<0$, the eigenstates of $V_\pm (x,A,B)$ are related by
    \begin{equation}
    \begin{array}{lll}
    \displaystyle
    \psi_n^{(+)}(x,a_0)=A(x,a_0) \psi_n^{(-)} (x,a_0)\\
    \displaystyle \psi_n^{(-)}(x,a_0)=A^\dagger (x,a_0)
    \psi_n^{(+)}(x,a_0),\\
    \displaystyle E_n^{(-)}(a_0)=E_n^{(+)}(a_0)
    \end{array}\label{bse1}
    \end{equation}
 Now we can show that the potentials in equation (\ref{bsp}) are shape invariant
 by two different relations between the parameters.\\

\noindent {\bf {Step 1}}\\
 The potentials of equation (\ref{bsp}) are shape invariant if we
 change $A\rightarrow A+\sqrt{|\lambda|}~ \mbox{and} ~ B \rightarrow
 B+\sqrt{|\lambda|}.$The shape invariant condition is given by
    \begin{equation}
    V_+(x,A,B)=V_-~\left(x,A+\sqrt{|\lambda|},B+\sqrt{|\lambda|}\right)+\left(A+B+2\sqrt{|\lambda|}\right)^2-(A+B)^2\label{e24}
    \end{equation}
 Now for $B<-\frac{1}{\sqrt{|\lambda|}}$ it is seen that the superpotential
  (\ref{sp}) resulting from change of parameters as above falls in the class of broken SUSY
problem for which $E_0^{(-)} \neq 0$. Though the potentials of
equation (\ref{bsp}) are shape invariant but we are unable to
determine the spectra for these potentials because of the absence
of zero energy ground state.\\
Another way of parameterizations $A\rightarrow A+\sqrt{|\lambda|}~
~\mbox{and}~~ B\rightarrow -B$ gives us
    \begin{equation}
        V_+(x,A,B)=V_-~\left(x,A+\sqrt{|\lambda|},-B\right)+\left(A-B+\sqrt{|\lambda|}\right)^2-(A+B)^2
        \label{sic1}
    \end{equation}
which shows that $V_-~\mbox{and}~ V_+$ are shape invariant. This
change of parameters $(A\rightarrow A+\sqrt{|\lambda|}~
~\mbox{and}~~ B\rightarrow -B)$ leads to a system with unbroken
SUSY since the parameter $B$ changes sign.
 Hence the ground state energy of the potential
$V_-(x,A+\sqrt{|\lambda|},-B)$ is zero. From the relation
(\ref{sic1}) we observe that $V_+(x,A,B) ~\mbox{and} ~
V_-\left(A+\sqrt{|\lambda|},-B\right)$ differ only by a constant,
hence we have
    \begin{equation}
    \begin{array}{ll}
    \displaystyle
    \psi_+(x,A,B)=\psi_-(x,A+\sqrt{|\lambda|},-B)\\
    \displaystyle
    E_n^{(+)}(A,B)=E_n^{(-)}(x,A+\sqrt{|\lambda|},-B)+\left(A-B+\sqrt{|\lambda|}\right)^2-(A+B)^2
    \end{array}\label{bse2}
    \end{equation}
Thus, if we can evaluate the spectrum and energy eigenfunctions of
unbroken SUSY $H_-(x,A+\sqrt{|\lambda|},-B)$, then we can
determine the spectrum and eigenfunctions $H_+(x,A,B)$ with broken SUSY. In the 2nd step we will do this.\\

\noindent {\bf{Step 2}}\\
With the help of shape invariant formalism in case of unbroken SUSY
discussed earlier, we obtain spectrum and eigenfunctions
for $V_-(x,A+\sqrt{|\lambda|},-B)$ as
    \begin{equation}\begin{array}{ll}\displaystyle
    E_n^{(-)}(A+\sqrt{|\lambda|},-B)=\left(A-B+\sqrt{|\lambda|}+2n\sqrt{|\lambda|}\right)^2-
    \left(A-B+\sqrt{|\lambda|}\right)^2\\
    \displaystyle
    \psi_n^{(-)}(x,A+\sqrt{|\lambda|},-B) \propto x^{\frac{B}{\sqrt{|\lambda|}}}(1+\lambda
    x^2)^{\frac{A}{2\sqrt{|\lambda|}}} P_n^{(\frac{B}{\sqrt{|\lambda|}}-\frac{1}{2},\frac{A}{\sqrt{|\lambda|}}-\frac{1}{2})}
    (1+2\lambda x^2)
    \end{array}\label{bse3}
    \end{equation}
Now using (\ref{bse3}) ,(\ref{bse2}) and (\ref{bse1}) we obtain
spectrum and eigenfunctions for $V^- (x,A,B)$ with broken SUSY as
    \begin{equation}
    \begin{array}{ll}
    \displaystyle
        E_n^{(-)}(A,B)=\left(A-B+\sqrt{|\lambda|}+2n\sqrt{|\lambda|}\right)^2-(A+B)^2\\
        \displaystyle
        \psi_n^{(-)}(x,A,B) \propto x^{\frac{1-B}{\sqrt{|\lambda|}}}(1+\lambda x^2)^{\frac{A}{2\sqrt{|\lambda|}}}
         P_n^{\left(\frac{1}{2}-\frac{B}{\sqrt{|\lambda|}},\frac{A}{\sqrt{|\lambda|}}-\frac{1}{2}\right)}(1+2\lambda x^2)
         \end{array}\label{e25}
         \end{equation}
Similar approach can be applied in case of $A<0~ \mbox{and}~ B>0$.
In this case we change $(A,B)$ into $(-A,B+\sqrt{|\lambda|})$ and
the shape invariance condition is
\begin{equation}
V_+(x,A,B)=V_-~\left(x,-A,B+\sqrt{|\lambda|}\right)+\left(B-A+\sqrt{|\lambda|}\right)^2-(A+B)^2\label{e26}
\end{equation}
And
\begin{equation}
\begin{array}{ll}
\displaystyle
E_n^{(-)}(A,B)=\left[B-A+\sqrt{|\lambda|}+2n\sqrt{|\lambda|}\right]^2-(A+B)^2\\
\displaystyle
\psi_n^{(-)}(x,A,B) \propto x^{\frac{1-A}{\sqrt{|\lambda|}}}(1+\lambda
x^2)^{\frac{B}{2\sqrt{|\lambda|}}}
P_n^{\left(\frac{B}{\sqrt{|\lambda|}}-\frac{1}{2},\frac{A}{\sqrt{|\lambda|}}-\frac{1}{2}\right)}(1+2\lambda
x^2)
\end{array}\label{e6.27}
\end{equation}

\section{Relation between $\Lambda$-deformed Hermite polynomial and Jacobi polynomial}
\noindent Here we shall obtain a relationship between the
$\Lambda$-deformed Hermite polynomials \cite{Ca+07} and Jacobi
polynomials. We recall that the solutions of the Hamiltonian for nonlinear
oscillator (\ref{e5.00}) is given by the equations (\ref{e5.01}) and (\ref{e5.02}) in terms of $\Lambda$-deformed Hermite
polynomial ${\cal {H}}_m(y,\Lambda)$ whose Rodrigues formula and generating function are
given in (\ref{rf}). On the other hand, putting $B =
0$ and $A = \frac{\alpha}{\sqrt{\lambda}}$ in the solution
(\ref{e6.8}) of Eqn.(\ref{e6.0}), the eigenfunctions of the QNLO
(equation (\ref{e5.03})) can be written in terms of Jacobi polynomial as
\begin{equation}
\psi_n(y) = N_n(1+\Lambda
y^2)^{-\frac{1}{2\Lambda}}P_n^{(-\frac{1}{2}-\frac{1}{\Lambda},-\frac{1}{2}-\frac{1}{\Lambda})}(iy\sqrt{\Lambda})~~,~~n
= 0,1,2\cdots <\frac{1}{\Lambda}~~~(\Lambda > 0) \label{eqn7}
\end{equation}
For $\Lambda < 0$, putting $B=0,
A=\frac{\alpha}{\sqrt{|\lambda|}}$ in the wavefunction of the 5th
entry of  Table \ref{t6.1} and using (\ref{e5.21}) we obtain
\begin{equation}
\psi_n(y) = N_n(1+\Lambda y^2)^{-\frac{1}{2\Lambda}}
P_n^{(-\frac{1}{2}-\frac{1}{\Lambda},-\frac{1}{2}-\frac{1}{\Lambda})}(y\sqrt{|\Lambda|})~~,
~~n = 0,1,2\cdots ~~~(\Lambda < 0) \label{eqn8}
\end{equation}
Comparing Eqns.(\ref{e5.02}) and (\ref{eqn8}) and also
Eqns.(\ref{e5.01}) and (\ref{eqn7}), it is possible to derive a
relation between $\Lambda$-deformed Hermite polynomial ${\cal
{H}}_n(y,\Lambda)$ and Jacobi polynomial $P_n^{(\alpha,
\beta)}(x)$ as
\begin{equation}
P_n^{(-\frac{1}{2}-\frac{1}{\Lambda},-\frac{1}{2}-\frac{1}{\Lambda})}(iy\sqrt{\Lambda})
= \frac{1}{n!}\left(\frac{1}{2i\sqrt{\Lambda}}\right)^n {\cal
{H}}_n(y,\Lambda),~~~~~~\forall~ \Lambda \label{e9}
\end{equation}
The Rodrigues formula and the generating function for the
$\Lambda$-deformed Hermite polynomial ${\cal {H}}_n(y,\Lambda)$
were given by {\cite{Ca+07}}
\begin{equation}
\begin{array}{lcl}
{\cal {H}}_n(y,\Lambda) &=& (-1)^nz_y^{\frac{1}{\Lambda}+\frac{1}{2}}\frac{d^n}{dy^n}\left[z_y^n z_y^{-(\frac{1}{\Lambda}+\frac{1}{2})}\right],~~~z_y = 1 + \Lambda y^2\\
{\cal {F}}(t,y,\Lambda) &=& (1+\Lambda (2ty -
t^2))^{\frac{1}{\Lambda}}
\end{array}\label{rf}
\end{equation}
It was shown {\cite{Ca+07}} that the polynomials obtained from the
generating function ${\cal {F}}(t,y,\Lambda)$ with those obtained
from Rodrigues formula are essentially the same and only differ in
the values of the global multiplicative coefficients. We have
observed that if the generating function ${\cal {F}}(t,y,\Lambda)$
is taken as
 \begin{equation}
    (1+\Lambda(2ty-t^2))^{\frac{1}{\Lambda}}=\sum_{n=0}^\infty
    \frac{1}{2^n}
      \frac{\left(-\frac{1}{\Lambda}\right)_n}{\left(\frac{1}{2}-\frac{1}{\Lambda}\right)_n}
      {\cal {H}}_n(y,\Lambda)\frac{t^n}{n!}\label{e10}
    \end{equation}
where $(a)_n$ represents P\"ochhammer symbol given by
$(a)_n=\frac{\Gamma(a+n)}{\Gamma(a)}$ then the polynomials
obtained from the above relation are exactly
same with those obtained from Rodrigues formula given in Eqn.(\ref{rf}). Correspondingly the recursion relations are obtained as
\begin{equation}
    (\Lambda(2n+1)-2)~ [2(1-n\Lambda) y {\cal {H}}_n(y,\Lambda)+(\Lambda(2n-1)-2)n
    {\cal {H}}_{n-1}(y,\Lambda)]=(n\Lambda-2){\cal
    {H}}_{n+1}(y,\Lambda)\label{rr1}
\end{equation}
and
\begin{equation}
\begin{array}{ll}
\displaystyle (\Lambda(n-2)-2) ~[2(\Lambda(2n-1)-2)n
    {\cal {H}}_n(y,\Lambda)-(\Lambda(n-1)-2){\cal {H}}'_n(y,\Lambda)]\\
    \displaystyle
    ~~~~~=n\Lambda(\Lambda(2n-1)-2)~[2(\Lambda(n-2)-2)y{\cal {H}}'_{n-1}(y,\Lambda)-(n-1)
    (\Lambda(2n-3)-2){\cal {H}}'_{n-2}(y,\Lambda)]
    \end{array}\label{rr2}
    \end{equation}
    where `prime' denotes differentiation with respect to $y$.
    For $\Lambda \rightarrow 0$ Eqns.(\ref{rr1}) and (\ref{rr2}) give the recursion relations for Hermite
    polynomial.
\newpage
\section{Summary}
  \noindent To summarize, we point out the following main results :
\begin{itemize}
 \item[$\blacktriangleright$]{
    We have constructed the coherent states for nonlinear oscillator via Gazeau-Klauder formalism.
    These coherent states are shown to satisfy the requirements of continuity of labeling, resolution of unity, temporal stability and action identity.
    The plots of the weighting distribution for these coherent states are almost Gaussian in nature. The Mandel parameter $Q$ is sub-Poissonian which indicates
    that the coherent states (\ref{e5.10}) exhibit squeezing for all values of $\mu$. The fractional revivals of the coherent states are evident from the figures \ref{f5.3}(b)
    and \ref{f5.3}(c) depicting squared modulus of the autocorrelation function. This is in contrast to the results obtained in ref \cite{Sc06} where the wave packet revival in
    an infinite well for the Schr\"odinger equation with position dependent mass was studied. In ref.\cite{Sc06}, it was found that though full revival takes place,
    there is no fractional revival in the usual sense. Instead, a very narrow wave packet is located near one wall of the well, when the mass is higher. In the present paper the spectral compositions and the intensities of the fractional revivals are determined by phase analysis.}

  \item[$\blacktriangleright$]{ We have studied various exactly solvable Hermitian, non-Hermitian $\mathcal{PT}$-symmetric and quasi-exactly solvable generalizations of the quantum nonlinear oscillator with the mass function
$\left(\frac{1}{1+\lambda x^2}\right)$. The eigenfunctions of the quantum nonlinear oscillator are previously obtained \cite{Ca+07} in terms of $\Lambda$-deformed Hermite polynomials. These $\Lambda$-deformed Hermite polynomials are actually related to classical Jacobi Polynomials as has been shown in equation (\ref{e9}). In the case of unbroken supersymmetry, it has been shown that all the generalizations of the quantum nonlinear oscillator listed in tables \ref{t6.1} and table \ref{t6.2} are shape invariant. In this context, the case of broken supersymmetry has also been discussed.}
\end{itemize}


\chapter{Future Directions}\label{c5}
 \noindent Here we point out some future issues related to the work presented in chapters \ref{c2}, \ref{c3} and \ref{c4}.

 \begin{itemize}
 \item[$\blacktriangleright$] {In section \ref{s2.1}, the modified factorization approach based on unbroken supersymmetry,
 has been extended to quantum systems with position dependent mass. It would be interesting to formulate such algorithm when supersymmetry is broken.
     }

 \item[$\blacktriangleright$] {Some exactly solvable 
 position-dependent mass Hamiltonians and quasi-Hermitian Hamiltonians have been obtained in section \ref{s2.2} and section \ref{s3.2} respectively. 
 The associated bound state wave functions are given in terms of $X_1$ Laguerre or Jacobi type EOPs. To look for solvable 
 position dependent mass and quasi-Hermitian Hamiltonians associated with exceptional orthogonal polynomials of higher co-dimension as well as multi-indexed ones 
 will be worthwhile.}

 \item[$\blacktriangleright$] {Formulation of higher order intertwining method for spectral modification
of a position-dependent mass Hamiltonian, by considering an $n$-th ($n\ge 3$) order differential operator as an intertwiner, 
is worth investigating in future. Also one can study the other possibilities of spectral
modifications apart form the ones reported in section \ref{s2.3}.}

 \item[$\blacktriangleright$] {In ref.\cite{Ma08,Mu08a}, the authors
have shown that it is possible to find solitons (with real eigenvalues)
of the nonlinear Schr\"{o}dinger equation $\frac{d^2\psi}{dx^2} + [4 \cos^2 x + 4 i V_0 \sin 2x] \psi(x) + |\psi|^2 \psi(x) = E \psi(x)$ above the $\mathcal{PT}$-threshold $V^{th}_0 =0.5$. This happens because the corresponding band structure remains real for some values of
bloch momentum $\nu$ even above the $\mathcal{PT}$-threshold. Stability analysis revealed that
 these solitons are unstable. However, we have shown, in section \ref{s3.1.2}, that no part of the band structure remains real beyond the second critical point $V_0^c \sim 0.888437$. Therefore an important direction of future work would be to study the existence as well as the stability of solitons (if any) beyond the second critical point $V_0^c$. Also it will be interesting to study
other related properties of a $\mathcal{PT}$-optical lattice (e.g. power oscillation, double refraction, non-reciprocity) near this second critical point.}

  \item[$\blacktriangleright$] {Generalized Swanson Hamiltonian is shown, in section \ref{s3.3}, to be non-isospectral 
  to the harmonic oscillator despite of $[\tilde{a},\tilde{a}^\dag] =$ Constant. The reason
  behind this anomaly is not clear till now. This requires further studies.}

  \item[$\blacktriangleright$] {The study of the revival dynamics of the quantum nonlinear oscillator using Renyi and Shanon information entropies and compare the results with the auto-correlation based analysis already done in section \ref{s4.1.2}, will be worth pursuing.}

  \item[$\blacktriangleright$] {The classical analogs of the Hamiltonians with generalized quantum nonlinear 
  oscillator potentials presented in tables \ref{t6.1} and \ref{t6.2}
  is another area of future investigation.}

 \end{itemize}

\vspace{2.5 cm}
\begin{center}
 $\divideontimes \divideontimes \divideontimes$
\end{center}


\singlespacing

\vspace{2 cm}
\begin{center}
 $\divideontimes \divideontimes \divideontimes$
\end{center}


\begin{thebibliography}{100}
\pagestyle{References}
\bibitem{Sc40} E. Schr\"{o}dinger, {\it Proc. R. Irish Acad.} A46 (1940) 9; A46 (1940) 183.
\bibitem{Da82} M. G. Darboux, {\it C. R. Acad. Sci. Paris} {94} (1882) 1456.
\bibitem{Cr55} M. Crum, {\it Quart. J. Math.} 6 (1955) 121.
\bibitem{IH51} L. Infeld and T. D. Hull, {\it Rev. Mod. Phys.} 23 (1951) 21.
\bibitem{Mi84} B. Mielnikh, {\it J. Math. Phys.}, 25 (1984) 3387.
\bibitem{Ni84} M.M. Nieto, {\it Phys. Lett.} B 145 (1984) 208.
\bibitem{Su85d} C.V. Sukumar {\it J.Phys. A} 18 (1985) 2937.
\bibitem{GL51} I.M. Gelfand and B.M. Levitan, {\it Am. Math. Soc. Transl.} 1 (1951) 253.
\bibitem{AM80} P.B. Abraham and H.E. Moses, {\it Phys. Rev.} A22 (1980) 1333.
\bibitem{Pu85} D.L. Pursey, {Phys. Rev.} D33 (1985) 1048.
\bibitem{Fl79} G.P. Flessas, {\it Phys. Lett. } A72 (1979) 289.
\bibitem{Ra79} M. Razavy, {\it Am. J. Phys.} 48 (1980) 285.
\bibitem{Du93a} A. de Souza Dutra, {\it Phys. Rev.} A47 (1993) R2437.
\bibitem{JR98} G. Junker and P. Roy, {\it Ann. Phys.} 270 (1998) 155.
\bibitem{Pa26} W. Pauli, {\it Zeitschrift fur Fizik} 36 (1926) 336.
\bibitem{Fo35} V. Fock, {\it Zeitschrift fur Fizik} 98 (1935) 145.
\bibitem{Tu88} A. V. Turbiner, {\it Commun. Math. Phys.} 118 (1988) 467.
\bibitem{Sh89} M. A. Shifman, {\it Int. J. Mod. Phys. } A126 (1989) 2897.
\bibitem{Us93} A.G. Ushveridze, {\it Quasi-Exactly Solvable Models in Quantum Mechanics} (IOP, Bristol), 1993.
\bibitem{Ra71} P. Ramond, {\it Phys. Rev.} D3 (1971) 2415.
\bibitem{NS71} A. Neveu and J. Schwarz, {\it Nucl. Phys.} B31, (1971) 86.
\bibitem{GL71} Y. A. Gelfand and E. P. Likhtman, {\it JETP Lett.} 13 (1971) 323.
\bibitem{VA73} D. Volkov and V. Akulov, {\it Phys. Lett.} B46 (1973) 109.
\bibitem{WZ74} J. Wess and B. Zumino, {\it Nucl. Phys.} B70 (1974) 39.
\bibitem{CM67} S. Coleman and J. Mandula, {\it Phys. Rev.} 159 (1967) 5.
\bibitem{Wi81} E. Witten, {\it Nucl. Phys.} B185 (1981) 513.
\bibitem{Wi82} E. Witten, {\it Nucl. Phys.} B202 (1982) 253.
\bibitem{CR83} M. de Crumbrugghe and V. Rittenberg, {\it Ann. Phys.} 151, (1983) 99.
\bibitem{CF83} F. Cooper and B. Freedman, {\it Ann. Phys.} 146 (1983) 262.
\bibitem{KNT85} V.A. Kostelecky, M.M. Nieto and D.R. Truax, {\it Phys. Rev.} D32 (1985) 2627.
\bibitem{So85} M.F. Sohnius, {\it Phys. Rep.} 128 (1985) 39.
\bibitem{HR84} R.W. Haymaker and A.R.P. Rau, {\it Am.J.Phys} 54 (1984) 928.
\bibitem{Ra86} A. R. P. Rau, {\it Phys. Rev. Lett.} 56, (1986) 95.
\bibitem{Su85e} C.V. Sukumar, {J. Phys.} A18 (1985) 2917.
\bibitem{AB+84} A.A. Andrianov, N.V. Borisov and M.V. Ioffe, {\it Phys. Lett.} A 105 (1984) 19.
\bibitem{CK00} F.Cooper, A.Khare and U.Sukhatme, {\it Supersymmetry in Quantum Mechanics}, (World Scientific, 2000).
\bibitem{Ba00} B. Bagchi, {\it Supersymmetry in Quantum and Classical Mechanics} (Chapman and Hall /CRC Monographs
and Surveys in Pure and Applied mathematics (2000)).
\bibitem{Ju96} G. Junker,{\it Supersymmetric methods in quantum mechanics and statistical physics}, Springer, 1996.
\bibitem{GMR11} A. Gangopadhyay, J.V. Mallow and C. Rasinariu, Supersymmetric quantum mechanics, (World SCintific, 2011).
\bibitem{CK95} F. Cooper, A Khare and U. Sukhatme, {\it Phys. Rep.} 251 (1995) 268.
\bibitem{Ro02} R.L. Rodrigues, The quantum mechanics SUSY algebra: An introductory review, arXiv: 0205017.
\bibitem{Ni76} H. Nicolai, {\it J.Phys.} A9 (1976) 1497.
\bibitem{Su86} C.V. Sukumar, {\it J. Phys.} A19 (1986) 2297.
\bibitem{AB+85}	A. Andrianov, N. Borisov, M. Eides and M. Ioffe, {\it Phys. Lett.} A109 (1985) 143.
\bibitem{Su87} C.V. Sukumar, {\it J. Phys.} A20 (1987) 2461.
\bibitem{Su85a} C. V. Sukumar, {\it J. Phys.} A18 (1985) L57.
\bibitem{DF98} G. Dunne and J. Feinberg, {\it Phys. Rev.} D 57 (1998) 1271.
\bibitem{CJ+08} F. Correa, V. Jakubsky, L.M. Nieto and M. Plyushchay, {\it Phys. Rev. Lett.} 101 (2008) 030403.
\bibitem{CJP08} F. Correa, V. Jakubsky, and M. Plyushchay {\it J.Phys. A} 41 (2008) 485303.
\bibitem{AICD95} A.A. Andrianov, M.V. Ioffe, F. Cannata, and J.P. Dedonder, {\it Int. J. Modern Phys.} A10 (1995) 2683.
\bibitem{AS03} A.A. Andrianov and A.V. Sokolov, {\it Nucl. Phys.} B660 (2003) 25.
\bibitem{AIS93} A.A. Andrianov, M.V. Iofe and V.P. Spiridonov, {\it Phys. Lett.} A174 (1993) 273.
\bibitem{AIN95} A.A. Andrianov, M.V. Ioffe, and D.N. Nishnianidze, {\it Phys. Lett.} A201 (1995) 103.
\bibitem{CNP07} F. Correa, L.M. Nieto and M. Plyushchay, {\it Phys. Lett. B} 644 (2007) 94.
\bibitem{Pl04} M.Plyushchay, {\it J.Phys. A} 37 (2004) 10375.
\bibitem{KP01} S.M. Klishevich and M.S. Plyushchay, {\it Nucl. Phys.} B606 (2001) 583.
\bibitem{AST01} H. Aoyama, M. Sato and T. Tanaka, {\it Nucl. Phys.} B619 (2001) 105.
\bibitem{AST01a} H. Aoyama, M. Sato and T. Tanaka, {\it Phys. Lett.} B503 (2001) 423.
\bibitem{FG04} D.J. Fernandez and N.F. Garcia, {\it AIP Conf. Proc.} 744 (2005) 236.
\bibitem{Fe10} D.J. Fernandez, {\it AIP Conf.Proc.} 1287 (2010) 3.
\bibitem{An+85} A.A. Andrianov et al., {\it Theor. Math. Phys.} 61 (1985) 965.
\bibitem{An+84} A.A. Andrianov et al., {\it Phys. Lett.} A105 (1984) 19.
\bibitem{AIN95a} A.A. Andrianov, M.V. Ioffe, D.N. Nishnianidze, {\it Theor. Math. Phys.} 104 (1995) 1129.
\bibitem{IGV06} M.V. Ioffe, J.Mateos Guilarte and P.A. Valinevich , {\it Ann. Phys.} 321 (2006) 2552.
\bibitem{KTV01} S. Kuru, A. Tegmen and A. Vercin,  {\it J. Math. Phys.} 42 (2001) 3344.
\bibitem{Io04} M.V. Ioffe, {\it J. Phys.} A37 (204) 10363.
\bibitem{AI88} A. A. Anderianov and M.V. Ioffe, {\it Phys. Lett.} B205 (1986) 507.
\bibitem{Ca+04a} F. Cannata et al., {\it J. Phys. A} 37 (2004) 10339.
\bibitem{IN03} M.V. Ioffe and A.I. Neelov, {\it J. Phys} A36 (2003) 2493.
\bibitem{Kh93} A. Khare, {\it J. Math. Phys.} 34 (1993) 1277.
\bibitem{DV93} S. Durand and L. Vinet, {\it J. Phys.} A23 (1990) 3661.
\bibitem{HS10} Y.R. Huang and W.C. Su, {\it J. Phys.} A43 (2010) 115302.
\bibitem{KMR93} A. Khare, A.K. Mishra and G. Rajasekara, {\it Int. J. Mod. Phys. } A8 (1993) 1245.
\bibitem{Du93} S. Durand, {\it Phys. Lett.} B312 (1993) 115.
\bibitem{Qu03} C. Quesne, {\it Mod. Phys. Lett.} A18 (2003) 515.
\bibitem{Ge83} L. Gendenshtein, {\it JETP Lett.} 38 (1983) 356.
\bibitem{DKS86} R. Dutt, A. Khare and U.P. Sukhatme, {\it Phys. Lett.} B181 (1986) 295.
\bibitem{DKS88} R. Dutt, A. Khare and U.P. Sukhatme, {\it Am. J. Phys.} 56 (1988) 163.
\bibitem{CGK87} F. Cooper, J.N. Ginocchio, and A. Khare, {\it Phys. Rev.} D36 (1987) 2458.
\bibitem{DKP88} J.W. Dabrowska, A. Khare and U. Sukhatme, {\it J. Phys.} A21 (1988) L195.
\bibitem{Gi84} J.N. Ginocchio, {\it Ann. Phys.} 152 (1984) 203.
\bibitem{Na79} G. A. Natanzon, {\it Theor. Mat. Fiz.} 38 (1979) 146.
\bibitem{Le89} G. Levai, {\it J.Phys.} A 22 (1989) 689.
\bibitem{KS93}  A. Khare, U.P. Sukhatme, {\it J. Phys. A} 26 (1993) L901.
\bibitem{Ba+93} D. Barclay et al., {\it Phys. Rev.} A48 (1993) 2786.
\bibitem{SRK97} U.P. Sukhatme, C. Rasinariu, A. Khare, {\it Phys. Lett.} A 234 (1997) 401.
\bibitem{EQ87} M.J. Englefield and C. Quesne, {\it J.Phys.} A24 (1987)3557.
\bibitem{WA90}J. Wu and Y. Alhassid, {\it J. Math. Phys.} 31 (1990) 557.
\bibitem{Ba98} A.B. Balantekin, {\it Phys. Rev.} D58 (1998) 013001.
\bibitem{Ch+98} S. Chaturvedi et al., {\it Phys. Lett.} A248 (1998) 109.
\bibitem{GRS99} A. Gangopadhyaaya, C. Rasinariu and U. Sukhatme, {\it Phys. Rev.} A60 (1999) 3482.
\bibitem{BRA99} A.B. Balantekin, M.A. Ribeiro and A.N.F. Aleixo, {\it J.Phys.} A32 (1999) 2785.
\bibitem{Su08} W.C. Su, {\it J.Phys.} A 41 (2008) 255307; {\it J. Phys.} A 42 (2009)  435301.
\bibitem{Du+93} R. Dutt et al., {\it Phys. Lett.} A174 (1993) 363.
\bibitem{GMS01} A. Gangopadhyaya, J.V. Mallow anmd U.P. Sukhatme, {\it Phys. Lett.} A283 (2001) 279.
\bibitem{BG+10} J. Bougie, A. Gangopadhyaya and J.V. Mallow, Phys.Rev.Lett. 105(2010)210402.
\bibitem{BGM11} J. Bougie, A. Gangopadhyaya and J.V. Mallow, {\it J. Phys.} A44 (2011) 275307.
\bibitem{Ra11} A. Ramos, {\it J. Phys. A} 44 (2011) 342001.
\bibitem{Le92} G. Levai, {\it J. Phys. A} 25 (1992) L521.
\bibitem{GM+99} A. Gangopadhyaya, J.V. Mallow, C. Rasinariu and U.P. Sukhatme, {\it Theo. Math. Phys.} 118 (1999) 285.
\bibitem{Le04}G. Levai, {\it Czech J. Phys.} 54 (2004) 1121.
\bibitem{TF01} V.M. Tkachuk and T.V. Fityo, {\it J. Phys.} A 34 (2001) 8673.
\bibitem{Tk99}V.M. Tkachuk, {\it J. Phys. A} 32 (1999) 1291.
\bibitem{QT03} C. Quesne and V.M. Tkachuk, {\it J. Phys. A}  36 (2003) 10373.
\bibitem{Tk01} V.M. Tkachuk, {\it J. Phys.} A34 (2001) 6339.
\bibitem{TV02} V.M. Tkachuk and O. Vozhyak, {\it Phys. Lett.} A301 (2002) 177.
\bibitem{KS99} A. Khare and U. Sukhatme, {\it J. Math. Phys.} 40 (1999) 5473.
\bibitem{LR99} G. Levai and P. Roy, {\it Phys. Lett.} A264 (1999) 117.
\bibitem{NRV94} N. Nag, R. Roychoudhury and Y.P. Varshini, {\it Phys. Rev.} A49 (1994) 5098.
\bibitem{ACD88} R.D. Amado, F. Cannata and J. Dedonder, {\it Phys. Rev.} A38 (1988) 3797.
\bibitem{Ba87b} D. Baye, {\it Phys. Rev. Lett.} 58 (1987) 2738.
\bibitem{Ba87a} D. Baye, {\it J. Phys.} A20 (1987) 5529.
\bibitem{KS89b} A. Khare and U. Sukhatme, {\it J. Phys.} A22 (1989) 2847.
\bibitem{SB97} J.M. Sparenberg and D. Baye, {\it Phys. Rev. Lett.} 79 (1997) 3802.
\bibitem{PSP93} J. Pappademos, U. Sukhatme and A. Pagnamenta, {\it Phys. Rev.} A48 (1999) 3535.
\bibitem{St95} A.A. Stahlhofen, {\it Phys. Rev.} A51 (1995) 934.
\bibitem{KS88} A. Khare and U.P. Sukhatme, {\it J. Phys.} A21 (1988) L501
\bibitem{ACDI95} A.A. Andrianov, F. Cannata, J.P. Dedonder, M.V. Ioffe, {\it Int.J.Mod.Phys.} A10 (1995) 2683.
\bibitem{KS04a} A. Khare, and U. Sukhatme, {\it J. Phys. A} 37 (2004) 10037.
\bibitem{IGV08} M.V. Ioffe, J.M. Guitarte, P.A. Valinevich, {\it Nucl. Phys.} B790 (2008) 414.
\bibitem{FNN00} D.J. Fernandez, J. Negro, L.M. Nieto {\it Phys Lett A} 275 (2000) 338.
\bibitem{SK02} U. Sukhatme and A Khare, {\it Phys. Atom. Nucl.} 65 (2002) 1122.
\bibitem{Fe+02} D.J. Fernandez et al, {\it J. Phys.} A 35 (2002) 4279.
\bibitem{IS85} T. Imbo and U. Sukhatme, {\it Phys. Rev. Lett.} 54 (1985) 2184.
\bibitem{RV88} R. Roychoudhury and Y.P. Varshni, {\it Phys. Rev.} A37 (1988) 2309.
\bibitem{CR90} F. Cooper and P. Roy, {\it Phys. Lett.} A143 (1990) 202.
\bibitem{GRT93} E. Gozzi, M. Reuter and W.D. Thacker, {\it Phys. Lett.} A183 (1993) 29.
\bibitem{CDS94} F. Cooper, J. Dawson and H. Shepard, {\it Phys. Lett.} A187 (1994) 140.
\bibitem{FR00} E.D. Filho and R.M. Ricotta, {\it Phys. Lett.} A269 (2000) 269.
\bibitem{Kh85} A. Khare, {\it Phys. Lett.} B161 (1988) 131.
\bibitem{DKS91} R. Dutt, A. Khare and U. Sukhatme, {\it Am. J. Phys.} 59 (1991) 723.
\bibitem{HKS97} M. Hruska, W. Keung and U. Sukhatme, {\it Phys. Rev.} A 55 (1997) 3345.
\bibitem{CCX06} G. Chen, Z. Chen and P. Xuan, {\it Phys. Lett.} A 352 (2006) 317.
\bibitem{BM91} D.T. Barclay and C.J. Maxwell, {it Phys. Lett.} A157 (1991) 357.
\bibitem{Ad+86} R. Adhikary et al., {\it Phys. Rev.} A38 (1986) 1679.
\bibitem{DKV87} R. Dutt, A. Khare and Y.V. Varshni, {\it Phys. Lett.} A123 (1987) 375.
\bibitem{Va92} Y.P. Varshni, {\it J.Phys. A} 25 (1992) 5761.
\bibitem{Fr+88} S. Fricke et al., {\it Phys. Rev.} A37 (1988) 1686.
\bibitem{SP90} U. Sukhatme and a. Pagnamenta, {\it Phys. Lett.} A 151 (1990) 7.
\bibitem{Du+99} R. Dutt et al., {\it Phys. Rev.} A48 (1999) 1845.
\bibitem{IJ94} A. Inomata and G. Junker, {\it Phys. Rev.} A50 (1994) 3638.
\bibitem{KKS88} W. Keung, E. Kovacs and U.P. Sukhatme, {\it Phys. Rev. Lett.} 60 (1988) 41.
\bibitem{GPS93} A. Gangopadhaya, P.K. Panigrahi and U.P. Sukhatme, {\it Phys. Rev.} A47 (1993) 2720.
\bibitem{HKN86} R.J. Hughes, V.A. Kostelecky and M.M. Nieto, {\it Phys. Rev.} D34 (1986) 1100.
\bibitem{NT93} Y. Nogami and F.M. Toyama, {\it Phys. Rev.} A 47 (1993) 1708.
\bibitem{Co+88} F. Cooper et al, {\it Ann. Phys.} 187 (1988) 1.
\bibitem{Su85f} C. V. Sukumar, {\it J. Phys.} A18 (1985) L697.
\bibitem{HV84} E. DHoker and L.Vinet {\it Phys. Lett.} B137 (1984) 72.
\bibitem{KM84b}  A. Khare and J. Maharana, {\it Nucl. Phys.} B244 (1984) 409.
\bibitem{Ui84} H. Ui, {\it Prog. Theor. Phys.} 72 (1984) 813.
\bibitem{HY10} R.L. Hall and O.Yesiltas, {\it Int. J. Mod. Phys.} E 19 (2010) 1923.
\bibitem{JR07} T. Jana and P. Roy, {\it Phys. Lett.} A 361 (2007) 55.
\bibitem{Le+01} P.T. Leung et al, {\it J. Math. Phys.} 42 (2001) 4802.
\bibitem{JR07a} T.K. Jana and P. Roy, {\it J. Phys.} A40 (2007) 5865.
\bibitem{FM90} D.Z. Freedman and P.F. Mende, {\it Nucl. Phys.} B 344 (1990) 317.
\bibitem{IN02} M.V. Ioffe and A.I. Neelov, {\it J. Phys. A} 35 (2002) 7613.
\bibitem{Gh12} P.K. Ghosh, {\it J.Phys.} A45 (2012) 183001.
\bibitem{GKS98} P.K. Ghosh, A. Khare and M. Shivakumar, {\it Phys. Rev.} A 58 (1998) 821.
\bibitem{SG95} C.V. Sukumar and P. Guha, {\it J.Math. Phys.} 36 (1995) 321.
\bibitem{GR94} A.K. Grant and J.L. Rosen, {\it J. Math. Phys.} 35 (1994) 2142.
\bibitem{HSK04} E. Harikumar, V. Sunil Kumar and A. Khare, {\it Phys. Lett.} B589 (2004) 155.
\bibitem{Gh05} P.K. Ghosh, {\it Eur. J. Phys.} C 42 (2005) 355.
\bibitem{GS09} J. Ben Geloun and F.G. Scholtz, {\it J. Phys.} A42 (2009) 165206.
\bibitem{PVZ11} S. Post, L. Vinet and A. Zhedanov, {\it J. Phys.} A 44 (2011) 435301.
\bibitem{Ta12} T. Tanaka, {\it Int. J. Mod. Phys. } A27 (2012) 1250102.
\bibitem{Fe84} D.J. Fernandez, {\it Lett. Math. Phys.} 8 (1984) 337.
\bibitem{Ro97} M.Robnik, {\it J. Phys.} {A30} (1997) 1287.
\bibitem{JR84} A. Jevicki and J.P. Rodrigues, {\it Phys. Lett.} B146 (1984) 55.
\bibitem{Ca95} J. Casahorran, {\it Physica} { A217} (1995) 429.
\bibitem{DP99} A. Das and S.A. Pernice, {\it Nucl.Phys.} B561 (1999) 357.
\bibitem{PS93} P.K. Panigrahi and U.P. Sukhatme, {\it Phys. Lett.} {A 178} (1993) 1993.
\bibitem{BU10a} M.S. Berger and N.S. Ussembayev, {\it Phys. Rev.} { A 82} (2010) 022121.
\bibitem{BU10b} M.S. Berger and N.S. Ussembayev,  {\it J. Phys.} {A43} (2010) 385309.
\bibitem{DR11} D. Datta and P. Roy, {\it Phys. Rev.} {A 83} (2011) 054102.
\bibitem{Qu08} C. Quesne, {\it J. Phys.} A41 (2008) 392001.
\bibitem{Qu09a} C. Quesne, {\it SIGMA} 5 (2009) 084.
\bibitem{Gr11a} Y. Grandati, {\it Ann. Phys.} 326 (2011) 2074.
\bibitem{Gr11b} Y. Grandati, {\it J. Math. Phys.} 52 (2011) 103505.
\bibitem{OS09a} S. Odake and R. Sasaki, {\it Phys. Lett.} B679 (2009) 414.
\bibitem{OS10} S. Odake and R. Sasaki, {\it J. Phys. A} 43 (2010) 335201.
\bibitem{Qu11} C. Quesne, {\it Mod. Phys. Lett.} A26 (2011) 1843.
\bibitem{BQ10} B. Bagchi and C. Quesne, {\it J. Phys. A} 43 (2010) 305301.
\bibitem{OS10d} S. Odake and R. Sasaki, {\it J. Math. Phys.} 51 (2010) 053513.
\bibitem{Qu11c}C. Quesne, {\it Int. J. Mod. Phys. A} 26 (2011) 5337.
\bibitem{UK09} D. Gomez-Ullate, N. Kamran and R. Milson, {\it J. Math. Anal. Appl.} 359 (2009) 352.
\bibitem{UK10} D. Gomez-Ullate, N. Kamran and R. Milson, {\it J. Approx. Theory} 162 (2010) 987.
\bibitem{BQR09} B. Bagchi, C. Quesne and R. Roychoudhury, {\it Pramana: J. Phys.} 73 (2009) 337.
\bibitem{HOS11} C-L. Ho, S. Odake and R. Sasaki, {\it SIGMA} 7 (2011) 107.
\bibitem{HS12} C-L. Ho and R. Sasaki, {\it ISRN Mathematical Physics}, 2012 (2012) 920475.
\bibitem{STZ10} R. Sasaki, S. Tsujimoto and A. Zhedanov, {\it J. Phys.} A43 (2010 ) 315204.
\bibitem{Ho11d} C.L. Ho, {\it J. Math. Phys.} 52 (2011) 122107.
\bibitem{Ho11a} C.L. Ho, {\it Prog. Theor. Phys.} 126 (2011) 185.
\bibitem{UKM12} D. Gomez-Ullate, N. Kamran and R. Milson, {\it J. Math. Anal. Appl.} 387 (2012) 410.
\bibitem{OS11a} S. Odake and R. Sasaki, {\it Phys. Lett.} B702 (2011) 164.
\bibitem{Ho11b} C.L. Ho, {\it Ann. Phys.} 326 (2011) 797.
\bibitem{DR10} D. Dutta and P. Roy, {J. Math. Phys.} 51 (2010 ) 042101.
\bibitem{Sr12} S. Sree Ranjani et al, {\it J. Phys. A} 45 (2012) 055210.
\bibitem{Ta10} T. Tanaka, {\it J. Math. Phys.} 51 (2010) 032101.
\bibitem{OS09b} S. Odake and R. Sasaki, {\it Phys. Lett.} B682 (2009) 130.
\bibitem{OS11c} S. Odake and R. Sasaki, {\it Prog. Theor. Phys.} 125 (2011) 851.
\bibitem{GG04} K. Goser, P. Gl\"{o}sek\"{o}tter and  J. Dienstuhl, {Nanoelectronics and Nanosystems. From
Transistors to Molecular and Quantum Devices}, Springer-Verlag, Berlin 2004.
\bibitem{Ph98} W.D. Phillips {\it Rev. Mod. Phys.} 70 (1998) 721.
\bibitem{WJP95} C. Zicovich-Wilson, W. Jaskilski and J.H. Planelles,{\it Int . J. Quant. Chem.} 54 (1995) 61.
\bibitem{To01} G. Todorovic et al., {\it Phys. Lett.} A279 (2001) 268.
\bibitem{In00} D. Indjin et al., {\it SuperLattices and Microstructures}, 28 (2000) 143.
\bibitem{BS95} V.G. Bagrov and B.F. Samsonov, {\it Theo. Math. Phys.} 104 (1995) 1051.
\bibitem{BS97} V.G. Bagrov and B.F. Samsonov,  {\it Phys. Part. Nucl.} 28 (1997) 374.
\bibitem{Ha11a} A. S. Halberg, {\it J. Math. Phys.} 52 (2011) 083505.
\bibitem{Ha10a} A. S. Halberg, {\it J. Math. Phys.} 51 (2010) 033521.
\bibitem{Ba93} D. Baye, {Phys. rev.} A48 (1993) 2040.
\bibitem{Am88} R.D. Amado, {\it Phys. Rev.} A37 (1988) 2277.
\bibitem{AC04} A.A. Andrianov and F. Canata, {\it J. Phys.} A37 (2004) 10297.
\bibitem{KS89} A. Khare and U.P. Sukhatme, {Phys. Rev.} A40 (1989) 6185.
\bibitem{LS75} M.B. Levitan and I.S. Sargsian, {\it Introduction to Spectral Theory} (AMS) 1975.
\bibitem{CRF01} J.F. Carinena, A. Ramos and D.J. Fernandez, {\it Ann. Phys.} 292 (201) 42.
\bibitem{Ca+98} F. Cannata et al., {\it J. Phys.} A32 (1999) 3583.
\bibitem{Or98} J.O. Rosas-Ortiz, {\it J. Phys.} A31 (1998) 10163.
\bibitem{FMR03} D.J. Fernandez, R. Munoz and A. Ramos, {\it Phys Lett} A 308 (2003) 11.
\bibitem{AF08} A.C. Astorga and D.J. Fernandez, {J. Phys.} A41 (2008) 475303.
\bibitem{FH05} D.J. Fernandez, E. Hernandez, {\it Phys.Lett.A} 338 (2005) 13.
\bibitem{KTP01} S. Kuru, A. Tegmen, and A. Vercin, {\it J. Math. Phys.} 42 (2001) 3344.
\bibitem{Ha10} A.S. Halberg, {\it J. Math. Phys.} 51 (2010) 033521.
\bibitem{HPS09} A. Schulze-Halberg, E. Pozdeeva and A. Suzko, {\it J. Phys.} A42 (2009) 115211.
\bibitem{An91} A. Anderson, {\it Phys. Rev. A} 43 (1991) 4602.
\bibitem{SP00} B.F. Samsonov and A.A. Pecheritsin, {\it Rus. Phys. J.} 43 (2000) 48.
\bibitem{NPS03} L.M. Nieto, A.A. Pecheritsin and B.F. Samsonov, {\it Ann. Phys.} 305 (2003) 151.
\bibitem{SH08} A. A. Suzko and A. S. Halberg, {\it Phys. Lett.} A{372} (2008) 5865.
\bibitem{JM94} G. Junker and S. Matthiesen, {\it J. Phys.} A 27 (1984) L751.
\bibitem{SWY00} W.M. Suen, C.W. Wong and K. Young, {\it Phys. Lett.} A271 (2000) 54.
\bibitem{KN08} S. Kuru and J. Negro, {\it Ann. Phys. }  323 (208) 413.
\bibitem{LN03} H.W. Lee and D.S. Novikov, {\it Phys. Rev. B} 68 (2003) 155402.
\bibitem{JP12} V. Jakubsky and M. Plyushchay , {\it Phys. Rev. D}  85 (2012) 045035.
\bibitem{MP06} J. Morales and J.J. Pena, {\it Phys. Scr.} 74 (2006) 71.
\bibitem{BS94} D. Baye, J-M. Sparenberg, {\it Phys. Rev. Lett.} 73 (1994) 2789.
\bibitem{SSB07} B.F. Samsonov, J-M. Sparenberg and D. Baye, {\it J. Phys.} A40 (2007) 4225.
\bibitem{LSSB00} H. Leeb, S.A. Sofianos, J.M.. Sparenberg and D. Baye, {\it Phys. Rev.} C62 (2000) 064003.
\bibitem{BCA99} A.B. Balantekin, M.A. Candido Ribeiro and A. N. Aleixo, {\it J. Phys.} A32 (1999) 2785.
\bibitem{ABC00a} A.N. Aleixo, A.B. Balantekin and M.A. Candido Ribeiro {\it J. Phys.} A33 (2000) 1503.
\bibitem{KN84} A. Kostelecky and M.M. Nieto, {\it Phys. Rev. Lett.} 53, (1984) 2285.
\bibitem{BB98} C.M. Bender and S. Boettcher, {\it Phys. Rev. Lett.} { 80} (1998) 5243.
\bibitem{Be05} C. M. Bender, {\it Contemp. Phys.} 46 (2005) 277.
\bibitem{Be07} C. M. Bender, {\it Rep. Prog. Phys.} 70 (2007) 947.
\bibitem{DDT01} P. Dorey, C. Dunning, and R. Tateo, {\it J. Phys.} A34 (2001) 5679.
\bibitem{BBJ02} C.M. Bender, D.C. Brody, and H.F. Jones, {\it Phys. Rev. Lett.} 89 (2002) 270401.
\bibitem{BBJ03} C.M. Bender, D.C. Brody and H.F. Jones, {\it Am. J. Phys.} 71 (2003) 1095.
\bibitem{BBRR04} C.M. Bender, J. Brod, A. Refig and M. Reuter, {\it J. Phys.} A37 (2004) 10139.
\bibitem{BMW03} C.M. Bender, P.N. Meisinger and Q. Wang, {\it J. Phys.} A36 (2003) 1973.
\bibitem{BJ04} C. M. Bender and H. Jones, {\it  Phys. Lett.} A 328 (2004) 102
\bibitem{BT06} C.M. Bender and B. Tan {\it J. Phys.} A39 (2006) 1945.
\bibitem{RR07a} R. Roychoudhury and P. Roy, {\it J. Phys.} A40 (2007) F617.
\bibitem{Wi03} S. Weigert, {\it Phys. Rev.} A 68 (2003) 062111.
\bibitem{BQZ01} B. Bagchi, C. Quesne and M. Znojil, {\it Mod. Phys.} A 16 (2001) 2047.
\bibitem{Be+10} C.M. Bender et al., {\it Ann. Phys.} 325 (2010) 2332.
\bibitem{Be+10a} C.M. Bender et al., {\it Phys. Rev. Lett.} 104 (2010) 061601.
\bibitem{BBM99} C.M. Bender, S. Boettcher, and P.N. Meisinger, {\it J. Math. Phys.} 40 (1999) 2201.
\bibitem{Sh++} K. C. Shin, {\it J. Math. Phys.} 42 (2001) 2513; {\it Commun. Math. Phys.} 229 (2002) 543; {\it J. Math. Phys.} 46 (2005) 082110.
\bibitem{PD98} F. Pham and E. Delabaere, {\it Phys. Lett.} A250 (1998) 25; E. Delabaere and F. Pham, {\it Phys. Lett. }A250 (1998) 29; E. Delabaere and D. T. Trinh, {\it J. Phys.} A33 (2000) 8771.
\bibitem{We06} S. Weigert, {\it J. Opt.} B5 (2003) S416; {\it J. Phys.} A39 (2006) 235; {\it J. Phys.} A39 (2006) 10239.
\bibitem{SG06} F.G. Scholtz and H.B. Geyer, {\it Phys. Lett.} B634 (2006) 84; {\it J. Phys.} A39 (2006) 10189.
\bibitem{Mo02a} A. Mostafazadeh, {\it J. Math. Phys.} { 43} (2002) 205.
\bibitem{Mo02b} A. Mostafazadeh  {\it J. Math. Phys.} { 43} ( 2002) 2814
\bibitem{Mo02c} A. Mostafazadeh  2002 {\it J. Math. Phys.} {\bf 43} 3944
\bibitem{Mo10}  A. Mostafazadeh, {\it Int. J. Geom. Meth. Mod. Phys.} 7 (2010) 1191.
\bibitem{SGH92} F.G. Scholtz, H.B. Geyer and J.W. Hahne, {\it Ann. Phys.}, { 213} (1992) 74.
\bibitem{KS04} Kretschmer R and Szymanowski L 2004 {\it Phys. Lett. A} {\bf 325} 112.
\bibitem{MB04} A. Mostafazadeh and A. Batal, {\it J.Phys.} A37 (2004) 11645.
\bibitem{So02} L. Solombrino, {\it J. Math. Phys.} 43 (2002) 5439.
\bibitem{BQ02} B. Bagchi and C. Quesne, {\it Phys. Lett.} A 301 (2002) 173.
 \bibitem{Mo05a} A. Mostafazadeh, {\it J. Math. Phys.} 46 (2005) 102108.
\bibitem{Mo06} A. Mostafazadeh, {\it J. Phys.} A39 (2006) 13506.
\bibitem{Jo05} H.F. Jones, {\it J. Phys.} A {38} (2005) 1741.
\bibitem{Mo05b} A. Mostafazadeh, {\it J. Phys.} A 38 (2005) 3213.
\bibitem{Ah01b} Z. Ahmed, {\it Phys. Lett.} A290 (2001) 19.
\bibitem{Ah02} Z. Ahmed, {\it Phys. Lett.} A294 (2002) 287.
\bibitem{Mo02d} A. Mostafazadeh, {\it Nucl. Phys.} B640 (2002) 419.
\bibitem{SR07a} A. Sinha and P. Roy, {\it J. Phys.} A39 (2007) L377.
\bibitem{Na54} M.A. Naimark, {\it Proc. Mosc. Math. Soc.} 3 (1954) 181.
\bibitem{Sa05} B.F. Samsonov, {\it J. Phys.} A38 (2005) L571.
\bibitem{Mo11} A. Mostafazadeh, {\it J. Phys.} A44 (2011) 375302.
\bibitem{Ah09} Z. Ahmed, {\it J. Phys.}, A42 (2009) 472005.
\bibitem{He04} W.D. Heiss, {\it J. Phys.} A 37 (2004) 2455.
\bibitem{Be04} M.V. Berry, {\it Czech.J.Phys.} 54 (2004) 1039.
\bibitem{Sm09} A.V. Smilga, {\it J. Phys. A} 42 (2009) 095301.
\bibitem{MR08} M. Muller and I. Rotter {\it J. Phys.} A41 (2008) 244018.
\bibitem{Mo09a} A. Mostafazadeh, {\it Phys. Rev. Lett.} 102 (2009) 220402; {Phys. Rev.} A80 (2009) 032711.
\bibitem{RDM05} A. Ruschhaupt, F. Delgado, and J. G. Muga, {\it J. Phys.} A38 (2005) L171.
\bibitem{MD09} A. Mostafazadeh and H.M. Dehnavi, {\it J. Phys.} A42(2009)125303.
\bibitem{MS11} A. Mostafazadeh and M. Sarisman, {\it Phys. Lett.} A375(2011) 3387.
\bibitem{Ah12} Z. Ahmed, {\it J. Phys.} A45 (2012) 032004.
\bibitem{Ka66} T. Kato, {\it Perturbation Theory of Linear Operators} (Berlin, Springer, 1966)
\bibitem{Ro09} I. Rotter, {\it J. Phys.} A42 (2009) 153001.
\bibitem{CMW07} H. Cartarius, J. Main, and G. Wunner, {\it Phys. Rev. Lett.} 99 (2007) 173003.
\bibitem{KGM08} S. Klaiman, U. Gunther and N. Moiseyev, {\it Phys. Rev. Lett.} 101 (2008) 080402.
\bibitem{Le+09} S. Leev et al., {\it Phys. Rev. Lett.} 103 (2009) 134101.
\bibitem{BQR08} B. Bagchi, C. Quesne and R. Roychoudhury, {\it J. Phys.} A41 (2008) 022001.
\bibitem{Zn99a} M. Znojil, {Phys. Lett.} A259 (1999) 220.
\bibitem{Zn99} M. Znojil, {\it Phys. Lett.} A264 (1999) 108.
\bibitem{Mo05d} A. Mostafazadeh, {\it J. Phys.} A38 (2005) 6557.
\bibitem{Zn00b} M. Znojil, {\it J. Phys.} A33(2000) 4561.
\bibitem{Zn01} M. Znojil, {\it Phys. Lett.} A285 (2001) 7.
\bibitem{BMQ02d} B. Bagchi, S. Mallik and C. Quesne, {\it Mod. Phys. Lett.} A17 (2002) 1651.
\bibitem{ZJ05} M. Znojil, V. Jakubsky, {\it J. Phys.} A38 (2005) 5041.
\bibitem{BR00} B. Bagchi and R. Roychoudhury {\it J. Phys.} {A33} (2000) L1.
\bibitem{Ah01f} Z. Ahmed, {\it Phys. Lett.} A 282 (2001) 343.
\bibitem{Le06} G. Levai, {\it J. Phys.} A39 (2006) 10161.
\bibitem{Le06a} G. Levai, {\it Czech. J. Phys.} 56 (2006) 453.
\bibitem{Le08} G. Levai, {\it Phys. Lett.} A372 (2008) 6484.
\bibitem{LM09} G. Levai and E. Magyari, {\it J. Phys.} A39 (2009) 195302.
\bibitem{LZ00} G. Levai and M. Znojil, {\it J. Phys.} A33 (2000) 7165.
\bibitem{LZ01} G. Levai and M. Znojil, {\it Mod. Phys. Lett.} A 30 (2001) 1973.
\bibitem{ZL00} M. Znojil and G. Levai, {\it Phys. Lett.} A 271 (2000) 327.
\bibitem{Zn06} M. Znojil, {\it J. Phys.} A 39 (2006) 4047.
\bibitem{Zn06a} M. Znojil, {\it J. Phys.} A39 (2006) 441.
\bibitem{BQ00} B. Bagchi and C. Quesne, {\it Phys. Lett.} A 273 (2000) 285.
\bibitem{BB98a} C.M. Bender and S. Boettcher, {\it J. Phys. } A 31 (1998) L273.
\bibitem{Zn00d} M. Znojil, {J. Phys.} A33 (2000) 4203.
\bibitem{BCQ00} B. Bagchi, F. Cannata and C. Quesne, {\it Phys. Lett.} A269 (2000) 79.
\bibitem{BM05} C.M. Bender, and M. Monou, {\it J. Phys.} A38 (2005) 2179.
\bibitem{KM00} A. Khare and B.P. Mandal, {\it Phys. Lett.} A272 (2000) 53.
\bibitem{Ba+00} B. Bagchi et al., {\it Phys. Lett.} A 289 (2000) 34.
\bibitem{KM09} A. Khare and B.P. Mandal, {\it Pramanna- J. Phys.} 73 (2009) 387.
\bibitem{BDM99} C. M. Bender, G. V. Dunne and P. N. Meisinger, {\it Phys. Lett.} A 252 (1999) 272.
\bibitem{Ce03} J.M. Cervero, {\it Phys. Lett.} A317 (2003) 26.
\bibitem{Sh04} K. C. Shin, {\it J. Phys.} A37 (2004) 8287.
\bibitem{Jo99} H.F. Jones, {\it Phys. Lett.} A 262 (1999) 242.
\bibitem{Ah01e} Z. Ahmed, {\it Phys. Lett.} A286 (2001) 231.
\bibitem{KS06} A. Khare and U. Sukhatme, {\it J. Phys.} A39 (2006) 10133.
\bibitem{KS04d} A. Khare and U. Sukhatme, {\it Phys. Lett.} A324(2004)406.
\bibitem{CDV07} F. Cannata, J.P. Dedonder and A. Ventura, {\it Annals of Physics} 322 (2007) 397.
\bibitem{Ah01i} Z. Ahmed, {\it Phys. Rev.} A64 (2001) 042716.
\bibitem{LSZ09} G. Levai, P. Siegl and M. Znojil, {\it J. Phys.} A42 (2009) 295201.
\bibitem{Jo08} H.F. Jones, {\it Phys. Rev.} D 78(2008)065032.
\bibitem{Zn01b} M. Znojil, {\it Czech. J. Phys.} 51 (2001) 420.
\bibitem{An+99b} A.A. Andrianov et al., {\it Int. J. Mod. Phys.} A14 (1999) 2675.
\bibitem{Zn02c} M. Znojil, {\it J. Phys.} A35 (2002) 2341.
\bibitem{ZC+00} M. Znojil, F. Cannata, B. Bagchi and R. Roychoudhury,{\it Phys. Lett.} B483 (2000) 284.
\bibitem{Le04a} G. Levai, {\it J. Phys.} A37 (2004) 10179.
\bibitem{LZ02} G. Levai and M. Znojil, {\it J. Phys.} A35 (2002) 8793.
\bibitem{ZL01d} M. Znojil and G. Levai, {\it Mod. Phys. Lett.} A16 (2001) 2273.
\bibitem{LCV02} G. Levai, F. Cannata and A. Ventura, {\it Phys. Lett.} A300 (2002) 271.
\bibitem{JZ05d} V. Jakubsky and M. Znojil, {\it Czech. J. Phys.} 55 (2005) 1113.
\bibitem{Le09} G. Levai, {\it Pramana-J. Phys.} 73 (2009) 329.
\bibitem{MK00}B.P. Mandol and A. Khare, {\it Phys. Lett.} A 272 (2000) 53.
\bibitem{Mo07c} A. Mostafazadeh, {\it Phys. Rev. Lett.} 99 (2007) 130502.
\bibitem{AF08b} P.E. Assis and A. Fring, {\it J. Phys.} A41 (2008) 244002.
\bibitem{GS08} U. Gunther and B.F. Samsonov, {\it Phys. Rev. Lett.} 101 (2008) 230404.
\bibitem{BBJB07} C.M. Bender, D.C. Brody, H.F. Jones and B.K. Meister, {\it Phys. Rev. Lett.} 98 (2007) 040403.
\bibitem{BM11} C.M. Bender, P.D. Mannheim, {\it Phys. Rev.} D84 (2011) 105038.
\bibitem{Na02} A. Nanayakkara, {\it Phys. Lett. }A304(2002)67.
\bibitem{Le07} G. Levai, {\it J. Phys.} A40 (2007) F273.
\bibitem{BDMS01} C.M. Bender, C.V. Dunne, P.N. Meisinger and M. Simsek, {\it Phys. Lett.} A311(2001)281.
\bibitem{ZT01} M. Znojil and M. Tater, {\it J. Phys.} A34 (2001) 1793.
\bibitem{FZ08} A. Fring and M. Znojil, {\it J. Phys.} A41 (2009) 194010.
\bibitem{Zn09} M. Znojil, {\it Phys. Rev.} D80 (2009) 105004;~ PT-symmetry and Quantum graph, arXiv: 1205.5211 (2012).
\bibitem{Zn11a} M. Znojil, {\it Phys. Lett.} A375 (2011) 3435.
\bibitem{Zn12} M. Znojil, {\it Ann. Phys.} 327 (2012) 893.
\bibitem{Gu09} A. Guo et al., {\it Phys. Rev. Lett.} 103 (2009) 093902.
\bibitem{Ru10} C.E. Ruter et al., {\it Nature Phys.} 6 (2010) 192.
\bibitem{Ma08} K.G. Makris et al, {\it Phys. Rev. Lett.} 100 (2008) 103904.
\bibitem{Be08} M. V. Berry, {\it J. Phys.} A41 (2008) 244007.
\bibitem{Mu08a} Z.H. Musslimani et.al., {\it Phys. Rev. Lett.} 100 (2008) 030402.
\bibitem{Ma10} K.G. Makris et al, {\it Phys. Rev.} A81 (2010) 063807.
 \bibitem{Lo09a} S. Longhi, {\it Phys. Rev. Lett.} 103 (2009) 123601.
\bibitem{Li+11} Z. Lin et al, {\it Phys. Rev. Lett.} 106 (2011) 213901.
\bibitem{Jo12} H.F. Jones, {\it J. Phys.} A45 (2012) 135306.
\bibitem{Lo11} S. Longhi, {\it J. Phys.} A 44(2011)485302.
\bibitem{CW12} H. Cartarius, G. Wunner,  {\it Phys. Rev.} A86 (2012) 013612.
\bibitem{Mu08b} Z.H. Musslimani et. al., {\it J. Phys.} A41 (2008) 244019.
\bibitem{AAG07} A.A. Abdullaev, A. Abdumalikov and R. Galimzyanov, {\it Phys. Lett.} A 367(2007)149.
\bibitem{BK10} Y.V. Bludov and V.V. Konotop, {\it Phys. Rev.} A 81(2010)013625.
\bibitem{DM11} R. Driben and B.A. Malomed, {\it Optics Letters} 36 (2011)4323.
\bibitem{GKN08} E. M. Graefe, H.J. Korsch and A.E. Niederle, {\it Phys. REv. Lett.} 101 (2008) 150408; {\it Phys. Rev.} A82(2010) 013629.
\bibitem{FF07} C.F.M. Faria and A. Fring, {\it Laser Physics} 17 (2007) 424.
\bibitem{BCM+05}C.M Bender, I.C.Pelaez, K.A. Milton and K.V.Shajesh, {\it Phys. Lett.} B613 (2005) 97.
\bibitem{BJR05} C.M. Bender, H.F. Jones and R.J. Rivers,{\it Physics Letters} B625 (2005) 333.
\bibitem{Be05a} C. M. Bender et al, {\it Physical Review} D71 (2005) 025014.
\bibitem{BM08} C.M. Bender and P.D. Mannheim, {\it Phys. Rev. Lett.} 100(2008)110402.
\bibitem{Mo05} J.W. Moffat, {\it Phys. Lett.} B627 (2005) 9.
\bibitem{Mo04a} A. Mostafazadeh, {\it Ann. Phys.} 309 (2004) 1.
\bibitem{Na04} A. Nanayakkara, {\it J. Phys.} A37 (2004) 4321.
\bibitem{Fa11} M. Fagoti et al, {\it Phys. Rev. } B83 (2011) 241406.
\bibitem{Ko10} T. Kottos, {\it Nature Physics} 6(2010)166.
\bibitem{AF09} O.C. Alvaredo and A. Fring, {\it J. Phys.} A42 (2009) 465211.
\bibitem{BM01} B. Basu-Mallick and B.P. Mandal, {\it Phys. Lett.} A 284 (2001) 231.
\bibitem{Ba02} B. Basu-Mallick, {\it Int. J. Mod. Phys.} B16 (2002) 1875.
\bibitem{Ba+04} B. Basu-Mallick et al., {\it Czech. J. Phys.} 54 (2004) 5.
\bibitem{BHH07} C.M. Bender, D.D. Holm and D.W. Hook, {\it J. Phys.} A40 (2007) F793.
\bibitem{CFB11} A. Cavagilla, A. Fring and B. Bagchi, {\it J. Phys.} A44 (2011) 325201.
\bibitem{Tr88} W. Trzeciakowski, {\it Phys. Rev.} B38 (1988) 12493.
\bibitem{GK93} M. R. Geller and W. Kohn,  {\it Phys. Rev. Lett.}  70 (1993) 3103.
\bibitem{Jo85} B.A. Joyce, {\it Rep. Prog. Phys.} 48 (1985) 1637.
\bibitem{Ba88} G. Bastard, {\it Wave Mechanics Applied to semiconductors Heterostructure} (Les Editions de Physique, Les Ulis, France 1988).
\bibitem{EHT90} G.T. Einevoll, P.C. Hemmer and J. Thomsen, {\it Phys. Rev.} B42 (1990) 3485.
\bibitem{Sh59} J.R. Shewell, {\it Am. J. Phys.} 27 (1959) 16.
\bibitem{GW69} T. Gora and F. Williams, {\it Phys. Rev.} {177} (1969) 1179.
\bibitem{Ba81} G. Bastard, {\it Phys. Rev.} B24 (1981) 5693.
\bibitem{ZK83}  Q.G. Zhu and H. Kroemer, {\it Phys. Rev.} B27 (1983) 3519.
\bibitem{LK83} T. Li and K.J. Kuhn, {\it Phys. Rev.} B27 (1983) 12760.
\bibitem{Ro83} O. Von Roos, {\it Phys. Rev.} B27 (1983) 7547.
\bibitem{BD66} D.J. BenDaniel and B.C. Duke, {\it Phys. Rev.} B152 (1966) 683.
\bibitem{Le95} J.M. Levy-Leblond, {\it Phys. Rev.} {A52} (1995) 1845.
\bibitem{MB84} R.A. Morrow and K. R.Brownstein, {\it Phys. Rev.} B30 (1984) 678.
\bibitem{ML01} N. Moiseyev and R. Lefebvre, {\it Phys. Rev.} A 64 (2001) 052711.
\bibitem{BBQT05} B. Bagchi, A. Banerjee, C. Quesne and V.M. Tkachuk, {\it J. Phys.} A38 (2005) 2929.
\bibitem{QT04} C. Quesne and V.M. Tkachuk, {\it J. Phys.} A 37 (2004) 4267.
\bibitem{DC99} L. Dekar, L. Chetouani and T. F. Hammann, {\it Phys.Rev.} A59 (1999) 107.
\bibitem{DCH98} L. Dekar, L. Chetouani and T. F. Hammann, {\it J. Math. Phys.} 39 (1998) 2551.
\bibitem{DA00} A. de Souza Dutra and C.A.S. Almeida, {\it Phys. Lett.} A275 (2000) 25.
\bibitem{BGQ06} B. Bagchi, P.S. Gorain and C. Quesne, {\it Mod. Phys. Lett} A21 (2006) 2703.
\bibitem{Ca+04} J.F. Carinena et al, {\it Nonlinearity} {17} (2004) 1941.
\bibitem{Ca+07} J.F. Carinena et al, {\it Ann.Phys.} 322 (2007) 434.
\bibitem{PR99} A. R. Plastino et al., {\it Phys.Rev.} A60 (1999) 4398.
\bibitem{DHA03} A. de Souza Dutra, M. Hott, C.A.S. Almeida, {\it Europhys. Lett.} 62 (2003) 8.
\bibitem{GO02} B. G\"{o}n\"{u}l, O. \"{O}zer, B. G\"{o}n\"{u}l, and F. \"{U}zg\"{u}n, {\it Mod.Phys.Lett.} A17 (2002) 2453.
\bibitem{GG02} B. G\"{o}n\"{u}l, B. G\"{o}n\"{u}l, D. Tutcu and O. \"{O}zer, {\it Mod.Phys.Lett.} A17 (2002) 2057.
\bibitem{GN07} A. Ganguly and L. M. Nieto, {\it J.Phys.} A40 (2007) 7265.
\bibitem{BGS10} B. Bagchi, A. Ganguly and A. Sinha, {\it Phys. Lett.} A 374 (2010) 2397.
\bibitem{SL03} K. Samani and F. Loran, Shape Invariant Potentials for Effective Mass Schr\"{o}dinger Equation, arXive:quant-ph/0302191.
\bibitem{MRT12} B. Midya, B. Roy and T. Tanaka, {\it J. Phys.} A45 (2012) 205303.
\bibitem{MI99} V. Milanovic and Z. Ikonic, {\it J.Phys.} A32 (1999) 7001.
\bibitem{Ta06} T. Tanaka, {\it J. Phys.} A39 (2006) 219.
\bibitem{Qu09b} C. Quesne, {\it SIGMA} { 5} (2009) 046.
\bibitem{BS62} A. Bhattacharjee and E. C. G. Sudarshan, {\it Nuovo Cimento} 25 (1962) 864.
\bibitem{Al02} A. D. Alhaidari, {\it Phys.Rev.} A66 (2002) 042116.
\bibitem{BQ05a} B. Bagchi, P. Gorain, C. Quesne and R. Roychoudhury, {\it Europhys. Lett.} 72 (2005) 155.
\bibitem{GIN06} A. Ganguly, M.V. Ioffe and L.M. Nieto, {\it J. Phys. A } {39} (2006) 14659.
\bibitem{Qu07e} C. Quesne, {\it SIGMA} 3 (2007) 067.
\bibitem{CC04} G. Chen and Z. Chen, {\it Phys. Lett.} A331 (2004) 312.
\bibitem{MM06} O. Mustafa and S.H. Mazharimousavi, {\it J. Phys.} A 39 (2006) 10537.
\bibitem{KKK02} R. Kock, E. Korcuk and M. Koca, {\it J. Phys.} A 35 (2002) L527.
\bibitem{BV95} M.S. de Bianchi and M.D. Ventra, {\it Eur. J. Phys.} 16 (1995) 260.
\bibitem{KSK05} R. Koc, G. Sahinoglu and M. Koca, {\it Eur. Phys. J.} B48 (2005) 583.
\bibitem{CDH95} L. Chetouani, L. Dekar and T.F. Hammann, {\it Phys. Rev.} A52 (1995) 82.
\bibitem{YY94} K.C. Yung and J.H. Yee, {\it Phys. Rev.} A50 (1994) 104.
\bibitem{Ha08} A.S. Halberg, {\it Int. J. Mod. Phys.} A23 (2008) 537.
\bibitem{CH11} H. Cobian and A.S> Halberg, {\it J. Phys.} A44 (2011) 285301.
\bibitem{RR02a} B. Roy and P. Roy, {\it J.Phys.} A35 (2002) 3961.
\bibitem{RR05} B. Roy and P. Roy, {\it Phys. Lett.} A 340 (2005) 70.
\bibitem{JR09} T.K. Jana and P. Roy, {\it Eurphys. Lett.} 87 (2009) 3003.
\bibitem{Al+07} A. D. Alhaidary et al., {\it Phys. Rev.} A75 (2007) 062711.
\bibitem{JYJ05} L. Jiyang, L. Yi and C. Jia, {\it Phys. Lett.} A345 (2005) 279.
\bibitem{BQR06} B. Bagchi, C. Quesne and R. Roychoudhury, {\it J. Phys.} A39 (2006) L127.
\bibitem{Sc26} E. Schr\"{o}dinger, {\it Naturwissenschaften} 14 (1926) 664.
\bibitem{Gl63} R.J. Glauber, {\it Phys. Rev.} 130 (1963) 2529; {\it Phys. Rev. Lett.} 10 (1963) 84.
\bibitem{Su63} E.C.G. Sudarshan, {\it Phys. Rev. Lett.} 10 (1963) 227.
\bibitem{Kl63} J.R. Klauder, {\it J. Math. Phys.} 4 (1963) 1055; {\it J. Math.
Phys.} 4 (1963) 1058.
\bibitem{Zh+90} W.M. Zhang et al, Rev.Mod.Phys. {\bf 62} 867 (1990).
\bibitem{Gi72} R. Gilmore, {\it Ann. Phys.} 74 (1972) 391.
\bibitem{Gi74} R. Gilmore, {\it Rev. Mev. de Fisica} 23 (1974) 142.
\bibitem{Pe72} A.M. Perelomov, {\it Commun. Math. Phys.} 26 (1972) 222.
\bibitem{Pe86} A.M. Perelomov, {\it Generalized Coherent States and Their Application}, (Springer-Verlag, Berlin,1986).
\bibitem{BG71} A.O. Barut and L. Girardello, {\it Commun.Math.Phys. }{21} (1971) 41.
\bibitem{NS78} M.M. Nieto and L.M. Simmons, {\it Phys. Rev. Lett.} 41 (1978) 207.
\bibitem{NS79} M.M. Nieto and L.M. Simmons, {\it Phys. Rev.} D20 (1979) 1321.
\bibitem{GNS80} V.P. Gutschick, M.M. Nieto and L.M. Simmons, {\it Phys. Lett.} A76 (1980) 15.
\bibitem{GNB79} V.P. Gutschick, M.M. Nieto, F. Baker, {\it Am. J. Phys.} 47 (1979) 755.
\bibitem{KS85} J. R. Klauder and B. S. Skagerstam, (1985){\it Coherent
states: Applications in Physics and Mathematical Physics} (World Scientific, Singapore).
\bibitem{Kl96} J.R.Klauder, {\it J.Phys.}  A29 (1996) L293.
\bibitem{GK99} J.P.Gazeau and J.R.Klauder, {\it J.Phys.} { A32} (1999) 123.
\bibitem{An+01} J.P.Antoine et al, {\it J.Math.Phys.} 42 (2001) 2349.
\bibitem{ED02} A.H. EI Kinani and M. Daoud, {\it J. Math. Phys.} 43 (2002) 714.
\bibitem{RR02} B. Roy and P. Roy, {\it Phys. Lett.} A296 (2002) 187.
\bibitem{Po03} D. Popov, {\it Phys. Lett.} A316 (2003) 369.
\bibitem{CF08} A. Chenaghlou and O. Faizy, {\it J. Math. Phys.} 49 (2008) 022014.
\bibitem{Ro03} P. Roy, {\it Opt. Comm.} 221 (2003) 145.
\bibitem{PSZ08} D. Popov, V. Sajfert and I. Zaharie, {\it Physica} A387 (2008) 4459.
\bibitem{NAH03} M. Novaes, M.A.M. Aguiar and J.E.M. Hornos, {\it J. Phys.} A36 (2003) 5773.
\bibitem{RR06} B. Roy and P. Roy, {\it Phys. Lett.} A 359 (2006) 110.
\bibitem{Ho01} J.M. Hollingworth et al, {\it J.Phys.A} 34 (2001) 9463.
\bibitem{CF02} A. Chenaghlou and H. Fakhri, {\it Mod. Phys. Lett.} A17 (202) 1701.
\bibitem{ED01a} A.H. EI Kinani and M. Daoud, {\it J. Phys.} A34 (2001) 5373.
\bibitem{NG03} M. Novaes and J.P. Gazeau, {\it J. Phys.} A36 (2003) 199.
\bibitem{AB05} S. Twareque Ali and F. Bagarello, {\it J. Math. Phys.} 46 (2005) 053518; {\it J. Math. Phys.} 49 (2008) 032110.
\bibitem{JKL01} H. Jeong, M.S. Kim and J. Lee, {\it Phys. Rev.} A64 (2001) 052308.
\bibitem{BL05} S.L. Braunstein and P. van Loock, {\it Rev. Mod. Phys.}  77 (2005) 513.
\bibitem{Sa12} B. Sanders, {\it J. Phys.} A 45 (2012) 244002.
\bibitem{Ro04} R.W. Robinett , {\it Phys. Rep.} 392 (2004) 1.
\bibitem{AP89} I. Sh. Averbukh and N. F. Perelman, {\it Phys. Lett.} A139 (1989)449.
\bibitem{YMS90} J.A. Yeazell, M. Mallalieu and C.R. Stroud, {\it Phys. Rev. Lett.} 64 (1990) 2007.
\bibitem{YS91} J.A. Yeazell, C.R. Stroud, {\it Phys. Rev.} A43 (1991) 5153.
\bibitem{Av92} I.Sh. Averbukh, {\it Phys. Rev.} A46 (1992) R2205.
\bibitem{AS97} D.L. Aronstein and C.R. Stroud, {\it Phys. Rev.} A55 (1997) 4526.
\bibitem{Na90} M. Nauenberg, {\it J. Phys.} B23 (1990) L385.
\bibitem{RS07} E. Romera and F. de los Santos, {\it Phys. Rev. Lett.} 99 (2007) 263601.
\bibitem{RS08} E. Romera and F. de los Santos, {\it Phys. Rev.} A79 (2008) 013837.
\bibitem{SS76} C.U. Segre and J.D. Sullival, {\it Am. J. Phys.} 44 (1976) 729.
\bibitem{BKT95} R. Bluhm, V.A. Kostelecky and B. Tudose, {\it Phys. Rev.} A52 (1995) 2234.
\bibitem{RA86} R.R. Puri and G.S. Agarwal, {\it Phys. Rev.} A33 (1986) 3610.
\bibitem{BMK92} V. Buzek, H. Moya-Cessa and P.L. Knight, {\it  Phys. Rev.} A45 (1992) 8190.
\bibitem{EVU94} V.V. Eryomin, S.I. Vetchinkin and I.M. Umanskii, {\it J. Chem. Phys.} 222 (1994) 394.
\bibitem{CM95} W.Y. Chen and G.J. Milburn, {\it Phys. Rev.} A51 (1995) 2328.
\bibitem{KK09} V. Krueckl and T. Kramer, {\it New. J. Phys.} 11 (2009) 093010.
\bibitem{VE94} S.L.Vetchinkin and V.V. Eryomin, {\it Chem. Phys. Lett.} 222 (1994) 394.
\bibitem{Sc06} A. Schmidt, {\it Phys. Lett.} A 353 (2006) 459.
\bibitem{Ro00} R.W. Robinett, {\it Am. J. Phys.} 68 (2000) 410.
\bibitem{RH02} R.W. Robinett and S. Heppelmann, {\it Phys. Rev.} A65 (2002) 062103.
\bibitem{RBP05} U. Roy, J. Banerji and P. K Panigrahi, {\it J. Phys.} A38 (2005) 9115.
\bibitem{BKP96} R. Bluhm, V.A. Kostelecky and J.F. Porter, {\it Am. J. Phys.} 64 (1996) 944.
\bibitem{BA97} J. Banerji and G.S. Agarwal, {\it Optics Express} 5 (1999) 220.
\bibitem{BKT97} R. Bluhm, V.A. Kostelecky and B. Tudose, {\it Phys. Rev.} A55 (1999) 819.
\bibitem{Wa+94} J. Wals et al., Phys. Rev. Lett. 72 (1994) 3783.
\bibitem{VVS96} M.J. Vrakking, D.M. Villeneuve and A. Stolow, {\it Phys. Rev.} A54 (1996) R37.
\bibitem{MRR10} B. Midya, B. Roy and R. Roychoudhury, {\it J. Math. Phys.} 51 (2010) 022109.
\bibitem{MN00} B. Mielnik, L.M. Nieto and O. Rosas-Ortiz, {\it Phys. Lett.} {\bf A269}(2000) 70.
\bibitem{Pe08} J.J. Pena et al., {\it Int. J. Quant. Chem.} 108 (2008) 2906.
\bibitem{BG04} B. Bagchi, P. Gorain, C. Quesne and R. Roychoudhury, {\it Mod.Phy.Lett.} A19 (2004) 2765.
\bibitem{AS65} M. Abramowitz and I. A. Stegun, {\it Handbook of Mathematical Functions}, (Dover publivcations, New York, 1965).
\bibitem{Ga80} M.G. Gasymov, {\it Funct. Anal. Appl.} 14 (1980) 11.
\bibitem{MG29} H. P. Mulholland and S. Goldstein, {\it Phil. Mag.} 8 (1929)834.
\bibitem{Bo48} C.J. Bouwkamp, {\it Kon. Nederl. Akad. Wetensch. Proc.} 51 (1948) 891.
\bibitem{KO90} N. Kamran and P.J. Olver, {\it J. Math. Ana. App.} 145 (1990) 342.
\bibitem{Fl83} G. Floquet, {\it Annales Scientifiques de l'E.N.S.} 12 (1883) 47.
\bibitem{Sw04} M.S. Swanson, {\it J. Math. Phys.} {45} (2004) 585.
\bibitem{BQR05} B.Bagchi, C.Quesne and R.Roychoudhury {\it J. Phys.} A38 (2005) L647.
\bibitem{MGH07} D.P. Musumbu, H.B. Geyer and W.D. Heiss,{\it J. Phys. A: Math. Theor.} { 40} (2007) F75.
\bibitem{Qu07} C. Quesne, {\it J. Phys.} A{40} (2007) F745.
\bibitem{BT08} B. Bagchi and T. Tanaka, {\it Phys. Lett.} A372 (2008) 5390.
\bibitem{SR07} A. Sinha and P. Roy, {\it J. Phys. A: Math. Theor.} { 40} (2007) 10599.
\bibitem{Ba07} B. Bagchi, {\it J. Phys.} A { 40} (2007) F1041.
\bibitem{ML74} P.M. Mathews and M. Lakshmanan, {\it Quart.Appl.Maths.} {32} (1974) 215.
\bibitem{LR03} M. Lakshmanan and S. Rajasekar, {\it Nonlinear dynamics,Integrability, Chaos and Patterns} (Advanced Texts in Physics, Springer, Berlin), 2003.
\bibitem{Ca+05} J.F. Carinena et al., { \it Regul. Caot. Dyn.} 10 (2005) 423.
\bibitem{Ve+93} S.I. Vetchinkin et al., {\it Chem. Phys. Let.} 215 (1993) 11.
\bibitem{Ve+94} V.V. Eryomin et al., {\it J. Chem. Phys.} 101 (1994) 10730.
\bibitem{Le01} G. Levai, {\it Czech. J. Phys.} 51 (2001) 1.
\bibitem{Ba+00a} B. Bagchi et al., {\it Phys. Lett.} A 269 (2000) 79.






\bibitem{Qu91} C. Quesne, {\it Int. J. Mod. Phys. A}  6 (1991) 1567.
\bibitem{TV97} V.M. Tkachuk, S.I. Vakarchuk, {\it Phys. Lett. A} 228 (1997) 141.
\bibitem{Zn04} M. Znojil, {\it J.Phys.A} 37 (2004) 9557.
\bibitem{Ho06} C.L. Ho, {\it Ann.Phys}. 321 (2006) 2170.
\bibitem{Me+99} A. Metz et al, {\it Phys. Rev. Lett.} 83(1999) 1542.
\bibitem{UH83} L. F. Urrutia and E. Hernandez, {\it Phys. Rev. Lett.} 51, (1983) 755.
\bibitem{GMS98} A. Gangopadhyay, J.V. Mallow and U.P. Sukhatme, {\it Phys. Rev.} A58 (1998) 4287.
\bibitem{ABC02} A.N. Aleixo, A.B. Balantekin and M.A. Candido Ribeiro, {\it J. Phys.} A35 (2002) 9063.
\bibitem{ABC00b} A.N. Aleixo, A.B. Balantekin and M.A. Candido Ribeiro, J. Phys. A 33 (2000) 3173.
\bibitem{ABC01} A.N. Aleixo, A.B. Balantekin and M.A. Candido Ribeiro, {\it J. Phys.} A34 (2001) 1109.
\bibitem{GKS98} P.K. Ghosh, A. Khare, and M. Sivakumar, {\it Phys. Rev. A} 58 (1998) 821.
\bibitem{CIN11} F. Cannata, M.V. Ioffe, and D. N. Nishnianidz, {\it J. Math. Phys.} 52 (2011) 022106.
\bibitem{Fe97} D.J. Fernandez, {\it Int. J. Mod. Phys.} {A12} (1997) 171.
\bibitem{FS03} D.J. Fernandez and E. Salinas-Hernandez {\it J. Phys.} A36 (2003) 2537.
\bibitem{MR04} B. Mielnik B and O. Rosas-Ortiz, {\it J. Phys.} A37 (2004) 10007.









\bibitem{Su05} A.A. Suzko, {\it Phys. Lett.} A335 (2005) 88.
\bibitem{NSS03} L.M. Nieto, B.F. Samsanov and A.A. Suzko, {\it J. Phys.} A36 (2003) 12293.











\bibitem{Ro05} B. Roy, {\it Europhys. Lett.} {72} (2005) 1.




\bibitem{Ha06} A.S. Halberg, {\it Int. J. Mod. Phys.} A21 (2006) 1359.






 \bibitem{Sh01} K.C. Shin, {\it J. Math. Phys.} {42} (2001) 2513.
 \bibitem{Ah03} Z. Ahmed, {Phys. Lett.} A310 (2003) 139.
\bibitem{BBM02} C.M. Bender, M.V. Berry and A. Mandilara, {\it J. Phys.} A35 (2002) L467.
\bibitem{ABB05} Z. Ahmed, C.M. Bender and M.V. Berry, {\it J. Phys.} A38 (2005) L627.


\bibitem{BM10} C.M. Bender and P.D. Mannheim, {\it Phys. Lett.} A374 (2010) 1616.

\bibitem{Mo03} A. Mostafazadeh, {\it J.Phys.} A36 (2003) 7081.


\bibitem{GSZ05} U. Guenther, F. Stefani, and M. Znojil, {\it J. Math. Phys.} 46 (2005) 063504.


\bibitem{Zn05} M. Znojil, {\it Phys. Lett.} A342 (2005) 36.
\bibitem{Zn11} M. Znojil, {\it SIGMA} 7 (2011), 018.
\bibitem{ZJ09} M. Znojil and V. Jakubsky, {\it Pramana- J. of Phys.} 73 (2009) 397.




\bibitem{Lo09b} S. Longhi, {\it Phys. Rev.} B 80 (2009) 165125.
\bibitem{WKP10} C.T. West, T. Kottos and T. Prosen, {\it Phys. Rev. Lett.} 104 (2010) 054102.
\bibitem{HT05} T. Hertog and G.T. Horowitz, {\it JHEP} 04 (2005) 005.
\bibitem{GW10} J.B. Gong and Q. Wang, {\it Phys. Rev.} A 82 (2010) 012103.






\bibitem{AT91} G.S. Agarwall and K. Tara, {\it Phys. Rev.} A43 (1991) 492.


\bibitem{FV96} R.L. de Matos Filho and W. Vogel, {\it Phys. Rev.} A54 (1996) 4560.
\bibitem{Ma82} L. Mandel, {\it Phys. Rev. Lett.} 49 (1982) 136.
\bibitem{MW65} L. Mandel and E. Wolf, {\it Rev. Mod. Phys.} 37 (1965) 231.
\bibitem{KM77} H.J. Kimble and L. Mandel, {\it Phys. Rev.} A15 (1977) 689.
\bibitem{Lo80} R. Loudon, {\it Rep. Prog. Phys.} 43 (1980) 913.


\bibitem{Sh04a} T.Shreecharan et al, {\it Phys.Rev.} { A69} (2004) 012102.


\bibitem{AB04} A.N.F. Aleixo and A.B. Balantekin, {\it J. Phys. A} 37 (2004) 8513.


\bibitem{PS86} J. Parker and C.R. Stroud, {\it Phys. Rev. Lett.} 56 (1986) 716.

\bibitem{Ba92} T Baumert et al, {\it Chem. Phys. Lett.} 191 (1992) 639.




\bibitem{SSI04} M. Spanner, E.A. Shapiro, and M. Ivanov, {\it Phys. Rev. Lett.} 92 (2004) 093001.
















%
%
%
\bibitem{CJT98} F. Cannata, G. Junker, J. Trost, {\it Phys. lett.} A246 (1998) 219.
\bibitem{St05} J. E. Strang, {\it Acad. Roy. Belg. Bull. Cl. Sci.} 16 (2005) 269.
\bibitem{DHS05} A. de Souza Dutra, M.B. Hott and V.G.C. Santos, {\it Eurphys. Lett.} 71 (2005) 166.
\bibitem{Yu04} C. Yuce, {\it Phys. Lett.} A 336(2004)290.
\bibitem{BJ11} C.M. Bender and H.F. Jones, {\it Phys. Rev.} A 34(2011)032103.
\bibitem{RR07} B. Roy and R. Roychoudhury, {\it J. Phys.} A 40 (2007)8479.
\bibitem{LSR03} G. Levai, A. Sinha and P. Roy, {\it J. Phys.} A36 (2003)7611.
\bibitem{Zn99d} M. Znojil, {\it J. Phys.} A32(2001)7419.
\bibitem{Fe+99} F.M. Fernandez et al., {\it J.Phys.} A32(1999)3105.
\bibitem{SR04} A. Sinha and P. Roy, {\it J. Phys.} A 37(2004)2509.

\bibitem{Moy11} N. Moiseyev, {\it Non-Hermitian Quantum Mechanics}, (Cambridge University press, 2011).
\bibitem{Zn03} M. Znojil, {\it J. Phys.} A 36 (2003) 7639.
\bibitem{GRS07} U. Guenther, I. Rotter and B.F. Samsonov, {\it J. Phys.} A 40 (2007) 8815.
\bibitem{BJ08g} C.M. Bender and H.F. Jones, {\it J. Phys.} A41 (2008) 244006.

\bibitem{Jo07} H.F. Jones, {\it Phys. Rev.} D 76 (2007) 125003.
\bibitem{CJP09} F. Correa, V. Jakubsky and M.S. Plyushchay, {\it Ann. Phys.} 324 (2009) 1078.
\bibitem{CP07} F. Correa and M.S. Plyushchay, {\it J. Phys.} A 40 (2007) 14403.
\bibitem{RQV88} P.Roy, R. Roychoudhury and Y.P. Varshni, {\it J. Phys.} A21 (1988) 1589.
\bibitem{SPB04} T. Shreecharan, P.K. Panigrahi and J. Banerji, {\it Phys. Rev.} A69 (2004) 012102.
\bibitem{DE02} M. Daoud and A.H. EI Kinani, {\it J. Phys.} A35 (2002) 2639.
\bibitem{GJ93} P.F. Gora and C. Jedrzejek, {\it Phys. Rev.} A58 (1993) 3291.
\bibitem{So94} A.I. Solomon, {\it Phys. Lett.} A196 (1994) 29.
\bibitem{Ma79} L. Mandel, {\it Opt. Lett.} 4 (1979) 205.
\bibitem{BA99} J. Banerji and G.S. Agarwal, {\it Phys. Rev.} A59 (1999) 4777.
\bibitem{ZF92} Z. Zhang and H. Fan, {\it Phys. Lett.} A165 (1992) 14.
\bibitem{Do+96} V.V. Dodonov et al., {\it Quant. Semiclassic. Opt} 8(1996) 413.
\end{thebibliography}
\end{document}